\documentclass[%
 reprint,
 amsmath,amssymb,
 aps,
]{revtex4-2}

\usepackage{graphicx}% Include figure files
\usepackage{dcolumn}% Align table columns on decimal point
\usepackage{bm}% bold math

\usepackage{xcolor}
\usepackage{comment}
\usepackage{nccmath}

\usepackage[unicode=true,
 bookmarks=true,bookmarksnumbered=true,bookmarksopen=false,
 breaklinks=true,pdfborder={0 0 0},backref=false,colorlinks=true]
 {hyperref}
\hypersetup{
citecolor={blue},
urlcolor={blue}
}

\usepackage[normalem]{ulem}

\newcommand{\beq}{\begin{equation}}
\newcommand{\eeq}{\end{equation}}
\newcommand{\bea}{\begin{eqnarray}}
\newcommand{\eea}{\end{eqnarray}}
\newcommand{\bwt}{\begin{widetext}}
\newcommand{\ewt}{\end{widetext}}

\newcommand{\bk}{\mathbf{k}}

\newcommand{\bA}{\mathbf{A}}

\newcommand{\bB}{\mathbf{B}}

\newcommand{\bK}{\mathbf{K}}
\newcommand{\bL}{\mathbf{L}}

\newcommand{\bq}{\mathbf{q}}
\newcommand{\bp}{\mathbf{p}}
\newcommand{\bg}{\mathbf{g}}
\newcommand{\bG}{\mathbf{G}}
\newcommand{\br}{\mathbf{r}}

\newcommand{\mbf}[1]{\mathbf{#1}}

\usepackage{babel}
\usepackage{physics}
\usepackage{amsmath}
\usepackage{xr}
\usepackage{hyperref}
\usepackage{amsmath}
\usepackage{amssymb}
\usepackage{stmaryrd}
\usepackage{graphicx}
\usepackage{esint}
\usepackage{subfigure}
\usepackage{multirow}
\usepackage{physics}
\usepackage{xcolor}
\usepackage{comment}

\newcommand{\kk}{{\bf{k}}}

\newcommand{\rr}{{\bf{r}}}
\newcommand{\RR}{{\bf{R}}}
\newcommand{\qqq}{{\bf{q}}}
\newcommand{\QQ}{{\bf{Q}}}

\newcommand{\GG}{{\bf{G}}}
\newcommand{\cmom}{{\boldsymbol \pi}}

\newcommand{\braketter}[1]{\langle {#1} \rangle}
\newcommand{\lp}{\left(}
\newcommand{\rp}{\right)}
\usepackage{nccmath}

\makeatother

\begin{document}

\preprint{APS/THF in B}

\title{Topological heavy fermions in magnetic field}

\author{
 \small Keshav Singh$^{*1,2}$~,
 \small Aaron Chew$^3$~,
 \small Jonah Herzog-Arbeitman$^3$~,
 \small B. Andrei Bernevig$^{3,4,5}$~ and
 \small Oskar Vafek$^{1,2}$~
\smallskip 
\\
$^1$\footnotesize \textit{National High Magnetic Field Laboratory, Tallahassee, Florida, 32310}
\\
$^2$\footnotesize \textit{Department of Physics, Florida State University, Tallahassee, Florida, 32306}
\\
$^3$\footnotesize \textit{Department of Physics, Princeton University, Princeton, NJ 08544}
\\
$^4$\footnotesize \textit{Donostia International Physics Center, P. Manuel de Lardizabal 4, 20018 Donostia-San Sebastian, Spain}
\\
$^5$\footnotesize \textit{IKERBASQUE, Basque Foundation for Science, Bilbao, Spain}
}

\begin{abstract}
The recently introduced topological heavy fermion model (THFM) provides a means for interpreting the low-energy electronic degrees of freedom of the magic angle twisted bilayer graphene as hybridization amidst highly dispersing topological conduction and weakly dispersing localized heavy fermions. 
In order to understand the Landau quantization of the ensuing electronic spectrum, a generalization of THFM to include the magnetic field $B$ is desired, but currently missing. Here we provide a systematic derivation of the THFM in $B$
and solve the resulting model to obtain the interacting Hofstadter spectra for single particle charged excitations. 
While naive minimal substitution within THFM fails to correctly account for the total number of magnetic subbands within the narrow band i.e. its total Chern number, our method --based on projecting the light and heavy fermions onto the irreducible representations of the magnetic translation group-- reproduces the correct total Chern number. 
Analytical results presented here offer an intuitive understanding of the nature of the (strongly interacting) Hofstadter bands.
%\clearpage
\end{abstract}

\maketitle
%\linenumbers
\section{Introduction} 
Since the discovery of the remarkable phase diagram of the magic angle twisted bilayer graphene (MATBG) \cite{cao2018unconventional,cao2018correlated}, substantial effort \cite{sharpe2019emergent,serlin2020intrinsic,PhysRevB.98.085435,PhysRevX.8.031087,PhysRevX.8.031089,PhysRevX.8.041041,PhysRevLett.121.257001,PhysRevLett.122.246401,PhysRevLett.122.106405,PhysRevLett.122.257002,PhysRevLett.124.046403,PhysRevLett.124.097601,PhysRevB.102.035161,PhysRevLett.124.166601,PhysRevX.10.031034,PhysRevLett.125.257602,PhysRevB.103.205414,khalaf2021charged,2021NatCo..12.5480C,PhysRevB.103.205411,PhysRevB.103.205412,PhysRevB.103.205413,PhysRevB.103.205414,PhysRevB.103.205415,PhysRevB.103.205416,PhysRevB.104.075143,PhysRevResearch.3.013033,PhysRevLett.127.266402,PhysRevX.11.041063,PhysRevB.103.205416,PhysRevLett.129.117602,2023PhRvB.107g5156X} has been devoted to understanding its rich physics. The presence of topological narrow bands within this system \cite{ahn2019failure,po2019faithful,song2019all,PhysRevB.103.205412} provides a novel platform to study the interplay between strong electron correlations and band topology. The recently introduced topological heavy fermion model (THFM) for MATBG \cite{song2022magic,cualuguaru2023twisted} bridges the contrary signatures of localized \cite{2019Natur.572..101X,rozen2021entropic} and delocalized physics\cite{choi2019electronic,kerelsky2019maximized} reported via STM and transport measurements\cite{lu2019superconductors,zondiner2020cascade}.
Within THFM the low energy electrons are viewed as a result of the
hybridization between heavy $p_x\pm ip_y$-like Wannier states, localized at the AA stacking sites, and topological conduction fermions, denoted by $f$ and $c$ respectively in analogy to heavy fermion materials\cite{song2022magic}. Among its other features, THFM allows for an intuitive explanation of the charged excitation spectra \cite{song2022magic} at integer fillings hitherto obtained via strong coupling expansion of projected models\cite{PhysRevLett.125.257602,PhysRevB.103.205415}. 

The large moir\'e period of $\sim$13nm in MATBG has revealed a sequence of broken symmetry Chern insulators yielding a plethora of finite magnetic field($\mathbf{B}$) induced phases at lower fluxes\cite{lu2019superconductors,yankowitz2019tuning,nuckolls2020strongly,2020Natur.582..198W,2021NatPh..17..710D,yu2022correlated,saito2021hofstadter,wu2021chern,xie2021fractional,yu2022correlated} and has showcased, for the first time, reentrant correlated Hofstadter states at magnetic fields as low as 31T \cite{PhysRevLett.128.217701}. Thus it becomes important to better understand the interplay of correlations and band topology in the presence of a perpendicular $\mathbf{B}$ field. Theoretical studies have previously focused on non-interacting \cite{hejazi2019landau,PhysRevB.84.035440,lian2020landau} and strong coupling \cite{wang2022narrow,PhysRevB.106.085140,PhysRevLett.129.076401} regimes. Although exact, each employed considerable numerical analysis, restricting a deeper physical understanding of the mechanism for Landau quantization. 

In this paper, we show how one can understand the Landau quantization of the strong coupling spectra in terms of hybridization amidst Landau levels (LLs) of $c$ fermions and hybrid Wannier states of $f$ fermions. We find that only a particular number of $f$ fermion momentum channels are allowed to hybridize to $c$ fermion LLs, with coupling strength which decreases with increasing $\mathbf{B}$ and increasing LL index $m$. Moreover, through our analysis we can clearly understand the reason why a naive minimal coupling is unable to recover the correct total Chern number of the narrow bands. In the flat band limit of THFM, our framework allows for an exact solution including the dominant interactions and analytically explains the total Chern number. Even for cases with a non-trivial Chern number, we explicitly demonstrate the dependence of total number of states on the magnetic field as is expected by the Streda formula\cite{Steda1982TheoryOQ}. Although going away from the flat band limit requires numerics, given the simple structure of our Hamiltonian, we are still able to compute the spectrum to unprecedentedly small fluxes and find it to be well captured by the analytical solution in the flat band limit, $M=0$, which can be taken all the way to $\mathbf{B}=0$ as shown in Figs.(\ref{fig:0fieldvsB field}),(\ref{fig: nu=-1 spec}),(\ref{fig:nu=-2}) at narrow band filling factors $\nu=0,-1,-2$ respectively.

The formulas we derive are general for any rational value of $\phi/\phi_0$, with $\phi$ being the flux through the unit cell and $\phi_0$ being the flux quantum $hc/e$, but we focus our analysis on the $1/q$ flux sequence and low $\mathbf{B}$ where the results become particularly transparent. Our analysis as well unveils the physical nature of the anomalous low energy mode which is seen to be almost $\mathbf{B}$-independent, also observed in previous numerics\cite{wang2022narrow}, as the anomalous zero-LL of a massless Dirac particle, a key ingredient of the topological heavy fermion picture of MATBG. Although this work deals directly with THFM, our methods apply more generally.
%%%%%%%%%%%%%%%%%%%%%%%%%%%%%%%%%%%%%%%%   FIGURE   %%%%%%%%%%%%%%%%%%%%%%%%%%%%%%%%%%%%%%%%%%%%%%%%%%%%%%

\begin{figure}[t]
    \centering
    \includegraphics[width=8.6cm]{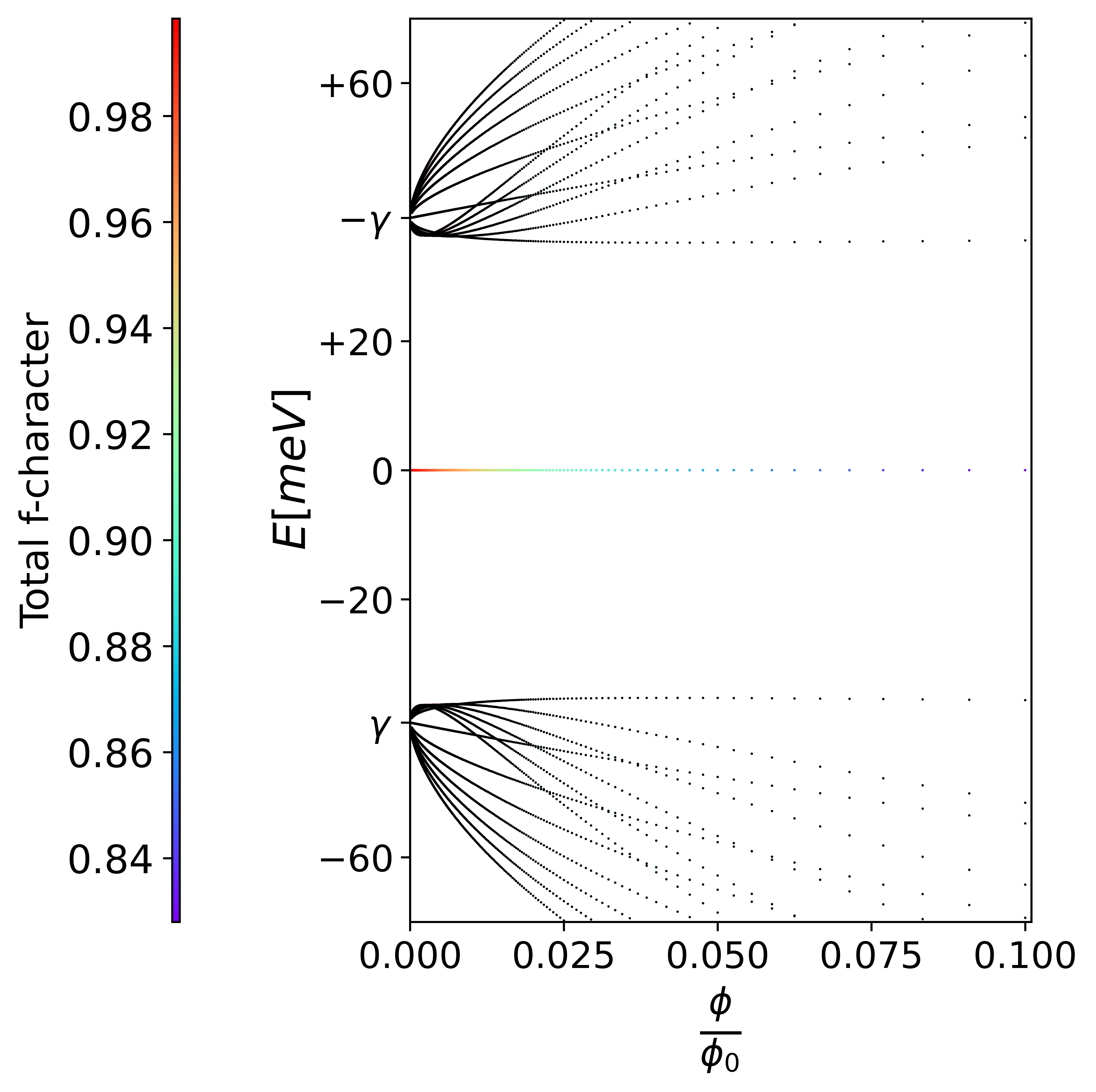}
    \caption{The spin-valley degenerate non-interacting Hofstadter spectra for THFM at $w_0/w_1=0.7$ in the flat band limit $M=0$. For illustration, we have fixed $m_\star=5$ so that the $\mathbf{B} \rightarrow 0$ energies for remote magnetic subbands, i.e. $\pm \gamma$, are tractable. The value of parameters used are $\gamma=-39.11$ meV, $v'_{\ast}=1.624$ $\mathrm{eV.\mathring{A}}$, $v_{\ast}=-4.483$ $\mathrm{eV.\mathring{A}}$ and $\lambda=0.3792L_m$. `Total-$f$ character' color labeling on left, unlike in the rest of the figures, represents the total $f$ weight of the flat bands composed of zero modes. We sum over the $f$-weights of each zero mode and normalize it by the total number of zero modes, i.e. $2q$. For the coupled modes obtained using ans\"atze in Eqs.(\ref{MT the ansatz 3})-(\ref{MT the ansatz 6}), the $f$-weight is obtained as $|c^{(\mu)}_{5}|^2$+$|c^{(\mu)}_{6}|^2$, after normalizing the eigenvector. The $f$-weight for the decoupled $f$-modes is 1, while that of the anomalous $c$ in Eq.(\ref{MT decoupled-c state})is zero. The remote bands in `black' do not correspond to the above color labeling.}
    \label{fig:Non-Interacting,flatband}
\end{figure}

%%%%%%%%%%%%%%%%%%%%%%%%%%%%%%%%%%%%%%%%   FIGURE   %%%%%%%%%%%%%%%%%%%%%%%%%%%%%%%%%%%%%%%%%%%%%%%%%%%%%%
\section{Results}
{\textbf{Review of THFM and the Key Challenge}}:
The THFM in momentum space is given by \cite{song2022magic} 
\begin{eqnarray}
&&\hat{H}_{0} = \sum_{\mid\mathbf{k} \mid< \Lambda_c} \sum_{\tau s}\sum_{aa'=1}^4H^{c,\tau}_{aa'}(\mathbf{k})\tilde{c}^{\dagger}_{\mathbf{k} a \tau s}\tilde{c}_{\mathbf{k} a'\tau s} + \nonumber \\
&&\!\!\sum_{\mid\mathbf{k}\mid < \Lambda_c}\!\!\sum_{\tau s}\sum_{a=1}^4 \sum_{b=1}^{2}\left(e^{-\frac{1}{2}\mathbf{k}^2 \lambda^2}H^{cf,\tau}_{ab}(\mathbf{k})\tilde{c}^{\dagger}_{\mathbf{k} a \tau s}\tilde{f}_{\mathbf{k} b \tau s} \!\!+\!\! \text{h.c.}\right).\label{zero field full}\nonumber\\
\end{eqnarray}
 Here $\Lambda_c$ is the momentum cutoff for $c$ fermions while $f$ fermions, whose bandwidth is negligibly small, reside in the entire moir\'e Brillouin zone (mBZ).  The tilde on the fermionic creation and annihilation operators indicates that they are at $\mathbf{B}=0$. The parameter $\lambda\approx 0.38L_m$ is a damping factor proportional to the spatial extent of the localized Wannier functions\cite{song2022magic}, with $L_m$ being the moir\'e period; $\tau=+1(-1)$ represents graphene valley $\mathbf{K}(\mathbf{K}')$ and $s$ spin $\uparrow,\downarrow$. 
The $c$-$c$ and $c$-$f$ couplings are
\begin{eqnarray}
H^{c,1}=\left(\begin{array}{cccc}
     0& 0& v_\ast k&0 \\
     0& 0& 0& v_\ast \bar{k} \\
    v_\ast \bar{k}&0 &0 &M \\
    0& v_\ast k& M& 0 \end{array}\right)\!\!\!\!\text{ ,}\; H^{cf,1}=\left(\begin{array}{cc}
     \gamma&v_\ast '\bar{k}  \\
    v_\ast ' k & \gamma\\
    0&0\\
    0&0
\end{array}\right),\label{Eq:Zero B, 6by6}\nonumber\\
\end{eqnarray}
where $k=k_x+ik_y$ and $\bar{k}=k_x-ik_y$. The explicit $\mathbf{k}$ dependence in $H^{c,1}(\mathbf{k})$ and $H^{cf,1}(\mathbf{k})$ above has been suppressed for brevity. The parameters $v_{\ast}$, $v'_{\ast}$, $M$ and $\Gamma$ were determined from the Bistritzer-MacDonald \cite{bistritzer2011moire} (BM) model in Ref.\cite{song2022magic}. 
The bandwidth of narrow bands is set by $2|M|$ and the gap between the narrow bands and the remote bands is $|\gamma|-|M|$. The couplings at the opposite graphene valley (i.e. at $\tau=-1$) can be obtained by replacing $k\leftrightarrow -\bar{k}$ in Eq.(\ref{Eq:Zero B, 6by6}).

In order to illustrate the key challenge to promoting the model to non-zero $B$, we consider a simplified case wherein we set the bandwidth and the spatial extent of the localized Wannier functions to zero, i.e. $M=\lambda=0$ in Eqs.(\ref{zero field full}) and (\ref{Eq:Zero B, 6by6}). As argued below and as shown in the supplementary note 8, the conclusions reached hold even for a general case without making this simplification.
Following a naive minimal substitution, we promote $k_x+ik_y\rightarrow -i\sqrt{2}\hat{a}/\ell$\cite{jain2007composite}, where the magnetic length is $\ell=\sqrt{\hbar c/(e B)}$ and $\hat{a}$ is the Landau level (LL) lowering operator. The eigenstates of thus minimally substituted Hamiltonian $\left(\begin{array}{cc}H^{c,1}&H^{cf,1}\\
{H^{cf,1}}^\dagger & 0\end{array}\right)$ can be obtained exactly. It can be readily verified that the zero modes take the form 
$\left(a_{1,m}\ket{m}, a_{2,m}\ket{m-1}, a_{3,m}\ket{m+1}, a_{4,m}\ket{m-2},\right. \nonumber \\
\left.b_{1,m}\ket{m}, b_{2,m}\ket{m-1}\right)^{T},$
where $|n\rangle$ denotes $n^{th}$-LL and
$a_{\alpha,m}$ and $b_{\beta,m}$ are complex coefficients with $\alpha\in\{1,\ldots,4\}$ and $\beta\in\{1,2\}$.
The LL index $m\in\{0,\ldots,m_{\ast}\}$ where $m_{\ast}$ denotes its upper cut-off. 
For $m=0$, the non-zero coefficients are 
$a_{3,0}=-i\gamma \ell$ and $b_{1,0}=\sqrt{2} v_\ast$, while for $m=1$ they are  $a_{3,1}=2v_\ast'^2-\gamma^2\ell^2$, $b_{1,1}=-2i\gamma \ell v_\ast$ and $b_{2,1}=2\sqrt{2}v_\ast v_\ast'$. For each $m\geq 2$, there are two zero modes whose non-zero coefficients are  
$a_{3,m}=\sqrt{2}\sqrt{m(m-1)}v_\ast'$, $a_{4,m}=i\gamma\ell\sqrt{m+1}$, $b_{2,m}=\sqrt{2}\sqrt{m^2-1}v_\ast$ and 
$a_{3,m}=-i\gamma\ell\sqrt{m-1}$, $a_{4,m}=\sqrt{2}\sqrt{m(m+1)}v_\ast'$, $b_{1,m}=\sqrt{2}\sqrt{m^2-1}v_\ast$, respectively. Including the anomalous zero-LL of the $c$-$c$ coupling, $\left(0,0,|0\rangle,0,0,0\right)^T$, we have a total of $2m_{\ast}+1$ zero modes within the narrow bands. 

These zero modes are well separated from the remote subbands by a gap that limits to $|\gamma|$ as $\mathbf{B}\rightarrow 0$. This gap cannot close in the stated limit even if we relax the above mentioned assumptions. 
The value of $m_{\ast}$ is typically determined by requiring that the LL spectrum converges in the energy range of interest. For us this energy window includes the narrow bands and perhaps several LLs from the remote bands. However, increasing $m_{\ast}$ results in an unbounded increase in the number of LLs within the narrow band energy range as seen from our exact result. In other words, since each LL contains $\phi/\phi_0$ states per moir\'e unit cell, the zero modes would accommodate $(2m_{\ast}+1)\phi/\phi_0$ states per moir\'e unit cell for each spin and valley. But the total Chern number of the narrow bands at $\mathbf{B}=0$ vanishes which means that the zero modes should accommodate precisely two states per moir\'e unit cell for each spin and valley {\it independent} of $\mathbf{B}$ \cite{Steda1982TheoryOQ}. 
This demonstrates that the naive minimal coupling is unable to account for the correct number of magnetic subbands within the narrow bands for an arbitrary $m_{\ast}$.

In the next sections we introduce the framework for systematically promoting THFM to non-zero $\mathbf{B}$ and naturally solve the problem illustrated above. Our approach also provides a deeper understanding of the nature of the Hofstadter subbands. 
This framework is also extended to include interactions at a mean-field (MF) level. We do so by illustrating the Landau quantization for the ``one-shot" Hartree-Fock (HF) bands obtained using the MF Hamiltonian for the parent valley polarized (VP) state\cite{song2022magic} at three different integer fillings of the narrow bands.
\\
\\
{\textbf{Basis states at non-zero magnetic field}}:
As illustrated in the previous section, the naive minimal coupling is inadequate. In order to develop a systematic framework for THFM at non-zero $B$, we begin by carefully constructing the basis states in the way that takes into account the nature of the $c$ and $f$ fermions. In addition, our construction takes advantage of the magnetic translation symmetry of the underlying Hamiltonian\cite{bistritzer2011moire}. This not only helps us to label our states using good quantum numbers but also allows us to transparently keep track of the total number of states at finite $\mathbf{B}$. 

We start by briefly reviewing the magnetic translation symmetry. In the presence of an out-of-plane magnetic field, employed via Landau gauge vector potential $\mathbf{A}=(0,Bx,0)$, the minimally coupled BM Hamiltonian\cite{bistritzer2011moire}, $H^\tau_{BM}(\mathbf{p}-\frac{e}{c}\mathbf{A})$, preserves the translational symmetry with respect to the primitive moir\'e lattice vector $\mathbf{L}_2=L_m(0,1)$ but translation with respect to the primitive moir\'e lattice vector $\mathbf{L}_1=L_m(\frac{\sqrt{3}}{2},\frac{1}{2})$ needs to be accompanied by a gauge transformation. In other words, if $f(\mathbf{r})$ is an eigenstate of $H^\tau_{BM}(\mathbf{p}-\frac{e}{c}\mathbf{A})$, then so is $\hat{t}_{\mathbf{L}_2}f(\mathbf{r}) = f(\mathbf{r}-\mathbf{L}_2)$ and $\hat{t}_{\mathbf{L}_1}f(\mathbf{r})=\exp\left(i\frac{L_{1x}y}{\ell^2}\right)f(\mathbf{r}-\mathbf{L}_1)$ with the same eigenvalue (also see supplementary note 1 for details). These operators do not commute as $\hat{t}_{\pm\mathbf{L}_2}\hat{t}_{\mathbf{L}_1} = \exp\left(\mp2\pi i \frac{\phi}{\phi_0}\right)\hat{t}_{\mathbf{L}_1}\hat{t}_{\pm \mathbf{L}_2}$, where the flux through the moir\'e unit cell is $\phi=B L_{1x}L_m$. However for $\phi/\phi_0=p/q$, with $p$ and $q$ being relatively prime integers, we have the commuting set of magnetic translation(MT) operators $[\hat{t}^q_{\mathbf{L}_2},\hat{t}_{\mathbf{L}_1}]=0$, which we use to label our basis states.  

We can now proceed to construct the non-zero $\mathbf{B}$ basis for $f$-fermions by utilizing MT. At $\mathbf{B}=0$ the basis for $f$s is composed of two Wannier functions $W_{\mathbf{R},b\tau}(\mathbf{r})=W_{\mathbf{0},b\tau}(\mathbf{r}-\mathbf{R})$ in each moir\'e unit cell which behave as $p_x \pm i p_y$ orbitals sitting on the AA stacking sites spanned by moir\'e triangular lattice vector $\mathbf{R}$. The highly localized nature of these states and the trivial topology of their bands allow us to construct a complete basis for the $f$s at $\mathbf{B}\neq 0$ using the hybrid Wannier method \cite{wang2022narrow,wang2023revisiting}. 
To this end we first construct hybrid Wannier states (hWs) out of $W_{\mathbf{0},b\tau}(\mathbf{r})$ by a repeated application of the translation operator $\hat{t}_{\mathbf{L}_2}$ (as seen in Eq.(\ref{eq:etas})). The hWs are Bloch extended in the $y$ direction, i.e. along $\mathbf{L}_2$, while localized in the $x$ direction with the localization center at the origin. Note that $\mathbf{A}$ respects the translational symmetry along $y$. Moreover, near the origin where $\mathbf{A}$ is small at small $B$ \cite{wang2023revisiting}, hWs must have a large overlap with the $\mathbf{B}\neq 0$ magnetic subbands that emanate out of $\mathbf{B}=0$ bands of $f$s i.e. the $\mathbf{B}\neq 0$ Hilbert space of $f$'s that we pursue. The rest of the basis is then generated similarly by projecting the hWs onto irreducible representations (irreps) of the magnetic translation group (MTG) as
\begin{equation}
\eta_{b\tau k_1 k_2}(\mathbf{r})=\frac{1}{\sqrt{\mathcal{N}}}
\sum_{s,n\in \mathbb{Z}} e^{2\pi i (s k_1+nk_2)}\hat{t}^s_{\mathbf{L}_1}\hat{t}^{n}_{\mathbf{L}_2} W_{\mathbf{0},b\tau}(\mathbf{r}).\label{eq:etas}
\end{equation}
Clearly, $\eta_{b\tau k_1 k_2}$ is a simultaneous eigenstate of $\hat{t}_{\mathbf{L}_1}$ and $\hat{t}^q_{\mathbf{L}_2}$ with eigenvalues $e^{-2\pi i k_1}$ and $e^{-2\pi i qk_2}$, respectively. Thus, $k_{1,2}$ labels the momentum along the primitive reciprocal lattice vectors $\mathbf{g}_{1,2}$, where $\mathbf{g}_{1}=\frac{4\pi}{\sqrt{3}L_m}(1,0)$ and $\mathbf{g}_{2}=\frac{2\pi}{\sqrt{3}L_m}(-1,\sqrt{3})$. For the $f$s $k_{1,2}\in [0,1)$ i.e. there are two $f$s per moire unit cell; the normalization factor $\mathcal{N} = s_{tot}n_{tot}$, where $s_{tot}$ and $n_{tot}$ denote the total count of $s$ and $n$ (see supplementary note 3 for details). The states with different $k_1$ and $[k_2]_{1/q}$ are guaranteed to be orthogonal which is apparent through their eigenvalues under the MTs, where $[b]_a$ denotes $b$ modulo $a$. In supplementary note 3, we prove that the overlap between states with $k_2$ differing by integral multiples of $1/q$ is very small i.e. to an excellent approximation, these states are also orthogonal. This stems from the extremely well localized nature of the $\mathbf{B}=0$ Wannier states.

In order to construct the non-zero $\mathbf{B}$ basis for $c$-fermions, we recall that the $\mathbf{B}=0$ basis for the $c$s is composed of four $\mathbf{k} \cdot \mathbf{p}$ Bloch states $e^{i{\bf k}\cdot{\bf \mathbf{r}}}\tilde{\Psi}_{\Gamma a \tau}(\mathbf{r})$, where $\tilde{\Psi}_{\Gamma a \tau}$ is the Bloch state at the $\Gamma$ point in mBZ\cite{song2022magic}. We can extend the $c$s to non-zero $B$ by multiplying $\tilde{\Psi}_{\Gamma a \tau}$ by LL wavefunctions, a result obtained when the $\mathbf{k}\cdot\mathbf{p}$ method is extended to $\mathbf{B}\neq0$ \cite{luttinger1955motion}. So as to use the same quantum numbers as for $f$s, we also project the $c$'s LL wavefunctions $\Phi_{m}$ onto the (orthonormal) irreps of the MTG as
\begin{equation}
\chi_{k_1 k_2 m}(\mathbf{r})=\frac{1}{\sqrt{\ell L_m }}\frac{1}{\sqrt{\mathcal{N}}}
\sum_{s\in \mathbb{Z}} e^{2\pi i s k_1}\hat{t}^s_{\mathbf{L}_1}\Phi_{m}(\mathbf{r},k_2\mathbf{g}_2).
\label{eq:chis}\end{equation}
Again, $\chi_{k_1 k_2 m}$ is a simultaneous eigenstate of $\hat{t}_{\mathbf{L}_1}$ and $\hat{t}^q_{\mathbf{L}_2}$ with eigenvalues $e^{-2\pi i k_1}$ and $e^{-2\pi i qk_2}$, respectively. Here $k_1 \in [0,1)$, but unlike in Eq.(\ref{eq:etas}), $k_2 \in [0,\frac{p}{q})$
i.e. there are $\phi/\phi_0=p/q$ states per moir\'e unit cell in each Landau level. (see supplementary note 3 for details on orthonormality and the domain of $k_{1(2)}$).
$\Phi_{j}(\mathbf{r},k_2\mathbf{g}_2) = e^{2\pi ik_2\frac{y}{L_m}}\varphi_j\left(x-k_2\frac{2\pi\ell^2}{L_m}\right)$, the harmonic oscillator (h.o.) wavefunctions
$\varphi_m(x)=e^{-x^2/2\ell^2}H_m(x/\ell)/\pi^{\frac{1}{4}}\sqrt{2^mm!}$ with Hermite polynomials $H_m$. The $k_2$ induced offset in the h.o. wavefunctions is $2\pi k_2\ell^2/L_m =qk_2L_{1x}/p$. Although the LL wavefunction $\Phi_m$ is an eigenstate of $\hat{t}_{\pm \mathbf{L}_2}$, the function $\chi_{k_1k_2 m}$ is not. Instead,  $\hat{t}_{\pm\mathbf{L}_2}\chi_{k_1k_2 m}(\mathbf{r})=e^{\mp2\pi ik_2}\chi_{[k_1\mp\frac{p}{q}]_1 k_2 m}(\mathbf{r})$. We utilize this identity in the proceeding sections. (see supplementary note 4B, Eq.s(92)-(94) for derivation).

Since the $q\mathbf{L}_2$ translations break up the $k_2$ domain into units of width $\frac{1}{q}$, from here on we use the label $k=(k_1,k_2)$ with $k_1\in[0,1)$ and we fix $k_2 \in [0,\frac{1}{q})$. The original $k_2$ domains are then accessed using labels $r' \in \{0,\ldots q-1\}$ and $r \in \{0,\ldots,p-1\}$, denoting the magnetic strip $[\frac{r'}{q},\frac{r'+1}{q})$ and $[\frac{r}{q},\frac{r+1}{q})$ along $\mathbf{g}_2$ for $\eta$ and $\chi$ respectively. We thus relabel the states as $\eta_{b\tau k r'}(\mathbf{r})$ and $\chi_{k r m}(\mathbf{r})$, respectively. Having assembled the low energy basis at $\mathbf{B}\neq 0$, we now expand the low energy field at a given spin $s$ and valley $\tau$ as 
\begin{eqnarray}\label{eqn:psiansatz}
\psi_{\tau,s}(\mathbf{r}) &=& \sum_{k\in[0,1)\otimes [0,\frac{1}{q})}\left(\sum_{b=1}^2\sum_{r'= 0}^{q-1} \eta_{b\tau k r'}(\mathbf{r}) f_{b\tau k r' s} +  \nonumber\right. \\ 
&&\left. \sum_{a=1}^4 \sum_{m=0}^{m_{a,\tau}}\sum_{r = 0}^{p-1} \Psi_{a\tau}(\mathbf{r})\chi_{k r m }(\mathbf{r})
c_{a\tau k r m s} \right), 
\end{eqnarray}
where $\Psi_{a\tau}(\mathbf{r})=\sqrt{A_{tot}}\tilde{\Psi}_{\Gamma a\tau }(\mathbf{r})$ with $A_{tot}$ being the total sample area, and $f_{b\tau k r' s}$ and $c_{a\tau k r m s}$ denote the annihilation operators for the $\mathbf{B}\neq 0$ $f$ and $c$ basis states, respectively.
Anticipating the appearance of anomalous Dirac LLs for the topological $c$ fermions, we allow for the $a$ dependence of the upper cutoff on the LL index at each valley $\tau$, denoted by $m_{a,\tau}$ above, with $m_{1,+1}=m_{2,-1}=m_\star+1$, $m_{2,+1}=m_{1,-1}=m_\star$, $m_{3,+1}=m_{4,-1}=m_\star+2$ and $m_{4,+1}=m_{3,-1}=m_\star-1$. As discussed in supplementary note 3, the choice of $m_\star$, although arbitrary, needs to be below an upper-bound to ensure the orthogonality amidst the $c$-states $\Psi_{a\tau}(\mathbf{r})\chi_{krm}(\mathbf{r})$. This is because it relies on the fact that their overlaps are exponentially small in $\ell^2\mathbf{g}^2$ as long as the LL index $m$ is held below an upper cutoff $m_\star\lesssim q/2$, where $\mathbf{g}$ is the reciprocal moir\'e lattice vector.
\\
\\
{\textbf{Non-interacting Hamiltonian at $\mathbf{B}\neq0$}}: The single particle THFM at $\mathbf{B}\neq 0$ can be computed using the low energy fields derived in the previous section (Eq. \ref{eqn:psiansatz}) 
\begin{eqnarray}\label{Single particle Hamiltonian in B}
\hat{H}^{B}_{0} &=& \sum_{\tau,s} \int d^2 \mathbf{r} \psi^{\dagger}_{\tau,s}(\mathbf{r})H^{\tau}_{BM}\left(\mathbf{p}-\frac{e}{c}\mathbf{A}\right)\psi_{\tau,s}(\mathbf{r}) \nonumber\\
&\approx& \sum_{\tau,s} H^{\tau,s}_{cc} + \sum_{\tau,s} \left(H^{\tau,s}_{cf}+ \text{h.c.}\right),
\end{eqnarray}
where the $f$-$f$ coupling is neglected in the last line because it is negligibly small (this is also the case at $\mathbf{B}=0$ in Eq.(\ref{zero field full})). The $c$-$c$ and $c$-$f$ couplings are
\begin{eqnarray}
H^{\tau,s}_{cc}&=&\sum_{k\in[0,1)\otimes [0,\frac{1}{q})} \sum_{a,a'=1}^4 \sum_{m=0}^{m_{a,\tau}}\sum_{m'=0}^{m_{a',\tau}}\sum_{r,\tilde{r}=0}^{p-1} \nonumber \\ 
&&\tilde{h}^{\tau}_{[amr],[a'm'\tilde{r}]}(k) c^\dagger_{a \tau k r m s}
c_{a'\tau k \tilde{r} m' s}\label{eq:fullcc2ndQnz},\\
H^{\tau,s}_{cf}&=&\sum_{k\in[0,1)\otimes [0,\frac{1}{q})}\sum_{a=1}^4 \sum_{b=1}^2 \sum_{m=0}^{m_{a,\tau}}
 \sum_{r=0}^{p-1}\sum_{r'=0}^{q-1}h^{\tau}_{[amr],[br']}(k)\nonumber \\ 
&&c^\dagger_{a\tau k r ms}
f_{b\tau k r's}.\label{eq:fullcf2ndQnz}
\end{eqnarray}
The matrix element for $c$-$c$ coupling  $\tilde{h}^{\tau}_{[amr],[a'm'\tilde{r}]}(k)=\langle\Psi_{a\tau}\chi_{krm}|H^{\tau}_{BM}(\mathbf{p}-\frac{e}{c}\mathbf{A})|\Psi_{a'\tau}\chi_{k\bar{r}m'}\rangle$ takes the same form as obtained by the direct minimal substitution in $c$-$c$ coupling in Eq.(\ref{Eq:Zero B, 6by6}) and expanding in LL basis, as is expected from $\mathbf{k}\cdot\mathbf{p}$\cite{luttinger1955motion}: 
\begin{eqnarray}\label{cc for U(4)}
\tilde{h}^{\tau}_{[amr],[a'm'\tilde{r}]}(k) = \delta_{r\tilde{r}}\left(\begin{array}{cc}
0_{2\times 2} & h^{\tau,c}_{mm'} \\
\sigma_x h^{\tau,c}_{mm'} \sigma_x & M\delta_{mm'}\sigma_x
\end{array}\right)_{aa'},
\end{eqnarray}
where the Pauli matrix $\sigma_x$ acts on the $c$ orbitals and
\begin{eqnarray}
h^{+1,c}_{mm'} = i\frac{\sqrt{2} v_{\ast}}{\ell}\left(\begin{array}{cc}
-\sqrt{m'}\delta_{m+1,m'} & 0 \\
0 & \sqrt{m}\delta_{m,m'+1}
\end{array}\right)\label{cc for U(4)2}
\end{eqnarray}
with $h^{-1,c}_{mm'}=-\sigma_xh^{+1,c}_{mm'}\sigma_x$. For $M=0$, we recover the LLs of two massless Dirac particles, with two zero LLs at each valley (see supplementary note $4$A for details of derivation). 

As discussed in the previous section, there are two $f$-states per moir\'e unit cell per valley for each spin projection. On the other hand, for each $c$-LL there are $p/q$ states per moir\'e unit cell per valley for each spin projection. In order to understand the hybridization between these states that, together with the $c$-$c$ coupling, gives rise to an isolated band of states whose total number is independent of $\mathbf{B}$ --because its total Chern number vanishes\cite{Steda1982TheoryOQ}-- we need to carefully analyze the $c$-$f$ coupling. 
%%%%%%%%%%%%%%%%%%%%%%%%%%%%%%%%%%%%%%%%   FIGURE   %%%%%%%%%%%%%%%%%%%%%%%%%%%%%%%%%%%%%%%%%%%%%%%%%%%%%%
\begin{figure}
    \centering
    \includegraphics[width=8.6cm]{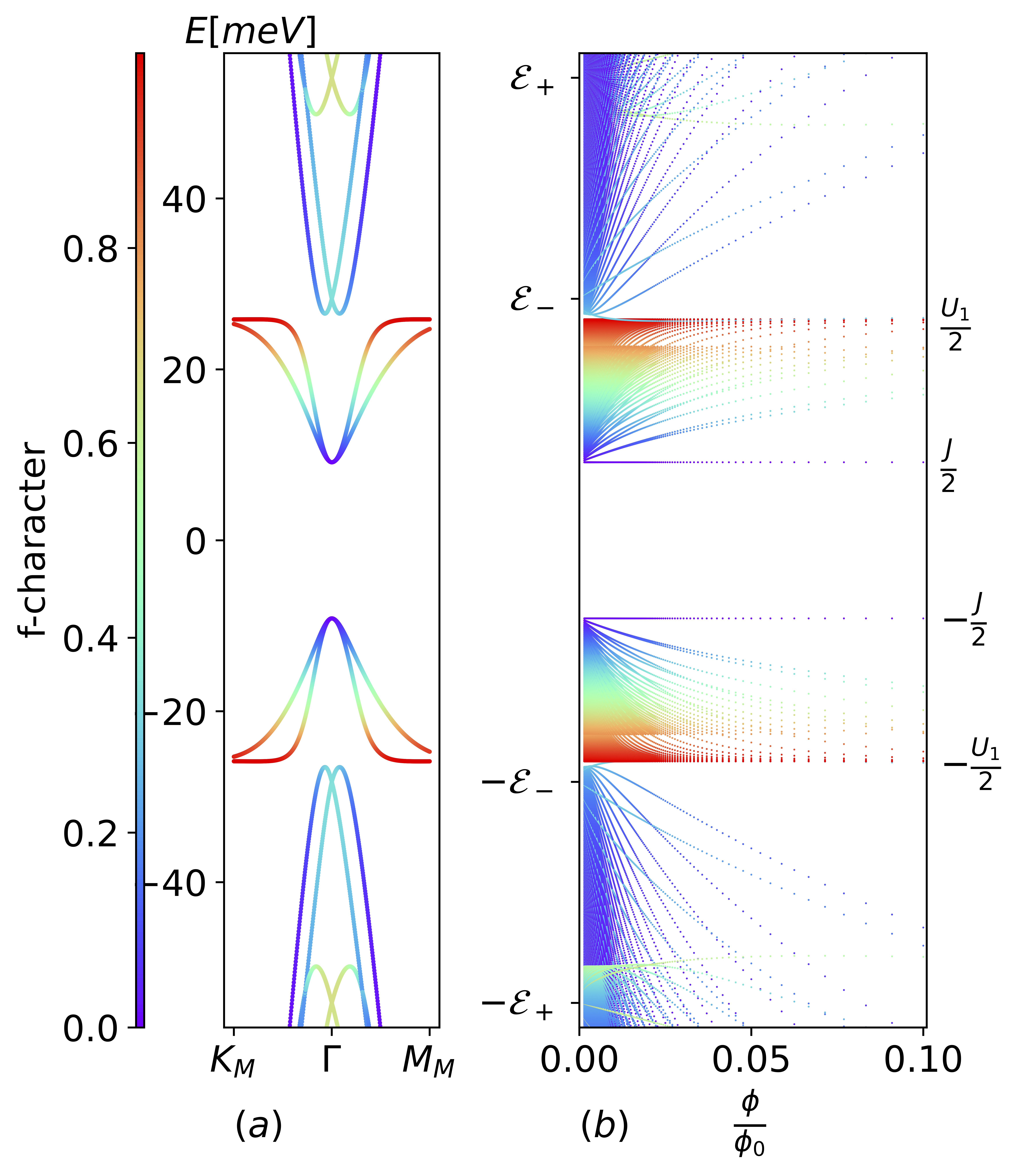}
    \caption{The spin degenerate interacting heavy fermion Hofstadter spectra (b) contrasted with zero field strong coupling spectra (a) in flat band ($M$=0) limit at $w_0/w_1=0.7$ for both valleys at CNP. We fix $m_\star$ = $\lceil \frac{q-3}{2} \rceil$. The $\mathbf{B}=0$ energies at $\Gamma$, labelled using $\mathcal{E}_\pm = \pm\frac{U_1}{4}+\sqrt{\frac{U^2_1}{16}+\gamma^2}$, are recovered in $\mathbf{B} \rightarrow 0$ limit of our theory. The value of parameters used are $J=18.27$ meV, $U_1=51.72$ meV, $\gamma=-39.11$ meV, $v'_{\ast}=1.624$ $ \mathrm{eV.\mathring{A}}$, $v_{\ast}=-4.483$ $\mathrm{\mathrm{eV.\mathring{A}}}$ and $\lambda=0.3792L_m$. Following \cite{song2022magic}, twist angle $\theta=1.05^{\circ}$ in this work and $L_m=134.218$ $ \mathrm{\mathring{A}}$. We set $k_{1,2}=0$, although inconsequential as magnetic subbands are Landau levels in this regime and thus do not disperse.}
    \label{fig:0fieldvsB field}
\end{figure}
%%%%%%%%%%%%%%%%%%%%%%%%%%%%%%%%%%%%%%%%   FIGURE   %%%%%%%%%%%%%%%%%%%%%%%%%%%%%%%%%%%%%%%%%%%%%%%%%%%%%%
Although formidable at first sight, it is actually possible to find an analytical expression for this matrix element $h^{\tau}_{[amr][br']}(k)=\langle\Psi_{a\tau}\chi_{krm}|H^{\tau}_{BM}(\mathbf{p}-\frac{e}{c}\mathbf{A})|\eta_{b\tau kr'}\rangle$ and thus determine  
the $c$-$f$ coupling at non-zero $\mathbf{B}$. The Methods-A provides the key steps for the derivation which we omit here for brevity. The result can be cast in a closed form expression which for $p=1$ and for $\ell\gg \lambda$ reads 
\begin{eqnarray} &&h^{\tau}_{[am0],[br']}(k) = \left(\begin{array}{c} \gamma \Upsilon_{m,r'}(k)\sigma_0 + h^{\tau,cf}_{m,r'}(k) \\ 0_{2\times 2} \end{array}\right)_{ab},\label{eq:cf for U(4)} \end{eqnarray} with $h^{+1,cf}_{m,r'}(k) =$ \begin{eqnarray} i\frac{\sqrt{2}v'_*}{\ell}\left(\begin{array}{cc} 0 & \sqrt{m}\Upsilon_{m-1,r'}(k) \\ -\sqrt{m+1}\Upsilon_{m+1,r'}(k)& 0 \end{array} \right),\label{cf for U(4) 2} \end{eqnarray}
and $h^{-1,cf}_{m,r'}(k) = -\sigma_x h^{+1,cf}_{m,r'}(k)\sigma_x$, where $\sigma_x$ acts in orbital space of $c$ and $f$ fermions. The matrix $\Upsilon_{m,r'}(k)$ is given as
\begin{eqnarray}\label{eq:I0}
\Upsilon_{m,r'}(k)= 
\sqrt{\frac{L_{1x}}{\ell}}
e^{i\pi r'_{q}\left(k_2-2k_1\right)}e^{i\pi r'_{q}(r'_{q} -1)\frac{1}{2q}} \nonumber \\ 
e^{-2\pi^2\frac{\lambda^2}{L^2_m}\left(k_2+\frac{r'_{q}}{q}\right)^2} \mathcal{F}_m\left(\lambda,(r'_{q}+qk_2)L_{1x}\right),
\end{eqnarray}
where  $r'_q=\text{sgn}_+\left(\frac{q}{2}-r'\right)\text{min}[r',q-r']$ with $\text{sgn}_+(x)$ being the usual sign function except at $0$ where it evaluates to $1$, and
\begin{eqnarray}
\mathcal{F}_m(\lambda,x_0)&=&\frac{1}{\pi^{\frac{1}{4}}\sqrt{2^mm!}}\sqrt{\frac{\ell^2}{\ell^2+\lambda^2}}e^{-\frac{x_0^2}{2(\ell^2+\lambda^2)}}\nonumber \\
&&\times \mathcal{H}_m\left(\frac{-2x_0\ell}{\ell^2+\lambda^2},-1+\frac{2\lambda^2}{\ell^2+\lambda^2}\right).
\end{eqnarray}
The two variable Hermite polynomials\cite{babusci2012integrals} are given by $\mathcal{H}_m(x,y)=m!\sum_{k=0}^{\lfloor\frac{m}{2} \rfloor} (x^{m-2k}y^k)/((m-2k)!k!)$,
where $\lfloor m \rfloor$ denotes the floor function at $m$. Their relation to the Hermite polynomials used above is $H_m(x) = \mathcal{H}_m(2x,-1)$. 

Although we can significantly simplify the form of $\hat{H}^{B}_{0}$ and gain a deeper analytical understanding of our solution as we do in next section, one can already use the above expressions to obtain the Hofstadter spectrum for THFM numerically.
Such numerical calculation recovers the correct total number of states within the narrow band energy window, i.e. 2 per moir\'e unit cell per valley for each spin projection regardless of the value of $m_\star$, thus solving the key challenge outlined earlier. As illustrated in Fig.(\ref{fig:Non-Interacting,flatband}) for $M=0$, these zero modes are well separated from the remote bands by a gap that limits to $|\gamma|$ as $\mathbf{B}\rightarrow 0$. The results are qualitatively the same for $M\neq 0$ as shown in supplementary Fig.(10b) except the zero modes split into a group of states with a width set by $2|M|$ as expected. In the following sections, we elucidate the nature of the Hofstadter subbands by carefully casting the $\mathbf{B}\neq 0$ Hamiltonian in terms of coupled and decoupled modes of $f$ fermions. This not only helps us to obtain an exact solution in the flat band limit but also to understand the total Chern number via straightforward analytical arguments.
\begin{figure}
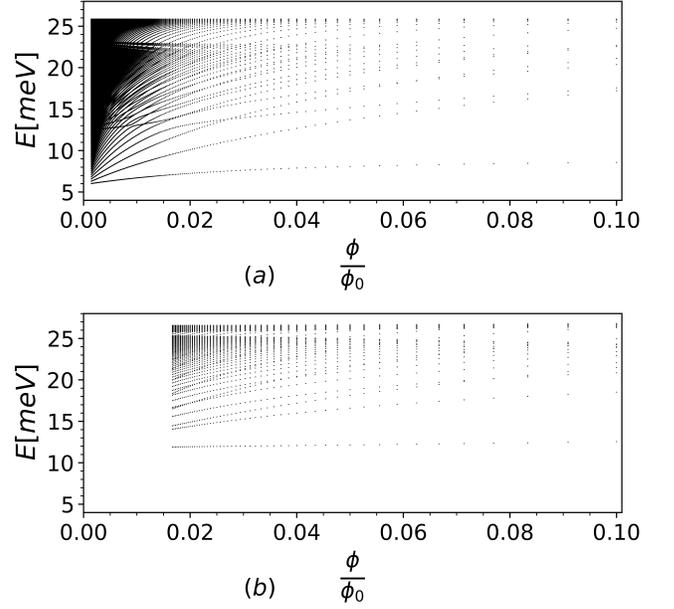

\includegraphics[width=8.6cm]{MainFigs/Figure3a.jpg}
\includegraphics[width=8.6cm]{MainFigs/Figure3b.jpg}
\caption{
(a) THFM Hofstadter spectrum at valley $\mathbf{K}'$ with $M=3.248$ meV (i.e. including dispersion of the flat bands) for $m_\star= \lceil \frac{q-3}{2} \rceil$ and $w_0/w_1=0.7$. (b) Strong coupling projected BM Hofstadter spectrum (i.e. in the flat band limit) at $w_0/w_1=0.7$ computed using the gauge-invariant basis of magnetic translation group irreps (see supplementary note 12 for details). The spectra above are spin degenerate.}\label{fig: M neq 0 spec}
\end{figure}
\\
\\
{\textbf{Analytical results for the non-interacting Hamiltonian at $\mathbf{B}\neq 0$}}: As mentioned, for $M=0$ we find two isolated zero modes per moir\'e unit cell per valley for each spin projection from numerical calculation. In order to obtain these zero modes analytically, we start by noting that the general form of our Hamiltonian at $\mathbf{B}\neq 0$ presented in the previous section immediately implies a certain lower bound on their number. Within each valley and for each spin projection, the Hamiltonian matrix at a given $k$ has the form 
 $\left(\begin{array}{cc} C & F\\ F^\dagger & 0\end{array}\right)$ where $C$ is a square matrix of dimension $4m_\star+6$ and $F$ is a $\left(4m_\star+6\right)\times 2q$ rectangular matrix; the last $2q\times2q$ block is filled with zeros. This automatically guarantees a lower bound on the number of zero modes equal to the difference in the number of $F$'s columns and rows, as is easily established by considering the singular value decomposition (SVD) of $F$ (see e.g. Ref\cite{cualuguaru2022general}). Moreover, as seen in the Eq.(\ref{eq:cf for U(4)}), $F$ has the form $\left(\begin{array}{cc} F'\\ 0_{\left(2m_\star+3\right)\times2q}\end{array}\right)$ where $F'$ is a $\left(2m_\star+3\right)\times2q$ rectangular matrix. Therefore,
half of the singular values of $F$ are guaranteed to vanish. This implies that we can readily obtain a (larger) lower bound of $2q-(2m_\star+3)$ zero modes. Physically, these zero modes are just linear combinations of different $f$s which decouple from $c$s. 
Clearly they do not account for the total number of zero modes in the spectrum, i.e. $2q$ at each $k$  or two per moir\'e unit cell per valley for each spin projection. As we go forth to show, the remaining $2m_\star+3$ zero modes are contributed by the coupled modes, which at $M\neq 0$ get split into a group of states with a width of $2|M|$ accounting for the bandwidth of magnetic subbands within the narrow bands.
Below we build a framework for analyzing them.

To that end, we define new fermion fields $\bar{f}$ by the canonical transformation  
\begin{eqnarray}
  \bar{f}_{b\tau k \bar{r} s}=\sum_{r'=0}^{q-1}V_{\bar{r}r'}f_{b\tau k r' s}\label{f-bar modes}, 
\end{eqnarray}
where the unitary matrix $V$ is defined via the SVD of matrix $\Upsilon_{mr'}=\sum_{m'=0}^{m_{a\tau}}\sum_{\bar{r}=0}^{q-1}U_{mm'}\Sigma_{m'\bar{r}}V_{\bar{r}r'}$. Here $U$ is a unitary matrix of dimensions $(m_{a,\tau}+1)\times(m_{a,\tau}+1)$ and $\Sigma$ is a rectangular matrix of dimensions $(m_{a,\tau}+1)\times q$ containing the singular values of the matrix $\Upsilon$ along the main diagonal and zeros elsewhere, i.e. $\Sigma_{mr}=\Sigma_m\delta_{mr}$. Moreover, using the closed form expression for $\Upsilon$ stated in the previous section, we find that the matrix $U$ above is extremely close to an identity matrix at low $\mathbf{B}$ (see supplementary note 5 for details). Substituting the SVD in Eq.(\ref{eq:cf for U(4)}) and using $U=\mathbb{I}$, we find that $2q-(2m_{max}+3)$ of the $\bar{f}$ modes decouple from the $c$'s for each valley, spin and $k$. For example at $\tau=+1$, the modes in Eq.(\ref{f-bar modes}) that decouple from the $c$s are the ones with $\bar{r}>m_\star +1$ and $\bar{r}>m_\star$ for $\bar{f}_{11k\bar{r}s}$ and $\bar{f}_{21k\bar{r}s}$, respectively (see supplementary note 6 and 7 for details at $\tau=+1$ and $\tau=-1$, respectively). We thus recover the $2q-(2m_\star+3)$ zero modes contributed by the decoupled $f$s as discussed earlier.  

For the remaining coupled modes, we note that the sum of the $c$-$c$ and $c$-$f$ couplings in Eqs.(\ref{eq:fullcc2ndQnz}) and (\ref{eq:fullcf2ndQnz}) at $\tau=+1$ can be rewritten in the $\bar{f}$ basis as
\begin{eqnarray}\label{MT xi-hamiltonian}
&&H^{+1,s}_{cc} + \left(H^{+1,s}_{cf}+\mbox{h.c.}\right) =\nonumber\\ &&\sum_k\sum_{\alpha,\alpha'=1}^6
\sum_{m=0}^{m_\alpha}
\sum_{m'=0}^{m_{\alpha'}}
\langle m|\hat{h}^{+1,s}_{\alpha,\alpha'} |m'\rangle  d^{\dagger}_{ms\alpha}(k)d_{m's\alpha'}(k),\nonumber\\
\end{eqnarray}
where $m_{\alpha=1,\ldots, 4}=m_{\alpha,+1}$, $m_{5}=m_\star+1$ and $m_6=m_\star$,  
\begin{eqnarray}
     d^{\dagger}_{m s\alpha}(k) = \left(c^{\dagger}_{11k0ms},c^{\dagger}_{21k0 ms},c^{\dagger}_{3  1k0ms},c^{\dagger}_{41k0ms},\right.\nonumber\\
     \left.\bar{f}^{\dagger}_{11kms},\bar{f}^{\dagger}_{21kms} \right)_\alpha.\label{MT 6-spinor}
\end{eqnarray}
The operator $\hat{h}^{+1,s}_{\alpha,\alpha'}$ is defined as
\begin{widetext}
\begin{eqnarray}\label{eq:MT hat h plus 1}
\hat{h}^{+1,s}_{\alpha,\alpha'} = \left(\begin{array}{cccccc}
0&0&-i\sqrt{2}\frac{v_{\ast}}{\ell}\hat{a}&0&\gamma \Sigma(\hat{a}^{\dagger}\hat{a})&i\sqrt{2}\frac{v'_{\ast}}{\ell}\hat{a}^{\dagger} \Sigma(\hat{a}^{\dagger}\hat{a})  \\
0&0&0&i\sqrt{2}\frac{v_{\ast}}{\ell}\hat{a}^{\dagger}&-i\sqrt{2}\frac{v'_{\ast}}{\ell}\hat{a} \Sigma(\hat{a}^{\dagger}\hat{a})  &\gamma \Sigma(\hat{a}^{\dagger}\hat{a}) \\
i\sqrt{2}\frac{v_{\ast}}{\ell}\hat{a}^{\dagger}&0&0&M&0&0 \\0&-i\sqrt{2}\frac{v_{\ast}}{\ell}\hat{a}&M&0&0&0\\
\gamma \Sigma(\hat{a}^{\dagger}\hat{a})    & i\sqrt{2}\frac{v'_{\ast}}{\ell}\Sigma(\hat{a}^{\dagger}\hat{a})\hat{a}^{\dagger}&0&0&0&0\\
-i\sqrt{2}\frac{v'_{\ast}}{\ell}\Sigma(\hat{a}^{\dagger}\hat{a})\hat{a} &  \gamma \Sigma(\hat{a}^{\dagger}\hat{a})&0&0&0&0
\end{array}
\right)_{\alpha,\alpha'}.
\end{eqnarray}
\end{widetext}
Here $\hat{a}$ is a simple h.o. lowering operator with eigenstate $|m\rangle$ and $\Sigma(m)=\Sigma_m$. 
In the $\mathbf{B}\rightarrow 0$ limit, up to the first order in flux, the singular values of $\Upsilon$ can be approximated as $\Sigma(m)\approx1 -\left(m+\frac{1}{2}\right)\frac{\lambda^2}{\ell^2}$.
We note in passing that we do not have to rely on this approximation and can find the full closed form expression for $\Sigma(m)$ as shown in the Eq.(\ref{Eq:Singular Values}).
  
A naive minimal substitution into Eq.(\ref{zero field full}) with $\lambda=0$ would reproduce Eq.(\ref{eq:MT hat h plus 1}) with unit singular values. However, the decoupling of $2-(2m_\star+3)/q$ modes per moir\'e unit cell per spin in each valley is completely overlooked by the naive minimal substitution. This is the reason why it fails to recover the correct count of subbands within the narrow bands as noted earlier.

While the decoupled $\bar{f}$ modes account for $2q-(2m_\star +3)$ zero modes, the remaining $2m_\star+3$ zero modes of the flat band limit (i.e. $M=0$) originate from the zero modes of the operator (\ref{eq:MT hat h plus 1}). This can be readily verified via exact solutions, the first of which is a pure $c$-mode 
\begin{equation}\label{MT decoupled-c state}
\theta_1=\left[0,0,\ket{0},0,0,0\right]^{T}.
\end{equation}
The above can be interpreted as the anomalous zero-LL of a massless Dirac particle coming from the $c$-$c$ coupling. 
The remaining spectrum can be solved using the ans\"atze:
\begin{eqnarray}
 \theta_3 &=& \left[c^{(3)}_1\ket{0},0, c^{(3)}_3\ket{1},  0,c^{(3)}_5\ket{0},0\right]^T, \label{MT the ansatz 3}  \\
 \theta_5 &=& \left[c^{(5)}_1\ket{1}, c^{(5)}_2\ket{0},c^{(5)}_3\ket{2},0,c^{(5)}_5\ket{1},c^{(5)}_6\ket{0}\right]^{T}, \label{MT the ansatz 5}\\
 \theta_{6_m} &=&\left[c^{(6_m)}_{1}\ket{m},   c^{(6_m)}_{2}\ket{m-1}, c^{(6_m)}_{3}\ket{m+1}, \right. \nonumber \\
&&\left.c^{(6_m)}_{4}\ket{m-2}, c^{(6_m)}_{5}\ket{m}, c^{(6_m)}_{6}\ket{m-1}\right]^{T}, \label{MT the ansatz 6}
\end{eqnarray}
where $m\in\{2,\ldots,m_\star+1\}$. $c^{(\mu)}_{\alpha}$ denotes the complex coefficient of the corresponding h.o state at index $\alpha$, and $\mu$ labels the ansatz index $\theta_{\mu}$ (with a slight abuse of notation we have $\mu=6$ for ans\"atze in Eq.(\ref{MT the ansatz 6}) $\forall m$). Using the above, we can set up the eigen-equation for each $\theta_{\mu}$, which yields a corresponding decoupled $\mu\times\mu$ Hermitian matrix with eigenvectors $c^{(\mu)}_{\alpha}$ (see supplementary note 6 for details).  

The anomalous $c$-mode $\theta_1$ offers one zero mode. The hermitian matrices obtained using ans\"atze $\theta_3$ and $\theta_5$ offer one zero mode each. The non-zero coefficients of these modes are $c^{(3)}_3=-i\Sigma_0\gamma\ell$, $c^{(3)}_5=\sqrt{2}v_\ast$ and $c^{(5)}_3=\Sigma_0\Sigma_1(2v_\ast'^2-\gamma^2\ell^2)$, $c^{(5)}_5=-2i\Sigma_0\gamma\ell v_\ast$, $c^{(5)}_6=2\sqrt{2}v_\ast v_\ast'\Sigma_1$, respectively. The hermitian matrix obtained using the ansatz $\theta_{6_m}$ is bipartite and offers two zero modes $\forall m$. The non-zero coefficients for these zero modes are $c^{(6_m)}_3=\sqrt{2}\sqrt{m(m-1)}v_\ast'\Sigma_{m-1}$, $c^{(6_m)}_4=i\gamma\ell\sqrt{m+1}\Sigma_{m-1}$, $c^{(6_m)}_6=\sqrt{2}\sqrt{m^2-1}v_\ast$ and $c^{(6_m)}_3=-i\gamma\ell\sqrt{m-1}\Sigma_{m}$, $c^{(6_m)}_4=\sqrt{2}\sqrt{m(m+1)}v_\ast'\Sigma_{m}$, $c^{(6_m)}_5=\sqrt{2}\sqrt{m^2-1}v_\ast$, respectively. The coupled modes thus offer a total of $2m_\star+3$ zero modes. Including the $2q-(2m_\star+3)$ of the decoupled $f$ modes, at each $k$ we recover sum total of $2q$ zero modes in the non-interacting case for each valley, independent of $m_{\star}$. This gives the total of $2$ states per moir\'e unit cell per spin independent of $\mathbf{B}$, i.e. the total Chern number $0$. Note that the magnetic subbands within the narrow band window remain separated by a gap from the remote subbands for $\mathbf{B}\neq 0$ because the remote bands emanate out of $\mathbf{B}\rightarrow 0$ energy eigenvalues $\pm \gamma$ obtained using above ans\"atze, as shown in Fig.(\ref{fig:Non-Interacting,flatband}). 

The analysis can straightforwardly be extended to include interactions using appropriate mean field parameters. In the next sections we illustrate it by discussing the strong coupling Hofstadter spectra at three integer fillings of the narrow bands.
\\
\\
\textbf{Electron-electron interactions at integer fillings of narrow bands at $\mathbf{B}\neq0$}: To understand the effect of interactions on single-particle Hofstadter spectra discussed above, we extend our formalism to illustrate the Landau quantization for ``one-shot" Hatree-Fock (HF) bands obtained using the mean-field (MF) Hamiltonian for a parent valley polarized (VP) state. The VP state at $\mathbf{B}=0$ is given by a product of valley polarized $\tilde{f}$-multiplets and the Fermi sea ($\ket{\text{FS}}$) of half-filled $\tilde{c}$-electron bands \cite{song2022magic}. The narrow band filling factor $\nu$ then determines the number of $f$-electrons to be filled per moir\'e unit cell above $\ket{\text{FS}}$. The $U(4)$-flavor of the $f$-electrons to be filled is further dictated by the $U(4)$ Hund's rule discussed in \cite{song2022magic}. Below, we start with the $\mathbf{B}\neq 0$ solution for single particle charge $\pm 1$ excitation at the charge neutrality point (CNP).
\\
{$\boldsymbol{\nu}\mathbf{=0}$}:
At CNP, the MF interactions for the parent VP state with $\tau=+1$ valley occupied by the $f$-electrons %in the matrix notation 
with respect to the spinor in Eq.(\ref{MT 6-spinor}) are taken to be \cite{song2022magic} 
\begin{eqnarray}
V^{+1,s,\nu=0}_{\alpha,\alpha'}=\left(\begin{array}{ccc}
         0& 0 & 0 \\
         0& -\frac{J}{2}\sigma_0 &0\\
        0& 0 & -\frac{U_1}{2}\sigma_0
    \end{array}\right)_{\alpha,\alpha'}\label{Eq:MF at CNP}.
\end{eqnarray}
Within this approximation we continue using the $\mathbf{B}=0$ MF parameters $J$ and $U_1$ obtained for the parent VP state in Ref\cite{song2022magic}. The MF parameter $U_1$ is the largest energy scale of the THFM as it corresponds to the strong onsite Coulomb repulsion amidst the localized Wannier states of the $f$-fermions. The MF parameter $J$ corresponds to the energy associated with the ferromagnetic exchange interaction between the $U(4)$ moments of $f$ and $c$ fermions with $a=\{3,4\}$.

The decoupled $f$ modes in the valley $\tau=+1$ now move to energy $-U_1/2$, while the spectrum of the coupled modes in the same valley can be obtained by solving the eigenvalues of the operator $\hat{h}^{+1,s}+V^{+1,s,\nu=0}$, where $\hat{h}^{+1,s}$ is defined in Eq.(\ref{eq:MT hat h plus 1}). The spectrum for sector $\tau=-1$ of the MF Hamitonian is the particle-hole symmetric partner of the spectrum for $\tau=+1$\cite{song2022magic}. 
Thus for a given valley quantum number $\tau$, the $2q-(2m_\star+3)$ decoupled $f$-modes now move to the energy $-\tau U_1/2$ forming the lower and upper bounds on the strong coupling energy window for narrow bands as shown in the Fig.(\ref{fig:0fieldvsB field}b). In order to understand the mode counting within the narrow band strong coupling energy window, we first discuss the solutions in the flat band limit $M=0$ which can be obtained using ans\"atze presented in Eq.(\ref{MT decoupled-c state})-Eq.(\ref{MT the ansatz 6}). 
The anomalous $c$-mode $\theta_1$ in Eq.(\ref{MT decoupled-c state}) forms the $\mathbf{B}$ independent level at $-J/2$ as shown in the Fig.(\ref{fig:0fieldvsB field}b). Using the remaining ans\"atze $\theta_{\mu}$, we can set up corresponding $\mu\times \mu$ hermitian matrices with eigenvectors $c_\alpha^\mu$. The three hermitian matrices in total offer $2m_\star+2$ modes within the strong coupling narrow band energy window. This can be understood by noting that these modes emanate out of the $2m_\star+2$ fold degenerate $\mathbf{B}\rightarrow 0$ energy eigenvalue $-J/2$ of the above hermitian matrices. Using the fact that $\Sigma_m\rightarrow 1$ and $\ell^{-1}\rightarrow 0$ as $\mathbf{B}\rightarrow 0$, it can be readily verified that the non-zero coefficients for these $\mathbf{B}\rightarrow 0$ $2m_\star+2$ degenerate modes at $\tau=+1$ are $c^{(3)}_3=1$ for $\theta_3$, $c^{(5)}_3=1$ for $\theta_5$, $c^{(6_m)}_3=1$ for $\theta_{6_m}$ and $c^{(6_m)}_4=1$ for $\theta_{6_m}$. Note that in this limit, we have three extra modes of $a=3$ than the $a=4$ $c$-fermion. Similarly at $\tau=-1$, we will have three extra modes of $a=4$ than the $a=3$ $c$-fermion in the $\mathbf{B}\rightarrow 0$ limit. This can be understood as a direct consequence of a winding number three at $\Gamma$, reported in \cite{PhysRevLett.127.266402}. We present an effective Hamiltonian of these modes in the next paragraph.
By the particle-hole symmetry\cite{song2022magic}, we similarly have $2m_\star+3$ modes emanating out of $+J/2$ for $\tau=-1$ (see supplementary note 7A for details). Including the $2q-(2m_\star+3)$ decoupled $f$s at energy $-\tau U_1/2$, we have a total of $2q$ magnetic subbands within the narrow band strong coupling window $\pm(J/2$ to $U_1/2)$, i.e. $2$ states per moir\'e unit cell per valley per spin. The remote magnetic subbands on the other hand emanate out of the $\mathbf{B}\rightarrow 0$ energies $-\tau \frac{U_1}{4}\pm \sqrt{\frac{U^2_1}{16}+\gamma^2}$, marked by $\pm\mathcal{E}_{\mp \tau}$ in Fig.(\ref{fig:0fieldvsB field}b). 
Note that all the $\mathbf{B} \rightarrow 0$ energies mentioned above correspond to the zero field THFM energies at $\Gamma\in$ mBZ at CNP, illustrated in Fig.(\ref{fig:0fieldvsB field}a).

The lowest energy single particle excitations at the CNP at $\mathbf{B}=0$ reside at the $\Gamma$ point, as can be seen in Fig.(\ref{fig:0fieldvsB field}a). The Landau quantization of these bands can be better understood through a simple effective Hamiltonian obtained by systematically projecting onto the subspace spanned by $a=\{3,4\}$ $c$-fermions. It qualitatively describes the modes emanating from energy eigenvalues $-\tau J/2$. Including both valleys, the effective Hamiltonian for each spin projection is 
$
H_{eff} = \left(\begin{array}{cc}
    H^{\tau=1}_{eff} & 0  \\
    0 & H^{\tau=-1}_{eff}
\end{array} \right)$, where 
\begin{eqnarray}
H^{\tau=1}_{eff} =  \left(\begin{array}{cc}
-\frac{J}{2} - \hbar \omega _c \hat{a}^\dagger \hat{a} & i\frac{A}{\ell^3}{\hat{a}}^{\dagger^3}  \\
    h.c. & -\frac{J}{2} - \hbar \omega _c \hat{a} \hat{a}^\dagger 
\end{array} \right)\label{Eq:Heff for CNP K},
\end{eqnarray}
and $H^{\tau=-1}_{eff}$ can be obtained by replacing $\hat{a}\leftrightarrow \hat{a}^\dagger$ and changing the overall sign of  $\omega_c$, $A$ and $J$ in $H^{\tau=1}_{eff}$. Also $h.c.$ in Eq.(\ref{Eq:Heff for CNP K}) represents hermitian conjugate. The values of coefficient $A$ and effective cyclotron frequency $\omega_c\sim \ell^{-2}$ are provided in the Methods-C. 
In the $\mathbf{B}\rightarrow 0$ limit we can drop the off-diagonal terms in $H_{eff}^{\tau}$ because they are $\mathcal{O}\left(\ell^{-3}\right)$. For each spin, the anomalous modes $(|0\rangle,0)^T$ and $(0,|0\rangle)^T$ at $\tau=+1$ and $-1$ respectively, are singly degenerate at energy $-\tau J/2$. All other modes are doubly degenerate (for each spin). The energies of these pairs are $-\tau (J/2 +n \hbar \omega _c)$ where $n=1,2,3,\ldots$. Including the spin degeneracy, this would result in a LL filling sequence of $0,\pm 2,\pm 6,\pm 10,\ldots$ in the asymptotic $\mathbf{B}\rightarrow 0$ limit. As $\mathbf{B}$ increases, however, the off-diagonal terms grow and cause the splitting of these pairs. For example, the splitting of the first pair, i.e. with $n=1$, is visible at $\phi/\phi_0\sim 0.025$  ($\sim0.63$Tesla) in the Fig.(\ref{fig:0fieldvsB field}b). Moreover, the $\mathbf{B}$ field required for the splitting of a given pair with an index $n$ decreases with increasing $n$ because each action of the $\hat{a}$ is accompanied by a square root of the LL index making the off-diagonal terms comparable with the diagonal terms at a lower $\mathbf{B}$. If we compare the Fig.(\ref{fig:0fieldvsB field}b) with the Hofstadter spectrum of the BM model in the strong coupling limit at CNP presented in Fig.(\ref{fig: M neq 0 spec}b), we see a qualitative agreement in the nature of the LL spectra for low $\mathbf{B}$. Note that the latter is computed by neglecting the band kinetic energy and using the gauge invariant formalism introduced in the Ref.\cite{PhysRevB.106.085140} without any recourse to the heavy fermion model.  For example, in the vicinity of $\phi/\phi_0 = 0.025$ we can see that the anomalous mode is followed by a nearly degenerate pair of LLs, an isolated LL, and another two nearly degenerate LLs in both Figs.(\ref{fig:0fieldvsB field}b) and (\ref{fig: M neq 0 spec}b). Through the effective model analysis presented above we understand that these features appear due to the splitting of asymptotic $\mathbf{B}\rightarrow 0$ degeneracy of non-anomalous modes by the $\mathcal{O}\left(\ell^{-3}\right)$ terms as $\mathbf{B}$ increases. The splitting amidst the first pair of LLs (after the anomalous mode) appears to grow faster with increasing $\mathbf{B}$ in Fig.(\ref{fig: M neq 0 spec}b) compared to that in Fig.(\ref{fig:0fieldvsB field}b). Thus although the LL sequence at CNP from both approaches is `$0,\pm 2,\pm 4$', the LL gap at
 $\pm4$ is significantly smaller in the $M=0$ THFM compared to that in the strong coupling Hofstadter spectrum of BM model when $\phi/\phi_0$ reaches $0.1$ (i.e. $2.5$ Tesla). Interestingly, the LL filling sequence `$0,\pm2,\pm 4$' at CNP was also reported in the experiment of Ref.\cite{lu2019superconductors}, on an MATBG device with a non-vanishing gap at the CNP at $\mathbf{B}=0$. We come back to the experimental comparison at $\nu=\pm2$ in the later section. 

%Thus the LL filling sequence changes to $0,\pm2,\pm4\ldots$, as $\mathbf{B}$ increases. 

%As $\mathbf{B}$ increases, the degeneracy of these pairs gets lifted resulting in the LL filling sequence $0,\pm2,\pm4,\pm6,\ldots$, as is seen when the off-diagonal terms are included. The leading LL filling sequence of '$0,\pm2,\pm 4$' is consistent with the results reported in \cite{lu2019superconductors}.

For $M\neq 0$, the numerically determined strong coupling Hofstadter spectrum for $\tau=-1$, is shown in Fig.(\ref{fig: M neq 0 spec}a) (see supplementary note 7A for details).
As we can see, the lowest mode stays decoupled from the rest of the spectrum. The effect of finite $M$ can be included by adding $M\left(1-\frac{M_c}{\ell^2}(2\hat{a}^\dagger \hat{a}+1)\right)\sigma_x \zeta_0$ to $H_{eff}$. The Pauli matrices $\sigma_x$ and $\zeta_0$ act in the $a=\{3,4\}$ $c$-orbital and valley space, respectively. The value of the coefficient $M_c$ is provided in Methods-C. For non-zero $M$ the double degeneracy of LLs which we saw at $M=0$ is lifted even in the $\mathbf{B}\rightarrow 0$ limit. This results in a LL sequence $0,\pm 2,\pm 4, \ldots$ for the parent VP state at CNP for a general $\mathbf{B}$.

\begin{figure}[t]
\includegraphics[width=8.6cm]{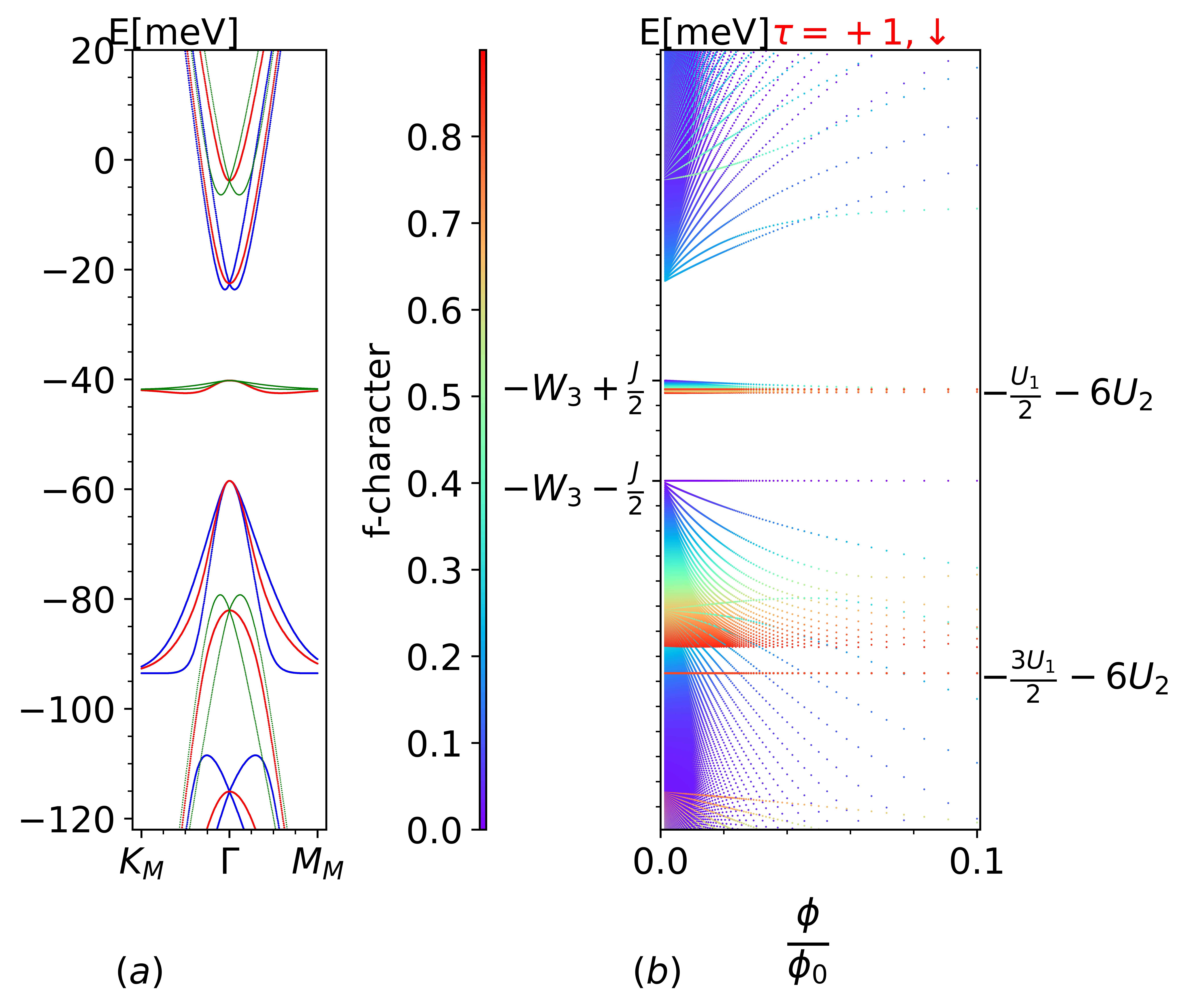}
\includegraphics[width=8.6cm]{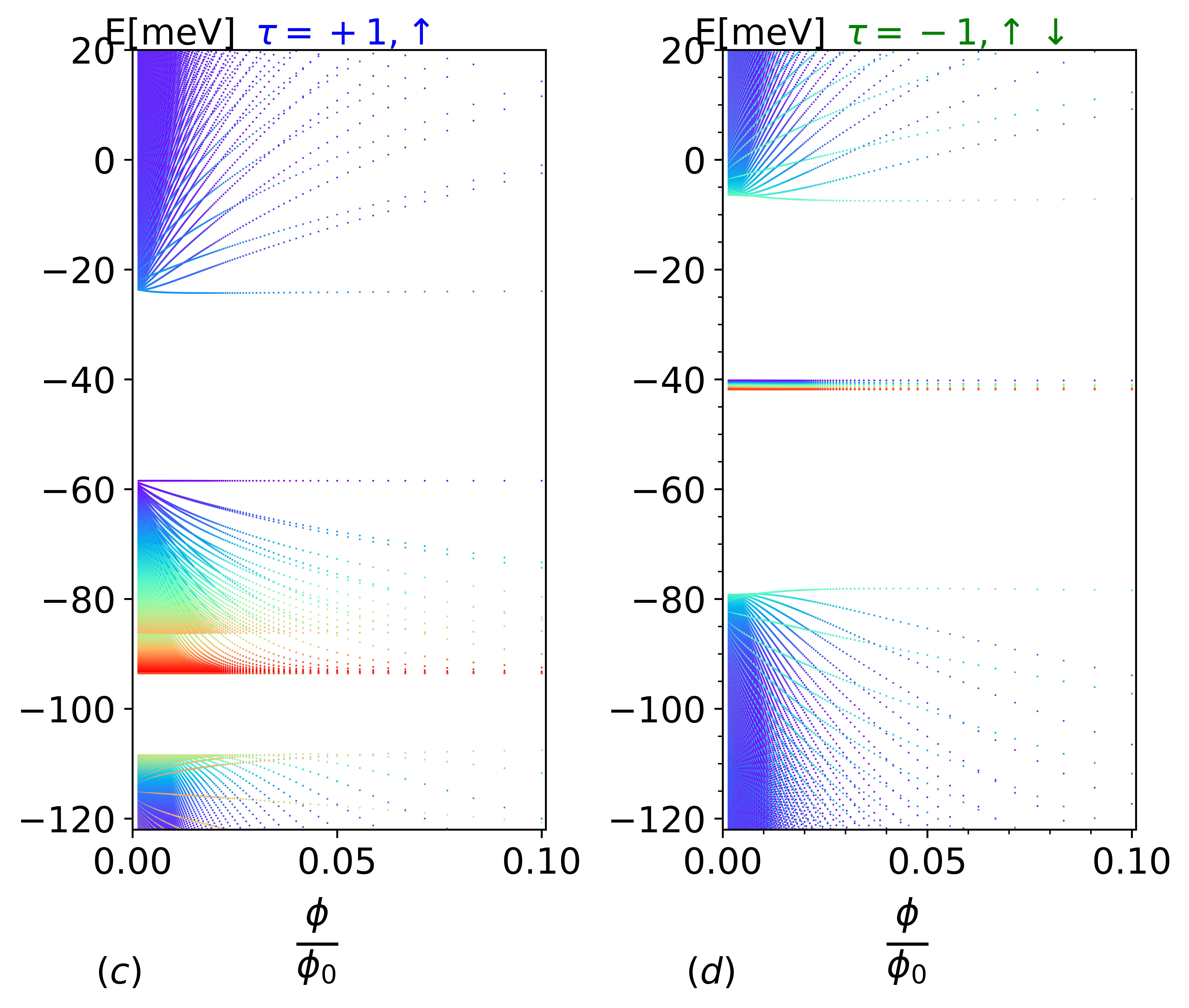}
\caption{Interacting heavy fermion Hofstadter spectra for sector (b) Valley \textcolor{red}{$\mathbf{K}$} spin $\textcolor{red}{\downarrow}$, (c)Valley \textcolor{blue}{$\mathbf{K}$} spin $\textcolor{blue}{\uparrow}$ and (d)Valley \textcolor{green}{$\mathbf{K}'$} spin $\textcolor{green}{\uparrow\downarrow}$(degenerate) contrasted with (a) zero field spectrum at filling $\nu=-1$ at $w_0/w_1=0.7$, with parameters $W_1=44.05$ meV, $W_3=49.33$ meV, $U_2=2.656$ meV in the flat band limit $M=0$ with $m_\star=\lceil \frac{q-3}{2} \rceil$.  The color on panel-(a) labels the spin and valley sector (\textcolor{red}{$\mathbf{K}$} $\textcolor{red}{\downarrow}$, 
 \textcolor{blue}{$\mathbf{K}$} $\textcolor{blue}{\uparrow}$ and \textcolor{green}{$\mathbf{K}'$} $\textcolor{green}{\uparrow\downarrow}$), while color labelling for panels-(b,c,d) denotes the f-character. The y-axis in panels (a) and (b) are aligned.}
\label{fig: nu=-1 spec}
\end{figure}

In the case of parent Kramers intervalley coherent state (KIVC), the effect of finite $M$ is included by adding $-M\left(1-\frac{M'_{c}}{\ell^2}(2\hat{a}^\dagger \hat{a} +1 )\right)\sigma_z\zeta_x$ to $H_{eff}$. The coefficient $M'_{c}$ is presented in the Methods-C. The LLs emanate out of the energy $\pm\sqrt{J^2/4+M^2}$ and are singly degenerate for each spin projection for a general $\mathbf{B}$. Similar to the flat band limit ($M=0$), the non-anomalous LLs occur in nearly degenerate pairs in the asymptotic $\mathbf{B}\rightarrow 0$ limit (see supplementary Fig.(14b)). These pairs of LLs split in energy with increasing $\mathbf{B}$. The splitting amidst the first pair of LLs (after the anomalous mode) is much weaker compared to other pairs (see supplementary Fig.(11b)). Thus, although the resulting LL sequence is `$0,\pm 2,\pm 4$', the LL filling gap $\pm4$ is much smaller compared to that for $0,\pm2$ similar to the $M=0$ case discussed earlier. More details for KIVC can be found in supplementary notes 9A and 10A2.  
\\
\\
{$\boldsymbol{\nu}\mathbf{=\pm 1}$}: In this section, we discuss the Landau quantization of the single particle excitation spectra at the narrow band filling factor of $\nu=- 1$ ($\nu=+ 1$ is related by particle-hole symmetry). Fig.(\ref{fig: nu=-1 spec}) shows the $\mathbf{B}=0$ spectrum and the Hofstadter spectra. We will show that all features of the spectrum can be analytically understood within the formalism, as well as through simple effective models. As in the case of CNP, we continue to use the $\mathbf{B}=0$ MF interactions for our analysis. The considered MF interactions are computed with respect to a partially spin- and completely valley-polarized parent state. For this state, the valley-spin flavor $\mathbf{K}\uparrow $ for both $b=1,2$ ($p_x\pm ip_y$) $f$-fermions and $\mathbf{K}\downarrow$ for $b=1$ $f$-fermion is occupied (at each unit cell) above the Fermi sea $|\text{FS}\rangle$ of half filled $c$-fermion bands (See also supplementary Eq.(S320) in Ref\cite{song2022magic}). 

For the sector valley \textcolor{red}{$\mathbf{K}$} spin \textcolor{red}{$\downarrow$}, the charge $\pm 1$ excitations occupy  Chern $\mp 1$ bands, which are separated from each other by a sizable gap: the Chern $-1$ and $+1$ bands, marked in red, can be seen in the energy windows $-30$ meV to $-50$ meV  and $-55$ meV to $-100$ meV of Fig(\ref{fig: nu=-1 spec}a), respectively. Below, we elucidate how our formalism captures the fact that the Chern $+(-)1$ bands gain(lose) states in presence of magnetic field $\mathbf{B}$, as they must to follow the Streda formula\cite{Steda1982TheoryOQ}. The MF interactions at sector \textcolor{red}{$\mathbf{K}\downarrow$} for the coupled modes with respect to the spinor in Eq.(\ref{MT 6-spinor}) read \cite{song2022magic} $V^{+1,\downarrow,\nu=-1}_{\alpha,\alpha'}=$
\begin{eqnarray}
    -\left(\begin{array}{ccc}
          W_1\sigma_0& 0 &0 \\
        0 &  W_3\sigma_0+\frac{J}{2}\sigma_z &0\\
        0& 0 & \left(U_1+6U_2\right)\sigma_0+\frac{U_1}{2}\sigma_z
    \end{array}\right)_{\alpha,\alpha'}.\label{eq:MF matrix for nu=-1}
\end{eqnarray}
The MF parameter $W_{a\in \{1,3\}}$ corresponds to the energy associated with the Coulomb repulsion between the $c$ and $f$ fermions, while $U_2$ corresponds to the energy associated with the next nearest neighbor Coulomb interaction of the $f$ fermions. The decoupled $f$-modes, i.e. the $\bar{f}_{11kr\downarrow}$ modes with $r\in\{m_\star+2,\ldots,q-1\}$ and $\bar{f}_{21k\bar{r}\downarrow}$ modes with $\bar{r}\in\{m_\star+1,\ldots,q-1\}$, are now at energies $-(\frac{3U_1}{2}+6U_2)=-93.5$ meV and $-(\frac{U_1}{2}+6U_2)=-41.8$ meV, respectively. The spectrum for the coupled modes can be obtained by solving the eigenvalues of the operator $\hat{h}^{+1,\downarrow}+V^{+1,\downarrow,\nu=- 1}$, where $\hat{h}^{+1,s}$ and $V^{+1,\downarrow,\nu=- 1}$ are defined in Eqs.(\ref{eq:MT hat h plus 1}) and (\ref{eq:MF matrix for nu=-1}) respectively. For $M=0$, the spectrum is exactly solvable. The anomalous $c$-mode in Eq.(\ref{MT decoupled-c state}) is an exact eigenstate which forms the $\mathbf{B}$ independent level at $-(W_3+\frac{J}{2})=-58.46$ meV. The remaining spectrum can be solved using the ans\"atze $\theta_3$, $\theta_5$ and $\theta_{6_m}$, presented in Eqs.(\ref{MT the ansatz 3})-Eq.(\ref{MT the ansatz 6}). The mode count can be understood as follows
\begin{enumerate}
    \item The spectrum for coupled modes includes $m_\star+2$ magnetic subbands emanating out of $\mathbf{B} \rightarrow 0$ energy eigenvalue $-(W_3+\frac{J}{2})$. The non-zero coefficients for these $\mathbf{B}\rightarrow 0$ eigenvectors are $c^{(3)}_3=1$ for $\theta_3$, $c^{(5)}_3=1$ for $\theta_5$ and $c^{(6_m)}_3=1$ for $\theta_{6_m}$. Including the anomalous $c$-mode in Eq.(\ref{MT decoupled-c state}), we have $m_\star+3$ modes emanating out of $-(W_3+\frac{J}{2})$. Moreover, accounting the $q-(m_\star+2)$ decoupled $f$ modes at energy $-(\frac{3U_1}{2}-6U_2)$, we have a total of $q+1$ magnetic subbands within the energy window of $-55$ meV to $-100$ meV. Recall that the isolated Chern +1 band resides in this same energy window at $\mathbf{B}=0$. We thus see that $q+1$ magnetic subbands emerge from the Landau quantization of the Chern +1 band.
    \item The spectrum for coupled modes includes $m_\star$ magnetic subbands emanating out of $\mathbf{B} \rightarrow 0$ energy eigenvalue $-(W_3-\frac{J}{2})=-40.19$ meV. The non-zero coefficients for these $\mathbf{B}\rightarrow 0$ eigenvectors are $c^{(6_m)}_3=1$ for $\theta_{6_m}$. Including the $q-(m_\star+1)$ decoupled $f$ modes at energy $-(\frac{U_1}{2}-6U_2)$, we have in total $q-1$ magnetic subbands in the energy window of $-30$ meV to $-50$ meV. Recall that the isolated Chern -1 band resides in this same energy window at $\mathbf{B}=0$. We thus see that $q-1$ magnetic subbands emerge from the Landau quantization of the Chern -1 band.  
    \end{enumerate}
Through the above mode count analysis, we see that the Chern $\pm 1$ bands Landau quantize into $q\pm 1$ magnetic subbands. Our formalism thus clearly shows that the total number of states per moir\'e unit cell for the Chern $\pm 1$ bands changes with magnetic field as $1\pm \frac{1}{q}=1\pm\frac{\phi}{\phi_0}$, as expected\cite{Steda1982TheoryOQ}. The mode count analysis for remaining MF valley-spin sectors, namely valley \textcolor{blue}{$\mathbf{K}$} spin \textcolor{blue}{$\uparrow$} and \textcolor{green}{$\mathbf{K}'$} spin \textcolor{green}{$\uparrow\downarrow$} (degenerate) can be found in supplementary note 6C,7B. The Hofstadter spectrum for each MF sector at $\nu=-1$ is shown in Fig.(\ref{fig: nu=-1 spec}-b,c,d) for $M=0$.
%%%%%%%%%%%%%%%%%%%%%%%%%%%%%%%%%%%%%%%%   FIGURE   %%%%%%%%%%%%%%%%%%%%%%%%%%%%%%%%%%%%%%%%%%%%%%%%%%%%%%
%%%%%%%%%%%%%%%%%%%%%%%%%%%%%%%%%%%%%%%%   FIGURE   %%%%%%%%%%%%%%%%%%%%%%%%%%%%%%%%%%%%%%%%%%%%%%%%%%%%%%

To better understand the Landau quantization of the dispersive (light mass) single particle excitations at $\nu=-1$, i.e. in vicinity of $\Gamma$, below we present an effective model analysis similar to that at CNP. As can be seen in Fig.(\ref{fig: nu=-1 spec}d)
the sectors \textcolor{green}{$\mathbf{K}'\uparrow\downarrow$} contribute magnetic subbands in the energy window $-30$ meV to $-50$ meV; adding a particle into any one of these subbands would move the filling towards CNP. Because we wish to focus on light mass excitations which move the filling away from CNP, we focus on the sectors \textcolor{blue}{$\mathbf{K}\uparrow$} and \textcolor{red}{$\mathbf{K}\downarrow$}. The effective Hamiltonian at the sector \textcolor{red}{$\mathbf{K}\downarrow$} takes the form 
\begin{eqnarray}
H^{\nu=-1,\mathbf{K}\downarrow}_{eff} &=& \left(\begin{array}{cc}
      -W_3-\frac{J}{2} -\hbar \bar{\omega}_c \hat{a}^\dagger \hat{a} & i\frac{\bar{A}}{\ell^3}{\hat{a}}^{\dagger^3} \\
      h.c.  &   -W_3+\frac{J}{2} -\hbar \tilde{\omega}_c \hat{a} \hat{a}^\dagger
   \end{array}\right)\nonumber\\
   &+&M\left(1-\frac{\bar{M}_c}{\ell^2} \hat{a} \hat{a}^\dagger-\frac{\tilde{M}_c}{\ell^2}\hat{a}^\dagger \hat{a} \right)\sigma_x \label{Eq:effective 2x2 nu=-1 VP, MT}
\end{eqnarray}
where Pauli matrix acts in the orbital space of the $a=\{3,4\}$ $c$-fermions. The coefficients $\bar{A}$, $\bar{M}_c$, $\tilde{M}_c$ and cyclotron frequencies $\tilde{\omega}_c\sim \ell^{-2}$, $\bar{\omega}_c\sim\ell^{-2}$ are provided in the Methods-C. The magnetic subbands of interest (the ones emerging from the Landau quantization of the light mass excitations) emanate out of the energy $-W_3- \sqrt{\frac{J^2}{4}+M^2}=-59.03$ meV, and are singly degenerate for a general $\mathbf{B}$. 

 The effective Hamiltonian at sector \textcolor{blue}{$\mathbf{K}\uparrow$} takes the same form as in Eq.(\ref{Eq:Heff for CNP K}). It can be obtained by replacing $-\frac{J}{2}$, $\omega_c$ and $A$ by $-(W_3+\frac{J}{2})$, $\bar{\omega}_c$ and $\bar{A}'$, respectively in Eq.(\ref{Eq:Heff for CNP K}). The effect of $M$ is included by adding $M\left(1-\frac{\bar{M}'_c}{\ell^2}(2\hat{a}^\dagger \hat{a}+1)\right)\sigma_x$ to the above obtained effective Hamiltonian. The coefficients $\bar{A}'$ and $\bar{M}'_c$ are provided in the Methods-C. The LLs emerge out of energies $-(W_3+J/2)+ M=-55.22$ meV and $-(W_3+J/2)- M=-61.71$ meV, and are singly degenerate for a general $\mathbf{B}$. 

In the $\mathbf{B}\rightarrow 0$ limit, we can drop the off-diagonal $\mathcal{O}\left(\ell^{-3}\right)$ term in both of the above effective Hamiltonians. Further taking the flat band limit $M=0$, we find that LL energies take the form $-(W_3+\frac{J}{2}+n\hbar\bar{\omega}_c)$, with $n\in\{0,1,2,\ldots\}$. The  $\mathbf{B}$ independent anomalous mode ($n=0$) is doubly degenerate as it is part of the spectrum at both sectors. The remaining modes ($n>0$) are triply degenerate each: singly degenerate at sector \textcolor{red}{$\mathbf{K}\downarrow$} and doubly degenerate at sector \textcolor{blue}{$\mathbf{K}\uparrow$}. This results in the LL sequence of $+1,-1,-4,-7,\ldots$ for $M=0$ in asymptotic $\mathbf{B}\rightarrow 0$ limit. The $+1$ gap in the sequence appears due to the Chern number $+1$ of the occupied band at sector \textcolor{red}{$\mathbf{K}\downarrow$} at $\mathbf{B}=0$. 

\begin{figure}[t]
    \centering\includegraphics[width=8.6cm]{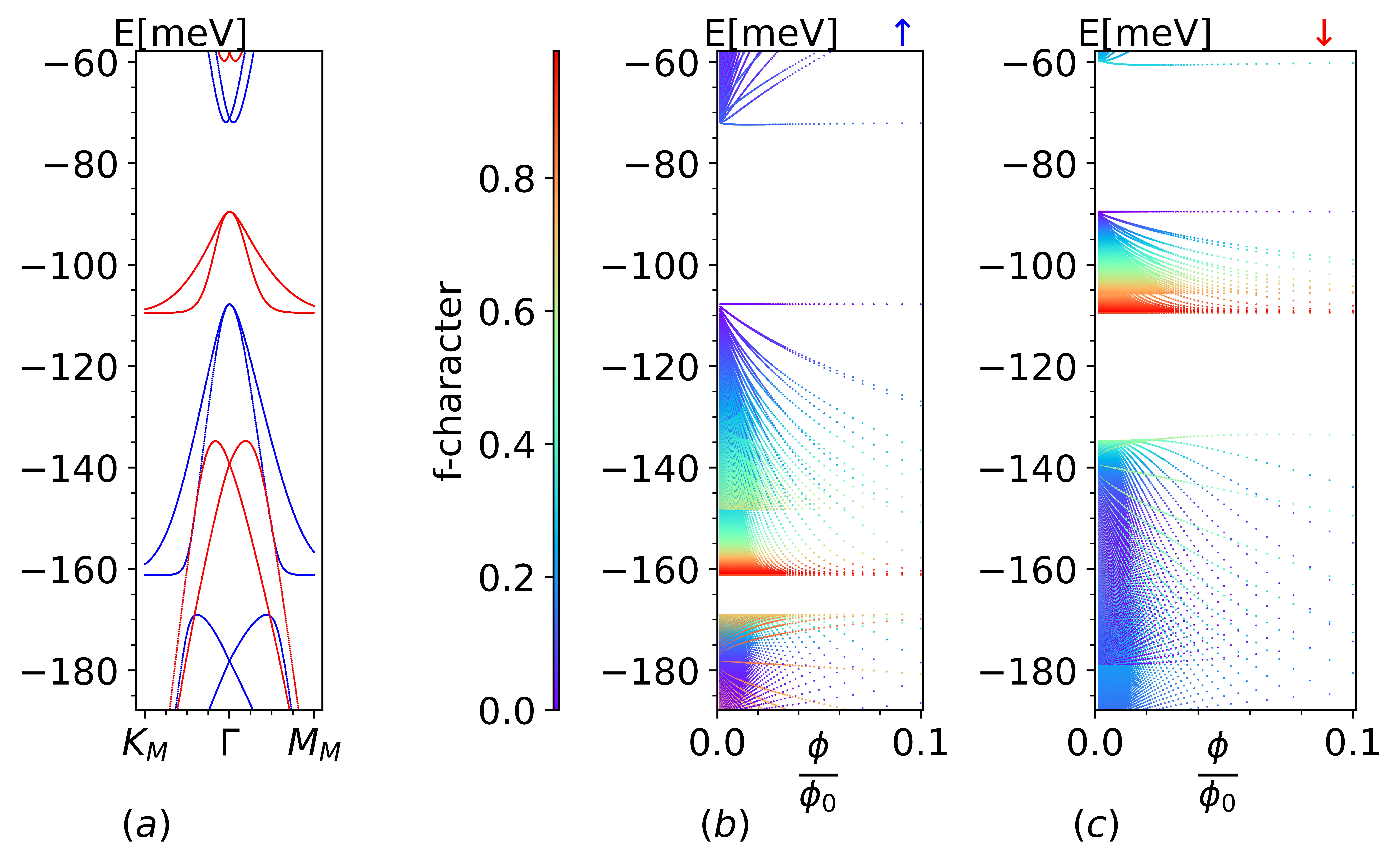}
    \caption{Interacting heavy fermion Hofstadter spectra for sectors (b) valley $\mathbf{K}$ spin (\textcolor{blue}{$\uparrow$}) and (c) valley $\mathbf{K}$ spin (\textcolor{red}{$\downarrow$}) contrasted with the (a) zero-field spectrum at filling $\nu=-2$ at $w_0/w_1=0.7$ in the flat band limit $M=0$ with $m_\star=\lceil \frac{q-3}{2} \rceil$. The color on panel-(a) labels the spin sector ($\textcolor{blue}{\uparrow}$, $\textcolor{red}{\downarrow}$) at valley $\mathbf{K}$, while color labelling for panels-(b,c) denotes the f-character.}
    \label{fig:nu=-2}
\end{figure}

Relaxing the $\mathbf{B}\rightarrow 0$ limit above, i.e. including the $\mathcal{O}\left(\ell^{-3}\right)$ terms in the above effective Hamiltonians (still $M=0$), we see that the LL energies change to: $E_m$, $E^{\downarrow}_n$ and $E_m$, $E^{\uparrow\pm}_n$ at sectors \textcolor{red}{$\mathbf{K}\downarrow$} and \textcolor{blue}{$\mathbf{K}\uparrow$}, respectively. Here $m\in\{0,1,2\}$,  $n\in\{0,1,2,\ldots\}$, $E_m=-W_3-J/2-m\hbar\bar{\omega_c}$, $E^{\downarrow}_n= -W_3-3\hbar\bar{\omega}_c/2-\hbar\tilde{\omega}_c/2-f_n- \sqrt{u^2_n+v_n\bar{A}^2}$ and $E^{\uparrow\pm}_n=-(W_3+J/2)-\hbar\bar{\omega}_c(n+2)\pm\sqrt{\hbar^2\bar{\omega}^2_c+v_n\bar{A}'^2}$. The coefficients  $f_n=n\hbar(\bar{\omega}_c+\tilde{\omega}_c)/2$, $u_n = J/2+3\hbar\bar{\omega}_c/2-\hbar\tilde{\omega}_c/2+n\hbar(\bar{\omega}_c-\tilde{\omega}_c)/2$ and $v_n=(n+1)(n+2)(n+3)/\ell^6$. Thus apart from the doubly degenerate levels at $E_m$, all others levels are singly degenerate. Since the LL energies are in the order : $E_0>E^{\uparrow,+}_0>E_1>E^{\uparrow,+}_1>\ldots$, the resulting LL sequence is $+1,-1,-2,-4,-5,\ldots$. %The LL gap at $-2$ is the smallest in this sequence, at least up to $\phi/\phi_0\sim 0.1$. 
Upon relaxing the flat band limit, i.e. for $M\neq 0$, all the above LL degeneracies get lifted. This results in LL sequence $+1,0,-1,\ldots$ for a general $\mathbf{B}$, with LL gaps at $+1,-1$ being the most dominant.
\\
\\
{$\boldsymbol{\nu}=\mathbf{=\pm 2}$:}
In this section, we discuss the Landau quantization of the single particle excitation spectra at the narrow band filling factor of $\nu=- 2$ ($\nu=+ 2$ is related by particle-hole symmetry). Fig.(\ref{fig:nu=-2}) shows the $\mathbf{B}=0$ spectrum and its continuation in field. As before, we will also derive the dominant LL sequence using an effective model that is simple enough for analytical solutions while capturing the low-energy features. As in the previous sections, we continue to use the $\mathbf{B}=0$ MF interactions for our analysis. The considered MF interactions are computed with respect to a spin and valley polarized parent state, for which the valley-spin flavor $\mathbf{K}\uparrow $ for both $b=1,2$ ($p_x\pm ip_y$) $f$-fermions are occupied (at each unit cell) above the Fermi sea $|\text{FS}\rangle$ of half filled $c$-fermion bands (See also supplementary Eq.(S333) in Ref\cite{song2022magic}). 

Below we discuss the valley sector $\mathbf{K}$ of the resulting MF Hamiltonian, for both spin $s=$\textcolor{blue}{$\uparrow$} and \textcolor{red}{$\downarrow$}. The single particle charge $+1$ excitations occupy the red band in energy window $-89.5$ meV to $-109.5$ meV of the Fig.(\ref{fig:nu=-2}a), which is part of the spectrum at sector $\mathbf{K}$\textcolor{red}{$\downarrow$}.  On the other hand, the single particle charge $-1$ excitations occupy the dispersive blue band in energy window $-107.8$ meV to $-161.2$ meV of the Fig.(\ref{fig:nu=-2}a), which is part of the spectrum at sector $\mathbf{K}$\textcolor{blue}{$\uparrow$}. The MF interactions for spin $s$, with respect to the spinor in Eq.(\ref{MT 6-spinor}) reads  \cite{song2022magic} $V^{+1,s,\nu=- 2}_{\alpha,\alpha'}=$
\begin{eqnarray}
    -2\left(\begin{array}{ccc}
          W_1\sigma_0&0  &0 \\
        0 & \left(W_3+\zeta_s\frac{J}{4}\right)\sigma_0 &0\\
        0& 0 & \left(\frac{4+\zeta_s}{4}U_1+6U_2\right)\sigma_0
\end{array}\right)_{\alpha,\alpha'},\label{eq:MF matrix for nu=-2}
\end{eqnarray}
where $\zeta_s=(+)-1$ for $s=$(\textcolor{blue}{$\uparrow$}) \textcolor{red}{$\downarrow$}. For the spin $s$, the $2q-(2m_\star+3)$ decoupled $f$ modes are at energy $-\left(\frac{4+\zeta_s}{2}U_1+12U_2\right)$, i.e. ($-161.17$ meV)$-109.45$ meV for $s=$(\textcolor{blue}{$\uparrow$}) \textcolor{red}{$\downarrow$}. The spectrum for the coupled modes can be obtained by solving the eigenvalues of the operator $\hat{h}^{+1,s}+V^{+1,s,\nu=- 2}$, where $\hat{h}^{+1,s}$ and $V^{+1,s,\nu=-2}$ are defined in Eqs.(\ref{eq:MT hat h plus 1}) and (\ref{eq:MF matrix for nu=-2}) respectively.
In the flat band limit $M=0$, the spectrum for coupled modes is exactly solvable. The anomalous $c$-mode in Eq.(\ref{MT decoupled-c state}) is an exact eigenstate which forms the $\mathbf{B}$ independent level at $-(2W_3+\zeta_s\frac{J}{2})$, i.e. ($-107.79$ meV)$-89.52$ meV for $s=$(\textcolor{blue}{$\uparrow$}) \textcolor{red}{$\downarrow$}. The remaining spectrum can be solved using the ans\"atze 
$\theta_3$, $\theta_5$ and $\theta_{6_m}$,
presented in Eqs.(\ref{MT the ansatz 3})-Eq.(\ref{MT the ansatz 6}). The spectrum of coupled modes include $2m_\star+2$ magnetic subbands emanating out of the $\mathbf{B}\rightarrow 0$ energy eigenvalue $-(2W_3+\zeta_s\frac{J}{2})$. The non-zero coefficients for the corresponding $\mathbf{B}\rightarrow 0$ eigenvectors are $c^{(3)}_3=1$ for $\theta_3$, $c^{(5)}_3=1$ for $\theta_5$, $c^{(6_m)}_3=1$ for $\theta_{6_m}$ and $c^{(6_m)}_4=1$ for $\theta_{6_m}$. We thus have a total of $2m_\star+3$ magnetic subbands emanating out of $\mathbf{B}\rightarrow 0$ energy $-(2W_3+\zeta_s\frac{J}{2})$. Including the $2q-(2m_\star+3)$ decoupled $f$ modes at $-\left(\frac{4+\zeta_s}{2}U_1+12U_2\right)$, we have a total of $2q$ magnetic subbands in the energy window $\left(-(2W_3+\zeta_s\frac{J}{2})\text{ to }-(\frac{4+\zeta_s}{2}U_1+12U_2)\right)$, i.e. $-89.52\text{ meV to }109.45$ meV and $-107.79\text{ meV to }161.17$ meV for s=\textcolor{red}{$\downarrow$} and \textcolor{blue}{$\uparrow$}, respectively. Recall that the red (\textcolor{red}{$\downarrow$}) and blue (\textcolor{blue}{$\uparrow$}) bands discussed earlier reside in the same energy window at $\mathbf{B}=0$. We thus have a total of $2q$ magnetic subbands emerging from the Landau quantization of single particle charge $\pm 1$ excitation bands at valley sector $\mathbf{K}$, i.e. two states per moir\'e unit cell for each. The discussion for sector $\mathbf{K'}\downarrow\uparrow$ can be found in supplementary note 7C.
%%%%%%%%%%%%%%%%%%%%%%%%%%%%%%%%%%%%%%%%   FIGURE   %%%%%%%%%%%%%%%%%%%%%%%%%%%%%%%%%%%%%%%%%%%%%%%%%%%%%%
%%%%%%%%%%%%%%%%%%%%%%%%%%%%%%%%%%%%%%%%   FIGURE   %%%%%%%%%%%%%%%%%%%%%%%%%%%%%%%%%%%%%%%%%%%%%%%%%%%%%%

To better understand the Landau quantization of the dispersive (light mass) single particle excitations at $\nu=-2$, i.e. in vicinity of $\Gamma$, below we present an effective model analysis similar to that in previous sections. As can be seen in the Fig.(\ref{fig:nu=-2}c), the sector $\mathbf{K}$\textcolor{red}{$\downarrow$} contributes to magnetic subbands in the energy window $-89.52\text{ meV to }-109.45$ meV; adding a particle into any one of these subbands would move the filling towards CNP. Same is true for the magnetic subbands contributed by sectors $\mathbf{K'}\uparrow\downarrow$, as can be seen in the supplementary Fig.(9). Because we wish to focus on light mass excitations which move the filling away from CNP, we focus only on the sector $\mathbf{K}$\textcolor{blue}{$\uparrow$}. 

The effective Hamiltonian at sector $\mathbf{K}$\textcolor{blue}{$\uparrow$} takes the same form as in Eq.(\ref{Eq:Heff for CNP K}). It can be obtained by replacing $-\frac{J}{2}$, $\omega_c$ and $A$ by $-(2W_3+\frac{J}{2})$, $\omega^{(\uparrow)}_c$ and $A^{(\uparrow)}$, respectively in Eq.(\ref{Eq:Heff for CNP K}). The effect of $M$ is included by adding $M\left(1-\frac{M^{(\uparrow)}_c}{\ell^2}(2\hat{a}^\dagger \hat{a}+1)\right)\sigma_x$ to the above obtained effective Hamiltonian. The coefficients $\omega^{(\uparrow)}_c$, $A^{(\uparrow)}$ and $M^{(\uparrow)}_c$ are provided in the Methods-C. The LLs emanate out of the energy $-(2W_3+ J/2)\pm M$ and are singly degenerate for a general $\mathbf{B}$. In the $\mathbf{B}\rightarrow 0$ limit, we can drop the off-diagonal $\mathcal{O}\left(\ell^{-3}\right)$ terms in the above effective Hamiltonian. Further setting $M=0$, we see that the LL energies take the form $-(2W_3+\frac{J}{2}+n\hbar\omega_c^{(\uparrow)})$, with $n\in\{0,1,2\ldots\}$. 
Similar to CNP, except the anomalous mode ($n=0$) at energy $-(2W_3+\frac{J}{2})$, every remaining mode ($n>0$) is doubly degenerate. This results in the LL sequence $-1,-3,-5,\ldots$ for $M=0$ in the asymptotic $\mathbf{B}\rightarrow 0$ limit. This asymptotic degeneracy of the non-anomalous modes is lifted as $\mathbf{B}$ increases, as is seen when $\mathcal{O}\left(\ell^{-3}\right)$ terms are included. Relaxing the flat band limit ($M\neq 0$) lifts the double degeneracy of the non-anomalous modes even in the asymptotic $\mathbf{B}\rightarrow 0$ limit. Thus for $M\neq 0$, we have the LL sequence $-1,-2,-3,\ldots$ for a general $\mathbf{B}$. Contrary to $\nu=0$, the LL sequence obtained at $\nu=-2$ differs from the LL filling sequence `$-2,-4,-6$' reported in the experiment of Ref.\cite{lu2019superconductors} at $\nu=-2$ by the appearance of the sizable gap at $-1$. Studying the origin of this difference will be a subject of future work.

\section{Discussion} We have put forward a generalization of THFM in finite \textbf{B}. Although the formalism applies to any rational value of $\frac{\phi}{\phi_0}$, the physical nature of hybridization amidst the heavy $f$ and topological $c$ fermions is particularly revealing for the $\frac{1}{q}$ sequence. The finite $\mathbf{B}$ analytical solution in the flat band limit provides an intuitive picture of the mechanism for Landau quantization of the strong coupling spectra of MATBG at integer fillings in terms of the decoupled $f$ modes and coupled $c$-$f$ modes, all the way to zero magnetic field. It also provides a deeper understanding of the nature of the $\pm \frac{J}{2}$ level at CNP, observed in numerics before\cite{wang2022narrow}, as the anomalous zero-LL of a massless Dirac particle, a key ingredient of the topological heavy fermion picture of MATBG. Although the number of the decoupled $f-$modes per unit cell per spin at CNP is dependent on the LL index upper cutoff, the total number of states in the narrow band strong coupling window remains pinned to 2 per unit cell per spin, independent of the upper cutoff, as expected for a total Chern number 0. Even though the full $M \neq 0$ problem requires numerical analysis, we are able to probe till fluxes at least as low as $1/700$, which was not possible through the the framework of strong coupling expansion. We moreover argue that the overall physical nature of the subbands should stay unchanged, as $M$ anyways is the smallest energy scale in the problem. 
Although we present the Landau quantization of one-shot HF bands in order to outline the theoretical procedure, in practice one can use the same methodology to Landau quantize the self-consistent HF bands. We argue that it would not drastically alter any of the interacting Hofstadter spectrum features because the one-shot states are adiabatically connected to the self-consistent states, owing to almost identical band structure features as the self-consistent state at every $\nu$ discussed \cite{song2022magic}. Throughout the text we neglected the spin Zeeman effect as it leads to a much smaller energy splitting than the orbital effect, the former is only a few Kelvin at the highest fields considered here while the latter is several meV, so at least an order of magnitude larger. The effect of renormalization of mean field parameters in magnetic field and heterostrain is yet to be incorporated in our framework. A full analysis for other integer fillings, translation symmetry broken candidate ground states and Hofstadter-scale fluxes where reentrant many-body and topological effects are at play \cite{PhysRevLett.128.217701,2020PhRvL.125w6804H,herzog2023hofstadter,fang2023symmetry}, is also left for the future work.

\bibliography{main}

\section{Methods}
\subsection{Evaluation of the \texorpdfstring{$c$-$f$}{cf} matrix elements at \texorpdfstring{$\mathbf{B}\neq0$.}{B not 0}}
Because $\eta_{b\tau kr'}(\mathbf{r})$ is constructed using repeated action of MT operators on the Wannier state $W_{\mathbf{0},b\tau}(\mathbf{r})$ as defined in Eq.(\ref{eq:etas}), and because the MT operators commute with $H^{\tau}_{BM}(\mathbf{p}-\frac{e}{c}\mathbf{A})$, we can reorder them so that $H^{\tau}_{BM}(\mathbf{p}-\frac{e}{c}\mathbf{A})$ acts directly on $W_{\mathbf{0},b\tau}(\mathbf{r})$. Since $H^{\tau}_{BM}(\mathbf{p}-\frac{e}{c}\mathbf{A})$ is linear in $\mathbf{p}-\frac{e}{c}\mathbf{A}$, the vector potential $\mathbf{A}$ now acts on the well localized state $W_{\mathbf{0},b\tau}(\mathbf{r})$ centered at the origin where $\mathbf{A}$ vanishes. Therefore, even though $\mathbf{A}$ is large at large $x$, we can safely neglect its contribution at low $\mathbf{B}$ (confirmed numerically in supplementary note 4C).  
Moreover, since $\Psi^\dagger_{a\tau}(\mathbf{r})$ is a Bloch state at $\Gamma$, it is invariant under the action of the moir\'e lattice translation operators. The $c$-$f$ coupling $h^{\tau}_{[amr],[br']}(k)$ at low $\mathbf{B}$ thus reduces to calculating the integral 
$\int d^2 \mathbf{r} \left(\hat{t}^{-n}_{\mathbf{L}_2}\hat{t}^{-s}_{\mathbf{L}_1}\chi_{krm}(\mathbf{r})\right)^\dagger\Psi^\dagger_{a\tau}(\mathbf{r})H^{\tau}_{BM}(\mathbf{p})W_{\mathbf{0},b\tau}(\mathbf{r})$, summed over all integer values of $s$ and $n$ and weighted by the factor of $e^{2\pi i\left(sk_1+n\left(k_2+r'/q\right)\right)}/\sqrt{\mathcal{N}}$ as follows from the Eq.(\ref{eq:etas}). 
The factor $\Psi^\dagger_{a\tau}(\mathbf{r})H^{\tau}_{BM}(\mathbf{p})W_{\mathbf{0},b\tau}(\mathbf{r})$ involves the Hamiltonian as well as the $c$ and $f$ wavefunctions strictly at $\mathbf{B}=0$. 
Its Fourier transform was calculated in Ref.\cite{song2022magic} and sets the $c$-$f$ coupling at $\mathbf{B}=0$ in momentum space, $e^{-\frac{1}{2}\mathbf{k}^2\lambda^2}H^{cf,\tau}(\mathbf{k})$, appearing in the Eq.(\ref{zero field full}).
Therefore, inverse Fourier transforming it gives 
$\Psi^\dagger_{a\tau}(\mathbf{r})H^{\tau}_{BM}(\mathbf{p})W_{\mathbf{0},b\tau}(\mathbf{r})=\sqrt{A_{uc}}H^{cf,\tau}(-i\boldsymbol{\nabla}_\mathbf{r})e^{-\frac{\mathbf{r}^2}{2\lambda^2}}/(2\pi\lambda^2)$. 
The factor $\hat{t}^{-n}_{\mathbf{L}_2}\hat{t}^{-s}_{\mathbf{L}_1}\chi_{krm}(\mathbf{r})$ can be computed by noting that $\chi_{krm}(\mathbf{r})$ is an eigenstate of $\hat{t}_{\mathbf{L}_1}$ and $\hat{t}^{-n}_{\mathbf{L}_2}\chi_{k r m}(\mathbf{r})=e^{2\pi in(k_2+r/q)}\chi_{[k_1+n\frac{p}{q}]_1 k_2 m}(\mathbf{r})$. Finally substituting the explicit expression of $\chi_{[k_1+n\frac{p}{q}]_1 k_2 m}(\mathbf{r})$ using Eq.(\ref{eq:chis}) reduces $h^{\tau}_{[amr],[br']}(k)$ to sum over integrals of a 2D gaussian, a plane wave factor along $y$, and shifted 1D h.o. wavefunctions along $x$. We thus have a standard gaussian integral in $y$, while the $x$-integral over the gaussian and and shifted 1D h.o. wavefunctions can be evaluated using results in Ref.\cite{babusci2012integrals}, (see supplementary note 4B for details).

It is particularly revealing to analyze the case $p=1$, i.e. the $\phi/\phi_0=1/q$ sequence. Since $r$ ranges from $0$ to $p-1$, the $1/q$ sequence is tantamount to setting $r=0$ in $h^{\tau}_{[amr][br']}$. Based on the results from the above discussion, after performing the summation over $n$, we find that $h^{\tau}_{[am0],[br']}(k)$ reduces to
$\int d^2\mathbf{r}\left(\hat{t}^{r'+jq}_{\mathbf{L}_1}\Phi_m(\mathbf{r},k_2\mathbf{g}_2)\right)^\ast H_{ab}^{cf,\tau}\left(-i\boldsymbol{\nabla}_{\mathbf{r}}\right)e^{-\frac{\mathbf{r}^2}{2\lambda^2}}$, summed over all integer values of $j$ and weighted by the factor $e^{-2\pi i(r'+jq)k_1}\sqrt{L_{1x}/\ell}/(2\pi\lambda^2)$.
For $a=b=1$, this integral can be visualized as an overlap between a 2D localized heavy state with size $\lambda$ sitting at the origin and a 1D h.o. shifted in the $x$-direction with a plane wave phase variation in the $y$-direction that depends on the shift (see supplementary Fig.(2)).
To understand for what choice of $m, r', j$ is this integral significant, note  that the h.o. wavefunction is localized in the $x$ direction about $\left(r' + jq + k_2 q\right)L_{1x}$, and its width is $\sim 2\sqrt{2m+1} \sqrt{q}$  unit cells. In addition, the combination $\frac{r'}{q}+j+k_2$ controls the period of oscillation in the $y$-direction set by $1/(\frac{r'}{q}+j+k_2)$ times the unit cell size.
The integer $r'+ jq$ thus determines the unit cell to which the h.o. is shifted, and, because $k_2 2\pi \ell^2/L_m = k_2 qL_{1x}$, the value of $k_2q\in [0,1)$ fine-tunes the shift within the unit cell. The index $j$ then determines $q$-unit-cell periodic revival of the h.o. states, also illustrated in supplementary Fig.(2). 
Consider the case $r'=j=0$. The h.o. is centered at the unit cell containing the localized heavy state and the period of oscillations in the $y$-direction is long compared to the unit cell, encompassing at least $q$ unit cells. The hybridization with the localized heavy state proportional to $\gamma$ is thus significant. The spatial extent of the h.o. state in the $x$-direction is $\sim2\sqrt{2m+1}\sqrt{q}$ unit cells, which at low $\mathbf{B}$ is much longer than the localized heavy state. Thus unless $m$ is close to $m_\star\lesssim q/2$, even though the h.o. state
oscillates and has $m$-nodes, the result of the integration will be approximately given by the value of the h.o. wavefunction at $-k_2qL_{1x}$, up to an overall phase. If we keep $j=0$ but increase $r'$ to $1$ then the h.o. is centered at the unit cell adjacent to the one containing the localized heavy state and the period of oscillations in the $y$-direction is still long, between $q/2$ and $q$ unit cells. The hybridization with the localized heavy state proportional to $\gamma$ will still be significant and the result of the integration will still be approximately given by the value of the h.o. wavefunction but now at $-(1+k_2q)L_{1x}$, up to an overall phase. However for values of $r'$ past $\sim \sqrt{2m+1}\sqrt{q}$, regardless of the value of $qk_2$, the contribution from the revival copy $j=0$ gets exponentially suppressed, due to the large off-set with the 2D localized state. So for values of $r'>q/2$, it is the $j=-1$ revival copy of the h.o. states which gives the dominant contribution. We thus neglect all other values of $j$ and only consider the contribution from h.o. state centered at the unit cell $r'+jq\rightarrow \text{sgn}_+\left(\frac{q}{2}-r'\right)\text{min}[r',q-r']$,  where $\text{sgn}_+(x)$ is the usual sign function except at $0$ where it evaluates to $1$. The $\mathcal{F}_m$ appearing in the compact expression of $c$-$f$ hybridization in Eq.(\ref{eq:I0}) comes from the $x$-integral, i.e. overlap of the 2D heavy localized state with harmonic oscillator wavefunction $\varphi_m$.  In the limit $\lambda \rightarrow 0$, the 2D heavy localized state becomes the Dirac $\delta$-function, and we recover $\mathcal{F}_m(\lambda\rightarrow 0, x_0)= \varphi_m(-x_0)$ as expected. We found that keeping the full form of $\mathcal{F}_m$ is needed in order to achieve accurate results even for the low $B$ range, therefore we do not take this limit when handling $\mathcal{F}_m$(see also Fig.(4) in supplementary note 5). The exponential suppression factor multiplying $\mathcal{F}_m$ in Eq.(\ref{eq:I0}) comes from the $y$-integral; its dependence on $r'_q$ is weaker than $\mathcal{F}_m$, which comes from the $x$-integral. 

The derivatives appearing in case of the matrix elements $h^{\tau}_{[1(2)m0],[2(1)r']}$, act on the localized heavy function to change its spatial symmetry from $s$ to $p_{x,y}$-like. Moreover an integration by parts and expressing the derivatives via h.o. raising and lowering operators allows us to relate these cases to the analysis without derivatives(see supplementary notes 4B2, 4B3 and 4C for details). 

\subsection{Closed form expression for the singular values of \texorpdfstring{$\Upsilon$}{Upsilon} appearing in \texorpdfstring{Eq.\ref{eq:MT hat h plus 1}}{Eqnum}}
The fact that $U$ is very close to an identity matrix allows us to obtain  an analytical expression for $\Sigma(m)$, which reads 
\begin{eqnarray}\label{Eq:Singular Values}
\Sigma(m) = \left(\frac{1}{\sqrt{\xi(\kappa)}}\frac{1}{2^m m!}\mathcal{H}_{mm}\left(0,\frac{\kappa^6}{\xi(\kappa)};0,\frac{\kappa^6}{\xi(\kappa)}\mid \frac{2}{\xi(\kappa)}\right) \right)^\frac{1}{2},
\end{eqnarray}
where $\kappa^2=\frac{\lambda^2}{\ell^2}=(\frac{\phi}{\phi_0}) 2\pi \lambda^2 /(L_{1x}L_m)$, $\xi(\kappa) = (1+\kappa^2 + \kappa^4)(1+\kappa^2)$ and 
\begin{eqnarray}\label{others4variablehermite}
&&\mathcal{H}_{mn}\left(x,y;w,z\vert \beta\right) = \nonumber \\
&&\sum^{\text{min}(m,n)}_{k=0} \frac{m!n!\beta^k}{(n-k)!(m-k)!k!} 
\mathcal{H}_{m-k}(x,y)\mathcal{H}_{n-k}(w,z). \nonumber\\
\end{eqnarray}
Further details of the derivation can be found in supplementary note 5.
\subsection{Effective Hamiltonian Coefficients}
The coefficients appearing in the effective Hamiltonian at CNP in flat band limit presented in Eq.(\ref{Eq:Heff for CNP K}) are $\ell^2 \hbar \omega_c=4.00\times10^5 \text{meV} \mathrm{\mathring{A}}^2$ and $A=4.27\times10^7$ $\text{meV} \mathrm{\mathring{A}}^3$. The coefficient appearing in the mass term for VP and KIVC states at CNP are $M_c= 1.28\times10^4 \mathrm{\mathring{A}}^2$ and $M'_c=2.09\times10^4 \mathrm{\mathring{A}}^2$, respectively. The coefficients appearing in the effective Hamiltonian given in Eq.(\ref{Eq:effective 2x2 nu=-1 VP, MT}) are $\ell^2 \hbar \bar{\omega}_c =6.92\times10^5 \text{meV} \mathrm{\mathring{A}}^2$, $\ell^2 \hbar \tilde{\omega}_c =4.22\times10^4 \text{meV} \mathrm{\mathring{A}}^2$ ,  $\bar{A}=6.69\times10^7$ $\text{meV} \mathrm{\mathring{A}}^3$, $\bar{M}_c= 1.54\times10^4 \mathrm{\mathring{A}}^2$ and $\tilde{M}_c= 2.54\times10^4 \mathrm{\mathring{A}}^2$. The coefficients appearing in the effective Hamiltonian for sector $\mathbf{K}\uparrow$ at $\nu=-1$ are $\bar{A}'=6.11\times10^7$$\text{meV} \mathrm{\mathring{A}}^3$ and $\bar{M}'_c=1.34\times10^4\mathrm{\mathring{A}}^2$. The coefficients appearing in the effective Hamiltonian for sector $\mathbf{K}\uparrow$ at $\nu=-2$ are $\ell^2 \hbar \omega_c^{(\uparrow)}= 8.31\times10^5 \text{meV} \mathrm{\mathring{A}}^2$, $A^{(\uparrow)}=5.79\times10^7 \text{meV} \mathrm{\mathring{A}}^3$ and $M_c^{(\uparrow)}=1.32\times 10^4
 \mathrm{\mathring{A}}^2$. The derivation of the effective Hamiltonians and full expressions for the coefficients can be found in supplementary note 10A, 10B and 10C.

\begin{acknowledgments} 
The authors thank
Xiaoyu Wang and Zhi-da Song for valuable conversations, and to Dumitru Călugăru for computational advice. J. H.-A. is supported by a Hertz Fellowship and by ONR Grant No. N00014- 20-1-2303. B.A.B is supported by  the DOE Grant No. DE-SC0016239 and by the EPiQS Initiative, Grant GBMF11070. A.C. was supported by Grant No.
GBMF8685 towards the Princeton theory program and by the Gordon and Betty Moore Foundation through
the EPiQS Initiative, Grant GBMF11070. Further sabbatical support for  A.C., J.H.A and B.A.B  was provided by the European Research Council (ERC) under the European Union’s Horizon 2020 research and innovation programme (grant agreement No. 101020833), the Schmidt Fund for Innovative Research, Simons Investigator Grant No. 404513. O.V. was supported by NSF Grant No. DMR-1916958 and is partially funded by the Gordon and Betty Moore Foundation’s EPiQS Initiative Grant GBMF11070, National High Magnetic Field Laboratory through NSF Grant No. DMR-1157490 and the State of Florida.
\end{acknowledgments}

%\section*{Author Contribution}
%All authors have contributed equally to this work.
%\section*{Competing Interests}
%The authors declare no competing interests.

\clearpage

\onecolumngrid
%\begin{center}
%\textbf{Supplementary Materials}
%\end{center}

\tableofcontents

\clearpage
%\appendix
\setcounter{figure}{0}
\setcounter{equation}{0}
\setcounter{section}{0}

\renewcommand{\figurename}{Supplementary Figure}
\renewcommand\thesection{Supplementary note \arabic{section}}

\section{Bistritzer-MacDonald model}
\label{apdx:BM-model}

The Bistritzer-Macdonald (BM) model of magic-angle twisted bilayer graphene\cite{bistritzer2011moire} in the valley $\bK$$(\tau=+1$) is 
\begin{eqnarray}\label{appx:BM model}
H_{BM}^{\bK}(\bp) = \left(\begin{array}{cc} v_F \sigma \cdot \bp & T(\br)e^{i\bq_1 \cdot\br} \\
e^{-i\bq_1 \cdot\br}T^{\dagger}(\br) & v_F \sigma \cdot (\bp+ \hbar\bq_1)
\end{array}\right),
\end{eqnarray}
 where $\sigma$ acts in sublattice space and $\bp=0$ denotes the moir\'e Dirac point $K_M$. The interlayer hopping functions are 
 \begin{eqnarray}
T(\br)=\sum_{j=1}^{3}T_{j}e^{-i\bq_1 \cdot \br}, 
 \end{eqnarray}
 where $\bq_1 = k_{\theta}(0,-1)$, $\bq_{2,3}=k_{\theta}(\pm \frac{\sqrt{3}}{2},\frac{1}{2})$, $k_{\theta}=\frac{8\pi}{3a_0}\sin\frac{\theta}{2}=\frac{4\pi}{3L_m}$, $a_0 \approx 0.246$nm, $L_m= \frac{a_0}{2\sin{\theta/2}}$ is moir\'e period. $\theta=1.05^{\circ}$ in this work.
 \begin{eqnarray}
T_{j+1}=w_0I_2+w_1\left(\cos\left(\frac{2\pi}{3}j\right)\sigma_x + \sin\left(\frac{2\pi}{3}j\right)\sigma_y\right) 
 \end{eqnarray}
where $I_n$ is the $n \times n$ unit matrix. We will present results for the case $w_0/w_1=0.7$. The BM model is invariant under translation by integer multiples of moir\'e lattice vectors $\bL_1=L_m(\frac{\sqrt{3}}{2},\frac{1}{2})$ and $\bL_2=L_m(0,1)$. Thus if $f(\br)$ is an eigenstate, then so is
\begin{eqnarray}
\hat{T}_{\bL_{1,2}}f(\br) = f(\br-\bL_{1,2})  
\end{eqnarray}
where $\hat{T}_{\bL_{1,2}}$ are the usual discrete translation operators.

\section{Magnetic Translation Operators}
At $\bB\neq 0$ in the Landau gauge $\bA=(0,Bx)$, the BM model in flux is found via minimally substitution: $H^{\bK}_{BM}\left(\bp-\frac{e}{c}\bA\right)=H^{\bK}_{BM}\left(p_x,p_y-\frac{eB}{c}x\right)$. The finite $\bB$ BM model at $\bK'$$(\tau=-1$) can be obtained similarly by applying time reversal to Eq.(\ref{appx:BM model}) followed by the minimal substitution. However at $\bB\neq 0$, $H^{\tau}_{BM}$ is still invariant under the translation by $\bL_2$, but a translation by $\bL_1$ needs to be accompanied by a gauge transformation as
\begin{eqnarray}
&&e^{i\frac{eB}{\hbar c}L_{1x}y}H^{\tau}_{BM}\left(p_x,p_y-\frac{eB}{c}x+\frac{eB}{c}L_{1x}\right)e^{-i\frac{eB}{\hbar c}L_{1x}y}=
H^{\tau}_{BM}\left(p_x,p_y-\frac{eB}{c}x\right).
\end{eqnarray}
In other words if $\tilde{f}(\br)$ is an eigenstate of $H^{\tau}_{BM}\left(p_x,p_y-\frac{eB}{c}x\right)$, then so is
\begin{eqnarray}
\tilde{f}(\br) \rightarrow \hat{t}_{\bL_1}\tilde{f}(\br)=e^{i\frac{eB}{\hbar c}L_{1x}y}\tilde{f}(\br-\bL_{1})
\end{eqnarray}
Thus $\hat{t}_{\bL_1}=e^{i\frac{eB}{\hbar c}L_{1x}y} \hat{T}_{\bL_1}$ is the generator of magnetic translation by $\bL_1$. Note that $\hat{t}_{L_1}$ can alternatively be presented as
\begin{eqnarray}
\hat{t}_{\bL_1} = e^{i\bq_{\phi}\cdot\br}\hat{T}_{\bL_1}
\end{eqnarray}
where the magnetic translation wavevector $\bq_{\phi}$ is defined as
\begin{eqnarray}
\bq_{\phi} = \frac{\phi}{\phi_{0}}\left(\frac{1}{2}\bg_1 + \bg_2 \right)
\end{eqnarray}
where $\bg_{1,2}$ are corresponding reciprocal lattice vectors to $\bL_{1,2}$ and are given as: $\bg_1=\frac{4\pi}{\sqrt{3}L_m}(1,0)$ and $\bg_2=\frac{4\pi}{\sqrt{3}L_m}(-\frac{1}{2},\frac{\sqrt{3}}{2})$. Hence 
\begin{equation}\label{Eq:Use in Identity}
 \hat{t}_{\bL_1}=e^{2\pi i \frac{\phi}{\phi_0}\frac{y}{L_2}}\hat{T}_{\bL_1},   
\end{equation}
where $L_2=\mid \bL_2 \mid = L_m$.
Magnetic translation by $\bL_2$ are generated by $\hat{t}_{\bL_2}\tilde{f}(\br)=\tilde{f}(\br-\bL_2)$. We thus have
\begin{eqnarray}
\hat{t}_{\bL_2}\hat{t}_{\bL_1}\tilde{f}(\br)&=&e^{i\frac{eB}{\hbar c}L_{1x}(y-L_2)}\tilde{f}(\br-\bL_1-\bL_2)\\
\hat{t}_{\bL_1}\hat{t}_{\bL_2}\tilde{f}(\br)&=&e^{i\frac{eB}{\hbar c}L_{1x}y}\tilde{f}(\br-\bL_1-\bL_2)\\
\Rightarrow \hat{t}_{\bL_2}\hat{t}_{\bL_1}&=&e^{-2\pi i\frac{\phi}{\phi_0}}\hat{t}_{\bL_1}\hat{t}_{\bL_2}
\end{eqnarray}
where $\phi_0=\frac{hc}{e}$ and the flux through the unit cell is $\phi=B L_{1x}L_2$.
If $\phi/\phi_0=p/q$, with $p$ and $q$ relatively prime integers,
\begin{eqnarray}
 \left[\hat{t}^q_{\bL_2},\hat{t}_{\bL_1}\right]=0
\end{eqnarray}
and 
\begin{eqnarray}
\left[\hat{t}_{\bL_{1,2}},H^{\tau}_{BM}\left(p_x,p_y-\frac{eB}{c}x\right)\right]=0.   
\end{eqnarray}
The simultaneous eigenstates of the magnetic translation operators $\hat{t}_{\bL_1}$ and $\hat{t}^{q}_{\bL_2}$ can thus be used to produce a complete and orthonormal set of basis states for solving the $\bB\neq 0$ BM model. As we will see below, depending on whether they are conduction or heavy fermions there will be $p$ states or $q$ states per momentum point. It will be helpful to derive an identity for $\hat{t}^s_{\bL_1}$ which relates it to $\hat{T}^s_{\bL_1}$.

\subsection{Landau gauge magnetic translation group identities}\label{apdx:mtg identities}
Using Eq.(\ref{Eq:Use in Identity}), we have
\begin{eqnarray}
\hat{t}^s_{\bL_1}&=&\left(e^{2\pi i \frac{\phi}{\phi_0}\frac{y}{L_2}}\hat{T}_{\bL_1}\right)^s=
\left(e^{2\pi i \frac{\phi}{\phi_0}\frac{y}{L_2}}\hat{T}_{\bL_1}\right)\ldots \left(e^{2\pi i \frac{\phi}{\phi_0}\frac{y}{L_2}}\hat{T}_{\bL_1}\right)\left(e^{2\pi i \frac{\phi}{\phi_0}\frac{y}{L_2}}\hat{T}_{\bL_1}\right)\left(e^{2\pi i \frac{\phi}{\phi_0}\frac{y}{L_2}}\hat{T}_{\bL_1}\right)\left(e^{2\pi i \frac{\phi}{\phi_0}\frac{y}{L_2}}\hat{T}_{\bL_1}\right)\\
&=&
\left(e^{2\pi i \frac{\phi}{\phi_0}\frac{y}{L_2}}\hat{T}_{\bL_1}\right)\ldots \left(e^{2\pi i \frac{\phi}{\phi_0}\frac{y}{L_2}}\hat{T}_{\bL_1}\right)\left(e^{2\pi i \frac{\phi}{\phi_0}\frac{y}{L_2}}\hat{T}_{\bL_1}\right)\left(e^{2\pi i \frac{\phi}{\phi_0}\frac{2y-L_{1y}}{L_2}}\hat{T}^2_{\bL_1}\right)\\
&=&
\left(e^{2\pi i \frac{\phi}{\phi_0}\frac{y}{L_2}}\hat{T}_{\bL_1}\right)\ldots \left(e^{2\pi i \frac{\phi}{\phi_0}\frac{y}{L_2}}\hat{T}_{\bL_1}\right)\left(e^{2\pi i \frac{\phi}{\phi_0}\frac{3y-2L_{1y}-L_{1y}}{L_2}}\hat{T}^3_{\bL_1}\right)\\
&=&
\left(e^{2\pi i \frac{\phi}{\phi_0}\frac{y}{L_2}}\hat{T}_{\bL_1}\right)\ldots \left(e^{2\pi i \frac{\phi}{\phi_0}\frac{y}{L_2}}\hat{T}_{\bL_1}\right)\left(e^{2\pi i \frac{\phi}{\phi_0}\frac{4y-3L_{1y}-2L_{1y}-L_{1y}}{L_2}}\hat{T}^4_{\bL_1}\right).
\end{eqnarray}
Now, $1+2+3+\ldots+(s-1)=\frac{1}{2}s(s-1)$. Therefore,
\begin{eqnarray}
\hat{t}^s_{\bL_1}&=&
e^{-\pi i s(s-1) \frac{p}{q}\frac{L_{1y}}{L_2}}
e^{2\pi i s\frac{p}{q}\frac{y}{L_2}}
\hat{T}^s_{\bL_1}
\end{eqnarray}
where $\phi/\phi_0=p/q$.
Although the above formula was derived for positive $s$, it also holds for negative $s$. This can be seen by noting that
\begin{eqnarray}
\left(e^{-\pi i s(s-1) \frac{\phi}{\phi_0}\frac{L_{1y}}{L_2}}
e^{i\frac{eB}{\hbar c}sL_{1x}y}
\hat{T}^s_{\bL_1}\right)
\left(e^{\pi i s(-s-1) \frac{\phi}{\phi_0}\frac{L_{1y}}{L_2}}
e^{-i\frac{eB}{\hbar c}sL_{1x}y}
\hat{T}^{-s}_{\bL_1}\right)
=1.
\end{eqnarray}
So, $\hat{t}^{-1}_{\bL_1}$ denotes the inverse of $\hat{t}_{\bL_1}$, and $\hat{t}^{-s}_{\bL_1}=\left(\hat{t}^{-1}_{\bL_1}\right)^s$.

\section{Finite \texorpdfstring{$\bB$}{bB} basis for the \texorpdfstring{$c$}{c} and \texorpdfstring{$f$}{f} fermions}\label{apdx:finite B Basis}
The $\bB=0$ basis for $f$ fermions is constituted by two AA-stacking localised Wannier states per valley per spin \cite{song2022magic}. We denote them as $W_{\mathbf{0},b\tau}(\br)$, for $b\in\{1,2\}$ and $\tau=\pm 1$. Using the recently introduced method \cite{wang2023revisiting}, we can construct the finite $\bB$ basis by first building hybrid Wannier states out of the zero field Wannier states followed by projection onto a representation of magnetic translation group(MTG). For given spin, the hybrid Wannier states read
\begin{eqnarray}
w_{b\tau}(\br,k_2\bg_2)&=&\sum_{n\in \mathbb{Z}}e^{ik_2\bg_2\cdot (n\bL_2)}\hat{t}^{n}_{\bL_2} W_{\mathbf{0},b\tau}(\br)=
\sum_{n\in \mathbb{Z}}e^{2\pi ik_2n}W_{0,b \tau}(\br-n\bL_2).
\end{eqnarray}
where $k_2\in [0,1)$. These hybrid Wannier states although localized along the $x$ direction are Bloch-like extended in the $y$ direction and thus eigenstates of $\hat{t}_{\bL_2}$ 
\begin{equation}
 \hat{t}_{\bL_2}w_{b\tau}(\br,k_2\bg_2)=w_{b\tau}(\br-\bL_2,k_2\bg_2)=e^{-2\pi i k_2}w_{b\tau}(\br,k_2\bg_2).
\end{equation}
We next construct normalized eigenstates of MTG out of the hybrid Wannier states as 
\begin{eqnarray}
\eta_{b\tau k_1 k_2}(\br)&=&\frac{1}{\sqrt{\mathcal{N}}}
\sum_{s\in \mathbb{Z}} e^{2\pi i s k_1}\left(\hat{t}^s_{\bL_1}w_{b\tau}(\br,k_2\bg_2)\right),
\end{eqnarray}
where $k_1\in[0,1)$ and normalization $\mathcal{N}=n_{tot}s_{tot}$. Here $s_{tot}$ and $n_{tot}$ denote the total count of $s$ and $n$. Note that although these summations should be unbounded, we require these cutoffs for defining the normalization of our states and intermediate steps. They eventually cancel in the final formulas are are not physical. The normalisation is justified in Eqs.(\ref{normalzf1})-(\ref{normalzf2}). The domains of $k_{1,2}$ can be understood via noting that $\eta_{b\tau k_1 k_2}$ is periodic under $k_{1,2}\rightarrow k_{1,2}+1$. Under magnetic translations $\hat{t}_{\bL_1}$ and $\hat{t}^{q}_{\bL_2}$, we then have
\begin{eqnarray}
\hat{t}_{\bL_1}\eta_{b \tau k_1k_2}&=&e^{-2\pi i k_1} \eta_{b \tau k_1k_2}\\
\hat{t}^q_{\bL_2}\eta_{b \tau k_1k_2}&=&e^{-2\pi i qk_2}\eta_{b \tau k_1k_2}.
\end{eqnarray}
Thus states with $k_2$ that differ by $1/q$ have the same eigenvalue under $\hat{t}^{q}_{\bL_2}$. This is because the $qL_2$ translations break up the $k_2$ domains into units of width $1/q$. To make it apparent from the quantum numbers, we relabel the states as
\begin{eqnarray}
\eta_{b\tau k r'}(\br) = \eta_{b\tau k_1 k_2+\frac{r'}{q}}(\br) \;\;
\text{,}\;\; k=(k_1,k_2)\in[0,1)\otimes[0,1/q) \;\;
\text{,}\;\; r'=0,1 \ldots q-1.
\end{eqnarray}
Thus $\eta_{b\tau k r'}(\br)$ with different $k$ quantum numbers are guaranteed to be orthogonal. Moreover we have $q$ states for given $k$, corresponding to $r'\in\{0 \ldots q-1\}$ for each $b\in\{1,2\}$, i.e. a total of $2q$ heavy fermion states for given $k$. For same $k$ and different $r'$ the overlaps for given valley are
\begin{eqnarray}
&&\int d^2\br \eta^*_{b\tau k r'_{1}}(\br)\eta_{b'\tau k r'_{2}}(\br)=\nonumber\\
&&\frac{1}{\mathcal{N}}
\sum_{s_1s_2\in \mathbb{Z}}\sum_{n_1n_2\in \mathbb{Z}}
e^{2\pi i (s_2-s_1) k_1}
e^{-2\pi i(k_2+\frac{r'_{1}}{q})n_1}
e^{2\pi i(k_2+\frac{r'_{2}}{q})n_2}
e^{\pi i s_1(s_1-1) \frac{p}{q}\frac{L_{1y}}{L_2}}
e^{-\pi i s_2(s_2-1) \frac{p}{q}\frac{L_{1y}}{L_2}}\nonumber\\
&&\times\int d^2\br
e^{2\pi i (s_2-s_1)\frac{p}{q}\frac{y}{L_2}}
W^*_{\mathbf{0}b\tau}(\br-s_1\bL_1-n_1\bL_2) W_{\mathbf{0}b'\tau}(\br-s_2\bL_1-n_2\bL_2).\label{normalzf1}
\end{eqnarray}
Now since the $\bB=0$ Wannier states are well localised and orthonormal, we have
\begin{eqnarray}
\int d^2\br
e^{2\pi i (s_2-s_1)\frac{p}{q}\frac{y}{L_2}}
W^*_{\mathbf{0}b\tau}(\br-s_1\bL_1-n_1\bL_2) W_{\mathbf{0}b'\tau}(\br-s_2\bL_1-n_2\bL_2) \approx \delta_{bb'}\delta_{s_1s_2}\delta_{n_1n_2}
\end{eqnarray}
and thus,
\begin{eqnarray}
&&\int d^2\br \eta^*_{b\tau k r'_{1}}(\br)\eta_{b'\tau k r'_{2}}(\br)\approx\delta_{bb'}\frac{1}{\mathcal{N}}
\sum_{s_1\in \mathbb{Z}}\sum_{n_1\in \mathbb{Z}}
e^{2\pi i\frac{r'_{2}-r'_{1}}{q}n_1}=
\delta_{r'_{1}r'_{2}}\delta_{bb'}\frac{s_{tot}n_{tot}}{\mathcal{N}}=\delta_{r'_{1}r'_{2}}\delta_{bb'}.\label{normalzf2}
\end{eqnarray}
Henceforth we have a complete orthonormal basis for the heavy fermions in finite $\bB$, constituted by $2q$ states for given $k$, i.e. 2 states per moir\'e unit cell per valley per spin, which clearly implies that these are total Chern 0 states. 

Now let us discuss the basis for $c$ fermions. Remember that at $\bB=0$, the basis for $c$ fermions is constituted by four $\bk \cdot \bp$ Bloch states at the $\Gamma$ point in the moir\'e Brillouin Zone(mBZ), per valley per spin. We denote them by $\tilde{\Psi}_{\Gamma a \tau}$ for $a\in\{1 \ldots 4\}$ and $\tau=\pm 1$. 

We promote the $\bB=0$ $c$ fermion basis to finite $\bB$ using a result obtained when the $\bk\cdot\bp$ method is extended to finite $\bB$\cite{luttinger1955motion}, which prescribes the ansatz for finite field $\bk\cdot\bp$ states to be Landau level(LL) coefficients on top of the zero field $\bk\cdot\bp$ states. Henceforth, we define the finite $\bB$ basis for $\bk\cdot\bp$ Bloch states as an expansion over product of LLs and the
$\bB=0$ $\bk\cdot\bp$ Bloch states at the $\Gamma$ point in mBZ. However in order to use the same quantum numbers label as finite $\bB$ $f$ fermion basis we first project the Landau gauge LL onto the representation of MTG as
\begin{eqnarray}
\chi_{k_1k_2m}(\br)&=&\frac{1}{\sqrt{\ell L_2}}\frac{1}{\sqrt{\mathcal{N}}}
\sum_{s\in \mathbb{Z}} e^{2\pi i s k_1}\left(\hat{t}^s_{\bL_1}e^{2\pi ik_2\frac{y}{L_2}}\varphi_m\left(x-k_2\frac{2\pi\ell^2}{L_2}\right)\right)
\label{eq:conduction_basis_def}
\end{eqnarray}
where $k_1\in[0,1)$ and unlike before, $k_2\in[0,p/q)$ as justified in Eqs.(\ref{k2 domain for chi 1})-(\ref{k2 domain for chi2}). The functions $\varphi_m$ are $m$th 1D harmonic oscillator states that determine the Landau gauge Landau levels
\begin{eqnarray}
\varphi_m(x)&=&\frac{1}{\pi^{\frac{1}{4}}}\frac{1}{\sqrt{2^mm!}}e^{-x^2/2\ell^2}H_m(x/\ell)
\end{eqnarray}
where $H_m$ is the Hermite polynomial and $\ell^2=\hbar c/(eB)$. The normalization of the LL MTG states is explained in Eqs.(\ref{normalzc1})-(\ref{normalzc2}).
The domains for the quantum numbers $k_{1,2}$ can be understood by noting that $\chi_{k_1k_2m}$ is periodic under $k_1\rightarrow k_1+1$ and up to a phase under $k_2\rightarrow k_2+p/q$ as shown below
\begin{eqnarray}
&&\chi_{k_1k_2+\frac{p}{q}m}=   \frac{1}{\sqrt{\ell L_2}}\frac{1}{\sqrt{\mathcal{N}}}
\sum_{s\in \mathbb{Z}} e^{2\pi i s k_1}\left(\hat{t}^s_{\bL_1}e^{2\pi i(k_2+\frac{p}{q})\frac{y}{L_2}}\varphi_m\left(x-(k_2+\frac{p}{q})\frac{2\pi\ell^2}{L_2}\right)\right),
\end{eqnarray}
now using the fact, $\frac{2\pi\ell^2}{L_2}\frac{p}{q}=L_{1x}$ we have
\begin{eqnarray}
 \chi_{k_1k_2+\frac{p}{q}m}&=& \frac{1}{\sqrt{\ell L_2}}\frac{1}{\sqrt{\mathcal{N}}}
\sum_{s\in \mathbb{Z}} e^{2\pi i s k_1}\left(\hat{t}^s_{\bL_1}e^{2\pi i(k_2+\frac{p}{q})\frac{y}{L_2}}\varphi_m\left(x- L_{1x}- k_2\frac{2\pi\ell^2}{L_2}\right)\right) \label{k2 domain for chi 1} \\
&=& e^{2\pi i k_2\frac{L_{1y}}{L_2}}\frac{1}{\sqrt{\ell L_2}}\frac{1}{\sqrt{\mathcal{N}}}
\sum_{s\in \mathbb{Z}} e^{2\pi i s k_1}\left(\hat{t}^s_{\bL_1} e^{2\pi i\frac{p}{q}\frac{y}{L_2}}\hat{T}_{\bL_1}\left(e^{2\pi ik_2\frac{y}{L_2}}\varphi_m\left(x- k_2\frac{2\pi\ell^2}{L_2}\right)\right)\right) \\
&=&e^{2\pi i k_2\frac{L_{1y}}{L_2}}\frac{1}{\sqrt{\ell L_2}}\frac{1}{\sqrt{\mathcal{N}}}
\sum_{s\in \mathbb{Z}} e^{2\pi i s k_1}\left(\hat{t}^{s+1}_{\bL_1}e^{2\pi ik_2\frac{y}{L_2}}\varphi_m\left(x- k_2\frac{2\pi\ell^2}{L_2}\right)\right)
\\
&=& e^{-2\pi i \left(k_1-k_2\frac{L_{1y}}{L_2}\right)}\chi_{k_1k_2m} \label{k2 domain for chi2}
\end{eqnarray}
The domain for $k_2$ can also be understood in the context of the fact that $\chi_{k_1k_2m}$ are essentially LLs and thus behave as Chern number $+1$ states in $\bB\neq 0$. In order to illustrate it, let us discuss the total number of available states for $\chi_{k_1k_2m}$. 
We start by noting that under magnetic translations $\hat{t}_{\bL_1}$ and $\hat{t}^{q}_{\bL_2}$
\begin{eqnarray}
\hat{t}_{\bL_1}\chi_{k_1k_2m}&=&e^{-2\pi i k_1} \chi_{k_1k_2m},\label{c in Mt 1}\\
\hat{t}^q_{\bL_2}\chi_{k_1k_2m}&=&e^{-2\pi i qk_2} \chi_{k_1k_2m}. \label{c in Mt 2}
\end{eqnarray}
If we have a system size $N_1\bL_1$ and $q N_2 \bL_2$, then 
\begin{eqnarray}
&&\hat{t}^{N_1}_{\bL_1}\chi_{k_1k_2m}=\chi_{k_1k_2m}\\
\Rightarrow &&e^{-2\pi i k_1 N_1}=1;\;\Rightarrow \;\; k_1=0,\frac{1}{N_1},\frac{2}{N_1},\ldots,1-\frac{1}{N_1}
\end{eqnarray}
Similarly,
\begin{eqnarray}
&&\hat{t}^{qN_2}_{\bL_2}\chi_{k_1k_2m}=\chi_{k_1k_2m}\\
\Rightarrow &&e^{-2\pi i k_2 qN_2}=1;\;\Rightarrow \;\; k_2=0,\frac{1}{qN_2},\frac{2}{qN_2},\ldots,\frac{p}{q}-\frac{1}{qN_2}
\end{eqnarray}
So, the total number of states is $N_1\frac{p}{q}qN_2=\frac{\Phi}{\Phi_0}N_1qN_2=\frac{\Phi}{\Phi_0}N_1qN_2\frac{L_{1x}L_2}{A_{uc}}=
\frac{\Phi}{\Phi_0}\frac{A_{tot}}{A_{uc}}=\frac{B}{\Phi_0}A_{tot}=\frac{\Phi_{tot}}{\Phi_0}$.
Here $A_{uc}=\hat{z}\cdot(\bL_1 \times \bL_2)$ denotes the area of moir\'e unit cell. Thus $qA_{uc}$ is the area of magnetic unit cell and total area $A_{tot}=N_1N_2qA_{uc}$. This is a well known result for the degeneracy of the Landau level. Let us now discuss the orthonormalization for the LL MTG eigenstates. The overlaps amidst the MTG LL are given as:
\begin{eqnarray}
&&\int d^2\br \chi^*_{k_1k_2m}(\br)\chi_{k_1'k_2'm'}(\br) = \label{normalzc1}\\
&&\frac{1}{\ell L_2\mathcal{N}}\sum_{s\in\mathbb{Z}}\sum_{s'\in\mathbb{Z}}
e^{-2\pi i s \left(k_1-\left(k_2-\frac{p}{q}\right)\frac{L_{1y}}{L_2}\right)}
e^{i\pi s(s+1)\frac{p}{q}\frac{L_{1y}}{L_2}}
 e^{2\pi i s' \left(k'_1-\left(k'_2-\frac{p}{q}\right)\frac{L_{1y}}{L_2}\right)}
e^{-i\pi s'(s'+1)\frac{p}{q}\frac{L_{1y}}{L_2}} \nonumber
\\
&&\int dy
e^{-iy \left(s\frac{p}{q}+k_2\right)\frac{2\pi}{L_2}}
e^{iy \left(s'\frac{p}{q}+k'_2\right)\frac{2\pi}{L_2}}
\int dx
\varphi_m\left(x-\left(s\frac{p}{q}+k_2\right)\frac{2\pi}{L_2}\ell^2\right)
\varphi_{m'}\left(x-\left(s'\frac{p}{q}+k'_2\right)\frac{2\pi}{L_2}\ell^2\right) \label{step1}\\
&&=\frac{1}{\ell L_2\mathcal{N}}\sum_{s\in\mathbb{Z}}\sum_{s'\in\mathbb{Z}}
e^{-2\pi i s \left(k_1-\left(k_2-\frac{p}{q}\right)\frac{L_{1y}}{L_2}\right)}
e^{i\pi s(s+1)\frac{p}{q}\frac{L_{1y}}{L_2}}
 e^{2\pi i s' \left(k'_1-\left(k'_2-\frac{p}{q}\right)\frac{L_{1y}}{L_2}\right)}
e^{-i\pi s'(s'+1)\frac{p}{q}\frac{L_{1y}}{L_2}}\nonumber \\
&& 
n_{tot}L_2\delta_{s,s'}\delta_{k_2,k_2'} \ell \delta_{m,m'} \label{step2}\\
&&= \frac{1}{L_2 \mathcal{N}} \sum_{s\in\mathbb{Z}}e^{2\pi i s(k_1-k_1')}n_{tot}L_2 \delta_{k_2,k_2'} \delta_{m,m'}\label{step3}
\\
&&=\frac{1}{ L_2\mathcal{N}}s_{tot}\delta_{k_1,k_1'}n_{tot}L_{2}\delta_{k_2,k_2'} \delta_{m,m'} \label{step 4}\\
&&= \delta_{k_1,k_1'}\delta_{k_2,k_2'}\delta_{m,m'}.\label{normalzc2}
\end{eqnarray}
where in Eq.(\ref{step1})-(\ref{step2}), we have used the fact that $y$-integral implies $k_2-k'_2=(s'-s)\frac{p}{q}$. Now since $k_2,k'_2 \in [0,\frac{p}{q})$, the integral evaluates to $n_{tot}L_2\delta_{k_2,k_2'}\delta_{s,s'}$. Moreover each harmonic oscillator function in the $x$-integral can be shifted by the same amount since the arguments coincide. Therefore the x-integral evaluates to $\ell\delta_{mm'}$. Eventually in Eq.(\ref{step3})-(\ref{step 4}), the $s$ summation implies $k_1-k'_1=\text{integer}$. But since $k_1,k'_1\in [0,1)$, we must have $k_1=k'_1$ and the summation evaluates to $s_{tot}\delta_{k_1,k'_1}$. Therefore $\chi$'s are orthogonal and normalized. To have the same quantum number label as that of $\bB\neq 0$ $f$ fermion states, we relabel the MTG LLs as
\begin{eqnarray}
\chi_{k rm}(\br) = \chi_{k_1 k_2+\frac{r}{q}m}(\br) \;\;
\text{,}\;\; k=(k_1,k_2)\in[0,1)\otimes[0,1/q) \;\;
\text{,}\;\; r=0,1 \ldots p-1.
\end{eqnarray}
Thus for given $k$, we have $p$ states, corresponding to $r\in\{0\ldots p-1\}$, i.e. $\frac{p}{q}$ states per moir\'e unit cell per valley per spin as expected. Finally we have the finite $\bB$ c fermion basis states as $\Psi_{a \tau}\chi_{krm}$, where 
\begin{equation}\label{Eq.our kp bloch}
 \Psi_{a \tau}=\sqrt{NA_{uc}}\tilde{\Psi}_{\Gamma a \tau}   
\end{equation}
with $N$ being the total number of moir\'e unit cells. The $\sqrt{NA_{uc}}$ factor is required due to normalization of $\chi$. Note that since $\Psi_{a\tau}$ is a Bloch state at $\Gamma$,
\begin{eqnarray}
&&\hat{T}_{\bL_{i}} \Psi_{a\tau}(\br) =  \Psi_{a\tau}(\br) \\
&& \implies \hat{t}_{\bL_i}\left(\Psi_{a\tau}(\br)\chi_{krm}\right)(\br)=\Psi_{a\tau}\hat{t}_{\bL_{i}}\chi_{krm}(\br),
\end{eqnarray}
where $i\in\{1,2\}$. Thus $\Psi_{a \tau}\chi_{krm}$ are MTG eigenstates with same eigenvalue as that of $\chi_{krm}$. Now let us discuss the orthonormalization for these states. Since in this work we focus on $\frac{1}{q}$ sequence, for the following discussion on orthonormality we set index $r$ in $\chi_{krm}$ to zero and work with $\chi_{k0m}$. The Bloch periodicity for $\Psi_{\Gamma a \tau}(\br)$ allows us to expand it as 
\begin{eqnarray}
\Psi_{\Gamma a \tau}(\br) = \frac{1}{\sqrt{NA_{uc}}}\sum_{\bG} e^{i\bG\cdot \br}C_{a\tau\bG},
\end{eqnarray}
where $\bG=n_1\bg_{1}+n_2\bg_2$, $n_{1,2}\in\mathbb{Z}$, denote moir\'e reciprocal lattice vectors and 
\begin{eqnarray}
 C_{a\tau\bG} = \frac{1}{\sqrt{NA_{uc}}}\int d^2 \br e^{-i\bG \cdot \br} \tilde{\Psi}_{\Gamma a \tau}(\br)   
\end{eqnarray}
can be found in \cite{cualuguaru2023twisted}. Now using Eq.(\ref{Eq.our kp bloch}), the overlap matrix can be given as
\begin{eqnarray}\label{Eq:cc OVP}
O^{(q,m_{max})}_{[m',a'][m,a]} = \int d^2 \br \left(\Psi_{a' \tau}\chi_{k'0 m'}\right)^{\ast}\Psi_{a \tau}\chi_{k0 m} = \sum_{\bG_1,\bG_2} C^{\ast}_{a'\tau\bG_1}C_{a\tau\bG_2} \int d^2 \br e^{i(\bG_2 - \bG_1)\cdot \br} \chi^{\ast}_{k'0 m'}\chi_{k 0 m},
\end{eqnarray}
where $1/q$ is the flux per unit cell per flux quantum $\Phi_0$ and $m_{max}$ denotes the upper cutoff on the indices $m$ and $m'$. The integral can be computed as:
\begin{eqnarray}
&&\int d^2\br e^{i\bg\cdot \br}\chi^*_{k'0m'}(\br)\chi_{k0 m}(\br) \\ \nonumber
&=& \frac{1}{\ell L_2 \mathcal{N}}\sum_{s,s'\in \mathbb{Z}}e^{2\pi i(k_1s - k'_1s')}\int d^2\br (e^{2\pi i k_2'\frac{y}{L_2} }\varphi_{m'}(x-k'_2\frac{2\pi \ell^2}{L_2}))^* t^{-s'}_{L_1} e^{i\bg\cdot \br} t^{s}_{L_1}(e^{2\pi i k_2\frac{y}{L_2} }\varphi_m(x-k_2\frac{2\pi \ell^2}{L_2}))
 \\ \nonumber
&=& \delta_{k_1,k'_1}\frac{s_{tot}}{\ell L_2\mathcal{N}}\sum_{s\in \mathbb{Z}}e^{2\pi i sk_1}\int d^2\br (e^{2\pi i k_2'\frac{y}{L_2} }\varphi_{m'}(x-k'_2\frac{2\pi \ell^2}{L_2}))^*  e^{i \bg\cdot \br} t^{s}_{L_1}(e^{2\pi i k_2\frac{y}{L_2} }\varphi_m(x-k_2\frac{2\pi \ell^2}{L_2})) \\ \nonumber
&=& 
\delta_{k_1,k'_1}\sum_{s\in \mathbb{Z}}\delta_{k_2+\frac{s}{q},k_2'-\frac{L_2g_y}{2\pi}}e^{2\pi i s(k_1-\frac{k_2}{2})}e^{-i\pi s(s-1)\frac{1}{2q}}\frac{1}{\ell}\int d x \varphi_{m'}(x-k'_2\frac{2\pi \ell^2}{L_2})  e^{i g_x x} \varphi_m(x-(k_2+\frac{s}{q})\frac{2\pi \ell^2}{L_2})
\end{eqnarray}
The delta function constraint implies
\begin{eqnarray}
\delta_{k_2+\frac{s}{q},k_2'-\frac{L_2g_y}{2\pi}} = \delta_{k_2,k'_2}\delta_{s,-qg_2},
\end{eqnarray}
where $g_2 = g_y\frac{L_2}{2 \pi} \in \mathbb{Z}$. Note that getting $\delta_{k_{1(2)},k_{1'(2')}}$ from the overlap is a direct consequence of Eqs.(\ref{c in Mt 1})-(\ref{c in Mt 2}) for $\frac{1}{q}$ sequence. The above integral can be evaluated as
\begin{figure}
\includegraphics[width=5.3cm]{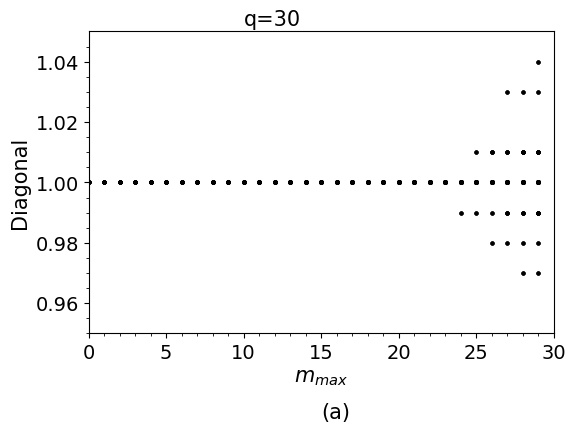}
\includegraphics[width=5.3cm]{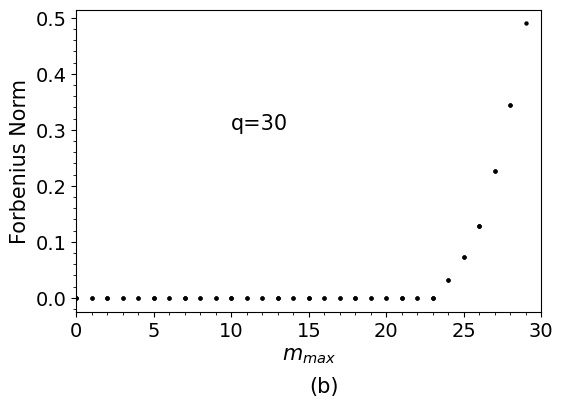}
\includegraphics[width=5.3cm]{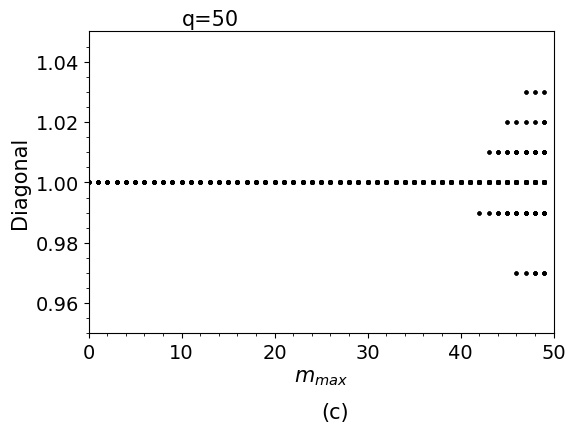}
\includegraphics[width=5.3cm]{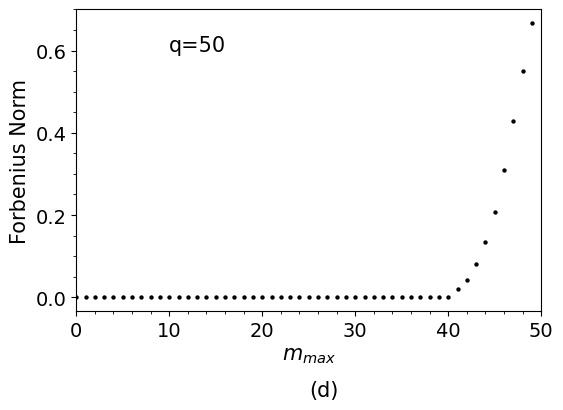}
\includegraphics[width=5.3cm]{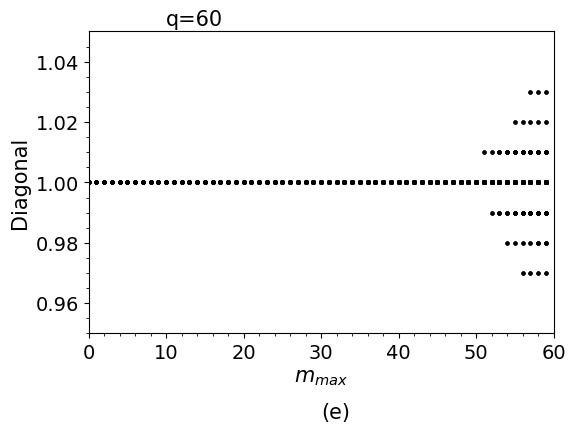}
\includegraphics[width=5.3cm]{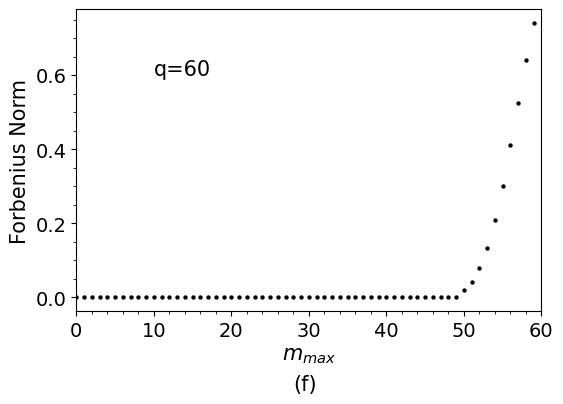}
\includegraphics[width=5.3cm]{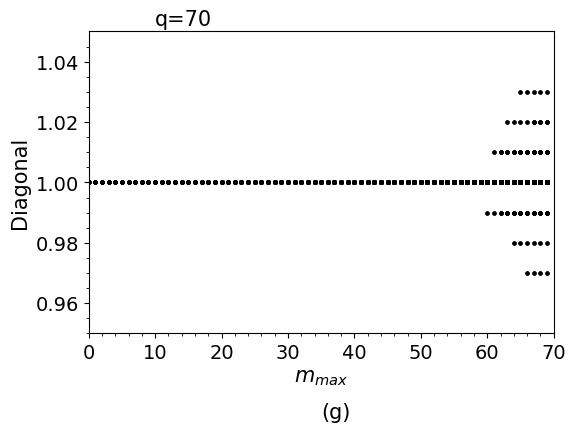}
\includegraphics[width=5.3cm]{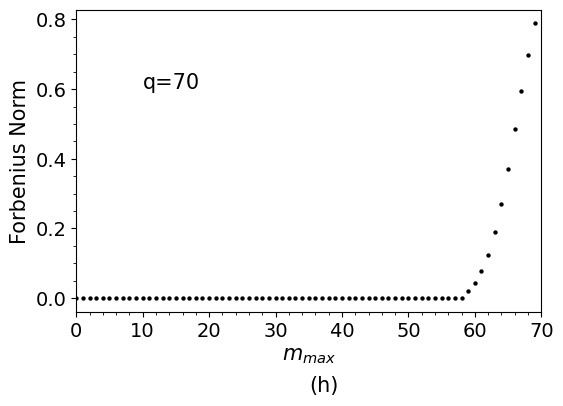}
\includegraphics[width=5.3cm]{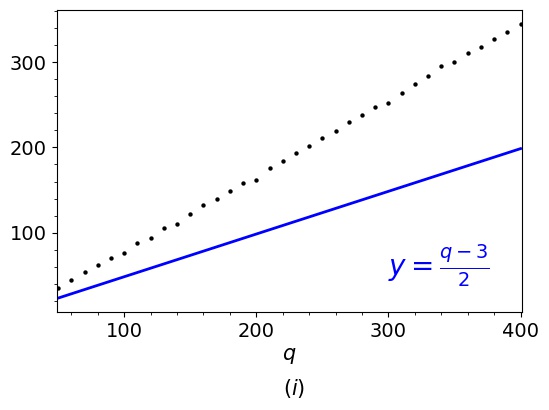}\label{Fig:mx q dependence plot}
\caption{Figures showing the diagonal values of Overlap matrix $O^{(q,m_{max})}_{[m',a'][m,a]}$ and frobenius norm of $O^{(q,m_{max})}-\mathbb{I}_{m_{max}}$ $\forall$ $m_{max}=0\ldots q-1$ at $w_0/w_1=0.8$ for (a,b) $q=30$, (c,d) $q=50$, (e,f) $q=60$ and (g,h) $q=70$ respectively. For higher values of $q$, the value of $m_{max}$ for which diagonal values start deviating from 1 have been (i) plotted using $\bullet$. }\label{fig:cc 0vp}
\end{figure}
\begin{eqnarray}
&=&\frac{1}{\ell}\int d x \varphi_{m'}(x-k_2\frac{2\pi \ell^2}{L_2})  e^{i g_x x} \varphi_m(x-(k_2-g_2)\frac{2\pi \ell^2}{L_2}) \\ \nonumber
&=&\frac{1}{\ell} e^{ig_xk_2\frac{2\pi \ell^2}{L_2}} \int d x \varphi_{m'}(x)  e^{i g_x x} \varphi_m(x+g_2\frac{2\pi \ell^2}{L_2}) \\ \nonumber
&=&\frac{1}{\ell} e^{ig_xk_2\frac{2\pi \ell^2}{L_2}} \int d x \varphi_{m'}(x)  e^{i g_x x} e^{i g_y\ell^2 p_x}\varphi_m(x) \\ \nonumber
&=&\frac{1}{\ell} e^{ig_xk_2\frac{2\pi \ell^2}{L_2}}e^{-\frac{i}{2}g_xg_y\ell^2} \int d x \varphi_{m'}(x)  e^{c_+a^\dagger + c_-a}\varphi_m(x) \\ \nonumber
&=& e^{ig_xk_2\frac{2\pi \ell^2}{L_2}}e^{-\frac{i}{2}g_xg_y\ell^2} \bra{m'} e^{c_+a^\dagger + c_-a}\ket{m},
\end{eqnarray} where $p_x$ is momentum operator along $x$ direction later changed to harmonic oscillator(h.o.) basis: $x=\frac{\ell}{\sqrt{2}}(a+a^{\dagger})$, and $p_x=\frac{i}{\sqrt{2\ell}}(a^{\dagger}-a)$. The h.o. kets are defined as $\langle m \mid x \rangle = \frac{1}{\sqrt{\ell}}\varphi_m(x)$, such that $\langle m \mid n \rangle = \delta_{m,n}$. Eventually we use the BCH formula, $e^Xe^Y=e^{X+Y+\frac{1}{2}[X,Y]+\ldots}$ , and
\begin{eqnarray}
c_{\pm} = \frac{i\ell}{\sqrt{2}}(g_x \pm i g_y).
\end{eqnarray}
We also know the identity
\begin{eqnarray}\label{small}
\bra{m'} e^{c_+a^\dagger + c_-a}\ket{m} = 
\begin{cases}
      e^{\frac{1}{2}c_+c_-}\sqrt{\frac{m!}{n!}}(c_+)^{n-m}L^{n-m}_{m}(-c_+c_-) & \text{for}~~ n \geq m,\\
      e^{\frac{1}{2}c_+c_-}\sqrt{\frac{n!}{m!}}(c_-)^{m-n}L^{m-n}_{n}(-c_+c_-) & \text{for}~~ n < m,
    \end{cases}   
\end{eqnarray}
where $L^{m-n}_{n}(x)$ is the associated Laguerre polynomial, 
\begin{eqnarray}
L^{m}_{N}(x) = \sum_{k=0}^{N} \frac{(N+m)!}{(N-k)!(m+k)!k!}(-x)^k.
\end{eqnarray}
As shown in Eq.(\ref{small}), the overlaps are exponentially small in $(c_+c_-)$, i.e. $(g\ell)^2$, and thus negligible. However, beyond an upper-bound on $m$, the orthonormality begins to fail. Supplementary Fig.(\ref{fig:cc 0vp}-a)-(\ref{fig:cc 0vp}-h) show that the diagonal values of overlap matrix in Eq.(\ref{Eq:cc OVP}) start deviating from $1$ beyond some upper-bound on $m$ exactly when the Frobenius norm of $O-\mathbb{I}$, with $\mathbb{I}$ being identity matrix of dimension same as that of $O$, starts deviating from $0$. Thus the deviation of diagonal values from $1$ is a good indicative of the breaking up of orthonormality.  In this work we choose $m_{max}=\lceil \frac{q-3}{2} \rceil$, which is well below the upper-bound as shown in Supplementary Fig.(\ref{fig:cc 0vp}-i).
\section{Computation of Matrix elements}
\subsection{c-c Coupling}
\label{apdx:cc matrix elements}
The $c-c$ coupling is given as
\begin{eqnarray}
H^{\tau}_{cc}&=&\sum_{k\in[0,1)\otimes [0,\frac{1}{q})} \sum_{a,a'=1}^4 \sum_{m=0}^{m_{a,\tau}}\sum_{m'=0}^{m_{a',\tau}}\sum_{r,\tilde{r}=0}^{p-1} \tilde{h}^{\tau}_{[amr],[a'm'\tilde{r}]}(k) c^\dagger_{a \tau k r m}
c_{a'\tau k \tilde{r} m'},
\end{eqnarray}
where
\begin{eqnarray}
\tilde{h}^{\tau}_{[amr],[a'm'r']}(k_1,k_2)&=&
\int d^2\br\Psi^*_{a \tau}(\br)\chi^*_{k r m}(\br) H^{\tau}_{BM}\left(p_x,p_y-\frac{eB}{c}x\right) \Psi_{a'\tau}(\br)\chi_{k r' m'}(\br).
\end{eqnarray}
Since the BM Hamiltonian is linear in $\bp-\frac{e}{c}\bA$, we can write it as
\begin{eqnarray}
H^{\tau}_{BM}\left(\bp-\frac{e}{c}\bA\right)&=&\mathcal{M}^{\tau}_\mu\left(p_\mu-\frac{e}{c}A_\mu(\br)\right)+\mathcal{T}^{\tau}(\br)
\end{eqnarray}
where $\mathcal{M}^{+1}_{\mu}=v_F\mathbb{I}_2 \otimes \mathbf{\sigma}$, $\mathcal{M}^{-1}_{\mu}=-v_F\mathbb{I}_2 \otimes \mathbf{\bar{\sigma}}$ with $\mathbf{\sigma}=(\sigma_x,\sigma_y)$ and $\bar{\sigma}=(\sigma_x,-\sigma_y)$. $\mathcal{T}^{\tau}$ denotes the remaining factors in BM Hamiltonian. So,
\begin{eqnarray}
&&H^{\tau}_{BM}\left(\bp-\frac{e}{c}\bA\right)\Psi_{a'\tau}(\br)\chi_{k r' m'}(\br)=
\mathcal{M}^{\tau}_\mu\left(p_\mu-\frac{e}{c}A_\mu(\br)\right)\Psi_{a'}(\br)\chi_{k_1,k_2+\frac{r'}{q},m'}(\br)+\mathcal{T}^{\tau}(\br)\Psi_{a'\tau}(\br)\chi_{kr'm'}(\br)\\
&&=\left(\mathcal{M}^{\tau}_\mu p_\mu\Psi_{a'\tau}(\br)+\mathcal{T}^{\tau}(\br)\Psi_{a'\tau}(\br)\right)\chi_{kr'm'}(\br)+
\mathcal{M}^{\tau}_\mu\Psi_{a'\tau}(\br)\left(\left(p_\mu-\frac{e}{c}A_\mu(\br)\right)\chi_{kr'm'}(\br)\right)\\
&&=\varepsilon^{\tau}_{a''a'}\Psi_{a''\tau}(\br)\chi_{kr'm'}(\br)+
\mathcal{M}^{\tau}_\mu\Psi_{a'\tau}(\br)\left(p_\mu-\frac{e}{c}A_\mu(\br)\right)\chi_{krm'}(\br),
\end{eqnarray}
where the matrix $\varepsilon^{\tau}$ is given in Eq.(\ref{the matrix factors from cc}) is $\bk$ independent because $\Psi_{a\tau}$ is defined at $\Gamma$. So, 
\begin{eqnarray}
&&\tilde{h}^{\tau}_{[amr],[a'm'r']}(k_1,k_2)=
\int d^2\br \Psi^*_{a\tau}(\br)\varepsilon^{\tau}_{a''a'}\Psi_{a''\tau}(\br)
\chi^*_{krm}(\br)\chi_{kr'm'}(\br)\\
&&+
\int d^2\br \Psi^*_{a\tau}(\br)
\mathcal{M}^{\tau}_\mu\Psi_{a'\tau}(\br)\chi^*_{krm}(\br)\left(p_\mu-\frac{e}{c}A_\mu(\br)\right)\chi_{kr'm'}(\br)
\end{eqnarray}
Since the factors $\varepsilon^{\tau}_{a''a'}\Psi^*_{a\tau}(\br)\Psi_{a''\tau}(\br)$ and $\Psi^*_{a\tau}(\br)
\mathcal{M}^{\tau}_\mu\Psi_{a'\tau}(\br)$ are periodic with respect to primitive moir\'e lattice vectors $\bL_1$ and $\bL_2$, we can perform a Fourier expansion as
\begin{eqnarray}
\varepsilon^{\tau}_{a''a'}\Psi^*_{a\tau}(\br)\Psi_{a''\tau}(\br)&=& \varepsilon^{\tau}_{a''a'}
\sum_\bg e^{i\bg\cdot\br}\left(\frac{1}{A_{tot}}\int d^2\br' e^{-i\bg\cdot\br'}\Psi^*_{a\tau}(\br')\Psi_{a''\tau}(\br') \right)\\
\Psi^*_{a\tau}(\br)
\mathcal{M}^{\tau}_\mu\Psi_{a'\tau}(\br)&=&
\sum_\bg e^{i\bg\cdot\br}\left(\frac{1}{A_{tot}}\int d^2\br' e^{-i\bg\cdot\br'}\Psi^*_{a\tau}(\br')
\mathcal{M}^{\tau}_\mu\Psi_{a'\tau}(\br')\right)
\end{eqnarray}
As shown in previous section, at small $B$ the overlaps of the MTG LL with $e^{i\bg\cdot\br}$ are suppressed by a factor of $\exp\left(-\bg^2\ell^2/4\right)$. So at small enough $B$ (large enough $\ell$) the overlaps with non-zero $\bg$ components are negligible. Keeping only the $\bg=0$ Fourier component then gives

\begin{eqnarray}
&&\tilde{h^{\tau}}_{[amr],[a'm'r']}(k_1,k_2)\approx
\left(\frac{1}{A_{tot}}\int d^2\br \Psi^*_{a\tau}(\br)\varepsilon^{\tau}_{a''a'}\Psi_{a''\tau}(\br)\right)
\left(\int d^2\br \chi^*_{km}(\br)\chi_{kr'm'}(\br)\right)+\nonumber\\
&&
\left(\frac{1}{A_{tot}}\int d^2\br \Psi^*_{a\tau}(\br)
\mathcal{M^{\tau}}_\mu\Psi_{a'\tau}(\br)\right)\left(\int d^2\br \chi^*_{krm}(\br)\left(p_\mu-\frac{e}{c}A_\mu(\br)\right)\chi_{kr'm'}(\br)\right)+
\mathcal{O}\left(\exp\left(-\bg^2\ell^2/4\right)\right)\\
&&=\varepsilon^{\tau}_{aa'}\delta_{rr'}\delta_{mm'}+
\mathcal{M}^{\tau,\mu}_{aa'}\left(\int d^2\br \chi^*_{krm}(\br)\left(p_\mu-\frac{e}{c}A_\mu(\br)\right)\chi_{kr'm'}(\br)\right)+
\mathcal{O}\left(\exp\left(-\bg^2\ell^2/4\right)\right),
\end{eqnarray}
where
\begin{eqnarray}
\mathcal{M}^{\tau,\mu}_{aa'}=\frac{1}{A_{tot}}\int d^2\br \Psi^*_{a\tau}(\br)
\mathcal{M}^{\tau}_\mu\Psi_{a'\tau}(\br)
\end{eqnarray}
The matrices $\epsilon^{\tau}_{a,a'}$ and $\mathcal{M}^{\tau,\mu}_{aa'}$ can be obtained from the zero field $cc$ coupling matrix 
\begin{eqnarray}\label{cc-matrix-zerofield}
&&H^{c,\tau}_{aa'}(\bk) = \left(\begin{array}{cc}
0_{2 \times 2} & v_{\ast}(\tau k_x\sigma_0 + ik_y\sigma_z) \\
v_{\ast}(\tau k_x\sigma_0 - ik_y\sigma_z) & M\sigma_x
\end{array}\right)
\end{eqnarray}
as
\begin{eqnarray}\label{the matrix factors from cc}
M^{\tau,x}_{aa'} =\frac{\tau v_{\ast}}{\hbar}\left(\begin{array}{cc}  
0&\sigma_{0}\\
\sigma_0&0
\end{array}\right)_{aa'} \;\;
\text{,}\;\;  M^{\tau,y}_{aa'} = \frac{iv_{\ast}}{\hbar}\left(\begin{array}{cc}  
0&\sigma_{z}\\
-\sigma_z&0
\end{array}\right)_{aa'} \;\;
\text{,}\;\; \varepsilon^{\tau}_{aa'} = M\left(\begin{array}{cc}  
0&0\\
0&\sigma_x
\end{array}\right)_{aa'}.
\end{eqnarray}
where the Pauli matrices $\sigma$ act in the orbital space of $c$ fermions. Since $\left[\hat{t}_{\bL_{1,2}},p_\mu-\frac{e}{c}A_\mu(\br)\right]=0$, we have

\begin{eqnarray}
\left(p_\mu-\frac{e}{c}A_\mu(\br)\right)\chi_{kr'm'}(\br)=
\frac{1}{\sqrt{\ell L_2 \mathcal{N}}}
\sum_{s\in \mathbb{Z}} e^{2\pi i s k_1}\left(\hat{t}^s_{\bL_1}
\left(p_\mu-\frac{e}{c}A_\mu(\br)\right)
e^{2\pi i\left(k_2+\frac{r'}{q}\right)\frac{y}{L_2}}\varphi_{m'}\left(x-\left(k_2+\frac{r'}{q}\right)\frac{2\pi\ell^2}{L_2}\right)\right).\nonumber\\
\end{eqnarray}

The action of the two components of $p_\mu-\frac{e}{c}A_\mu(\br)$ in Landau gauge can be calculated as
\begin{eqnarray}
&&\frac{\hbar}{i}\frac{\partial}{\partial x}e^{2\pi i\left(k_2+\frac{r'}{q}\right)\frac{y}{L_2}}\varphi_{m'}\left(x-\left(k_2+\frac{r'}{q}\right)\frac{2\pi\ell^2}{L_2}\right)=
\frac{\hbar}{i}\frac{1}{\ell}e^{2\pi i\left(k_2+\frac{r'}{q}\right)\frac{y}{L_2}}
\ell\frac{\partial}{\partial x}\varphi_{m'}\left(x-\left(k_2+\frac{r'}{q}\right)\frac{2\pi\ell^2}{L_2}\right)\nonumber\\
&&
=
\frac{\hbar}{i}\frac{1}{\ell}e^{2\pi i\left(k_2+\frac{r'}{q}\right)\frac{y}{L_2}}
\frac{\hat{a}-\hat{a}^\dagger}{\sqrt{2}}\varphi_{m'}\left(x-\left(k_2+\frac{r'}{q}\right)\frac{2\pi\ell^2}{L_2}\right)\nonumber\\
&&
=
\frac{\hbar}{i}\frac{1}{\sqrt{2}\ell}e^{2\pi i\left(k_2+\frac{r'}{q}\right)\frac{y}{L_2}}
\left(\sqrt{m'}
\varphi_{m'-1}\left(x-\left(k_2+\frac{r'}{q}\right)\frac{2\pi\ell^2}{L_2}\right)
-
\sqrt{m'+1}
\varphi_{m'+1}\left(x-\left(k_2+\frac{r'}{q}\right)\frac{2\pi\ell^2}{L_2}\right)
\right)\nonumber\\
\end{eqnarray}
and
\begin{eqnarray}
&&\left(\frac{\hbar}{i}\frac{\partial}{\partial y}-\frac{eB}{c}x\right)e^{2\pi i\left(k_2+\frac{r'}{q}\right)\frac{y}{L_2}}\varphi_{m'}\left(x-\left(k_2+\frac{r'}{q}\right)\frac{2\pi\ell^2}{L_2}\right)=\nonumber\\
&&\frac{\hbar}{\ell}
\left(\left(k_2+\frac{r'}{q}\right)\frac{2\pi\ell}{L_2}-\frac{x}{\ell}\right)e^{2\pi i\left(k_2+\frac{r'}{q}\right)\frac{y}{L_2}}\varphi_{m'}\left(x-\left(k_2+\frac{r'}{q}\right)\frac{2\pi\ell^2}{L_2}\right)=\\
&&-\frac{\hbar}{\ell}
e^{2\pi i\left(k_2+\frac{r'}{q}\right)\frac{y}{L_2}}\frac{a+a^\dagger}{\sqrt{2}}\varphi_{m'}\left(x-\left(k_2+\frac{r'}{q}\right)\frac{2\pi\ell^2}{L_2}\right)=\\
&&-\frac{\hbar}{\sqrt{2}\ell}
e^{2\pi i\left(k_2+\frac{r'}{q}\right)\frac{y}{L_2}}
\left(\sqrt{m'}\varphi_{m'-1}\left(x-\left(k_2+\frac{r'}{q}\right)\frac{2\pi\ell^2}{L_2}\right)+
\sqrt{m'+1}\varphi_{m'+1}\left(x-\left(k_2+\frac{r'}{q}\right)\frac{2\pi\ell^2}{L_2}\right)\right).
\end{eqnarray}
Substituting back to $\chi$, and using the orthogonality of $\chi$'s and Eq.(\ref{the matrix factors from cc}), we finally have,
\begin{eqnarray}
&&\tilde{h}^{+1}_{[amr],[a'm'\bar{r}]}(k_1,k_2)\approx
\delta_{r\bar{r}}\left(\begin{array}{cccc}
0 & 0 & -i\frac{\sqrt{2} v_{\ast}}{\ell}\sqrt{m'}\delta_{m+1,m'} & 0 \\
0 & 0 & 0 & i\frac{\sqrt{2} v_{\ast}}{\ell}\sqrt{m}\delta_{m,m'+1}\\
i\frac{\sqrt{2} v_{\ast}}{\ell}\sqrt{m}\delta_{m,m'+1} & 0 & 0 & M\delta_{mm'}\\
0 & -i\frac{\sqrt{2} v_{\ast}}{\ell}\sqrt{m'}\delta_{m+1,m'} & M\delta_{mm'} & 0
\end{array}\right)_{aa'}, \\
&&\tilde{h}^{-1}_{[amr],[a'm'\bar{r}]}(k_1,k_2)\approx
\delta_{r\bar{r}}\left(\begin{array}{cccc}
0 & 0 & -i\frac{\sqrt{2} v_{\ast}}{\ell}\sqrt{m}\delta_{m,m'+1} & 0 \\
0 & 0 & 0 & i\frac{\sqrt{2} v_{\ast}}{\ell}\sqrt{m'}\delta_{m+1,m'}\\
i\frac{\sqrt{2} v_{\ast}}{\ell}\sqrt{m'}\delta_{m+1,m'} & 0 & 0 & M\delta_{mm'}\\
0 & -i\frac{\sqrt{2} v_{\ast}}{\ell}\sqrt{m}\delta_{m,m'+1} & M\delta_{mm'} & 0
\end{array}\right)_{aa'}.\nonumber\\
\end{eqnarray}
Note that a straightforward canonical substitution in Eq.(\ref{cc-matrix-zerofield}), $k_x+ik_y\rightarrow \sqrt{2B}\hat{a}$, would yield us the same finite field $c-c$ coupling, where $\hat{a}$ denotes the LL lowering operator for the corresponding $c$ fermion LLs.
\subsection{c-f Coupling}\label{c-f matrix element calculation}\label{apdx:cf matrix elements}
The $c-f$ coupling at finite $\bB$ is given as
\begin{eqnarray}
H^{\tau}_{cf}&=&\sum_{k\in[0,1)\otimes [0,\frac{1}{q})}\sum_{a=1}^4 \sum_{b=1}^2 \sum_{m=0}^{m_{a,\tau}}
 \sum_{r=0}^{p-1}\sum_{r'=0}^{q-1}h^{\tau}_{[amr],[br']}(k)c^\dagger_{a\tau k r m}
f_{b\tau k r'},
\end{eqnarray}
where
\begin{eqnarray}
&&h^{\tau}_{[amr],[br']}(k_1,k_2) =
\int d^2\br\Psi^*_{a\tau}(\br)\chi^*_{krm}(\br) H^{\tau}_{BM}\left(p_x,p_y-\frac{eB}{c}x\right) \eta_{b\tau k r'}(\br) \\
&&=\frac{1}{\sqrt{\mathcal{N}}}
\sum_{s\in \mathbb{Z}} e^{2\pi i s k_1}
\int d^2\br\Psi^*_{a\tau}(\br)\chi^*_{k_1,k_2+\frac{r}{q},m}(\br) H^{\tau}_{BM}\left(p_x,p_y-\frac{eB}{c}x\right)
\hat{t}^s_{\bL_1}w_{b\tau}(\br,\left(k_2+\frac{r'}{q}\right)\bg_2)\\
&&=\frac{1}{\sqrt{\mathcal{N}}}
\sum_{s\in \mathbb{Z}} e^{2\pi i s k_1}
\int d^2\br\Psi^*_{a\tau}(\br)\chi^*_{krm}(\br) \hat{t}^s_{\bL_1} H^{\tau}_{BM}\left(p_x,p_y-\frac{eB}{c}x\right)
w_{b\tau}(\br,\left(k_2+\frac{r'}{q}\right)\bg_2)\\
&&=\frac{s_{tot}}{\sqrt{\mathcal{N}}}
\int d^2\br\Psi^*_{a\tau}(\br)\chi^*_{krm}(\br) H^{\tau}_{BM}\left(p_x,p_y-\frac{eB}{c}x\right)
w_{b\tau}\left(\br,\left(k_2+\frac{r'}{q}\right)\bg_2\right)\\
&&=\frac{s_{tot}}{\sqrt{\mathcal{N}}}\sum_{n\in \mathbb{Z}}e^{2\pi i\left(k_2+\frac{r'}{q}\right) n}
\int d^2\br\Psi^*_{a\tau}(\br)\chi^*_{krm}(\br) H^{\tau}_{BM}\left(p_x,p_y-\frac{eB}{c}x\right)
\hat{t}^{n}_{\bL_2} W_{\mathbf{0},b\tau}(\br)\\
&&=\frac{s_{tot}}{\sqrt{\mathcal{N}}}\sum_{n\in \mathbb{Z}}e^{2\pi i\left(k_2+\frac{r'}{q}\right) n}
\int d^2\br\Psi^*_{a\tau}(\br)\left(\hat{t}^{-n}_{\bL_2}\chi_{krm}(\br)\right)^* H^{\tau}_{BM}\left(p_x,p_y-\frac{eB}{c}x\right)
 W_{\mathbf{0},b\tau}(\br)\\
&&= \frac{s_{tot}}{\sqrt{\mathcal{N}}}\sum_{n\in \mathbb{Z}}e^{2\pi i\left(k_2+\frac{r'}{q}\right) n}
\int d^2\br\Psi^*_{a\tau}(\br)\left(\hat{t}^{-n}_{\bL_2}\chi_{krm}(\br)\right)^* H^{\tau}_{BM}\left(p_x,p_y\right)W_{\mathbf{0},b\tau}(\br) \nonumber \\ 
&& -\frac{s_{tot}}{\sqrt{\mathcal{N}}}\sum_{n\in \mathbb{Z}}e^{2\pi i\left(k_2+\frac{r'}{q}\right) n}
\int d^2\br\Psi^*_{a\tau}(\br)\left(\hat{t}^{-n}_{\bL_2}\chi_{krm}(\br)\right)^* \left(\frac{eBx}{c}\mathcal{M}^{\tau}_y\right)
W_{\mathbf{0},b\tau}(\br)
\end{eqnarray}

Note that the $B$-field term eventually turns out negligible because it acts on a well localized function and so at small $B$ it is exponentially suppressed in the region where the vector potential is appreciable.

Using the fact $\hat{t}^{-1}_{\bL_2}\hat{t}_{\bL_1}=e^{2\pi i\frac{p}{q}}\hat{t}_{\bL_1}\hat{t}^{-1}_{\bL_2}$, we have
\begin{eqnarray}
&&\hat{t}^{-n}_{\bL_2}\chi_{krm}(\br)=
\frac{1}{\sqrt{\ell L_2\mathcal{N}}}
\sum_{s\in \mathbb{Z}} e^{2\pi i s k_1}\hat{t}^{-n}_{\bL_2}\hat{t}^s_{\bL_1}e^{2\pi i\left(k_2+\frac{r}{q}\right)\frac{y}{L_2}}\varphi_m\left(x-\left(k_2+\frac{r}{q}\right)\frac{2\pi\ell^2}{L_2}\right)\\
&&=e^{2\pi i\left(k_2+\frac{r}{q}\right)n}\frac{1}{\sqrt{\ell L_2 \mathcal{N}}}
\sum_{s\in \mathbb{Z}} e^{2\pi i s k_1}e^{2\pi i sn\frac{p}{q}}\hat{t}^s_{\bL_1}
e^{2\pi i\left(k_2+\frac{r}{q}\right)\frac{y}{L_2}}\varphi_m\left(x-\left(k_2+\frac{r}{q}\right)\frac{2\pi\ell^2}{L_2}\right)\\
&&=e^{2\pi i\left(k_2+\frac{r}{q}\right)n}\chi_{\text{mod}\left(k_1+n\frac{p}{q},1\right) k_2 r m}(\br)
\end{eqnarray}
Therefore, 
\begin{eqnarray}
&&h^{\tau}_{[amr],[br']}(k_1,k_2) = \frac{s_{tot}}{\sqrt{\mathcal{N}}}\sum_{n\in \mathbb{Z}}e^{2\pi i n\frac{r'-r}{q}}
\int d^2\br\chi^*_{\text{mod}\left(k_1+n\frac{p}{q},1\right),k_2 r m}(\br) \Psi^*_{a\tau}(\br)H\left(p_x,p_y\right)
W_{\mathbf{0},b\tau}(\br) \\
&&- \left(\frac{eB}{c}\right)\frac{s_{tot}}{\sqrt{\mathcal{N}}}\sum_{n\in \mathbb{Z}}e^{2\pi i n\frac{r'-r}{q}}
\int d^2\br \left(x\chi^*_{\text{mod}\left(k_1+n\frac{p}{q},1\right),k_2 r m}(\br)\right) \Psi^*_{a\tau}(\br)\mathcal{M}^{\tau}_y
W_{\mathbf{0},b\tau}(\br)
\end{eqnarray}
In order to proceed we use the fact that \cite{song2022magic}
\begin{eqnarray}
&&
\int d^2\br e^{-i\bk'\cdot\br}\Psi^*_{a\tau}(\br) H^{\tau}_{BM}\left(p_x,p_y\right)
W_{\mathbf{0}b\tau}(\br)\approx \sqrt{A_{uc}} e^{-\bk'^2\lambda^2/2}H^{cf,\tau}_{ab}(\bk'),
\end{eqnarray}
which is $\sqrt{NA_{uc}}$ bigger than in \cite{song2022magic} because of choice of normalisation of $\chi$ discussed in previous section. Using
\begin{eqnarray}\label{zerocf}
H^{cf,\tau}_{ab}(\bk')=\left(\begin{array}{c}
\gamma\sigma_0+v'_\ast(\eta_{v}k_x\sigma_x+k_y\sigma_y)\\
0_{2\times 2}
\end{array}\right)_{ab},
\end{eqnarray}
we have 
\begin{eqnarray}
\Psi^*_{a\tau}(\br) H^{\tau}_{BM}\left(p_x,p_y\right)
W_{\mathbf{0},b\tau}(\br)&=&\int\frac{d^2\bk'}{(2\pi)^2}e^{i\bk'\cdot\br}
\int d^2\br' e^{-i\bk'\cdot\br'}\Psi^*_{a\tau}(\br') H^{\tau}_{BM}\left(p'_x,p'_y\right)
W_{\mathbf{0},b\tau}(\br')\\
&\approx& \sqrt{A_{uc}}\int\frac{d^2\bk}{(2\pi)^2}e^{i\bk\cdot\br}e^{-\bk^2\lambda^2/2}H^{cf,\tau}_{ab}(\bk)\\
&=&\sqrt{A_{uc}}H^{cf,\tau}_{ab}\left(\frac{1}{i}\frac{\partial}{\partial \br}\right)\int\frac{d^2\bk}{(2\pi)^2}e^{i\bk\cdot\br}e^{-\bk^2\lambda^2/2} \\
&=&\sqrt{A_{uc}}H^{cf,\tau}_{ab}\left(\frac{1}{i}\frac{\partial}{\partial \br}\right)
\frac{1}{2\pi \lambda^2}e^{-\br^2/(2\lambda^2)}.
\end{eqnarray}
Moreover since $H^{\tau}_{BM}(\bp)$ is linear in $\bp$, we can approximate
\begin{equation}
    \int d^2 \br e^{-i\bk'\cdot\br} \Psi^*_{a\tau}(\br)\mathcal{M}^{\tau}_y W_{\mathbf{0},b\tau}(\br) \approx \sqrt{A_{uc}} e^{-\bk'^2\lambda^2/2} M^{\tau}_{ab},
\end{equation}
where $M^{\tau}_{ab}$ can be read of from Eq.(\ref{zerocf}) to be
\begin{equation}
M^{\tau}_{ab} = -\frac{iv_*'}{\hbar}
\begin{pmatrix}
0 & 1\\
-1 & 0\\
0 & 0\\
0 & 0  
\end{pmatrix}_{(ab)} = M_{ab}.
\end{equation}
Thus we have
\begin{eqnarray}
\Psi^*_{a\tau}(\br)\mathcal{M}^{\tau}_y W_{\mathbf{0},b\tau}(\br) &=& \int \frac{d^2\bk'}{(2\pi)^2} e^{-i\bk' \cdot \br} \int d^2\br' e^{-i\bk'\cdot\br'} \Psi^*_{a\tau}(\br')\mathcal{M}^{\tau}_y W_{\mathbf{0},b\tau}(\br') \\
&\approx& \sqrt{A_{uc}}\int \frac{d^2\bk'}{(2\pi)^2}e^{-i\bk' \cdot \br} e^{-\frac{{\bk'^2}\lambda^2}{2}} M_{ab} \\
&=& \sqrt{A_{uc}}e^{-\br^2/(2\lambda^2)} \frac{1}{2\pi \lambda^2}M_{ab}.
\end{eqnarray}
Therefore,
\begin{eqnarray}
&&h^{\tau}_{[amr],[br']}(k_1,k_2)\approx \frac{s_{tot}\sqrt{A_{uc}}}{\sqrt{\mathcal{N}}}\sum_{n\in \mathbb{Z}}e^{2\pi i n\frac{r'-r}{q}}
\int d^2\br\chi^*_{\text{mod}\left(k_1+n\frac{p}{q},1\right)k_2rm}(\br)
H^{cf,\tau}_{ab}\left(\frac{1}{i}\frac{\partial}{\partial \br}\right)
\frac{1}{2\pi \lambda^2}e^{-\br^2/(2\lambda^2)}\nonumber \\
&&-\frac{eB}{c}\frac{s_{tot}\sqrt{A_{uc}}}{\sqrt{\mathcal{N}}}\sum_{n\in \mathbb{Z}}e^{2\pi i n\frac{r'-r}{q}}
\int d^2\br x \chi^*_{\text{mod}\left(k_1+n\frac{p}{q},1\right)k_2rm}(\br)
M_{ab} \frac{1}{2\pi \lambda^2}e^{-\br^2/(2\lambda^2)}\\
&&=\frac{s_{tot}\sqrt{A_{uc}}}{\sqrt{\mathcal{N}}}\sum_{n\in \mathbb{Z}}e^{2\pi i n\frac{r'-r}{q}}
\int d^2\br\left(H^{cf,\tau}_{ab}\left(\frac{1}{i}\frac{\partial}{\partial \br}\right)\chi_{\text{mod}\left(k_1+n\frac{p}{q},1\right)k_2rm}(\br)\right)^*
\frac{1}{2\pi \lambda^2}e^{-\br^2/(2\lambda^2)} \nonumber \\
&&-\frac{eB}{c}\frac{s_{tot}\sqrt{A_{uc}}}{\sqrt{\mathcal{N}}}\sum_{n\in \mathbb{Z}}e^{2\pi i n\frac{r'-r}{q}}
\int d^2\br\left(x\chi_{\text{mod}\left(k_1+n\frac{p}{q},1\right)k_2rm}(\br)\right)^*
\frac{1}{2\pi \lambda^2}e^{-\br^2/(2\lambda^2)}M_{ab}.\label{substitute into 1}
\end{eqnarray}
Using $\hat{t}^s_{\bL_1} =
e^{-\pi i s(s-1) \frac{p}{q}\frac{L_{1y}}{L_2}}
e^{2\pi i s\frac{p}{q}\frac{y}{L_2}}
\hat{T}^s_{\bL_1}$, derived in the \ref{apdx:mtg identities}, we have
\begin{eqnarray}
&&\chi_{\text{mod}\left(k_1+n\frac{p}{q},1\right),k_2rm}(\br)=\frac{1}{\sqrt{\ell L_2\mathcal{N}}}
\sum_{s\in \mathbb{Z}} e^{2\pi i s n\frac{p}{q}}e^{2\pi i s k_1}\hat{t}^s_{\bL_1}e^{2\pi i(k_2+\frac{r}{q})\frac{y}{L_2}}\varphi_m\left(x-(k_2+\frac{r}{q})\frac{2\pi\ell^2}{L_2}\right)\\
&=&\frac{1}{\sqrt{\ell L_2\mathcal{N}}}
\sum_{s\in \mathbb{Z}} e^{2\pi i s n\frac{p}{q}}e^{2\pi i s k_1}e^{-\pi i s(s-1) \frac{p}{q}\frac{L_{1y}}{L_2}}
e^{2\pi i s\frac{p}{q}\frac{y}{L_2}}
\hat{T}^s_{\bL_1}
e^{2\pi i(k_2+\frac{r}{q})\frac{y}{L_2}}\varphi_m\left(x-(k_2+\frac{r}{q})\frac{2\pi\ell^2}{L_2}\right)\\
&=&\frac{1}{\sqrt{\ell L_2\mathcal{N}}}
\sum_{s\in \mathbb{Z}} e^{2\pi i s n\frac{p}{q}}e^{2\pi i s k_1}e^{-\pi i s(s-1) \frac{p}{q}\frac{L_{1y}}{L_2}}
e^{2\pi i s\frac{p}{q}\frac{y}{L_2}}
e^{2\pi i(k_2+\frac{r}{q})\frac{y-sL_{1y}}{L_2}}\varphi_m\left(x-sL_{1x}-(k_2+\frac{r}{q})\frac{2\pi\ell^2}{L_2}\right)\\
&=&\frac{1}{\sqrt{\ell L_2\mathcal{N}}}
\sum_{s\in \mathbb{Z}} e^{2\pi i s n\frac{p}{q}}e^{2\pi i s k_1}e^{-2\pi i s(k_2+\frac{r}{q})\frac{L_{1y}}{L_2}}e^{-\pi i s(s-1) \frac{p}{q}\frac{L_{1y}}{L_2}}
e^{2\pi i\left(k_2+\frac{r}{q}+ s\frac{p}{q}\right)\frac{y}{L_2}}
\varphi_m\left(x-sL_{1x}-(k_2+\frac{r}{q})\frac{2\pi\ell^2}{L_2}\right)\label{substitution1}
\end{eqnarray} 
Substituting Eq.({\ref{substitution1}}) back in Eq.(\ref{substitute into 1}), we have
\begin{eqnarray}
&&h^{\tau}_{[amr],[br']}(k_1,k_2)\approx
\frac{s_{tot}\sqrt{A_{uc}}}{\mathcal{N}\sqrt{\ell L_2}}\sum_{n\in \mathbb{Z}}\sum_{s\in \mathbb{Z}}e^{2\pi i n\frac{r'-r}{q}}
 e^{-2\pi i s n\frac{p}{q}}e^{-2\pi i s k_1}e^{2\pi i s\left(k_2+\frac{r}{q}\right)\frac{L_{1y}}{L_2}}e^{\pi i s(s-1) \frac{p}{q}\frac{L_{1y}}{L_2}}
\nonumber\\
\bigg[ &&\int d^2\br e^{-2\pi i\left(k_2+\frac{r}{q}+ s\frac{p}{q}\right)\frac{y}{L_2}}
\varphi_m\left(x-sL_{1x}-\left(k_2+\frac{r}{q}\right)\frac{2\pi\ell^2}{L_2}\right)
H^{cf,\tau}_{ab}\left(\frac{1}{i}\frac{\partial}{\partial \br}\right)\frac{1}{2\pi \lambda^2}e^{-\br^2/(2\lambda^2)}\nonumber \\
&& - \frac{eB}{c}\int d^2\br e^{-2\pi i\left(k_2+\frac{r}{q}+ s\frac{p}{q}\right)\frac{y}{L_2}} x
\varphi_m\left(x-sL_{1x}-\left(k_2+\frac{r}{q}\right)\frac{2\pi\ell^2}{L_2}\right) M_{ab} \frac{1}{2\pi \lambda^2}e^{-\br^2/(2\lambda^2)} \bigg]\\
&&\approx
\frac{s_{tot}n_{tot}\sqrt{A_{uc}}}{\mathcal{N}\sqrt{\ell L_2}}\sum_{j\in \mathbb{Z}}\sum_{s\in \mathbb{Z}}\delta_{r'-r,jq+sp}
e^{-2\pi i s k_1}e^{2\pi i s\left(k_2+\frac{r}{q}\right)\frac{L_{1y}}{L_2}}e^{\pi i s(s-1) \frac{p}{q}\frac{L_{1y}}{L_2}}
\nonumber\\
&&\bigg[\int d^2\br e^{-2\pi i\left(k_2+\frac{r}{q}+ s\frac{p}{q}\right)\frac{y}{L_2}}
\varphi_m\left(x-sL_{1x}-\left(k_2+\frac{r}{q}\right)\frac{2\pi\ell^2}{L_2}\right)
H^{cf,\tau}_{ab}\left(\frac{1}{i}\frac{\partial}{\partial \br}\right)\frac{1}{2\pi \lambda^2}e^{-\br^2/(2\lambda^2)}\nonumber \\
&& - \frac{eB}{c}\int d^2\br e^{-2\pi i\left(k_2+\frac{r}{q}+ s\frac{p}{q}\right)\frac{y}{L_2}} x
\varphi_m\left(x-sL_{1x}-\left(k_2+\frac{r}{q}\right)\frac{2\pi\ell^2}{L_2}\right) M_{ab} \frac{1}{2\pi \lambda^2}e^{-\br^2/(2\lambda^2)} \bigg]
\end{eqnarray}
where the sum over $n$ above led to the Diophantine equation
\begin{eqnarray}
\frac{r'-r-sp}{q}=\text{integer} \equiv j.
\end{eqnarray} 

Plugging in the normalization factors results in
\begin{eqnarray}\label{cfinB}
&&h^{\tau}_{[amr],[br']}(k_1,k_2)
\approx
\frac{\sqrt{L_{1x}}}{\sqrt{\ell}}\sum_{j\in \mathbb{Z}}\sum_{s\in \mathbb{Z}}\delta_{r'-r,jq+sp}
e^{-2\pi i s k_1}e^{2\pi i s\left(k_2+\frac{r}{q}\right)\frac{L_{1y}}{L_2}}e^{\pi i s(s-1) \frac{p}{q}\frac{L_{1y}}{L_2}}
\nonumber\\
&&\frac{1}{2\pi \lambda^2}\bigg[ \int d^2\br e^{-2\pi i\left(k_2+\frac{r}{q}+ s\frac{p}{q}\right)\frac{y}{L_2}}
\varphi_m\left(x-sL_{1x}-\left(k_2+\frac{r}{q}\right)\frac{2\pi\ell^2}{L_2}\right)
H^{cf,\tau}_{ab}\left(\frac{1}{i}\frac{\partial}{\partial \br}\right)e^{-\br^2/(2\lambda^2)} \nonumber \\
&& - \frac{\hbar}{\ell^2}\int d^2\br e^{-2\pi i\left(k_2+\frac{r}{q}+ s\frac{p}{q}\right)\frac{y}{L_2}}
x\varphi_m\left(x-sL_{1x}-\left(k_2+\frac{r}{q}\right)\frac{2\pi\ell^2}{L_2}\right)
M_{ab}e^{-\br^2/(2\lambda^2)} \bigg].
\end{eqnarray}
\begin{figure}
    \centering
\includegraphics[width=12cm]{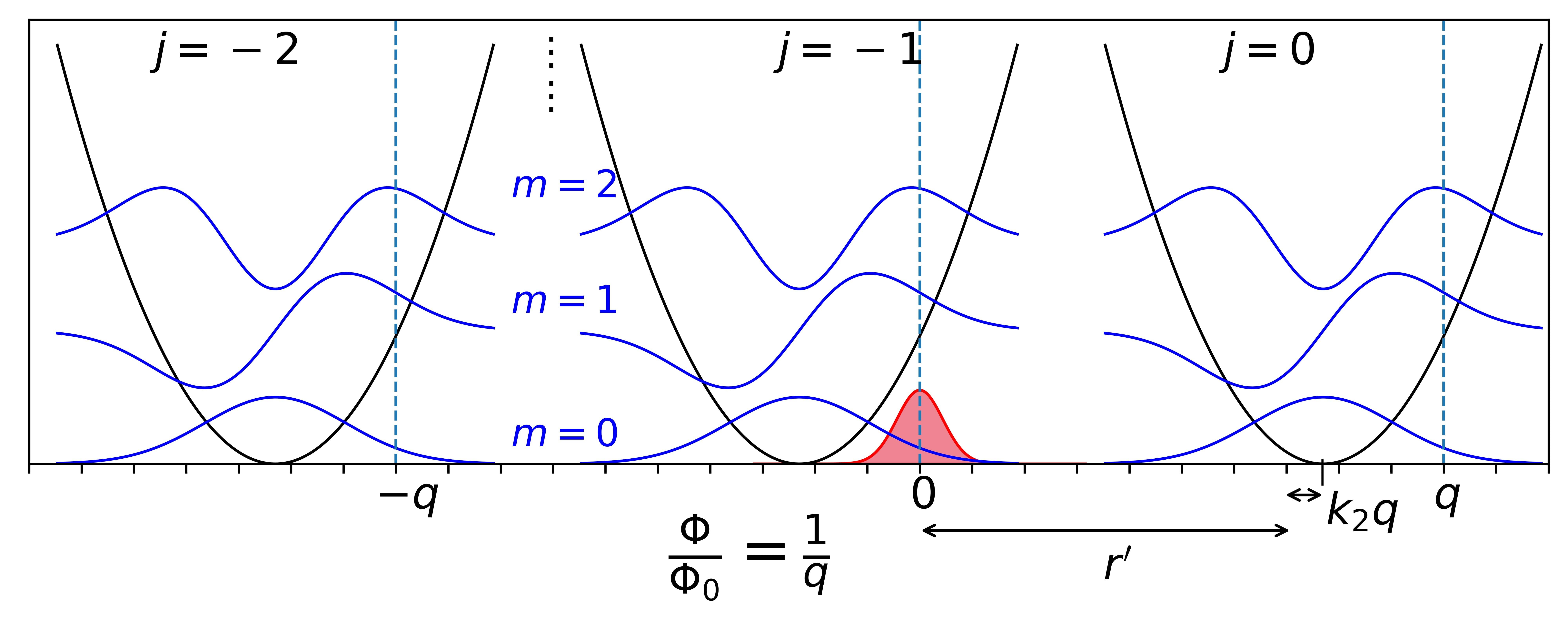}
    \caption{Schematic representation of the $cf$ coupling for $\phi/\phi_0=1/q$ discussed in the methods section A of main text. Each tick represents (the 1D projection of) a moire unit cell, illustrating
 the overlap between a 2D localized heavy state  with size $\lambda$ siting at the origin (red) and a Landau level (LL) i.e. a 1D harmonic oscillator (h.o.) shifted in the $x$-direction, its wavefunctions sketched by blue, with a plane wave phase variation in the $y$-direction (not shown) that depends on the shift. The $r'$ determines the momentum absorbed by the LL as well as the unit cell to which the h.o. is shifted and $k_2q$ fine tunes the shift within the unit cell. The index $j$ then determines $q$-unit-cell periodic revival of the h.o. The black parabolas mimic a quadratic potential to accompany the h.o. wavefunctions.}
    \label{fig:cf-schematic}
\end{figure}
Note that the $y$ integrals are elementary. At small $\bB$ and not too large $m$, the $x$ integrals we be performed by taking the advantage of the mismatch in length scales i.e. $\ell\gg \lambda$ and Taylor expand the harmonic oscillator functions about the origin. However at larger $\bB$ we may need to obtain the integrals with higher accuracy.
To this end we will find it useful to employ two-variable Hermite polynomial $\mathcal{H}_n(x,y)$ result from \cite{babusci2012integrals}. It will be useful to first discuss the integrals required for casting Eq.(\ref{cfinB}) into a closed form. 

We start by defining the two-variable Hermite polynomial as
\begin{eqnarray}\label{2varHermite}
\mathcal{H}_n(x,y)=n!\sum_{k=0}^{\left[\frac{n}{2}\right]}\frac{x^{n-2k}y^k}{(n-2k)!k!}, 
\end{eqnarray}
where $[n]$ denotes the floor function at $n$. The two-variable Hermite polynomials are related to the Hermite polynomial in harmonic oscillator wavefunction as $H_{m}(x) = \mathcal{H}_m(2x,-1)$. The required $x$ integral in Eq.(\ref{cfinB}) then reads
\begin{eqnarray}
&&A_m\left(\lambda,x_0\right)=\int_{-\infty}^{\infty} dx e^{-x^2/(2\lambda^2)} \varphi_m\left(x-x_0\right)=
\frac{1}{\pi^{\frac{1}{4}}}\frac{1}{\sqrt{2^mm!}}\int_{-\infty}^{\infty} dx e^{-x^2/(2\lambda^2)}
e^{-(x-x_0)^2/2\ell^2}H_m((x-x_0)/\ell)
\end{eqnarray}
Let $y=x/\ell$ and $y_0=x_0/\ell$. Then,
\begin{eqnarray}
A_m\left(\lambda,\ell y_0\right)&&=
\frac{\ell}{\pi^{\frac{1}{4}}}\frac{1}{\sqrt{2^mm!}}\int_{-\infty}^{\infty} dy e^{-\frac{\ell^2}{2\lambda^2}y^2}
e^{-\frac{1}{2}(y-y_0)^2}H_m(y-y_0)\\
&&=
\frac{\ell}{\pi^{\frac{1}{4}}}\frac{e^{-\frac{1}{2}y_0^2}}{\sqrt{2^mm!}}\int_{-\infty}^{\infty} dy e^{-\frac{1}{2}\left(\frac{\ell^2}{\lambda^2}+1\right)y^2}
e^{y_0y}H_m(y-y_0)\\
&&=
\frac{\ell}{\pi^{\frac{1}{4}}}\frac{e^{-\frac{1}{2}y_0^2}}{\sqrt{2^mm!}}\int_{-\infty}^{\infty} dy e^{-\frac{1}{2}\left(\frac{\ell^2}{\lambda^2}+1\right)y^2}
e^{y_0y}\mathcal{H}_m(2y-2y_0,-1).
\end{eqnarray}
Now, using the formula in \cite{babusci2012integrals}
\begin{eqnarray}
&&\int_{-\infty}^{\infty} dx \mathcal{H}_n(ax+b,y)e^{-cx^2+\alpha x}=\sqrt{\frac{\pi}{c}}e^{\frac{\alpha^2}{4c}}
\mathcal{H}_n\left(b+\frac{\alpha}{2}\frac{a}{c},y+\frac{a^2}{4c}\right)
\end{eqnarray}
we finally have
\begin{eqnarray}
A_m\left(\lambda,x_0\right)&&=
\frac{\pi^{\frac{1}{4}}}{\sqrt{2^{m-1}m!}}\sqrt{\frac{\lambda^2\ell^2}{\ell^2+\lambda^2}}
e^{-\frac{1}{2}x_0^2/(\ell^2+\lambda^2)}
\mathcal{H}_m\left(-2x_0\frac{\ell}{\ell^2+\lambda^2},\frac{2\lambda^2}{\ell^2+\lambda^2}-1\right).
\end{eqnarray}

Note that
\begin{eqnarray}
A_m\left(\lambda\rightarrow 0,x_0\right)&&=\sqrt{2\pi}\lambda\frac{\pi^{-\frac{1}{4}}}{\sqrt{2^m m!}}
e^{-\frac{1}{2}x_0^2/\ell^2}
\mathcal{H}_m\left(-2\frac{x_0}{\ell},-1\right)
=
\sqrt{2\pi}\lambda\varphi_m(-x_0)
\end{eqnarray}
as expected because in the limit $\lambda\rightarrow 0$, the harmonic oscillator wavefunction does not vary significantly within the spatial extent of the heavy fermion gaussian, which is set by $\lambda$.

Now, let
\begin{eqnarray}\label{apdx:F}
\mathcal{F}_m(\lambda,x_0)&=&\frac{1}{\sqrt{2\pi}\lambda}A_m(\lambda,x_0)\\
&=&\frac{1}{\pi^{\frac{1}{4}}\sqrt{2^{m}m!}}\sqrt{\frac{\ell^2}{\ell^2+\lambda^2}}
e^{-\frac{1}{2}x_0^2/(\ell^2+\lambda^2)}
\mathcal{H}_m\left(-2x_0\frac{\ell}{\ell^2+\lambda^2},\frac{2\lambda^2}{\ell^2+\lambda^2}-1\right) \label{xintegral}\\
\mathcal{F}_m(\lambda\rightarrow 0,x_0)&=&\phi_m\left(-x_0\right).
\end{eqnarray}
The y integrals in Eq.(\ref{cfinB}) can be performed using
\begin{eqnarray}
&&\int_{-\infty}^\infty dy e^{-2\pi i k\frac{y}{L_2}}e^{-y^2/2\lambda^2}=\sqrt{2\pi} \lambda e^{-2\pi^2k^2 \lambda^2/L_2^2}.\label{yintegral}
\end{eqnarray}

Using the integrals provided in Eq.(\ref{xintegral}) and Eq.(\ref{yintegral}), we now proceed to evaluate Eq.(\ref{cfinB}).

\subsubsection{Non-derivative or the zeroth-order coupling \texorpdfstring{$\gamma$}{gamma}} \label{apdx:Non-derivative coupling}

\begin{eqnarray}
&&I^0_{[mr],[r']}(k_1,k_2)= \frac{\sqrt{L_{1x}}}{\sqrt{\ell}}\sum_{j\in \mathbb{Z}}\sum_{s\in \mathbb{Z}}\delta_{r'-r,jq+sp}
e^{-2\pi i s k_1}e^{2\pi i s\left(k_2+\frac{r}{q}\right)\frac{L_{1y}}{L_2}}e^{\pi i s(s-1) \frac{p}{q}\frac{L_{1y}}{L_2}}
\nonumber\\
&&\frac{1}{2\pi \lambda^2}\int d^2\br e^{-2\pi i\left(k_2+\frac{r}{q}+ s\frac{p}{q}\right)\frac{y}{L_2}}
\varphi_m\left(x-sL_{1x}-\left(k_2+\frac{r}{q}\right)\frac{2\pi\ell^2}{L_2}\right)
e^{-\br^2/(2\lambda^2)}=\\
&&
 \frac{\sqrt{L_{1x}}}{\sqrt{\ell}}\sum_{j\in \mathbb{Z}}\sum_{s\in \mathbb{Z}}\delta_{r'-r,jq+sp}
e^{-2\pi i s k_1}e^{2\pi i s\left(k_2+\frac{r}{q}\right)\frac{L_{1y}}{L_2}}e^{\pi i s(s-1) \frac{p}{q}\frac{L_{1y}}{L_2}}
e^{-2\pi^2 \left(k_2+\frac{r}{q}+ s\frac{p}{q}\right)^2\frac{\lambda^2}{L^2_2}}
\mathcal{F}_m\left(\lambda,\left(s+\frac{r}{p}+k_2\frac{q}{p}\right)L_{1x}\right)\nonumber\\ \label{I0}
\end{eqnarray}
\subsubsection{\texorpdfstring{$k_x$}{kx} coupling}\label{apdx:kx coupling}
\begin{eqnarray}
&&I^x_{[mr],[r']}(k_1,k_2)=
\frac{\sqrt{L_{1x}}}{\sqrt{\ell}}\sum_{j\in \mathbb{Z}}\sum_{s\in \mathbb{Z}}\delta_{r'-r,jq+sp}
e^{-2\pi i s k_1}e^{2\pi i s\left(k_2+\frac{r}{q}\right)\frac{L_{1y}}{L_2}}e^{\pi i s(s-1) \frac{p}{q}\frac{L_{1y}}{L_2}}
\nonumber\\
&&\frac{1}{2\pi \lambda^2}\int d^2\br e^{-2\pi i\left(k_2+\frac{r}{q}+ s\frac{p}{q}\right)\frac{y}{L_2}}
\varphi_m\left(x-sL_{1x}-\left(k_2+\frac{r}{q}\right)\frac{2\pi\ell^2}{L_2}\right)
\left(\frac{1}{i}\frac{\partial}{\partial x}\right)e^{-\br^2/(2\lambda^2)}\\
&&=\frac{\sqrt{L_{1x}}}{\sqrt{\ell}}\sum_{j\in \mathbb{Z}}\sum_{s\in \mathbb{Z}}\delta_{r'-r,jq+sp}
e^{-2\pi i s k_1}e^{2\pi i s\left(k_2+\frac{r}{q}\right)\frac{L_{1y}}{L_2}}e^{\pi i s(s-1) \frac{p}{q}\frac{L_{1y}}{L_2}} \nonumber\\
&&\frac{1}{2\pi \lambda^2}\frac{i}{\ell}\int d^2\br e^{-2\pi i\left(k_2+\frac{r}{q}+ s\frac{p}{q}\right)\frac{y}{L_2}}
\left(\ell\frac{\partial}{\partial x}\varphi_m\left(x-sL_{1x}-\left(k_2+\frac{r}{q}\right)\frac{2\pi\ell^2}{L_2}\right)\right)
e^{-\br^2/(2\lambda^2)}
\end{eqnarray}
\begin{eqnarray}
&&=\frac{\sqrt{L_{1x}}}{\sqrt{\ell}}\sum_{j\in \mathbb{Z}}\sum_{s\in \mathbb{Z}}\delta_{r'-r,jq+sp}
e^{-2\pi i s k_1}e^{2\pi i s\left(k_2+\frac{r}{q}\right)\frac{L_{1y}}{L_2}}e^{\pi i s(s-1) \frac{p}{q}\frac{L_{1y}}{L_2}}
\nonumber\\
&&\frac{1}{2\pi \lambda^2}\frac{i\sqrt{m}}{\sqrt{2}\ell}\int d^2\br e^{-2\pi i\left(k_2+\frac{r}{q}+ s\frac{p}{q}\right)\frac{y}{L_2}}
\varphi_{m-1}\left(x-sL_{1x}-\left(k_2+\frac{r}{q}\right)\frac{2\pi\ell^2}{L_2}\right)
e^{-\br^2/(2\lambda^2)}\nonumber \\
&&-\frac{\sqrt{L_{1x}}}{\sqrt{\ell}}\sum_{j\in \mathbb{Z}}\sum_{s\in \mathbb{Z}}\delta_{r'-r,jq+sp}
e^{-2\pi i s k_1}e^{2\pi i s\left(k_2+\frac{r}{q}\right)\frac{L_{1y}}{L_2}}e^{\pi i s(s-1) \frac{p}{q}\frac{L_{1y}}{L_2}}
\nonumber\\
&&\frac{1}{2\pi \lambda^2}\frac{i\sqrt{m+1}}{\sqrt{2}\ell}\int d^2\br e^{-2\pi i\left(k_2+\frac{r}{q}+ s\frac{p}{q}\right)\frac{y}{L_2}}
\varphi_{m+1}\left(x-sL_{1x}-\left(k_2+\frac{r}{q}\right)\frac{2\pi\ell^2}{L_2}\right)
e^{-\br^2/(2\lambda^2)}\\
&&=\frac{\sqrt{L_{1x}}}{\sqrt{\ell}}\sum_{j\in \mathbb{Z}}\sum_{s\in \mathbb{Z}}\delta_{r'-r,jq+sp}
e^{-2\pi i s k_1}e^{2\pi i s\left(k_2+\frac{r}{q}\right)\frac{L_{1y}}{L_2}}e^{\pi i s(s-1) \frac{p}{q}\frac{L_{1y}}{L_2}}
\nonumber\\
&&\frac{i\sqrt{m}}{\sqrt{2}\ell}e^{-2\pi^2 \left(k_2+\frac{r}{q}+ s\frac{p}{q}\right)^2\frac{\lambda^2}{L^2_2}}
\mathcal{F}_{m-1}\left(\lambda,\left(s+\frac{r}{p}+k_2\frac{q}{p}\right)L_{1x}\right)\nonumber \\
&&-\frac{\sqrt{L_{1x}}}{\sqrt{\ell}}\sum_{j\in \mathbb{Z}}\sum_{s\in \mathbb{Z}}\delta_{r'-r,jq+sp}
e^{-2\pi i s k_1}e^{2\pi i s\left(k_2+\frac{r}{q}\right)\frac{L_{1y}}{L_2}}e^{\pi i s(s-1) \frac{p}{q}\frac{L_{1y}}{L_2}}
\nonumber\\
&&\frac{i\sqrt{m+1}}{\sqrt{2}\ell}e^{-2\pi^2 \left(k_2+\frac{r}{q}+ s\frac{p}{q}\right)^2\frac{\lambda^2}{L^2_2}}
\mathcal{F}_{m+1}\left(\lambda,\left(s+\frac{r}{p}+k_2\frac{q}{p}\right)L_{1x}\right)
\end{eqnarray}
So, finally
\begin{eqnarray}
&&I^x_{[mr],[r']}(k_1,k_2)=\frac{i}{\sqrt{2}\ell}\frac{\sqrt{L_{1x}}}{\sqrt{\ell}}\sum_{j\in \mathbb{Z}}\sum_{s\in \mathbb{Z}}\delta_{r'-r,jq+sp}
e^{-2\pi i s k_1}e^{2\pi i s\left(k_2+\frac{r}{q}\right)\frac{L_{1y}}{L_2}}e^{\pi i s(s-1) \frac{p}{q}\frac{L_{1y}}{L_2}}
e^{-2\pi^2 \left(k_2+\frac{r}{q}+ s\frac{p}{q}\right)^2\frac{\lambda^2}{L^2_2}}\nonumber\\
&&\times\left(
\sqrt{m}\mathcal{F}_{m-1}\left(\lambda,\left(s+\frac{r}{p}+k_2\frac{q}{p}\right)L_{1x}\right)
-\sqrt{m+1}\mathcal{F}_{m+1}\left(\lambda,\left(s+\frac{r}{p}+k_2\frac{q}{p}\right)L_{1x}\right)
\right)\label{Ix}
\end{eqnarray}

\subsubsection{\texorpdfstring{$k_y$}{ky} coupling}\label{apdx:ky-coupling}
\begin{eqnarray}
&&I^y_{[mr],[r']}(k_1,k_2)= \frac{\sqrt{L_{1x}}}{\sqrt{\ell}}\sum_{j\in \mathbb{Z}}\sum_{s\in \mathbb{Z}}\delta_{r'-r,jq+sp}
e^{-2\pi i s k_1}e^{2\pi i s\left(k_2+\frac{r}{q}\right)\frac{L_{1y}}{L_2}}e^{\pi i s(s-1) \frac{p}{q}\frac{L_{1y}}{L_2}}
\nonumber\\
&&\frac{1}{2\pi \lambda^2}\int d^2\br e^{-2\pi i\left(k_2+\frac{r}{q}+ s\frac{p}{q}\right)\frac{y}{L_2}}
\varphi_m\left(x-sL_{1x}-\left(k_2+\frac{r}{q}\right)\frac{2\pi\ell^2}{L_2}\right)
\left(\frac{1}{i}\frac{\partial}{\partial y}\right)e^{-\br^2/(2\lambda^2)}\\
&&= \frac{\sqrt{L_{1x}}}{\sqrt{\ell}}\sum_{j\in \mathbb{Z}}\sum_{s\in \mathbb{Z}}\delta_{r'-r,jq+sp}
e^{-2\pi i s k_1}e^{2\pi i s\left(k_2+\frac{r}{q}\right)\frac{L_{1y}}{L_2}}e^{\pi i s(s-1) \frac{p}{q}\frac{L_{1y}}{L_2}}
\left(k_2+\frac{r}{q}+ s\frac{p}{q}\right)\frac{2\pi}{L_2}
\nonumber\\
&&\times e^{-2\pi^2 \left(k_2+\frac{r}{q}+ s\frac{p}{q}\right)^2\frac{\lambda^2}{L^2_2}}
\mathcal{F}_m\left(\lambda,\left(s+\frac{r}{p}+k_2\frac{q}{p}\right)L_{1x}\right) \label{Iy}
\end{eqnarray}

\subsubsection{Minimal coupling}\label{apdx:Minimal Coupling}
The minimal coupling term, $e\mathcal{M}_{u}A_\mu/c$, can be expressed in terms of the non-derivative coupling $I^0_{[mr],[r']}(k_1,k_2)$ and $k_y$ coupling $I^y_{[mr],[r']}(k_1,k_2)$ as shown in this section. Let us call this term $I_A$, given as
\begin{eqnarray}
&&I_A = -\frac{\sqrt{L_{1x}}}{\sqrt{\ell}}\sum_{j\in \mathbb{Z}}\sum_{s\in \mathbb{Z}}\delta_{r'-r,jq+sp}
e^{-2\pi i s k_1}e^{2\pi i s\left(k_2+\frac{r}{q}\right)\frac{L_{1y}}{L_2}}e^{\pi i s(s-1) \frac{p}{q}\frac{L_{1y}}{L_2}}
\nonumber\\
&&\times \frac{1}{2\pi \lambda^2} \frac{\hbar}{\ell^2}\int d^2\br e^{-2\pi i\left(k_2+\frac{r}{q}+ s\frac{p}{q}\right)\frac{y}{L_2}}
x\varphi_m\left(x-sL_{1x}-\left(k_2+\frac{r}{q}\right)\frac{2\pi\ell^2}{L_2}\right)
e^{-\br^2/(2\lambda^2)} \label{Substitute here Harmonic}
\end{eqnarray}
 We can re-express the factor $x\varphi(x-x_0)$ as
\begin{eqnarray}
 x\varphi_m(x-x_0) &&=
\frac{1}{\pi^{\frac{1}{4}}}\frac{1}{\sqrt{2^mm!}}
e^{-(x-x_0)^2/2\ell^2}xH_m((x-x_0)/\ell) \nonumber\\
&& = \ell\frac{1}{\pi^{\frac{1}{4}}}\frac{1}{\sqrt{2^mm!}}
e^{-(x-x_0)^2/2\ell^2}(\frac{x-x_0}{\ell})H_m((x-x_0)/\ell) + x_0\phi_m(x-x_0).
\end{eqnarray}
Moreover, using recursion relation for Hermite polynomials 
\begin{eqnarray}
    xH_m(x)= \frac{1}{2}H_{m+1}(x) + mH_{m-1}(x)
\end{eqnarray}
we have
\begin{eqnarray}\label{Substitute this harmonic osc}
x\varphi_m(x-x_0) &&= \ell\frac{1}{\pi^{\frac{1}{4}}}\frac{1}{\sqrt{2^mm!}}
e^{-(x-x_0)^2/2\ell^2} \left(\frac{1}{2}H_{m+1}((x-x_0)/\ell) + mH_{m-1}((x-x_0)/\ell)\right)+ x_0\varphi_m(x-x_0) \nonumber \\
&& = \ell\sqrt{\frac{m+1}{2}} \varphi_{m+1}(x-x_0) + \ell \sqrt{\frac{m}{2}}\phi_{m-1}(x-x_0)  + x_0\varphi_{m}(x-x_0)
\end{eqnarray}

Substituting Eq.(\ref{Substitute this harmonic osc}) into Eq.(\ref{Substitute here Harmonic}), we have:
\begin{eqnarray}
&&I_A = -\frac{\sqrt{L_{1x}}}{\sqrt{\ell}}\sum_{j\in \mathbb{Z}}\sum_{s\in \mathbb{Z}}\delta_{r'-r,jq+sp}
e^{-2\pi i s k_1}e^{2\pi i s\left(k_2+\frac{r}{q}\right)\frac{L_{1y}}{L_2}}e^{\pi i s(s-1) \frac{p}{q}\frac{L_{1y}}{L_2}}
\nonumber\\
&&\times \frac{1}{2\pi \lambda^2}\frac{\hbar}{\ell}\int d^2\br e^{-2\pi i\left(k_2+\frac{r}{q}+ s\frac{p}{q}\right)\frac{y}{L_2}}\bigg[\sqrt{\frac{m+1}{2}}\varphi_{m+1}(x-sL_{1x}-\left(k_2+\frac{r}{q}\right)\frac{2\pi\ell^2}{L_2}) \nonumber \\
&&+\sqrt{\frac{m}{2}}\varphi_{m-1}(x-sL_{1x}-\left(k_2+\frac{r}{q}\right)\frac{2\pi\ell^2}{L_2}) \nonumber  \\
&&+\frac{1}{\ell}(sL_{1x}+\left(k_2+\frac{r}{q}\right)\frac{2\pi\ell^2}{L_2})\varphi_m(x-sL_{1x}-\left(k_2+\frac{r}{q}\right)\frac{2\pi\ell^2}{L_2})\bigg]
e^{-\br^2/(2\lambda^2)}.
\end{eqnarray}
Using the fact $\ell^2 = \frac{qL_{1x}L_2}{2\pi p}$, note that
\begin{equation}
    \frac{1}{\ell^2}(sL_{1x}+\left(k_2+\frac{r}{q}\right)\frac{2\pi\ell^2}{L_2}) = \frac{2\pi}{L_2}(k_2 + \frac{r}{p} + s\frac{p}{q}).
\end{equation}
Thus we have
\begin{eqnarray}
&&I_A = -\hbar\bigg[ \frac{1}{\sqrt{2}\ell}\bigg( \sqrt{m}I^0_{[m-1r],[r']}(k_1,k_2)  + \sqrt{m+1}I^0_{[m+1r],[r']}(k_1,k_2)\bigg) +I^y_{[mr],[r']}(k_1,k_2)\bigg]\label{minimal coupling term}
\end{eqnarray}

\subsubsection{Closed form expression for \texorpdfstring{$c$}{c}-\texorpdfstring{$f$}{f} coupling at finite field}
The closed form expression for the $c$-$f$ coupling at finite $\bB$ then reads
\begin{eqnarray}
H^{cf,\tau}&\approx&\sum_{k\in[0,1)\otimes [0,1/q)]}
\sum_{a=1}^4 \sum_{b=1}^2 \sum_{m=0}^{m_{a,\tau}}
 \sum_{r=0}^{p-1}\sum_{r'=0}^{q-1}\bigg[
\left(\begin{array}{c}
\gamma I^0_{[mr],[r']}(k) \sigma_0+v'_\ast(\tau I^x_{[mr],[r']}(k)\sigma_x+I^y_{[mr],[r']}(k)\sigma_y)\\
0_{2\times 2}
\end{array}\right)_{ab} + \nonumber\\
\end{eqnarray}
\begin{eqnarray}
\left(\begin{array}{c}   
-v_*'\frac{1}{\sqrt{2}\ell}\left( \sqrt{m}I^0_{[m-1r],[r']}(k)  + \sqrt{m+1}I^0_{[m+1r],[r']}(k)\right)\sigma_y -v_*'I^y_{[mr],[r']}(k) \sigma_y\\
0_{2\times 2}
\end{array}\right)_{ab}\bigg]c^\dagger_{a\tau krm}
f_{b\tau k r}.\label{cfmatrixinB}
\end{eqnarray}

\subsection{Low field Analysis of \texorpdfstring{$c$}{c}-\texorpdfstring{$f$}{f} coupling}\label{apdx:Low-field cf coupling}

For analyzing the low $\bB$ structure for $c$-$f$ hybridization matrix, given in Eq.(\ref{cfmatrixinB}), we will find it useful to first discuss recursion relations for $\mathcal{F}_m(\lambda,x_0)$ defined in Eq.(\ref{xintegral}). We start by considering the generating function for $\mathcal{H}_m(x,y)$ (defined in Eq.(\ref{2varHermite}))
\begin{eqnarray}
e^{xt+yt^2}&=&\sum_{n=0}^{\infty}\frac{t^n}{n!}\mathcal{H}_n\left(x,y\right).
\end{eqnarray}

Using the generating function, we can derive the following recursion relations
\begin{eqnarray}
\frac{\partial}{\partial t}e^{xt+yt^2}&=&\left(x+2yt\right)e^{xt+yt^2}\Rightarrow
\sum_{n=1}^{\infty}\frac{t^{n-1}}{(n-1)!}\mathcal{H}_n\left(x,y\right)=\sum_{n=0}^{\infty}\frac{t^n}{n!}x
\mathcal{H}_n\left(x,y\right)+\sum_{n=0}^{\infty}\frac{t^{n+1}}{n!}2y\mathcal{H}_n\left(x,y\right)\\
&&\sum_{n=0}^{\infty}\frac{t^{n}}{n!}\mathcal{H}_{n+1}\left(x,y\right)-\sum_{n=1}^{\infty}\frac{t^{n}}{(n-1)!}2y
\mathcal{H}_{n-1}\left(x,y\right)=\sum_{n=0}^{\infty}\frac{t^n}{n!}x\mathcal{H}_n\left(x,y\right)\\
&&n=0:\; \mathcal{H}_{1}\left(x,y\right)=x\mathcal{H}_{0}\left(x,y\right)\\
&&n>0:\;\mathcal{H}_{n+1}\left(x,y\right)-2yn \mathcal{H}_{n-1}\left(x,y\right)=x\mathcal{H}_{n}\left(x,y\right)
\end{eqnarray}
Therefore
\begin{eqnarray}
&&n=0:\; \sqrt{2}\mathcal{F}_{1}\left(\lambda,x_0\right)=-2x_0\frac{\ell}{\ell^2+\lambda^2} \mathcal{F}_{0}\left(\lambda,x_0\right)\\
&&n>0:\;\sqrt{2^{n+1}(n+1)!}\mathcal{F}_{n+1}\left(\lambda,x_0\right)+2\left(1-\frac{2\lambda^2}{\ell^2+\lambda^2}\right)n \sqrt{2^{n-1}(n-1)!}\mathcal{F}_{n-1}\left(\lambda,x_0\right)=-2x_0\frac{\ell}{\ell^2+\lambda^2}\sqrt{2^nn!} \mathcal{F}_{n}\left(\lambda,x_0\right)\nonumber\\
&&\implies \left(1+\frac{\lambda^2}{\ell^2}\right)\frac{1}{\sqrt{2}}\left(\sqrt{n+1}\mathcal{F}_{n+1}\left(\lambda,x_0\right)+
\left(1-\frac{2\lambda^2}{\ell^2+\lambda^2}\right)\sqrt{n} \mathcal{F}_{n-1}\left(\lambda,x_0\right)\right)=-\frac{x_0}{\ell} \mathcal{F}_{n}\left(\lambda,x_0\right) \nonumber\\
&&\implies \left(1+\frac{\lambda^2}{\ell^2}\right)\sqrt{\frac{n+1}{2}}\mathcal{F}_{n+1}\left(\lambda,x_0\right)+
\left(1-\frac{\lambda^2}{\ell^2}\right)\sqrt{\frac{n}{2}} \mathcal{F}_{n-1}\left(\lambda,x_0\right)=-\frac{x_0}{\ell} \mathcal{F}_{n}\left(\lambda,x_0\right).\label{required recursion}
\end{eqnarray}
From Eq.(\ref{Iy}), we have
\begin{eqnarray}
&& I^{y}_{[mr],[r']}= \frac{\sqrt{L_{1x}}}{\sqrt{\ell}}\sum_{j\in \mathbb{Z}}\sum_{s\in \mathbb{Z}}\delta_{r'-r,jq+sp}
e^{-2\pi i s k_1}e^{2\pi i s\left(k_2+\frac{r}{q}\right)\frac{L_{1y}}{L_2}}e^{\pi i s(s-1) \frac{p}{q}\frac{L_{1y}}{L_2}}
\left(k_2+\frac{r}{q}+ s\frac{p}{q}\right)\frac{2\pi}{L_2}
\nonumber\\
&&\times e^{-2\pi^2 \left(k_2+\frac{r}{q}+ s\frac{p}{q}\right)^2\frac{\lambda^2}{L^2_2}}
\mathcal{F}_m\left(\lambda,\left(s+\frac{r}{p}+k_2\frac{q}{p}\right)L_{1x}\right) \\
&&= \frac{p}{q} \frac{2\pi \ell^2}{\ell L_{1x}L_2}\frac{\sqrt{L_{1x}}}{\sqrt{\ell}}\sum_{j\in \mathbb{Z}}\sum_{s\in \mathbb{Z}}\delta_{r'-r,jq+sp}
e^{-2\pi i s k_1}e^{2\pi i s\left(k_2+\frac{r}{q}\right)\frac{L_{1y}}{L_2}}e^{\pi i s(s-1) \frac{p}{q}\frac{L_{1y}}{L_2}}
\left(qk_2+r+sp\right)\frac{L_{1x}}{p}\frac{1}{\ell}
\nonumber\\
&&\times e^{-2\pi^2 \left(k_2+\frac{r}{q}+ s\frac{p}{q}\right)^2\frac{\lambda^2}{L^2_2}}
\mathcal{F}_m\left(\lambda,\left(sp+r+qk_2\right)\frac{L_{1x}}{p}\right).
\end{eqnarray}. Now using the fact $\frac{2\pi p \ell^2}{q L_{1x}L_2}= 1$ and the recursion relation given in Eq.(\ref{required recursion}), we have
\begin{eqnarray}
&&I^{y}_{[mr],[r']}(k)=\frac{-1}{\ell}\frac{\sqrt{L_{1x}}}{\sqrt{\ell}}\sum_{j\in \mathbb{Z}}\sum_{s\in \mathbb{Z}}\delta_{r'-r,jq+sp}
e^{-2\pi i s k_1}e^{2\pi i s\left(k_2+\frac{r}{q}\right)\frac{L_{1y}}{L_2}}e^{\pi i s(s-1) \frac{p}{q}\frac{L_{1y}}{L_2}}
\nonumber\\
&&\times e^{-2\pi^2 \left(k_2+\frac{r}{q}+ s\frac{p}{q}\right)^2\frac{\lambda^2}{L^2_2}}
\bigg(\left(1+\frac{\lambda^2}{\ell^2}\right)\sqrt{\frac{m+1}{2}}\mathcal{F}_{m+1}\left(\lambda,\left(sp+r+qk_2\right)\frac{L_{1x}}{p}\right) \nonumber \\
&&+ \left(1-\frac{\lambda^2}{\ell^2}\right)\sqrt{\frac{m}{2}}\mathcal{F}_{m-1}\left(\lambda,\left(sp+r+qk_2\right)\frac{L_{1x}}{p}\right)\bigg) \\
&&=-\frac{1}{\sqrt{2}\ell} \left(\left(1+\frac{\lambda^2}{\ell^2}\right)\sqrt{m+1}I^0_{[m+1r],[r']}(k)
+ \left(1-\frac{\lambda^2}{\ell^2}\right)\sqrt{m}I^0_{[m-1r],[r']}(k)\right).\label{Iy reduced to I0}
\end{eqnarray}
where $I^0_{[mr],[r']}$ is defined in Eq.(\ref{I0}). Using the above derived form for $I^y_{[mr],[r']}$ in Eq(\ref{Iy reduced to I0}), the minimal coupling term, given in Eq.(\ref{minimal coupling term}), can be rewritten as
\begin{eqnarray}
&&-\frac{I_A}{\hbar} = \frac{\lambda^2}{\ell^2}\frac{1}{\sqrt{2}\ell}\bigg( \sqrt{m}I^0_{[m-1r],[r']}(k)  -\sqrt{m+1}I^0_{[m+1r],[r']}(k)\bigg)
\end{eqnarray}
\begin{figure}
\includegraphics[width=8.2cm]{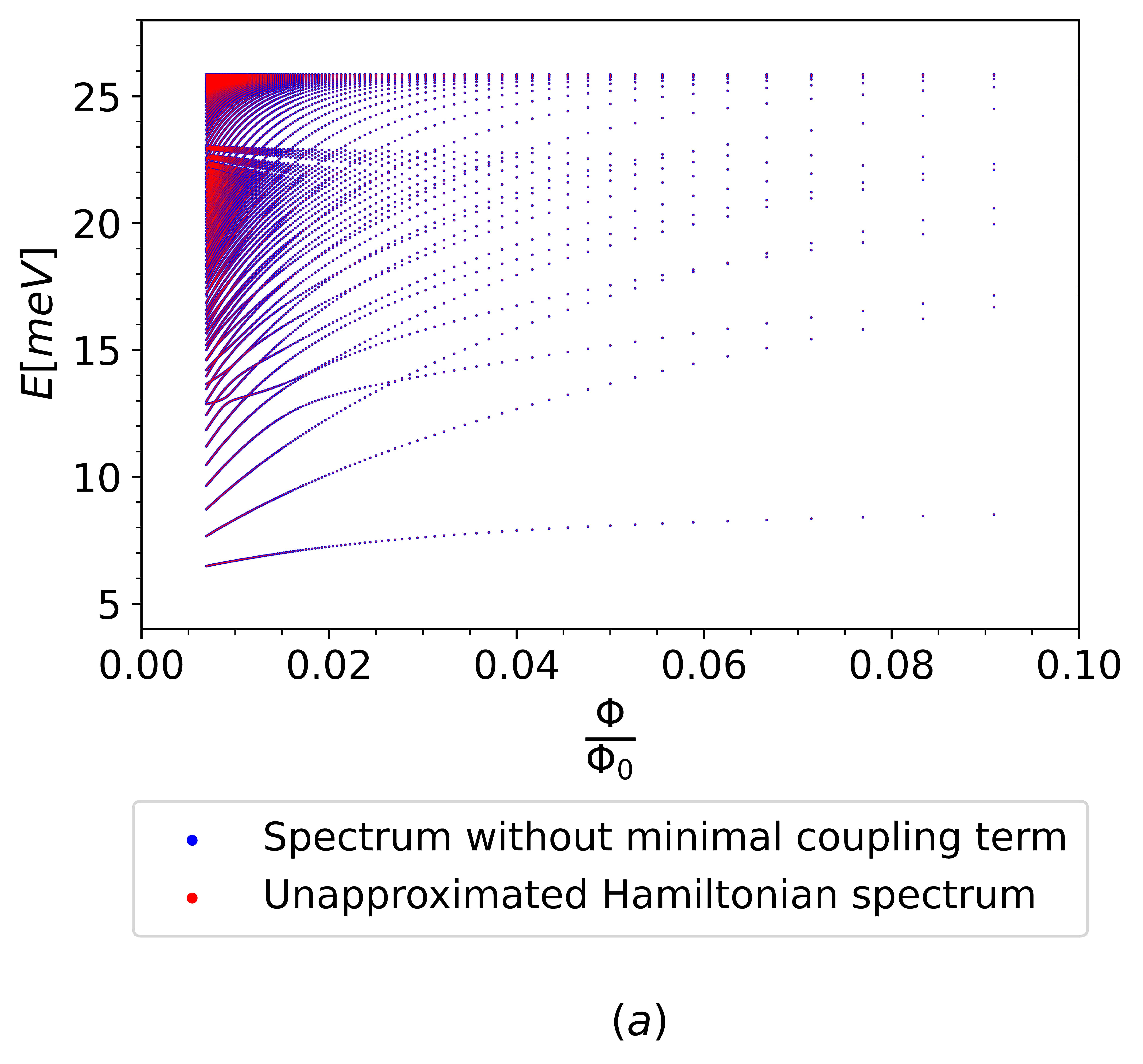}
\includegraphics[width=8.2cm]{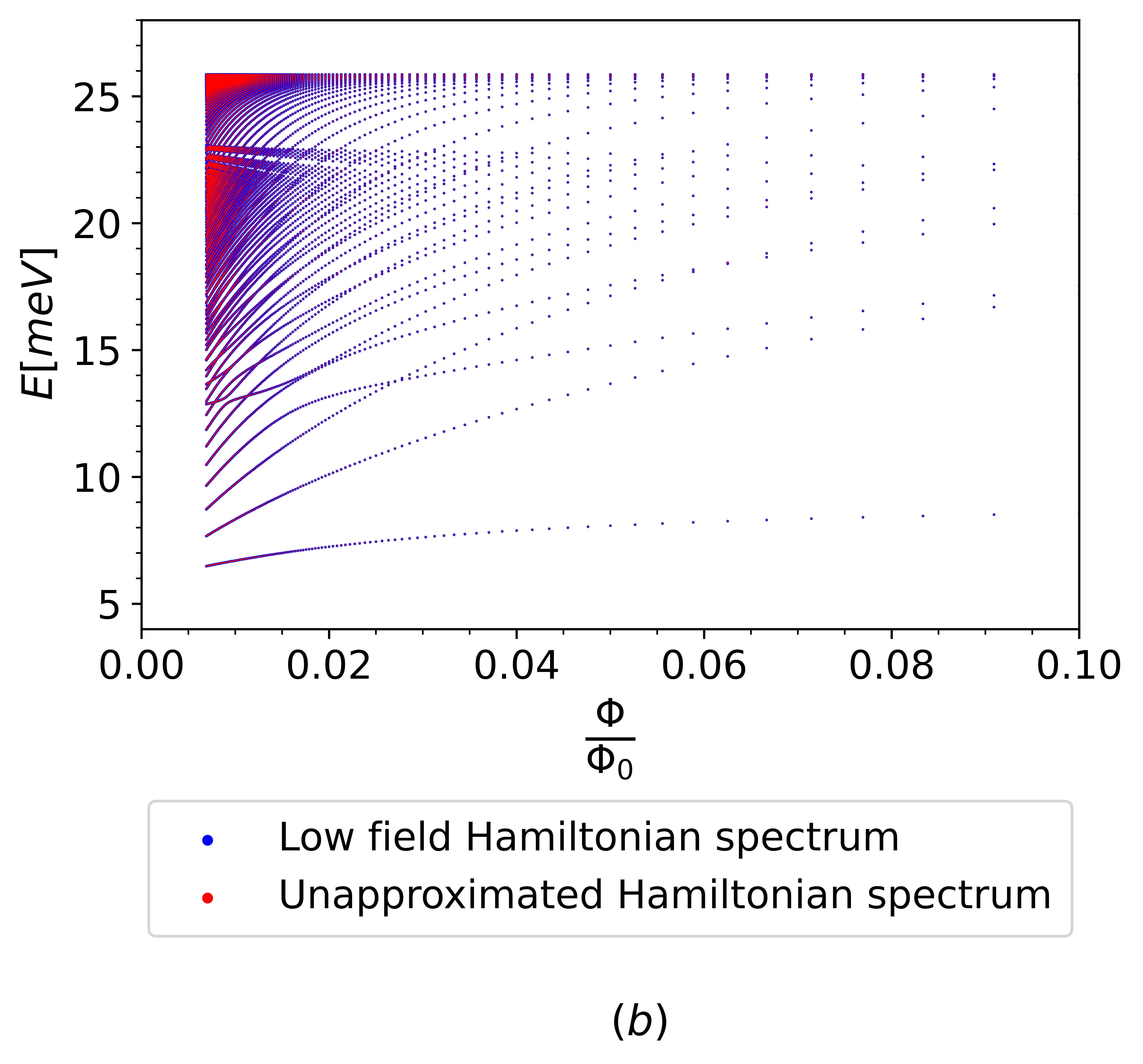}
\caption{These figures justify dropping $O(\frac{\lambda^2}{\ell^2})$ in the prefactor of $I^{0,x,y}$ in the $c$-$f$ hybridization matrix.(a)Comparison of the unapproximated Hamiltonian spectrum which includes the minimal coupling to the one without minimal coupling at valley $\bK'$ at CNP.(b)Comparison of the unapproximated Hamiltonian spectrum to the one obtained using low field $c$-$f$ coupling at valley $\bK'$ at CNP. These figures correspond to parameters $w_0/w_1=0.7$,$J=18.27meV$, $U_1=51.72 meV$, $\gamma=-39.11meV$, $v'_{\ast}=1.624eV.\mathring{A}$, $v_{\ast}=-4.483eV.\mathring{A}$, $M=3.248meV$ and $\lambda=0.3792L_m$. We set $k_{1,2}=0$ and $m_{max}=\lceil \frac{q-3}{2} \rceil$.}
\label{fig:Justifying low field cf}
\end{figure}

Note that this is an $O(\frac{\lambda^2}{\ell^2})$ term. At low $\bB$, i.e $\lambda^2\ll \ell^2$, this term is negligible as anticipated. This is confirmed numerically as shown in Supplementary Fig.(\ref{fig:Justifying low field cf}a). Note that we however still need to keep the full $\frac{\lambda^2}{\ell^2}$ dependence within $\mathcal{F}_m$ for accuracy in low $\bB$ range as shown discussed in Supplementary Fig.(\ref{fig:Justifying lambda in singular vals}). Keeping this in mind, let us now dwell into a low $\bB$ limit wherein we can drop the $O(\frac{\lambda^2}{\ell^2})$ term in the recursion relation Eq.(\ref{required recursion}), i.e.
\begin{eqnarray}\label{low field recursion}
\sqrt{\frac{n+1}{2}}\mathcal{F}_{n+1}\left(\lambda,x_0\right)+
\sqrt{\frac{n}{2}} \mathcal{F}_{n-1}\left(\lambda,x_0\right)=-\frac{x_0}{\ell} \mathcal{F}_{n}\left(\lambda,x_0\right).
\end{eqnarray}
Moreover using Eq.(\ref{Iy reduced to I0}), $I^y_{[mr],[r']}$ in this limit reads
\begin{eqnarray}
I^y_{[mr],[r']}(k) =-\frac{1}{\sqrt{2}\ell} \left(\sqrt{m+1}I^0_{[m+1r],[r']}(k)
+ \sqrt{m}I^0_{[m-1r],[r']}(k)\right).   
\end{eqnarray}
Now using the above and Eq.(\ref{Ix}), we have
\begin{eqnarray}
I^x_{[mr],[r']}(k)+i I^y_{[mr],[r']}(k)&\approx& -i\frac{\sqrt{2}}{\ell} \sqrt{m+1}I^0_{[m+1r],[r']}(k)\\
I^x_{[mr],[r']}(k)-i I^y_{[mr],[r']}(k)&\approx& i\frac{\sqrt{2}}{\ell} \sqrt{m}I^0_{[m-1r],[r']}(k).
\end{eqnarray}
Then at low field we can re-express the $c$-$f$ coupling as
\begin{eqnarray}
h^{+1}_{[amr][br']}(k) = \left(\begin{array}{cc}
\gamma I^0_{[mr],[r']}(k) & i\frac{\sqrt{2}v'_*}{\ell}\sqrt{m}I^0_{[m-1r],[r']}(k) \\
-i\frac{\sqrt{2}v'_*}{\ell}\sqrt{m+1}I^0_{[m+1r],[r']}(k)& \gamma I^0_{[mr],[r']}(k)\\
0 & 0 \\
0 & 0
\end{array}
\right),\label{lowfieldcf1}
\end{eqnarray}
\begin{eqnarray}
h^{-1}_{[amr][br']}(k) = \left(\begin{array}{cc}
\gamma I^0_{[mr],[r']}(k) & i\frac{\sqrt{2}v'_*}{\ell}\sqrt{m+1}I^0_{[m+1r],[r']} (k)\\
-i\frac{\sqrt{2}v'_*}{\ell}\sqrt{m}I^0_{[m-1r],[r']}(k)   & \gamma I^0_{[mr],[r']}(k)\\
0 & 0 \\
0 & 0
\end{array}
\right).\label{lowfieldcf2}
\end{eqnarray}
Supplementary Fig.(\ref{fig:Justifying low field cf}b) shows that dropping $O(\frac{\lambda^2}{\ell^2})$ in Eq.(\ref{required recursion}) is indeed an excellent approximation. It is thus justified to study the low field $c$-$f$ hybridization matrix, given in Eq.(\ref{lowfieldcf1}) and Eq.(\ref{lowfieldcf2}), in pursuit of obtaining an analytical solution.
\section{Singular Values for \texorpdfstring{$I^{0}$}{I0}}\label{apdx:Singular values}
In this section, we study the singular values of $I^{0}_{m,r'}(k)=I^{0}_{[m0],[r']}(k)$, where $I^{0}_{[m0],[r']}(k)$ is defined in Eq.(\ref{I0}). As discussed in the main text, these singular values controls the hybridization strength between the heavy and conduction fermions in finite $\bB$. Let us start by considering the singular value decomposition of $I^{0}_{m,r'}(k)$: 
\begin{equation}\label{eq:SVD}
I^{0}_{m,r'}(k)= U_{mm'}\Sigma_{m'\tilde{r}}V_{\tilde{r}r'},  
\end{equation} where $U,V$ are are $(m_{a,\tau}+1) \times (m_{a,\tau}+1)$ and $q \times q$ unitary matrices respectively, and the $(m_{a,\tau}+1)\times q$ rectangular matrix $\Sigma$ contains the singular values along the main diagonal and zeros elsewhere, where summation convention on repeated indices is implied. The columns of matrix $U$ are the eigenvectors of the following matrix 
\begin{eqnarray}
\Lambda^{0}_{mm'}(k) = \sum_{r'=0}^{q-1}I^{0}_{mr'}(k)I^{0^{\ast}}_{m'r'}(k).
\end{eqnarray}
 Since $r'$ is being summed over, we are allowed to shift its range to $-\lfloor \frac{q}{2}\rfloor, -\lfloor \frac{q}{2}\rfloor+1, \ldots, q-1-\lfloor \frac{q}{2}\rfloor$ and thus set $j=0$ in Eq.(\ref{I0}) based on the discussion in main text. Since $\mathcal{F}_m$ in Eq.(\ref{I0}) decays exponentially at $\pm \lfloor \frac{q}{2} \rfloor$, in the low field limit, we can replace the bounds of this sum by $\pm \infty$ and have
 \begin{eqnarray}
\Lambda^{0}_{mm'}(k) \approx \frac{L_{1x}}{\ell}\sum_{r'=-\infty}^{\infty}e^{-4\pi^2 \frac{\lambda^2}{L^2_2}\left(k_2+\frac{r'}{q}\right)^2} \mathcal{F}_m\left(\lambda,\left(r'+k_2q\right)L_{1x}\right)\mathcal{F}_{m'}\left(\lambda,\left(r'+k_2q\right)L_{1x}\right).
\end{eqnarray} 
Using the Dirac comb identity, $\sum_{r'\in \mathbb{Z}}\delta(\rho-r') = \sum_{t\in\mathbb{Z}}e^{2\pi i t\rho}$, we can convert this summation to an integral as
\begin{eqnarray}\label{Sigma-m^2}
\Lambda^{0}_{mm'}(k) &\approx& \frac{L_{1x}}{\ell}\sum_{t\in\mathbb{Z}}\int_{-\infty}^{\infty}d\rho e^{i 2\pi t\rho}
e^{-4\pi^2 \frac{\lambda^2}{L^2_2}\left(k_2+\frac{\rho}{q}\right)^2} \mathcal{F}_m\left(\lambda,\left(\rho+k_2q\right)L_{1x}\right)\mathcal{F}_{m'}\left(\lambda,\left(\rho+k_2q\right)L_{1x}\right) \nonumber \\
&=& \frac{1}{\ell}\sum_{t\in\mathbb{Z}}e^{-i 2\pi t k_2q}\int_{-\infty}^{\infty}d\rho e^{i 2\pi t\rho/L_{1x}}
e^{-\frac{\lambda^2}{\ell^2}\frac{\rho^2}{\ell^2}} \mathcal{F}_m\left(\lambda,\rho \right)\mathcal{F}_{m'}\left(\lambda,\rho \right).
\end{eqnarray}
Because $2\pi t/L_{1x}$ is a large wavevector compared to the length scale at which the h.o. wavefunctions vary, $\sim \ell$, i.e. $\frac{L_{1x}}{2\pi t} \ll \ell$ at low field, the overlaps are exponentially suppressed in $(\frac{2\pi t\ell}{L_{1x}})^2$. We can thus set $t=0$ in the above sum. Moreover since the exponential factor contains $\frac{\lambda^2}{\ell^2}$, the off-diagonal terms can be neglected and thus $\Lambda^{0}_{mm'}$ is diagonal up to a good approximation. This implies that the matrix $U$ is close to identity and the low $\bB$ singular values can be approximated as $\Sigma_m=\sqrt{\Lambda^0_{mm}}$. Note that the information of magnetic quantum number $k$ gets dissolved automatically due to the low field limit where the magnetic sub-bands are not $k$ dispersive. We thus have
\begin{eqnarray}
   \Sigma^{2}_m &&= \frac{1}{\ell}\int_{-\infty}^{\infty}d\rho 
e^{-\frac{\lambda^2}{\ell^2}\frac{\rho^2}{\ell^2}}
\mathcal{F}_m\left(\lambda,\rho \right)\mathcal{F}_{m}\left(\lambda,\rho \right) \label{integral for singular values}
\end{eqnarray}
$\mathcal{F}_m(\lambda,\rho)$ defined in Eq.(\ref{xintegral}) can re-expressed as:
\begin{eqnarray}
\mathcal{F}_m(\lambda,\rho)&=&
\frac{1}{\pi^{\frac{1}{4}}\sqrt{2^{m}m!}}\sqrt{\frac{1}{1+(\frac{\lambda}{\ell})^2}}
e^{-\frac{1}{2} (\frac{\rho}{\ell})^2/(1+(\frac{\lambda}{\ell})^2)}
\mathcal{H}_m\left(\frac{-2(\frac{\rho}{\ell})}{1+(\frac{\lambda}{\ell})^2},\frac{2(\frac{\lambda}{\ell})^2}{1+(\frac{\lambda}{\ell})^2}-1\right)
\end{eqnarray}
Naively one would expect that dropping $O(\frac{\lambda^2}{\ell^2})$ in $\mathcal{F}_m(\lambda,x_0)$, i.e. replacing $\mathcal{F}_m(\lambda,x_0)$ by $\varphi(-x_0)$, to be good enough approximation at low field. However as discussed in main text, the fairly strong $m$ dependence of singular values gets compromised under such an approximation, as shown in Supplementary Fig.(\ref{fig:Justifying lambda in singular vals}a).
\begin{figure}
\includegraphics[width=7cm]{Images_for_supplementary/Wrong_SVD_arxiv.jpg}
\includegraphics[width=9.8cm]{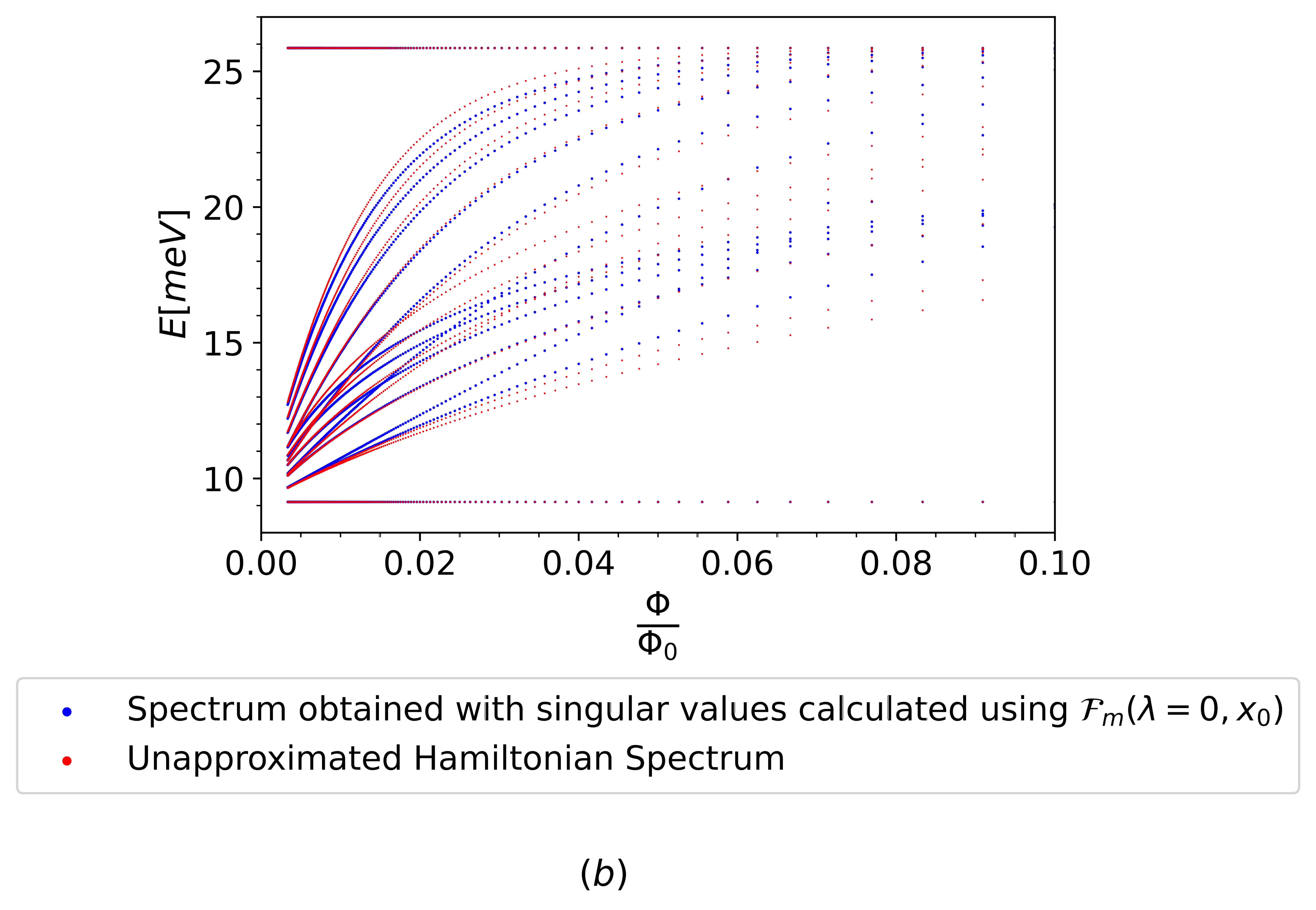}
\caption{These figures illustrate the importance of including the $O(\frac{\lambda^2}{\ell^2})$ in $\mathcal{F}_m(\lambda,x_0)$ for obtaining the correct interacting heavy fermion Hofstadter spectrum. (a) The comparison of numerically obtained singular values for $I^{0}$ with the ones obtained numerically post setting $\frac{\lambda}{\ell}=0$ in $\mathcal{F}_m(\lambda,x_0)$ . (b) The figure comparing unapproximated Hamiltonian spectrum with the one calculated using singular values marked in \textcolor{red}{red} from the left panel at $w_0/w_1=0.7$ in flat band limit at valley $\bK'$ at CNP. $m_{max}$ is set to 5 for illustration.
}\label{fig:Justifying lambda in singular vals}
\end{figure}

Let $\rho/\ell = x$ and $\kappa = \lambda/\ell$. Then

\begin{eqnarray}
\Sigma^{2}_m &&= \frac{1}{\sqrt{\pi}}\frac{1}{2^m m!}\frac{1}{1+\kappa^2}\int_{-\infty}^{\infty}dx
e^{-\left(\kappa^2 + \frac{1}{1+\kappa^2}\right)x^2}
\mathcal{H}^2_m\left(\frac{-2}{1+\kappa^2}x,\frac{\kappa^2 -1}{\kappa^2 +1} \right)
\end{eqnarray}
Now, using the result from \cite{babusci2012integrals}, we have
\begin{eqnarray}
&&\int_{-\infty}^{\infty} dx \mathcal{H}_m(ax+b,y)\mathcal{H}_n(cx+d,y)e^{-fx^2+\alpha x}=\sqrt{\frac{\pi}{f}}e^{\frac{\alpha^2}{4f}}
\mathcal{H}_{m,n}\left(b + \frac{a\alpha}{2f},y +\frac{a^2}{2f};d + \frac{c\alpha}{2f},z + \frac{c^2}{2f}\mid \frac{ac}{2f}\right)
\end{eqnarray}
where 
\begin{eqnarray}{\label{eqn2inxHermite}}
    \mathcal{H}_{m,n}\left(x,y;w,z\vert \beta\right) = \sum^{min(m,n)}_{k=0} \frac{m!n!}{(n-k)!(m-k)!k!}\beta^k\mathcal{H}_{m-k}(x,y)\mathcal{H}_{n-k}(w,z),
\end{eqnarray}
and the two variable Hermite polynomial $\mathcal{H}_n(x,y)$ is defined in Eq.(\ref{2varHermite}),
we have
\begin{eqnarray}\label{Sigma_sq}
 \Sigma^{2}_m &&= \frac{1}{\sqrt{\xi(\kappa)}}\frac{1}{2^m m!}\mathcal{H}_{mm}(0,\frac{\kappa^6}{\xi(\kappa)};0,\frac{\kappa^6}{\xi(\kappa)}\mid \frac{2}{\xi(\kappa)}),
\end{eqnarray}
where $\xi(\kappa)$ is given as 
\begin{eqnarray}
    \xi(\kappa) = (1+\kappa^2 + \kappa^4)(1+\kappa^2).
\end{eqnarray}
Thus
\begin{eqnarray}{\label{apdxSingularvalue}}
\Sigma_m = \left(\frac{1}{\sqrt{\xi(\kappa)}}\frac{1}{2^m m!}\mathcal{H}_{mm}(0,\frac{\kappa^6}{\xi(\kappa)};0,\frac{\kappa^6}{\xi(\kappa)}\mid \frac{2}{\xi(\kappa)}) \right)^\frac{1}{2}.
\end{eqnarray}
Note that 
\begin{eqnarray}
    \kappa^2 = \frac{\lambda^2}{\ell^2} = \frac{2\pi\lambda^2}{A_{uc}}\frac{p}{q} = \frac{2\pi\lambda^2}{A_{uc}}\frac{\phi}{\phi_0}
\end{eqnarray}
where $A_{uc}$ is moire unit cell area. Since $\Sigma^2_m$ is an analytic function in $\kappa^2$, we have an expression for the singular values continuously down to zero flux. Also note that as $\frac{\phi}{\phi_0} \rightarrow 0 \implies \kappa^2 \rightarrow 0 \implies \xi(\kappa) \rightarrow 1 \implies \mathcal{H}_{mm}(0,\frac{\kappa^6}{\xi(\kappa)};0,\frac{\kappa^6}{\xi(\kappa)}\mid \frac{2}{\xi(\kappa)}) \rightarrow 2^mm! \implies \Sigma_m \rightarrow 1$, consistent with numerical results. At extremely low flux, we can approximate the singular values as 
\begin{eqnarray}
   lt_{\kappa\rightarrow 0} \Sigma_m(\kappa) = 1- \left(m+\frac{1}{2}\right)\kappa^2,\label{Eq:limit for singular values}
\end{eqnarray}
i.e. approaching $1$ at zero flux with a negative slope slope of $\left(m+\frac{1}{2}\right)\frac{2\pi\lambda^2}{A_{uc}}$ as shown in Supplementary Fig.(\ref{fig:Singular vals analytical fig}c). Note that even at extremely low flux, the $m$ dependence of singular values grow linearly.

It would prove feasible to discuss the asymptote of $\Sigma^{2}_m$ for large values of m.
Let us start expanding the four variable Hermite polynomial in Eq.(\ref{Sigma_sq}).
\begin{eqnarray}
\Sigma^2_m &=& \frac{1}{\sqrt{\xi(\kappa)}}\frac{1}{2^mm!}\Big[\sum_{k=0}^{m} \frac{m!m!\tau^k}{(m-k)!(m-k)!k!}\mathcal{H}_{m-k}(0,y)\mathcal{H}_{m-k}(0,y)\Big]
\end{eqnarray}
where for brevity we denote $\frac{\kappa^6}{\xi(\kappa)}$ and $\frac{2}{\xi(\kappa)}$ by $y$ and $\beta$ respectively. Then
\begin{eqnarray}
\Sigma^2_m &=& \frac{1}{\sqrt{\xi(\kappa)}}\frac{1}{2^m}\Big[\frac{1}{m!}\mathcal{H}_{m}(0,y)\mathcal{H}_{m}(0,y)\Big] \label{kis0} \\
&+&\frac{1}{\sqrt{\xi(\kappa)}}\frac{1}{2^m}\Big[\sum_{k=1}^{m-1} \frac{m!\beta^k}{(m-k)!(m-k)!k!}\mathcal{H}_{m-k}(0,y)\mathcal{H}_{m-k}(0,y)\Big] \label{k in sum} \\
&+& \frac{1}{\sqrt{\xi(\kappa)}}\frac{1}{2^m}\beta^{m} \label{k is m},
\end{eqnarray}

\begin{figure}
\includegraphics[width=8.4cm]{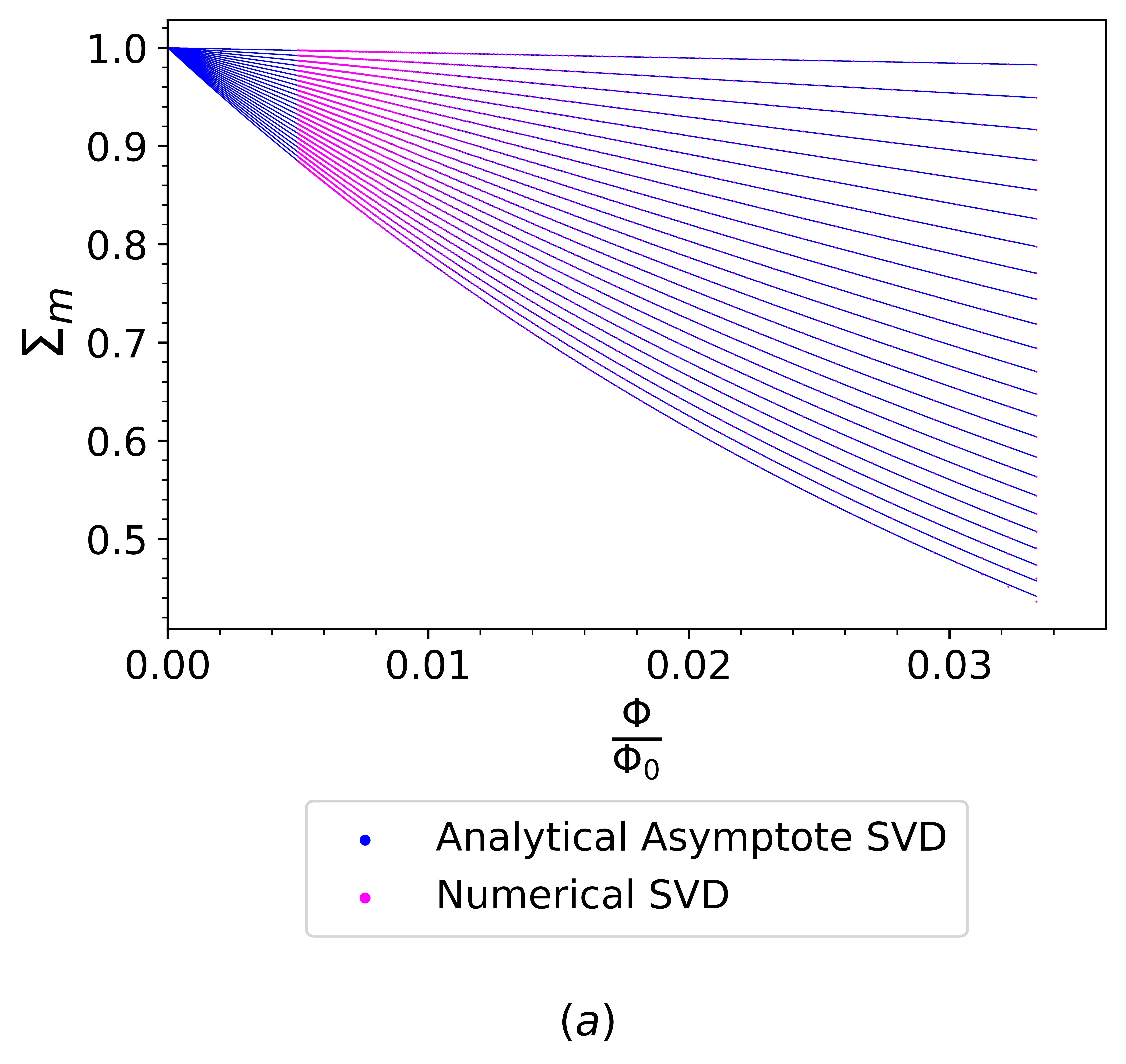}
\includegraphics[width=8.4cm]{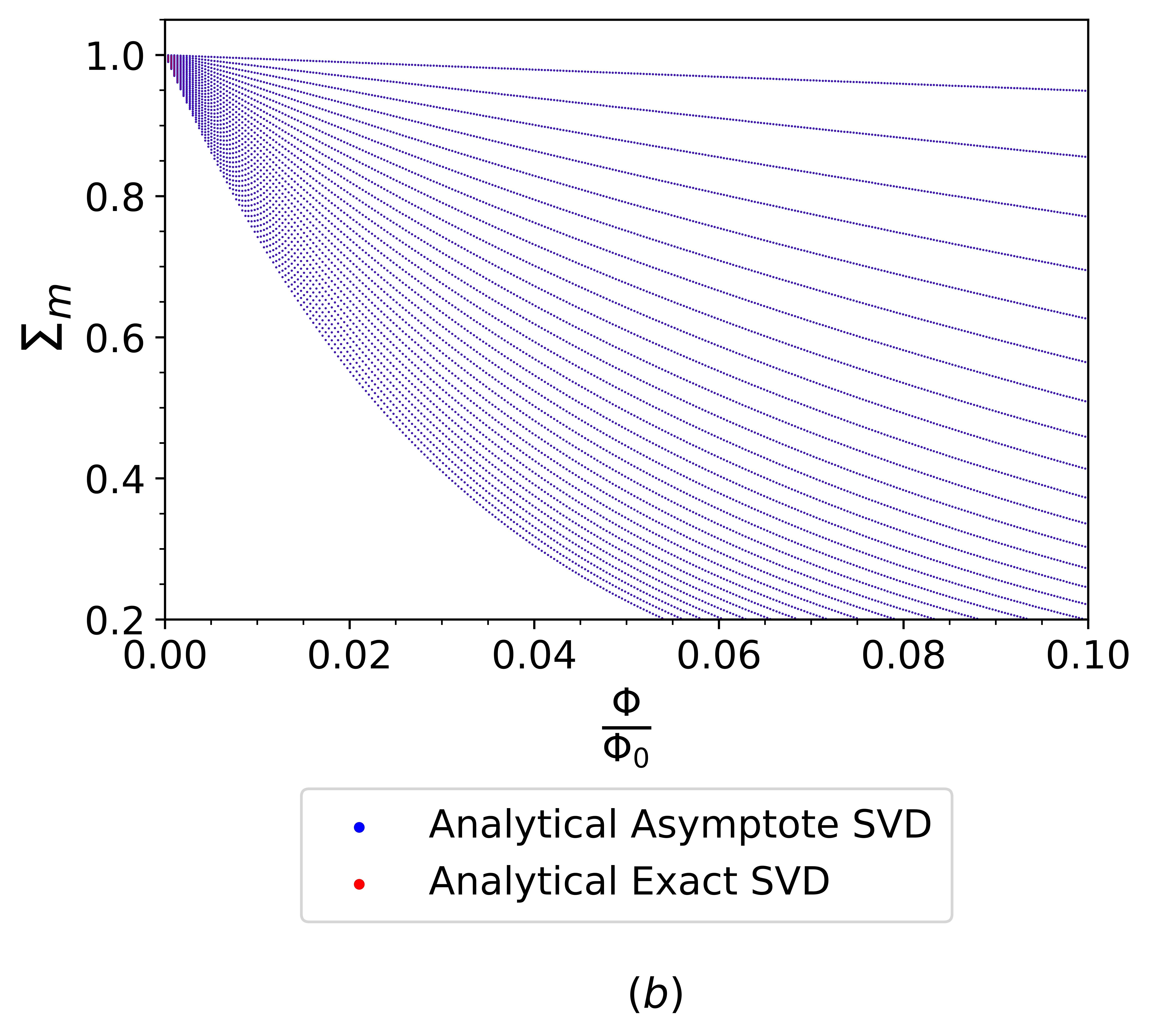}
\includegraphics[width=8.5cm]{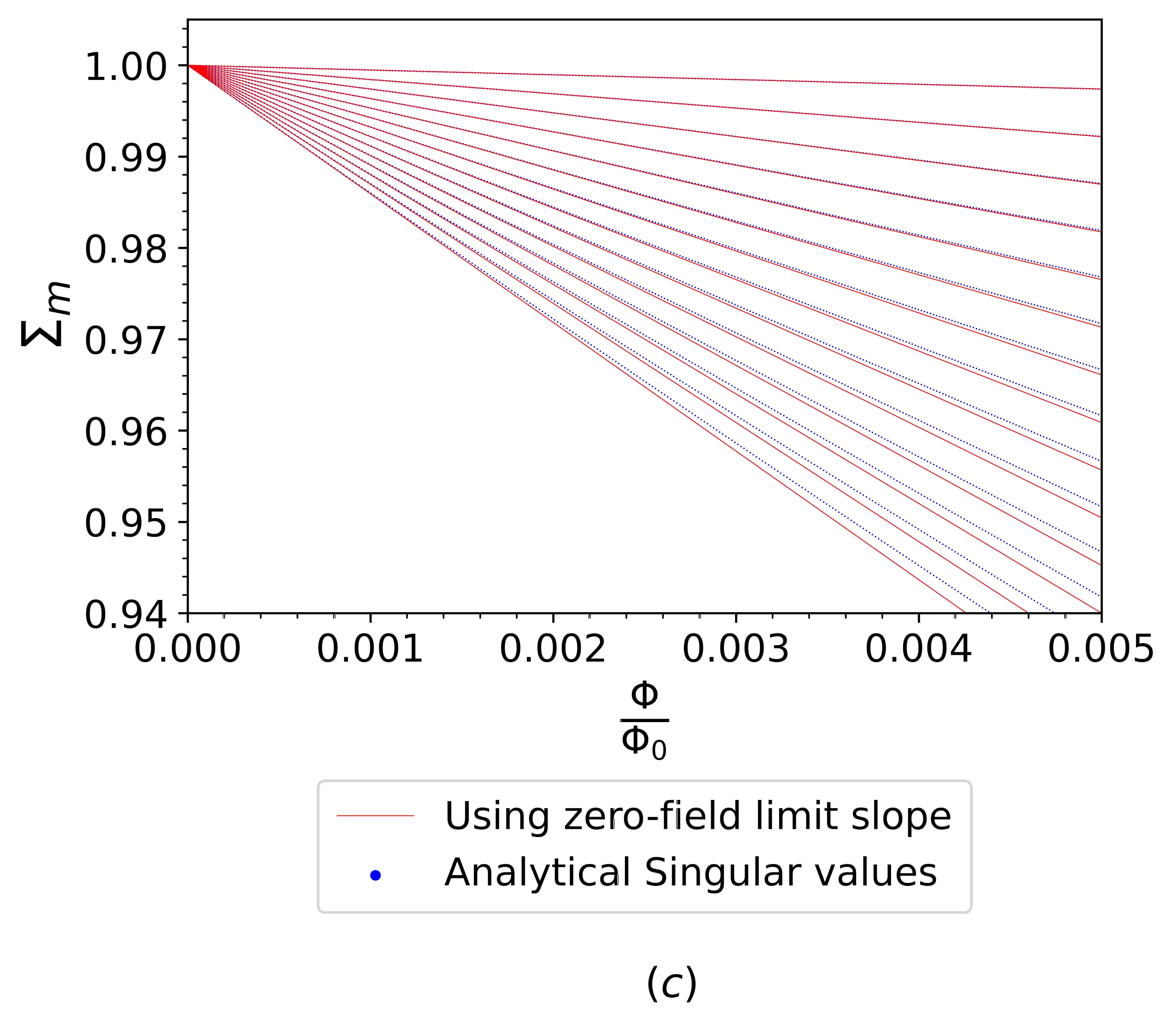}

\caption{(a) The comparison of first 25 numerically obtained singular values with the ones obtained using the analytical asymptotes for singular values. (b) The comparison of asymptotes with exact expression for singular values. (c) Comparing $1-(m+\frac{1}{2})\kappa^2$ with analytically computed singular values at extremely low field.}
\label{fig:Singular vals analytical fig}
\end{figure}
where we divided the sum over $k$ into $k=0$, $k=1\ldots m-1$ and $k=m$, shown in Eq.(\ref{kis0}), Eq.(\ref{k in sum}) and Eq.(\ref{k is m}) respectively. Now note that $\mathcal{H}_m(0,y)$ is non-zero only if $m$ is even, so
\begin{eqnarray}
\mathcal{H}_{m}(0,y) = \left(1-\text{mod}(m,2)\right)m!\frac{y^{\frac{m}{2}}}{\frac{m}{2}!} \\
\implies \mathcal{H}_{m}^{2}(0,y) = \left(1-\text{mod}(m,2)\right)\frac{m!m!y^m}{\frac{m}{2}!\frac{m}{2}!}.\label{fate of Hm(0,y)}
\end{eqnarray}
Using Eq.(\ref{fate of Hm(0,y)}), we have 
\begin{eqnarray}
\Sigma^2_m &=& \frac{1}{\sqrt{\xi(\kappa)}} \left( \left(1-\text{mod}(m,2)\right)\left(\frac{y}{2}\right)^m \frac{m!}{\frac{m}{2}!\frac{m}{2}!} + \left({\frac{\beta} {2}}\right)^m\right) \nonumber \\
&+& \frac{1}{\sqrt{\xi(\kappa)}}\frac{1}{2^m}\Big[ \sum_{k=1}^{m-1} \left(1-\text{mod}(m-k,2)\right) \frac{m!\beta^ky^{m-k}}{(\frac{m-k}{2})!(\frac{m-k}{2})!k!} \Big] \label{Sigma_sq_substitute here}.
\end{eqnarray}
Now consider 
\begin{eqnarray}
 &&\bar{A}(m) = \frac{m!}{(\frac{m}{2})!(\frac{m}{2})!} \\
 &&\implies ln(\bar{A}(m)) = ln(m!) - 2ln\left(\left(\frac{m}{2}\right)!\right) \approx m ln(2) \label{strirling1}\\
 && \implies \bar{A}(m) = 2^m \label{barA},
\end{eqnarray}
where in Eq.(\ref{strirling1}), we use the Stirling's approximation for large $m$, i.e. $ln(m!)=mln(m)-m$. Similarly consider
\begin{eqnarray}
&&\bar{B}_k(m) = \frac{m!}{(\frac{m-k}{2})!(\frac{m-k}{2})!} \\
&&\implies ln(\bar{B}_k(m)) = ln(m!) - 2ln\left(\left( \frac{m-k}{2}!\right) \right) - ln(k!) 
\end{eqnarray}
\begin{eqnarray}
&& \approx mln(m) - mln\left(\frac{m-k}{2}\right) + kln\left(\frac{m-k}{2}\right) - kln(k) \label{Stirling 2}\\
&& = mln\left(\frac{2m}{m-k}\right) + kln\left(\frac{m-k}{2k}\right) \\
&&\implies \bar{B}_k(m) = 2^{m-k} e^{mln\left(\frac{m}{m-k}\right)+kln\left(\frac{m-k}{k}\right)} \label{barB}
\end{eqnarray}
where in Eq.(\ref{Stirling 2}) we again have used the Stirling's approximation. Now substituting back Eq.(\ref{barA}) and Eq.(\ref{barB}) in Eq.(\ref{Sigma_sq_substitute here}) we have the following asymptote:\\
\begin{itemize}
\item If m is even
\begin{eqnarray}\label{Eq:even m}
 \Sigma_m^2 &=& \frac{1}{\sqrt{\xi(\kappa)}}\bigg[\left(\frac{1}{\xi(\kappa)}\right)^m + \left(\frac{\kappa^6}{\xi(\kappa)}\right)^m + \sum_{k=2,4,6,\ldots}^{m-2}e^{mln(\frac{m}{m-k})+kln(\frac{m-k}{k})}\left(\frac{1}{\xi(\kappa)}\right)^k \left(\frac{\kappa^6}{\xi(\kappa)}\right)^{m-k}\bigg].
\end{eqnarray}

\item If m is odd
\begin{eqnarray}\label{Eq:odd m}
 \Sigma_m^2 &=& \frac{1}{\sqrt{\xi(\kappa)}}\bigg[\left(\frac{1}{\xi(\kappa)}\right)^m + \sum_{k=1,3,5,\ldots}^{m-1}e^{mln(\frac{m}{m-k})+kln(\frac{m-k}{k})}\left(\frac{1}{\xi(\kappa)}\right)^k \left(\frac{\kappa^6}{\xi(\kappa)}\right)^{m-k}\bigg].
\end{eqnarray}

\end{itemize}

Supplementary Fig.(\ref{fig:Singular vals analytical fig}b) shows that the asymptotic expression for singular values, given in Eq.(\ref{Eq:even m}) and Eq.(\ref{Eq:even m}) , exactly matches the analytical singular values, given in Eq.(\ref{apdxSingularvalue}), for all $m\neq 0$ (since the asymptotic formulation does not hold for $m=0$). In the Supplementary Fig.(\ref{fig:Singular vals analytical fig}b), $m=0$ singular value on the asymptotic branch has been calculated using the analytical value, i.e. $\Sigma_0=\frac{1}{\sqrt{\sqrt{\xi(\kappa)}}}$.
\section{Discussion For Valley \texorpdfstring{$\mathbf{K}$}{tau = +1}}\label{apdx:Valley K discussion}
In this section we discuss finite $\bB$ THFM at valley $\mathbf{K}$, i.e. $\tau=+1$. Note that summation convention on repeated indices is implied unless explicitly stated.
Using the SVD decomposition discussed we can re-write the matrix elements as 
\begin{eqnarray}
h^{1}_{[1m0],[1r']}c^{\dagger}_{1 1 k 0 ms}f_{1 1 k r's} = 
\gamma \Sigma_{m',\bar{r}} c^\dagger_{1 1 k 0 ms}U_{mm'} V_{\bar{r},r'}f_{11kr's}= \sum_{m=0}^{m_{1,+1}}\sum_{\bar{r}=0}^{m_{1,+1}}\gamma \Sigma_m \delta_{m\bar{r}} c^\dagger_{1 1 k 0 ms} \bar{f}_{11k\bar{r}s},
\label{11K}
\end{eqnarray}
where we used the fact that $U$ is an identity matrix and the rectangular matrix $\Sigma_{m,r'}$ is non-zero only along its main diagonal.
\begin{eqnarray}
h^{1}_{[2m0],[1r']}c^{\dagger}_{2 1 k 0 m}f_{1 1 k r's} &=&
-i \sqrt{2}\frac{v'_{\ast}}{\ell} \sum_{m=0}^{m_{2,+1}} \sqrt{m+1}c^{\dagger}_{2 1 k 0 m}  \Sigma_{m+1,\bar{r}}V_{\bar{r},r'}f_{1 1 k r's}\nonumber \\
&=&-i\sqrt{2}\frac{v'_{\ast}}{\ell}\sum_{m=0}^{m_{2,+1}}\sum_{\bar{r}=0}^{m_{2,+1}+1}\sqrt{m+1}\Sigma_{m+1} \delta_{m+1,\bar{r}} c^{\dagger}_{2 1 k 0 ms} \bar{f}_{11k\bar{r}s}
\label{21K}
\end{eqnarray}
Because $m_{2,+1}+1=m_{1,+1}=m_{max}+1$, the upper bound on the $\bar{r}$ summation is the same in (\ref{11K}) and (\ref{21K}). 
Similarly
\begin{eqnarray}
h^{1}_{[1m0],[2r']}c^{\dagger}_{1 1 k 0 ms}f_{2 1 k r's} = i\sqrt{2} \sqrt{m}\frac{v'_{\ast}}{\ell} \sum_{m=0}^{m_{1,+1}}\sum_{\bar{r}=0}^{m_{1,+1}-1} \Sigma_{\bar{r}}\delta_{m-1,\bar{r}}c^{\dagger}_{11k0ms}\bar{f}_{21k\bar{r}s}, \label{eq:12k} 
\end{eqnarray}
\begin{eqnarray}
h^{1}_{[2m0],[2r']}c^{\dagger}_{2 1 k 0 ms}f_{2 1 k r's}= \gamma\sum_{m=0}^{m_{2,+1}}\sum_{\bar{r}=0}^{m_{2,+1}}\Sigma_{m}\delta_{m,\bar{r}} c^\dagger_{2 1 k 0 ms}\bar{f}_{2 1 k\bar{r}s}.\label{eq:22k}
\end{eqnarray}
Because $m_{1,+1}-1=m_{2,+1}=m_{max}$, the upper bound on $\bar{r}$ summation is the same in Eq.(\ref{eq:12k}) and Eq.(\ref{eq:22k}). Thus for a given $k$, out of the available $2q$ $f-$modes, only $2m_{max}+3$ couple; recall that $m_{max}\lesssim q/2$.  
\subsection{CNP}\label{apdx:KCNP}
The mean-field interactions for $f$-fermion modes at CNP in $\bar{f}$ basis is given as 
\begin{equation}
V^{f,\tau=+1,s} = \sum_{k\in[0,1)\otimes [0,\frac{1}{q})}V^{f,\tau=+1,s}_{coupled} + V^{f,\tau=+1,s}_{decoupled},
\end{equation}
\begin{equation}
V^{f,\tau=+1,s}_{coupled} =-\frac{U_1}{2}\left(\sum_{m=0}^{m_{max}+1}\bar{f}^{\dagger}_{11kms}\bar{f}_{11kms} + \sum_{m=0}^{m_{max}}\bar{f}^{\dagger}_{21kms}\bar{f}_{21kms} \right)
\end{equation}
\begin{equation}
V^{f,\tau=+1,s}_{decoupled}=- \frac{U_1}{2}\sum_{b=1}^{2}\sum_{m'=m_{max}+\bar{b}}^{q-1}\bar{f}^{\dagger}_{b1km's}\bar{f}_{b1km's},
\end{equation}
where $\bar{1}(\bar{2})=2(1)$.
Note that there are $2q-(2m_{max}+3)$ decoupled $f$ modes for each $k$. The coupled modes  

The coupled modes can then be described as
\begin{equation}\label{xi-hamiltonian}
H^{\tau=+1,s}_{coupled} = \sum_k\sum_{\alpha,\alpha'=1}^6
\sum_{m=0}^{m_\alpha}
\sum_{m'=0}^{m_{\alpha'}}
\Xi_{m\alpha,m'\alpha'}d^{\dagger}_{m\alpha s}(k)d_{m'\alpha' s}(k),
\end{equation}
where $m_{\alpha=1,\ldots, 4}=m_{\alpha,+1}$, $m_{5}=m_{max}+1$ and $m_6=m_{max}$, and
\begin{eqnarray}\label{6-spinor}
d^{\dagger}_{m\alpha s}(k) = \left(c^{\dagger}_{11k0ms},c^{\dagger}_{21k0 ms},c^{\dagger}_{3  1k0ms},c^{\dagger}_{41k0ms},\bar{f}^{\dagger}_{11kms},\bar{f}^{\dagger}_{21kms} \right)_\alpha.
\end{eqnarray}
We now define an operator 

\begin{eqnarray}\label{eq:hat h plus 1}
\hat{h}^{+1}_{\alpha,\alpha'} = \left(\begin{array}{cccccc}
0&0&-i\sqrt{2}\frac{v_{\ast}}{\ell}\hat{a}&0&\gamma \Sigma(\hat{a}^{\dagger}\hat{a})&i\sqrt{2}\frac{v'_{\ast}}{\ell}\hat{a}^{\dagger} \Sigma(\hat{a}^{\dagger}\hat{a})  \\
0&0&0&i\sqrt{2}\frac{v_{\ast}}{\ell}\hat{a}^{\dagger}&-i\sqrt{2}\frac{v'_{\ast}}{\ell}\hat{a} \Sigma(\hat{a}^{\dagger}\hat{a})  &\gamma \Sigma(\hat{a}^{\dagger}\hat{a}) \\
i\sqrt{2}\frac{v_{\ast}}{\ell}\hat{a}^{\dagger}&0&-\frac{J}{2}&M&0&0 \\0&-i\sqrt{2}\frac{v_{\ast}}{\ell}\hat{a}&M&-\frac{J}{2}&0&0\\
\gamma \Sigma(\hat{a}^{\dagger}\hat{a})    & i\sqrt{2}\frac{v'_{\ast}}{\ell}\Sigma(\hat{a}^{\dagger}\hat{a})\hat{a}^{\dagger}&0&0&-\frac{U_1}{2}&0\\
-i\sqrt{2}\frac{v'_{\ast}}{\ell}\Sigma(\hat{a}^{\dagger}\hat{a})\hat{a} &  \gamma \Sigma(\hat{a}^{\dagger}\hat{a})&0&0&0&-\frac{U_1}{2}
\end{array}
\right)_{\alpha,\alpha'},
\end{eqnarray}

where $\hat{a}$ is a simple h.o. lowering operator in terms of which
matrix $\Xi_{m\alpha,m'\alpha'}$
can be expressed as
\begin{equation}
\Xi_{m\alpha,m'\alpha'} = \langle m|\hat{h}^{+1}_{\alpha,\alpha'} |m'\rangle.
\end{equation}
Here $|m\rangle$ is a simple h.o. eigenstate and $\Sigma(m)=\Sigma_m$. The non-interacting Hofstadter spectrum can be obtained by solving Eq.(\ref{eq:hat h plus 1}) after setting the mean field terms $J,U_1$ to zero. The spectrum for $M\neq 0$ including the $2q-(2m_{max}+3$ zero modes (contributed by decoupled $f$s) is shown in Fig.(\ref{fig:Non-Interacting with M}) with $2q$ modes at each $k$, i.e. 2 modes per moir\'e unit cell per valley for each spin projection.  

We now discuss the exact solutions to the eigenstates of the operator in Eq.(\ref{eq:hat h plus 1}) in the flat band limit $M=0$. The $\bB$ field independent $-J/2$ Landau level energy shown in main-text Fig.(1) comes from the anomalous $c$-mode 
\begin{equation}\label{decoupled-c state}
\theta_1=\left[0,0,\ket{0},0,0,0\right]^{T}.
\end{equation} 
The rest of the problem can be solved using the following ans\"atze:
\begin{eqnarray}
 \theta_3 &=& \left[c^{(3)}_1\ket{0},0, c^{(3)}_3\ket{1},  0,c^{(3)}_5\ket{0},0\right]^T, \label{the ansatz 3}  \\
 \theta_5 &=& \left[c^{(5)}_1\ket{1}, c^{(5)}_2\ket{0},c^{(5)}_3\ket{2},0,c^{(5)}_5\ket{1},c^{(5)}_6\ket{0}\right]^{T},
 \label{the ansatz 5}
 \end{eqnarray}
 \begin{eqnarray}
 \theta_{6_m} &=&\left[c^{(6_m)}_{1}\ket{m},   c^{(6_m)}_{2}\ket{m-1}, c^{(6_m)}_{3}\ket{m+1}, c^{(6_m)}_{4}\ket{m-2}, c^{(6_m)}_{5}\ket{m}, c^{(6_m)}_{6}\ket{m-1}\right]^{T}, \label{the ansatz 6}
\end{eqnarray}
where $m\in\{2,\ldots,m_{max}+1\}$. $c^{(\beta)}_{\alpha}$ denotes the coefficient of corresponding h.o state at index $\alpha$ in the 6-component spinor in Eq.(\ref{6-spinor}) and $\beta$ labels the ansatz index $\theta_{\beta}$. Using the above, we can set up the eigen-equation and solve for the corresponding coefficients.  

The ans\"atze $\theta_3$ and $\theta_5$ yield the $3\times 3$ and $5\times 5$ Hermitian matrices, whose eigenvectors are $c^{(3)}_\alpha$ and $c^{(5)}_\alpha$, respectively:
\begin{eqnarray}\label{3x3}
&&h^{+1}_3=\left(\begin{array}{ccc}
0  & -i\frac{\sqrt{2}v_*}{\ell} & \gamma \Sigma_0 \\
i\frac{\sqrt{2}v_*}{\ell} & -\frac{J}{2} & 0\\
\gamma \Sigma_0 & 0 & -\frac{U_1}{2}
\end{array}
\right),
\end{eqnarray}

\begin{equation}\label{5x5}
h^{+1}_5 = \left(\begin{array}{ccccc}
0 & 0 & -i\frac{2v_*}{\ell} & \gamma \Sigma_1 & i\frac{\sqrt{2}v_*'}{\ell} \Sigma_0  \\
0 & 0 & 0 &-i\frac{\sqrt{2}v_*'}{\ell}\Sigma_1 & \gamma \Sigma_0 \\
i\frac{2v_*}{\ell} & 0 & -\frac{J}{2} & 0 & 0\\
\gamma \Sigma_1& i\frac{\sqrt{2}v_*'}{\ell}\Sigma_1& 0 & -\frac{U_1}{2} & 0\\
-i\frac{\sqrt{2}v_*'}{\ell}\Sigma_0 & \gamma \Sigma_0 & 0 & 0& -\frac{U_1}{2} 
\end{array}
\right).
\end{equation}

Similarly, the ansatz $\theta^{m}_6$ yields the following $6\times 6$ Hermitian matrix for each $m$, whose eigenvectors are $c^{(6,m)}_{\alpha}$: 
\begin{eqnarray}\label{decoupledblock66}
h^{+1,m}_6&=&\left(\begin{array}{cccccc}
0 & 0 & -i\sqrt{2m+2}\frac{v_*}{\ell} & 0 &\gamma \Sigma_m & i\sqrt{2m}\frac{v_*'}{\ell} \Sigma_{m-1}  \\
0 & 0 & 0 & i\sqrt{2m-2}\frac{v_*}{\ell} & -i\sqrt{2m}\frac{v_*'}{\ell}\Sigma_m &\gamma \Sigma_{m-1}\\
+i\sqrt{2m+2}\frac{v_*}{\ell} & 0 & -\frac{J}{2} & 0& 0& 0\\
0 & -i\sqrt{2m-2}\frac{v_*}{\ell} & 0 & -\frac{J}{2} & 0 & 0\\
\gamma \Sigma_m & i\sqrt{2m}\frac{v_*'}{\ell}\Sigma_m  & 0  & 0 & -\frac{ U_1}{2} & 0\\
-i\sqrt{2m}\frac{v_*'}{\ell}\Sigma_{m-1} & \gamma \Sigma_{m-1} & 0 & 0 & 0& -\frac{U_1}{2} 
\end{array}
\right).
\end{eqnarray}
The magnetic subbands within the narrow bands from the coupled modes emanate out of the $\bB\rightarrow 0$ energy eigenvalue of the above decoupled matrices, $-\frac{J}{2}$, which is $2$ fold degenerate for matrix in Eq.(\ref{decoupledblock66})$\forall m$ and singly degenerate for matrices in Eq.(\ref{3x3}) and Eq.(\ref{5x5}). Including the decoupled $c$ mode, we have $2m_{max}+3$ magnetic modes emanating out of this $2m_{max}+3$ fold degenerate $\bB\rightarrow 0$ energy eigenvalue. Now recall that we have $2q-(2m_{max}+3)$ decoupled $f$ modes with energy $-\frac{U_1}{2}$. Thus in total we have $2q$ magnetic modes within the narrow bands, which corresponds to 2 states per moir\'e unit cell per spin. The spectrum for flat band limit has been shown in the main text Fig.(2). Note that the $\bB\rightarrow 0$ energies recovered by the decoupled matrices are the corresponding zero field energies of THFM at $\Gamma$ in mBZ. 
\subsection{\texorpdfstring{$\nu=\pm 1$}{nu=pm 1}, Spin \texorpdfstring{$\uparrow$}{uparrow}}\label{apdx:K nu=-1 spin-up}
Here we discuss the finite $\bB$ THFM for valley $\bK$ spin $\uparrow$. The interactions read as \cite{song2022magic} 
\begin{eqnarray}
V^{\tau=+1,s=\uparrow}_{\nu=\pm 1} &=& \nu \sum_{k}\bigg( \sum_{a=1}^{4}\sum_{m=0}^{m_{a,+1}} \sum_{r=0}^{p-1}W_{a}c^{\dagger}_{a1krm\uparrow}c_{a1krm\uparrow} + \sum_{a=3,4}\sum_{m=0}^{m_{a,-1}}\sum_{r=0}^{p-1}\frac{J}{2} c^{\dagger}_{a1krm\uparrow} c_{a1krm\uparrow} \nonumber \\
&+& \sum_{b=1,2}\sum_{r'=0}^{q-1}\left(\frac{3U_1}{2} + 6U_2\right)f^{\dagger}_{b1kr'\uparrow}f_{b1kr'\uparrow} \bigg).
\end{eqnarray}
The interaction for $f$ modes in the $\bar{f}$ basis can then be given as
\begin{eqnarray}
V^{f,\tau=+1,\uparrow}_{\nu=\pm1} &=& \sum_{k\in[0,1)\otimes [0,\frac{1}{q})}V^{f,\tau=+1,\uparrow,\nu=\pm 1}_{coupled} + V^{f,\tau=+1,\uparrow,\nu=\pm 1}_{decoupled}
\end{eqnarray}
where
\begin{eqnarray}
&& V^{f,\tau=+1,\uparrow,\nu=\pm 1}_{coupled} = \nu\left(\frac{3U_1}{2} + 6U_2 \right) \left(\sum_{m=0}^{m_{max}+1}\bar{f}^{\dagger}_{11kms}\bar{f}_{11km\uparrow} + \sum_{m=0}^{m_{max}}\bar{f}^{\dagger}_{21km\uparrow}\bar{f}_{21km\uparrow}\right),\\
&& V^{f,\tau=+1,s,\nu= \pm 1}_{decoupled}= \nu\sum_{b=1}^{2}\sum_{m'=m_{max}+\bar{b}}^{q-1} \left(\frac{3U_1}{2} + 6U_2\right) \bar{f}^{\dagger}_{b1km'\uparrow}\bar{f}_{b1km'\uparrow}.\end{eqnarray}
where $\bar{1}(\bar{2})=2(1)$. Yet again there are $2q-(2m_{max}+3)$ decoupled $f$ modes for each $k$. Physically this corresponds to $2-(2m_{max}+3)/q$ per moir\'e unit cell. The coupled modes can then be described by
\begin{equation}\label{xi-hamiltonian-1,spin up}
H^{\tau=+1,\uparrow,\nu=\pm 1}_{coupled} = \sum_k\sum_{\alpha,\alpha'=1}^6
\sum_{m=0}^{m_\alpha}
\sum_{m'=0}^{m_{\alpha'}}
\Xi^{\uparrow,\nu=\pm 1}_{m\alpha,m'\alpha'}d^{\dagger}_{m\alpha \uparrow}(k)d_{m'\alpha' \uparrow}(k),
\end{equation}
where $m_{\alpha=1,\ldots, 4}=m_{\alpha,+1}$, $m_{5}=m_{max}+1$ and $m_6=m_{max}$, and
\begin{equation}\label{6-spinor-1,spin up}
d^{\dagger}_{m\alpha \uparrow}(k) = \left(c^{\dagger}_{11k0m\uparrow},c^{\dagger}_{21k0 m\uparrow},c^{\dagger}_{3  1k0m\uparrow},c^{\dagger}_{41k0m\uparrow},\bar{f}^{\dagger}_{11km\uparrow},\bar{f}^{\dagger}_{21km\uparrow} \right)_\alpha,
\end{equation}
with
\begin{eqnarray}
\Xi^{\uparrow,\nu=\pm 1}_{m\alpha,m'\alpha'} = \langle m|\hat{h}^{+1,\uparrow,\nu=\pm 1}_{\alpha,\alpha'} |m'\rangle,
\end{eqnarray}
where the operators $\hat{h}^{+1,\uparrow,\nu=\pm 1}_{\alpha,\alpha'}$ are given as
\begin{eqnarray}\label{eq:-1 spin up}
\hat{h}^{+1,\uparrow,\nu=\pm 1}_{\alpha,\alpha'} = \left(\begin{array}{cccccc}
\nu W_1&0&-i\sqrt{2}\frac{v_{\ast}}{\ell}\hat{a}&0&\gamma \Sigma(\hat{a}^{\dagger}\hat{a})&i\sqrt{2}\frac{v'_{\ast}}{\ell}\hat{a}^{\dagger} \Sigma(\hat{a}^{\dagger}\hat{a})  \\
0& \nu W_1&0&i\sqrt{2}\frac{v_{\ast}}{\ell}\hat{a}^{\dagger}&-i\sqrt{2}\frac{v'_{\ast}}{\ell}\hat{a} \Sigma(\hat{a}^{\dagger}\hat{a})  &\gamma \Sigma(\hat{a}^{\dagger}\hat{a}) \\
i\sqrt{2}\frac{v_{\ast}}{\ell}\hat{a}^{\dagger}&0&\nu (W_3 + \frac{J}{2})&M&0&0 \\0&-i\sqrt{2}\frac{v_{\ast}}{\ell}\hat{a}&M&\nu (W_3 + \frac{J}{2})&0&0\\
\gamma \Sigma(\hat{a}^{\dagger}\hat{a})    & i\sqrt{2}\frac{v'_{\ast}}{\ell}\Sigma(\hat{a}^{\dagger}\hat{a})\hat{a}^{\dagger}&0&0&\nu(\frac{3U_1}{2} + 6U_2)&0\\
-i\sqrt{2}\frac{v'_{\ast}}{\ell}\Sigma(\hat{a}^{\dagger}\hat{a})\hat{a} &  \gamma \Sigma(\hat{a}^{\dagger}\hat{a})&0&0&0& \nu(\frac{3U_1}{2} + 6U_2)
\end{array}
\right)_{\alpha,\alpha'},
\end{eqnarray}
where $\hat{a}$ is a simple h.o. lowering operator with $|m\rangle$ being a simple h.o. eigenstate and $\Sigma(m)=\Sigma_m$. The exact eigenstates for the above operator are exactly solvable in flat band limit, $M=0$.
The $\bB$ field independent level $\nu(W_3+J/2)$ is formed by the anomalous $c$-mode in Eq.(\ref{decoupled-c state}). The rest of the problem can be solved using the ans\"atze: Eq.(\ref{the ansatz 3})-Eq.(\ref{the ansatz 6}).
The corresponding coefficients $c^{(3)}_{\alpha}$, $c^{(5)}_{\alpha}$ and $c^{(6,m)}_{\alpha}$ can thus be solved as eigenvectors of the following $3\times 3$, $5\times 5$ and $m_{max}$ $6\times 6$ Hermitian matrices respectively:
\begin{eqnarray}\label{3x3,-1, K spin up}
&&h^{+1, \nu=\pm1}_3=\left(\begin{array}{ccc}
\nu W_1  & -i\frac{\sqrt{2}v_*}{\ell} & \gamma \Sigma_0 \\
i\frac{\sqrt{2}v_*}{\ell} & \nu(W_3 + \frac{J}{2}) & 0\\
\gamma \Sigma_0 & 0 & \nu(\frac{3}{2}U_1+6U_2)
\end{array}
\right),
\end{eqnarray}
\begin{equation}\label{5x5,-1, K spin up}
h^{+1,\nu=\pm 1}_5 = \left(\begin{array}{ccccc}
\nu W_1 & 0 & -i\frac{2v_*}{\ell} & \gamma \Sigma_1 & i\frac{\sqrt{2}v_*'}{\ell} \Sigma_0  \\
0 & \nu W_1 & 0 &-i\frac{\sqrt{2}v_*'}{\ell}\Sigma_1 & \gamma \Sigma_0 \\
i\frac{2v_*}{\ell} & 0 & \nu(W_3+\frac{J}{2}) & 0 & 0\\
\gamma \Sigma_1& i\frac{\sqrt{2}v_*'}{\ell}\Sigma_1& 0 & \nu(\frac{3U_1}{2}+6U_2) & 0\\
-i\frac{\sqrt{2}v_*'}{\ell}\Sigma_0 & \gamma \Sigma_0 & 0 & 0& \nu(\frac{3U_1}{2}+6U_2) 
\end{array}
\right),
\end{equation}
\begin{eqnarray}\label{decoupledblock66,-1, K spin up}
h^{+1,m,\nu=\pm 1}_6&=&\left(\begin{array}{cccccc}
\nu W_1 & 0 & -i\sqrt{2m+2}\frac{v_*}{\ell} & 0 &\gamma \Sigma_m & i\sqrt{2m}\frac{v_*'}{\ell} \Sigma_{m-1}  \\
0 & \nu W1 & 0 & i\sqrt{2m-2}\frac{v_*}{\ell} & -i\sqrt{2m}\frac{v_*'}{\ell}\Sigma_m &\gamma \Sigma_{m-1}\\
+i\sqrt{2m+2}\frac{v_*}{\ell} & 0 & \nu(W_3+\frac{J}{2}) & 0& 0& 0\\
0 & -i\sqrt{2m-2}\frac{v_*}{\ell} & 0 & \nu(W_3+\frac{J}{2}) & 0 & 0\\
\gamma \Sigma_m & i\sqrt{2m}\frac{v_*'}{\ell}\Sigma_m  & 0  & 0 & \nu(\frac{3U_1}{2}+6U_2) & 0\\
-i\sqrt{2m}\frac{v_*'}{\ell}\Sigma_{m-1} & \gamma \Sigma_{m-1} & 0 & 0 & 0& \nu(\frac{3U_1}{2}+6U_2) 
\end{array}
\right).
\end{eqnarray}
The magnetic subbands within the narrow bands from the coupled modes emanate out of the $\bB\rightarrow 0$ energy eigenvalue of the above decoupled matrices, $\nu(W_3+\frac{J}{2})$, which is $2$ fold degenerate for matrix in Eq.(\ref{decoupledblock66,-1, K spin up})$\forall m$ and singly degenerate for matrices in Eq.(\ref{3x3,-1, K spin up}) and Eq.(\ref{5x5,-1, K spin up}). Including the decoupled $c$ mode, we have $2m_{max}+3$ magnetic modes emanating out of this $\bB\rightarrow 0$ energy eigenvalue. Now recall that we have $2q-(2m_{max}+3)$ decoupled $f$ modes with energy $\nu(\frac{3U_1}{2}+6U_2)$. Thus in total we have $2q$ magnetic modes within the narrow bands, which corresponds to 2 states per moir\'e unit cell. The spectrum for flat band limit has been shown in Supplementary Fig.(\ref{fig:nu=-1 for Supp}b). Note that the $\bB\rightarrow 0$ energies recovered by the decoupled matrices are the corresponding zero field energies of THFM at $\Gamma$ in mBZ.

\subsection{\texorpdfstring{$\nu=\pm 1$}{nu=pm 1}, Spin \texorpdfstring{$\downarrow$}{downarrow}}\label{apdx:K nu=-1 spin-down}
Here we discuss the finite $\bB$ THFM at $\nu=\pm 1$ for $\tau=+1$ and spin $\downarrow$. The interactions read as\cite{song2022magic} 
\begin{eqnarray}
&&V^{\tau=+1,\downarrow}_{\nu=\pm 1} = \nu \sum_{k}\bigg( \sum_{a=1}^{4}\sum_{m=0}^{m_{a,+1}} \sum_{r=0}^{p-1}W_{a}c^{\dagger}_{a1krm\downarrow}c_{a1krm\downarrow}  +\sum_{a=3,4}\sum_{m=0}^{m_{a,+1}}\sum_{r=0}^{p-1}(-1)^{a+1}\frac{J}{2} c^{\dagger}_{a1krm\downarrow} c_{a1krm\downarrow} \nonumber \\
&&+ \sum_{b=1,2}\sum_{r'=0}^{q-1}\left(\frac{2+(-1)^{b+1}}{2}U_1 + 6U_2 \right)f^{\dagger}_{b1kr'\downarrow}f_{b1kr'\downarrow} \bigg).
\end{eqnarray}
Here $W_{a\in \{1\ldots 4\}}$ and $U_2$ are mean field coefficients with $W_1=W_2$ and $W_3=W_4$ \cite{song2022magic}.
In the $\bar{f}$ basis we can re-write the interaction for $f$-fermion modes as
\begin{eqnarray}
V^{f,\tau=+1,\downarrow}_{\nu=\pm1} &=& \sum_{k\in[0,1)\otimes [0,\frac{1}{q})}V^{f,\tau=+1,\downarrow,\nu=\pm 1}_{coupled} + V^{f,\tau=+1,\downarrow,\nu=\pm 1}_{decoupled}
\end{eqnarray}
where
\begin{eqnarray}
&& V^{f,\tau=+1,\downarrow,\nu=\pm 1}_{coupled} = \nu\left(\frac{3}{2}U_1 + 6U_2 \right) \left(\sum_{m=0}^{m_{max}+1}\bar{f}^{\dagger}_{11km\downarrow}\bar{f}_{11km\downarrow} \right)+ \nu\left(\frac{1}{2}U_1 + 6U_2 \right)\left(\sum_{m=0}^{m_{max}}\bar{f}^{\dagger}_{21kms}\bar{f}_{21km\downarrow}\right), \\
&& V^{f,\tau=+1,s,\nu= \pm 2}_{decoupled}= \nu\sum_{b=1}^{2}\sum_{m'=m_{max}+\bar{b}}^{q-1} \left(\frac{2+(-1)^{b+1}}{2}U_1 + 6U_2\right) \bar{f}^{\dagger}_{b1km'\downarrow}\bar{f}_{b1km'\downarrow},
\end{eqnarray}
where $\bar{1}(\bar{2})=2,1$. Note that out of the available $2q$ $f$ modes, $q-(m_{max}+2)$ are decoupled with energy $\nu \left(\frac{3}{2}U_1+ 6U_2\right)$ and $q-(m_{max}+1)$ are decoupled with energy $\nu \left(\frac{1}{2}U_1+6U_2\right)$, i.e. a total of $2-(2m_{max}+3)/q$ decoupled $f$ modes per moir\'e unit cell. 
The coupled modes can then be described by
\begin{equation}\label{xi-hamiltonian-1}
H^{\tau=+1,\downarrow,\nu=\pm 1}_{coupled} = \sum_k\sum_{\alpha,\alpha'=1}^6
\sum_{m=0}^{m_\alpha}
\sum_{m'=0}^{m_{\alpha'}}
\Xi^{\downarrow,\nu=\pm 1}_{m\alpha,m'\alpha'}d^{\dagger}_{m\alpha \downarrow}(k)d_{m'\alpha' \downarrow}(k),
\end{equation}
where $m_{\alpha=1,\ldots, 4}=m_{\alpha,+1}$, $m_{5}=m_{max}+1$ and $m_6=m_{max}$, and
\begin{equation}\label{6-spinor-1}
d^{\dagger}_{m\alpha \downarrow}(k) = \left(c^{\dagger}_{11k0m\downarrow},c^{\dagger}_{21k0 m\downarrow},c^{\dagger}_{3  1k0m\downarrow},c^{\dagger}_{41k0m\downarrow},\bar{f}^{\dagger}_{11km\downarrow},\bar{f}^{\dagger}_{21km\downarrow} \right)_\alpha,
\end{equation}
with
\begin{eqnarray}
\Xi^{\downarrow,\nu=\pm 1}_{m\alpha,m'\alpha'} = \langle m|\hat{h}^{+1,\downarrow,\nu=\pm 1}_{\alpha,\alpha'} |m'\rangle,
\end{eqnarray}
where the operators $\hat{h}^{+1,\downarrow,\nu=\pm 1}_{\alpha,\alpha'}$ are given as
\begin{eqnarray}\label{eq:-1 spin down}
\hat{h}^{+1,\downarrow,\nu=\pm 1}_{\alpha,\alpha'} = \left(\begin{array}{cccccc}
\nu W_1&0&-i\sqrt{2}\frac{v_{\ast}}{\ell}\hat{a}&0&\gamma \Sigma(\hat{a}^{\dagger}\hat{a})&i\sqrt{2}\frac{v'_{\ast}}{\ell}\hat{a}^{\dagger} \Sigma(\hat{a}^{\dagger}\hat{a})  \\
0& \nu W_1&0&i\sqrt{2}\frac{v_{\ast}}{\ell}\hat{a}^{\dagger}&-i\sqrt{2}\frac{v'_{\ast}}{\ell}\hat{a} \Sigma(\hat{a}^{\dagger}\hat{a})  &\gamma \Sigma(\hat{a}^{\dagger}\hat{a}) \\
i\sqrt{2}\frac{v_{\ast}}{\ell}\hat{a}^{\dagger}&0&\nu (W_3 + \frac{J}{2})&M&0&0 \\0&-i\sqrt{2}\frac{v_{\ast}}{\ell}\hat{a}&M&\nu (W_3 - \frac{J}{2})&0&0\\
\gamma \Sigma(\hat{a}^{\dagger}\hat{a})    & i\sqrt{2}\frac{v'_{\ast}}{\ell}\Sigma(\hat{a}^{\dagger}\hat{a})\hat{a}^{\dagger}&0&0&\nu(\frac{3}{2}U_1 + 6U_2)&0\\
-i\sqrt{2}\frac{v'_{\ast}}{\ell}\Sigma(\hat{a}^{\dagger}\hat{a})\hat{a} &  \gamma \Sigma(\hat{a}^{\dagger}\hat{a})&0&0&0& \nu(\frac{1}{2}U_1 + 6U_2)
\end{array}
\right)_{\alpha,\alpha'}.
\end{eqnarray}

The eigenstates of the above operator are exactly solvable for the flat band limit, i.e. $M=0$. The decoupled $c$ fermion given in Eq.(\ref{decoupled-c state}) forms the field independent $\nu(W_3+\frac{J}{2})$ level. The remaining eigenstates can be obtained using the ans\"atze given in Eqs.(\ref{the ansatz 3})-(\ref{the ansatz 6}). Setting up eigen-equation for these ans\"atze yield us the following $3\times 3$, $5\times 5$ and $m_{max}$ $6\times 6$ matrices: 
\begin{eqnarray}\label{3x3,-1}
&&h^{+1, \nu=\pm1}_3=\left(\begin{array}{ccc}
\nu W_1  & -i\frac{\sqrt{2}v_*}{\ell} & \gamma \Sigma_0 \\
i\frac{\sqrt{2}v_*}{\ell} & \nu(W_3 + \frac{J}{2}) & 0\\
\gamma \Sigma_0 & 0 & \nu(\frac{3}{2}U_1+6U_2)
\end{array}
\right),
\end{eqnarray}
\begin{equation}\label{5x5,-1}
h^{+1,\nu=\pm 1}_5 = \left(\begin{array}{ccccc}
\nu W_1 & 0 & -i\frac{2v_*}{\ell} & \gamma \Sigma_1 & i\frac{\sqrt{2}v_*'}{\ell} \Sigma_0  \\
0 & \nu W_1 & 0 &-i\frac{\sqrt{2}v_*'}{\ell}\Sigma_1 & \gamma \Sigma_0 \\
i\frac{2v_*}{\ell} & 0 & \nu(W_3+\frac{J}{2}) & 0 & 0\\
\gamma \Sigma_1& i\frac{\sqrt{2}v_*'}{\ell}\Sigma_1& 0 & \nu(\frac{3U_1}{2}+6U_2) & 0\\
-i\frac{\sqrt{2}v_*'}{\ell}\Sigma_0 & \gamma \Sigma_0 & 0 & 0& \nu(\frac{U_1}{2}+6U_2)
\end{array}
\right),
\end{equation}
\begin{eqnarray}\label{decoupledblock66,-1}
h^{+1,m,\nu=\pm 1}_6&=&\left(\begin{array}{cccccc}
\nu W_1 & 0 & -i\sqrt{2m+2}\frac{v_*}{\ell} & 0 &\gamma \Sigma_m & i\sqrt{2m}\frac{v_*'}{\ell} \Sigma_{m-1}  \\
0 & \nu W1 & 0 & i\sqrt{2m-2}\frac{v_*}{\ell} & -i\sqrt{2m}\frac{v_*'}{\ell}\Sigma_m &\gamma \Sigma_{m-1}\\
+i\sqrt{2m+2}\frac{v_*}{\ell} & 0 & \nu(W_3+\frac{J}{2}) & 0& 0& 0\\
0 & -i\sqrt{2m-2}\frac{v_*}{\ell} & 0 & \nu(W_3-\frac{J}{2}) & 0 & 0\\
\gamma \Sigma_m & i\sqrt{2m}\frac{v_*'}{\ell}\Sigma_m  & 0  & 0 & \nu(\frac{3U_1}{2}+6U_2) & 0\\
-i\sqrt{2m}\frac{v_*'}{\ell}\Sigma_{m-1} & \gamma \Sigma_{m-1} & 0 & 0 & 0& \nu(\frac{U_1}{2}+6U_2) 
\end{array}
\right),
\end{eqnarray}
where $m\in\{2\ldots m_{max}+1\}$. The magnetic subbands within the narrow bands from the coupled modes emanate out of the $\bB\rightarrow 0$ energy eigenvalue of the above decoupled matrices, $\nu (W_3 \pm \frac{J}{2})$. $\nu (W_3 + \frac{J}{2})$ here is singly degenerate for each of the above matrices and thus in total $m_{max}+2$ fold degenerate. On the other hand $\nu (W_3 - \frac{J}{2})$ is in total $m_{max}$ fold degenerate for the matrix in Eq.(\ref{decoupledblock66,-1}) including all values of $m$. Including the anomalous $c$ mode and $2q-(2m_{max}+3)$ decoupled $f$ modes, we have a total of $2q$ magnetic modes within the narrow bands, which corresponds to 2 states per moir\'e unit cell per spin. The spectrum for flat band limit has been shown in the main text Fig.(5-b). Note that the $\bB\rightarrow 0$ energies recovered by the decoupled matrices are the corresponding zero field energies of THFM at $\Gamma$ in mBZ.

\subsection{\texorpdfstring{$\nu=\pm 2$}{nu=pm 2}, Spin \texorpdfstring{$\uparrow\downarrow$}{uparrow downarrow}}\label{apdx:nu=-2 Valley K}
In this section we discuss the finite $\bB$ THFM at $\nu=\pm 2$, Valley $\bK$ and spin $\uparrow\downarrow$ sectors. The mean field interaction at filling $\nu = \pm 2$ for VP state at $\tau=+1$ for spin sector $s$ reads\cite{song2022magic} 
\begin{figure}
\includegraphics[width=6cm]{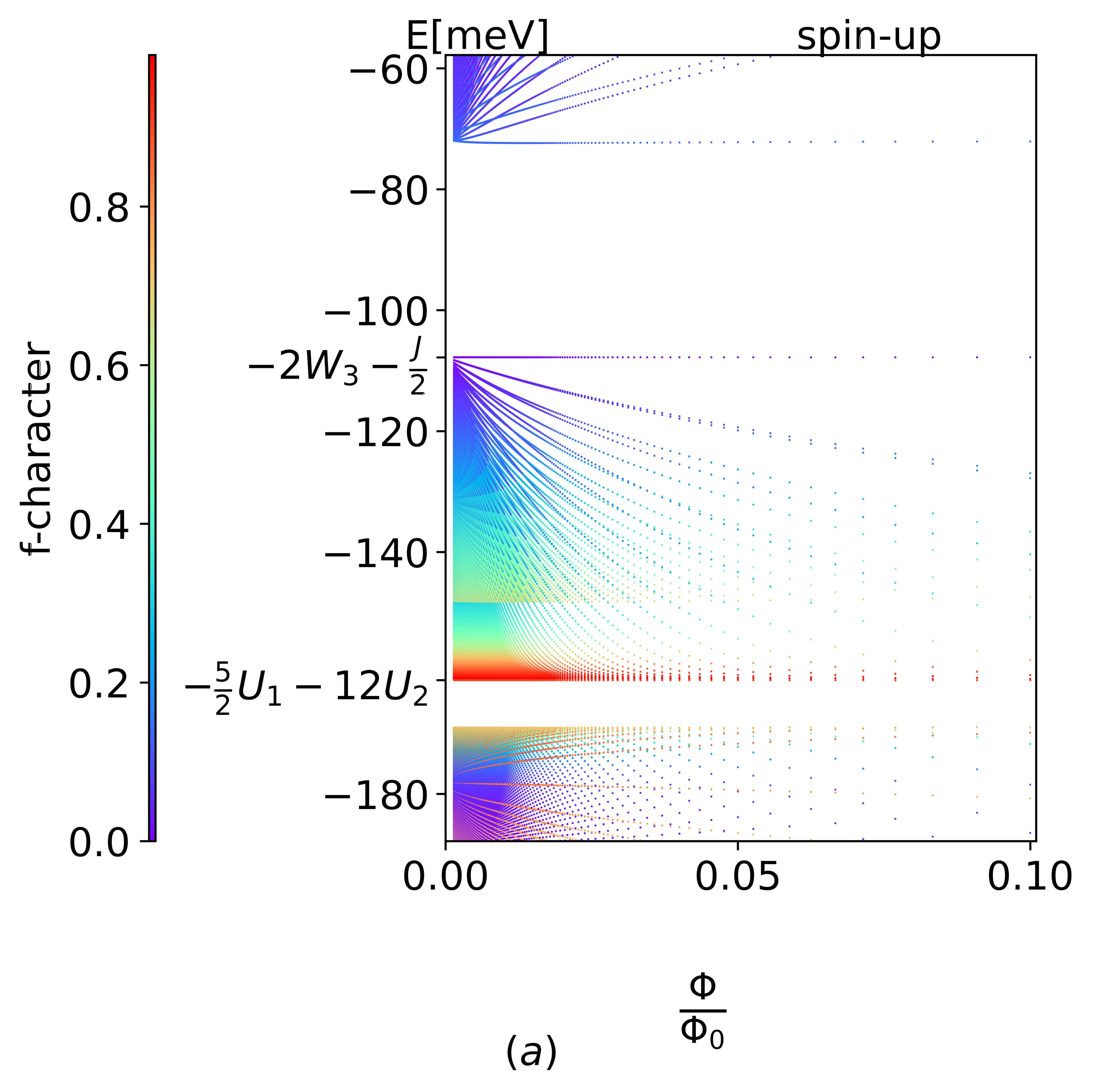}
\includegraphics[width=6cm]{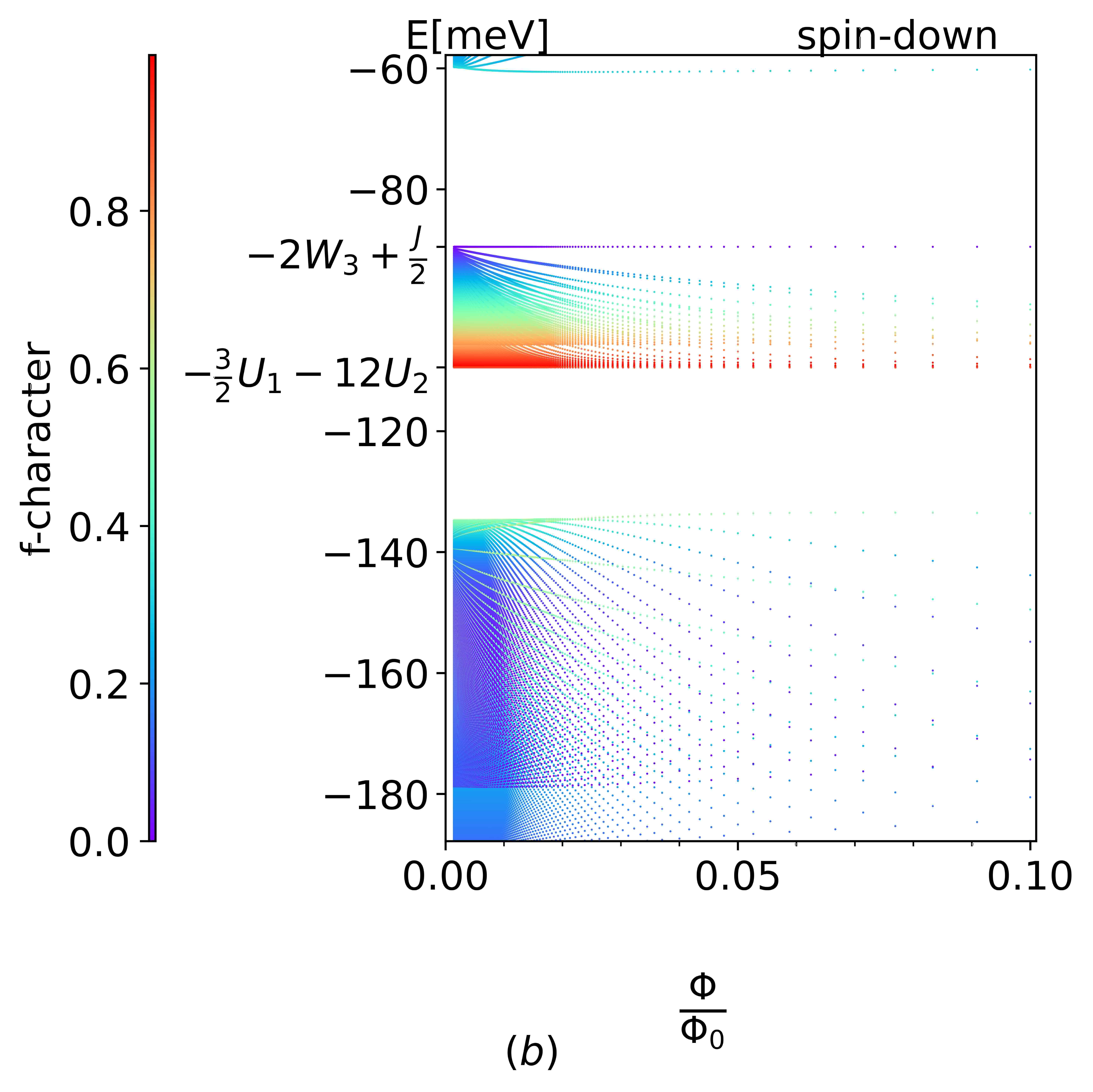}
\caption{Interacting heavy fermion Hofstadter spectra for valley $\bK$ (a) spin $\uparrow$ and (b) spin $\downarrow$ at filling $\nu=-2$ at $w_0/w_1=0.7$ in the flat band limit $M=0$.  $m_{max}=\lceil \frac{q-3}{2} \rceil$.}\label{fig:nu=-2 valley K for Supp}
\end{figure}

\begin{eqnarray}
&&V^{\tau=+1,s}_{\nu=\pm 2} = \nu \sum_{k}\bigg( \sum_{a=1}^{4}\sum_{m=0}^{m_{a,+1}} \sum_{r=0}^{p-1}W_{a}c^{\dagger}_{a1krms}c_{a1krms} +\sigma_s \sum_{a=3,4}\sum_{m=0}^{m_{a,+1}}\sum_{r=0}^{p-1}\frac{J}{4} c^{\dagger}_{a1krms} c_{a1krms} \nonumber \\
&&+ \sum_{b=1,2}\sum_{r'=0}^{q-1}\left(\frac{4+\sigma_s}{4}U_1 + 6U_2 \right)f^{\dagger}_{b1kr's}f_{b1kr's} \bigg),
\end{eqnarray}
where $\sigma_s=\pm 1$ for $s=\uparrow \downarrow$ respectively. In the $\bar{f}$ basis we can re-write the interaction for $f$-fermion modes as
\begin{eqnarray}
V^{f,\tau=+1,s}_{\nu=\pm2} &=& \sum_{k\in[0,1)\otimes [0,\frac{1}{q})}V^{f,\tau=+1,s,\nu=\pm 2}_{coupled} + V^{f,\tau=+1,s,\nu=\pm 2}_{decoupled}
\end{eqnarray}
where
\begin{eqnarray}
&& V^{f,\tau=+1,s,\nu=\pm 2}_{coupled} = \nu\left(\frac{4+\sigma_s}{4}U_1 + 6U_2 \right) \left(\sum_{m=0}^{m_{max}+1}\bar{f}^{\dagger}_{11kms}\bar{f}_{11kms} + \sum_{m=0}^{m_{max}}\bar{f}^{\dagger}_{21kms}\bar{f}_{21kms}\right), \\
&& V^{f,\tau=+1,s,\nu= \pm 2}_{decoupled}= \nu\left(\frac{4+\sigma_s}{4}U_1 + 6U_2\right) \sum_{b=1}^{2}\sum_{m'=m_{max}+\bar{b}}^{q-1}\bar{f}^{\dagger}_{b1km's}\bar{f}_{b1km's},
\end{eqnarray}
where $\bar{1}(\bar{2})=2,1$.
Note that we have $2-(2m_{max}+3)/q$ decoupled $f$ states per moir\'e unit cell for each spin projection. The coupled modes can then be described by
\begin{equation}\label{xi-hamiltonian pm 2}
H^{\tau=+1,s,\nu=\pm 2}_{coupled} = \sum_k\sum_{\alpha,\alpha'=1}^6
\sum_{m=0}^{m_\alpha}
\sum_{m'=0}^{m_{\alpha'}}
\Xi^{s,\nu=\pm 2}_{m\alpha,m'\alpha'}d^{\dagger}_{m\alpha s}(k)d_{m'\alpha' s}(k),
\end{equation}
where $m_{\alpha=1,\ldots, 4}=m_{\alpha,+1}$, $m_{5}=m_{max}+1$ and $m_6=m_{max}$, and
\begin{equation}\label{6-spinor pm 2}
d^{\dagger}_{m\alpha s}(k) = \left(c^{\dagger}_{11k0ms},c^{\dagger}_{21k0 ms},c^{\dagger}_{3  1k0ms},c^{\dagger}_{41k0ms},\bar{f}^{\dagger}_{11kms},\bar{f}^{\dagger}_{21kms} \right)_\alpha,
\end{equation}
with
\begin{eqnarray}
\Xi^{s,\nu=\pm 2}_{m\alpha,m'\alpha'} = \langle m|\hat{h}^{+1,s,\nu=\pm 2}_{\alpha,\alpha'} |m'\rangle,
\end{eqnarray}
where the operators $\hat{h}^{+1,s,\nu=\pm 2}_{\alpha,\alpha'}$ are given as

\begin{eqnarray}\label{eq:pm 2 spin updown}
\hat{h}^{+1,s,\nu=\pm 2}_{\alpha,\alpha'} = \left(\begin{array}{cccccc}
\nu W_1&0&-i\sqrt{2}\frac{v_{\ast}}{\ell}\hat{a}&0&\gamma \Sigma(\hat{a}^{\dagger}\hat{a})&i\sqrt{2}\frac{v'_{\ast}}{\ell}\hat{a}^{\dagger} \Sigma(\hat{a}^{\dagger}\hat{a})  \\
0& \nu W_1&0&i\sqrt{2}\frac{v_{\ast}}{\ell}\hat{a}^{\dagger}&-i\sqrt{2}\frac{v'_{\ast}}{\ell}\hat{a} \Sigma(\hat{a}^{\dagger}\hat{a})  &\gamma \Sigma(\hat{a}^{\dagger}\hat{a}) \\
i\sqrt{2}\frac{v_{\ast}}{\ell}\hat{a}^{\dagger}&0&\nu (W_3 + \sigma_s \frac{J}{4})&M&0&0 \\0&-i\sqrt{2}\frac{v_{\ast}}{\ell}\hat{a}&M&\nu (W_3 + \sigma_s \frac{J}{4})&0&0\\
\gamma \Sigma(\hat{a}^{\dagger}\hat{a})    & i\sqrt{2}\frac{v'_{\ast}}{\ell}\Sigma(\hat{a}^{\dagger}\hat{a})\hat{a}^{\dagger}&0&0&\nu(\frac{4+\sigma_s}{4}U_1 + 6U_2)&0\\
-i\sqrt{2}\frac{v'_{\ast}}{\ell}\Sigma(\hat{a}^{\dagger}\hat{a})\hat{a} &  \gamma \Sigma(\hat{a}^{\dagger}\hat{a})&0&0&0& \nu(\frac{4+\sigma_s}{4}U_1 + 6U_2)
\end{array}
\right)_{\alpha,\alpha'},
\end{eqnarray}
where $\hat{a}$ is a simple h.o. lowering operator, $|m\rangle$ is a simple h.o. eigenstate and $\Sigma(m)=\Sigma_m$. The exact eigenstates for the above operator are exactly solvable in flat band limit, $M=0$. The field independent $\pm(2W_3 + \frac{J}{2})$ and $\pm(2W_3 - \frac{J}{2})$ levels at fillings $\nu = \pm 2$ and $\tau=+1$ for spin $\uparrow \downarrow$ sectors respectively is formed by the anomalous $c$-mode 
in Eq.(\ref{decoupled-c state}). The rest of the problem can be solved using the ans\"atze: Eq.(\ref{the ansatz 3})-Eq.(\ref{the ansatz 6}).
The corresponding coefficients $c^{(3)}_{\alpha}$, $c^{(5)}_{\alpha}$ and $c^{(6,m)}_{\alpha}$ can thus be solved as eigenvectors of the following $3\times 3$, $5\times 5$ and $m_{max}$ $6\times 6$ Hermitian matrices respectively:
\begin{eqnarray}\label{3x3,-2, K}
&&h^{+1,s,\nu=\pm 2}_3=\left(\begin{array}{ccc}
\nu W_1  & -i\frac{\sqrt{2}v_*}{\ell} & \gamma \Sigma_0 \\
i\frac{\sqrt{2}v_*}{\ell} & \nu (W_3 + \sigma_s \frac{J}{4}) & 0\\
\gamma \Sigma_0 & 0 & \nu(\frac{4+\sigma_s}{4}U_1 + 6U_2)
\end{array}
\right),
\end{eqnarray}
\begin{equation}\label{5x5,-2, K}
h^{+1,s,\nu=\pm 2}_5 = \left(\begin{array}{ccccc}
\nu W_1 & 0 & -i\frac{2v_*}{\ell} & \gamma \Sigma_1 & i\frac{\sqrt{2}v_*'}{\ell} \Sigma_0  \\
0 & \nu W_1 & 0 &-i\frac{\sqrt{2}v_*'}{\ell}\Sigma_1 & \gamma \Sigma_0 \\
i\frac{2v_*}{\ell} & 0 & \nu (W_3 + \sigma_s \frac{J}{4}) & 0 & 0\\
\gamma \Sigma_1& i\frac{\sqrt{2}v_*'}{\ell}\Sigma_1& 0 & \nu(\frac{4+\sigma_s}{4}U_1 + 6U_2) & 0\\
-i\frac{\sqrt{2}v_*'}{\ell}\Sigma_0 & \gamma \Sigma_0 & 0 & 0& \nu(\frac{4+\sigma_s}{4}U_1 + 6U_2) 
\end{array}
\right),
\end{equation}
\begin{eqnarray}\label{decoupledblock66,-2, K}
h^{+1,m,s,\nu=\pm 2}_6&=&\left(\begin{array}{cccccc}
\nu W_1 & 0 & -i\sqrt{2m+2}\frac{v_*}{\ell} & 0 &\gamma \Sigma_m & i\sqrt{2m}\frac{v_*'}{\ell} \Sigma_{m-1}  \\
0 & \nu W1 & 0 & i\sqrt{2m-2}\frac{v_*}{\ell} & -i\sqrt{2m}\frac{v_*'}{\ell}\Sigma_m &\gamma \Sigma_{m-1}\\
+i\sqrt{2m+2}\frac{v_*}{\ell} & 0 & \nu (W_3 + \sigma_s \frac{J}{4}) & 0& 0& 0\\
0 & -i\sqrt{2m-2}\frac{v_*}{\ell} & 0 & \nu (W_3 + \sigma_s \frac{J}{4}) & 0 & 0\\
\gamma \Sigma_m & i\sqrt{2m}\frac{v_*'}{\ell}\Sigma_m  & 0  & 0 & \nu(\frac{4+\sigma_s}{4}U_1 + 6U_2) & 0\\
-i\sqrt{2m}\frac{v_*'}{\ell}\Sigma_{m-1} & \gamma \Sigma_{m-1} & 0 & 0 & 0& \nu(\frac{4+\sigma_s}{4}U_1 + 6U_2)
\end{array}
\right).
\end{eqnarray}
The magnetic subbands within the narrow bands from the coupled modes emanate out of the $\bB\rightarrow 0$ energy eigenvalue of the above decoupled matrices, $\nu (W_3 + \sigma_s \frac{J}{4})$, which is $2$ fold degenerate for matrix in Eq.(\ref{decoupledblock66,-2, K})$\forall m$ and singly degenerate for matrices in Eq.(\ref{3x3,-2, K}) and Eq.(\ref{5x5,-2, K}), for each spin projection. Including the decoupled $c$ mode, we have $2m_{max}+3$ magnetic modes emanating out of this $\bB\rightarrow 0$ energy eigenvalue. Now recall that we have $2q-(2m_{max}+3)$ decoupled $f$ modes with energy $\nu(\frac{4+\sigma_s}{4}U_1 + 6U_2)$. Thus in total we have $2q$ magnetic modes within the narrow bands, which corresponds to 2 states per moir\'e unit cell for each spin projection. The spectrum for flat band limit has been shown in Supplementary Fig.(\ref{fig:nu=-2 valley K for Supp}). Note that the $\bB\rightarrow 0$ energies recovered by the decoupled matrices are the corresponding zero field energies of THFM at $\Gamma$ in mBZ.

\section{Discussion For Valley \texorpdfstring{$\mathbf{K'}$}{tau = -1}}\label{apdx:Valley K' discussion}
In this section we discuss finite $\bB$ THFM at valley $\mathbf{K'}$, i.e $\tau=-1$. Note that summation convention on repeated indices is implied unless explicitly stated.
Using the SVD decomposition discussed we can re-write the matrix elements as 
\begin{eqnarray}
h^{-1}_{[1m0],[1r']}c^{\dagger}_{1 -1 k 0 ms}f_{1 -1 k r's} &=& 
\gamma \Sigma_{m',\bar{r}} c^\dagger_{1 -1 k 0 ms}U_{mm'} V_{\bar{r},r'}f_{1-1kr's} =\sum_{m=0}^{m_{1,-1}}\sum_{\bar{r}=0}^{m_{1,-1}}\gamma \Sigma_m \delta_{m\bar{r}} c^\dagger_{1 -1 k 0 ms} \bar{f}_{1-1k\bar{r}s},
\label{Supp:11}
\end{eqnarray}
where we used the fact that $U$ is an identity matrix and the rectangular matrix $\Sigma_{m,r'}$ is non-zero only along its main diagonal. Similarly
\begin{eqnarray}
h^{-1}_{[2m0],[1r']}c^{\dagger}_{2 -1 k 0 ms}f_{1 -1 k r's} &=&
-i \sqrt{2}\frac{v'_{\ast}}{\ell} \sum_{m=0}^{m_{2,-1}} \sqrt{m}c^{\dagger}_{2 -1 k 0 ms}  \Sigma_{m-1,\bar{r}}V_{\bar{r},r'}f_{1 -1 k r's} \nonumber \\
&=& -i\sqrt{2}\frac{v'_{\ast}}{\ell}\sum_{m=0}^{m_{2,-1}}\sum_{\bar{r}=0}^{m_{2,-1}-1}\sqrt{m}\Sigma_{\Bar{r}} \delta_{m-1,\bar{r}} c^{\dagger}_{2 -1 k 0 ms} \bar{f}_{1-1k\bar{r}s}.
\label{Supp:21}
\end{eqnarray}
Because $m_{2,-1}-1=m_{1,-1}=m_{max}$, the upper bound on the $\bar{r}$ summation is the same in Eq.(\ref{Supp:11}) and Eq.(\ref{Supp:21}). 
Similarly
\begin{eqnarray}
h^{-1}_{[1m0],[2r']}c^{\dagger}_{1 -1 k 0 ms}f_{2 -1 k r's}&=& i \sqrt{2}\frac{v'_{\ast}}{\ell} \sum_{m=0}^{m_{1,-1}} \sqrt{m+1}c^{\dagger}_{1 -1 k 0 ms}\Sigma_{m+1,\bar{r}}V_{\bar{r},r'}f_{2 -1 k r's}\nonumber \\
&=& i \sqrt{2}\frac{v'_{\ast}}{\ell} \sum_{m=0}^{m_{1,-1}}\sum_{\bar{r}=0}^{m_{1,-1}+1} \sqrt{m+1} c^{\dagger}_{1 -1 k 0 ms}\bar{f}_{2-1k\bar{r}s}\label{Supp:12}
\end{eqnarray}

\begin{eqnarray}
h^{1}_{[2m0],[2r']}c^{\dagger}_{2 -1 k 0 ms}f_{2 -1 k r's} &=& \sum_{m=0}^{m_{2,-1}}\sum_{\bar{r}=0}^{m_{2,-1}}\gamma\Sigma_{m}\delta_{m,\bar{r}} c^\dagger_{2 -1 k 0 ms}\bar{f}_{2 -1 k\bar{r}s}.\label{Supp:22}
\end{eqnarray}
Because $m_{1,-1}+1=m_{2,-1}=m_{max}+1$, the upper bound on $\bar{r}$ summation is the same in Eq.(\ref{Supp:12}) and Eq.(\ref{Supp:22}). Thus for a given $k$, out of the available $2q$ $f-$modes, only $2m_{max}+3$ couple. 
\subsection{CNP}\label{apdx:K'CNP}

The interactions in $\bar{f}$ basis at CNP \cite{song2022magic} can be given as 
\begin{eqnarray}
&&V^{f,\tau=-1,s} = \sum_{k\in[0,1)\otimes [0,\frac{1}{q})}V^{f,\tau=-1,s}_{coupled} + V^{f,\tau=-1,s}_{decoupled},  \\
&& V^{f,\tau=-1,s}_{coupled} = \frac{U_1}{2}\left(\sum_{m=0}^{m_{max}}\bar{f}^{\dagger}_{1-1kms}\bar{f}_{1-1kms} + \sum_{m=0}^{m_{max}+1}\bar{f}^{\dagger}_{2-1kms}\bar{f}_{2-1kms} \right), \nonumber \\
\\
&& V^{f,\tau=-1,s}_{decoupled}= \frac{U_1}{2}\sum_{b=1}^{2}\sum_{m'=m_{max}+b}^{q-1}\bar{f}^{\dagger}_{b-1km's}\bar{f}_{b-1km's},
\end{eqnarray}
Note that there are $2q-(2m_{max}+3)$ decoupled $f$ modes for each $k$. Physically this corresponds to $2-(2m_{max}+3)/q$ states per moir\'e unit cell. 
\begin{figure}
    \centering
    \includegraphics[width=8cm]{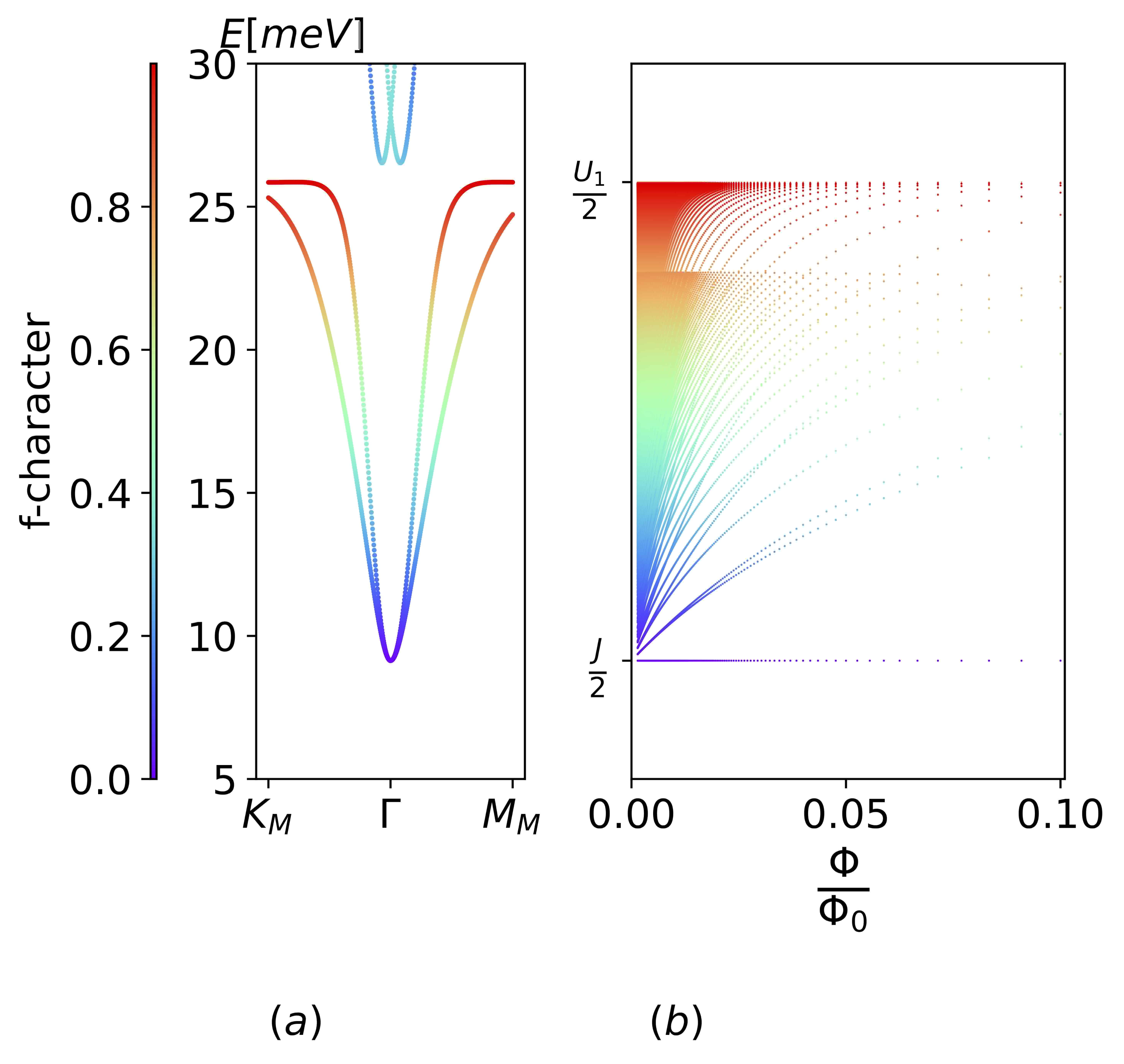}
    \caption{Interacting heavy fermion Hofstadter spectra for (b) CNP at $\bK'$ contrasted with (a) corresponding zero-field spectra at $w_0/w_1=0.7$ in the flat band limit $M=0$. $m_{max}=\lceil \frac{q-3}{2} \rceil$.}
    \label{fig:K' CNP}
\end{figure}
The coupled modes can then be described as
\begin{equation}\label{Supp:xi-hamiltonian}
H^{\tau=-1,s}_{coupled} = \sum_k\sum_{\alpha,\alpha'=1}^6
\sum_{m=0}^{m_\alpha}
\sum_{m'=0}^{m_{\alpha'}}
\bar{\Xi}_{m\alpha,m'\alpha'}\bar{d}^{\dagger}_{m\alpha s}(k)\bar{d}_{m'\alpha's}(k),
\end{equation}
where $m_{\alpha=1,\ldots, 4}=m_{\alpha,-1}$, $m_{5}=m_{max}$ and $m_6=m_{max}+1$, and
\begin{eqnarray}\label{6-spinor,K'}
\bar{d}^{\dagger}_{m\alpha s}(k) = \left(c^{\dagger}_{1-1k0ms},c^{\dagger}_{2-1k0 ms},c^{\dagger}_{3  -1k0ms},c^{\dagger}_{4-1k0ms},\bar{f}^{\dagger}_{1-1kms},\bar{f}^{\dagger}_{2-1kms} \right)_\alpha.
\end{eqnarray}
The following operator can be defined for $\tau=-1$ 
\begin{eqnarray}\label{eq:hat h -1}
\hat{h}^{-1}_{\alpha,\alpha'} = \left(\begin{array}{cccccc}
0&0&-i\sqrt{2}\frac{v_{\ast}}{\ell}\hat{a}^{\dagger}&0&\gamma \Sigma(\hat{a}^{\dagger}\hat{a})&i\sqrt{2}\frac{v'_{\ast}}{\ell}\hat{a} \Sigma(\hat{a}^{\dagger}\hat{a})  \\
0&0&0&i\sqrt{2}\frac{v_{\ast}}{\ell}\hat{a}&-i\sqrt{2}\frac{v'_{\ast}}{\ell}\hat{a}^{\dagger} \Sigma(\hat{a}^{\dagger}\hat{a})  &\gamma \Sigma(\hat{a}^{\dagger}\hat{a}) \\
i\sqrt{2}\frac{v_{\ast}}{\ell}\hat{a}&0&\frac{J}{2}&M&0&0 \\0&-i\sqrt{2}\frac{v_{\ast}}{\ell}\hat{a}^{\dagger}&M&\frac{J}{2}&0&0\\
\gamma \Sigma(\hat{a}^{\dagger}\hat{a})    & i\sqrt{2}\frac{v'_{\ast}}{\ell}\Sigma(\hat{a}^{\dagger}\hat{a})\hat{a}&0&0&\frac{U_1}{2}&0\\
-i\sqrt{2}\frac{v'_{\ast}}{\ell}\Sigma(\hat{a}^{\dagger}\hat{a})\hat{a}^{\dagger} &  \gamma \Sigma(\hat{a}^{\dagger}\hat{a})&0&0&0&\frac{U_1}{2}
\end{array}
\right)_{\alpha,\alpha'},
\end{eqnarray}
where $\hat{a}$ is a simple h.o. lowering operator in terms of which
matrix $\bar{\Xi}_{m\alpha,m'\alpha'}$
can be expressed as
\begin{equation}
\bar{\Xi}_{m\alpha,m'\alpha'} = \langle m|\hat{h}^{-1}_{\alpha,\alpha'} |m'\rangle.
\end{equation}
Here $|m\rangle$ is a simple h.o. eigenstate and $\Sigma(m)=\Sigma_m$. We now discuss the the exact solutions to the eigenstates of the operator in Eq.(\ref{eq:hat h -1}) in flat band limit, $M=0$.  The $\bB$ field independent $J/2$ Landau level comes from the anomalous $c$-mode 
\begin{equation}\label{anam c at K'}
\theta_1=\left[0,0,0,\ket{0},0,0\right]^{T}.
\end{equation} 
The rest of the problem can be solved using the following ans\"atze:
\begin{eqnarray}
 \bar{\theta}_3 &=& \left[0,c^{(3)}_2\ket{0},0, c^{(3)}_4\ket{1},  0,c^{(3)}_6\ket{0}\right]^T, \label{ansatz 3 K'}  \\
 \bar{\theta}_5 &=& \left[c^{(5)}_1\ket{0}, c^{(5)}_2\ket{1},0,c^{(5)}_4\ket{2},c^{(5)}_5\ket{0},c^{(5)}_6\ket{1}\right]^{T}, \label{ansatz 5 K'}\nonumber\\
 \\
 \bar{\theta}_{6_m} &=&\left[c^{(6_m)}_{1}\ket{m-1},   c^{(6_m)}_{2}\ket{m}, c^{(6_m)}_{3}\ket{m-2},c^{(6_m)}_{4}\ket{m+1}, c^{(6_m)}_{5}\ket{m-1}, c^{(6_m)}_{6}\ket{m}\right]^{T},\label{ansatz 6 K'}
\end{eqnarray}
where $m\in\{2,\ldots,m_{max}+1\}$. $c^{(\beta)}_{\alpha}$ denotes the coefficient of corresponding h.o state at index $\alpha$ in the 6-component spinor in Eq.(\ref{6-spinor,K'}) and $\beta$ labels the ansatz index $\theta_{\beta}$. Using the above, we can set up the eigen-equation and solve for the corresponding coefficients. 

The ans\"atze $\bar{\theta}_3$ and $\bar{\theta}_5$ yield the $3\times 3$ and $5\times 5$ Hermitian matrices, whose eigenvectors are $c^{(3)}_\alpha$ and $c^{(5)}_\alpha$, respectively:
\begin{eqnarray}\label{3x3,K'}
&&h^{-1}_3=\left(\begin{array}{ccc}
0  & i\frac{\sqrt{2}v_*}{\ell} & \gamma \Sigma_0 \\
-i\frac{\sqrt{2}v_*}{\ell} & \frac{J}{2} & 0\\
\gamma \Sigma_0 & 0 & \frac{U_1}{2}
\end{array}
\right),
\end{eqnarray}

\begin{equation}\label{5x5,K'}
h^{-1}_5 = \left(\begin{array}{ccccc}
0 & 0&0& \gamma\Sigma_0& i\frac{\sqrt{2}v_*'}{\ell}\Sigma_1 \\
0 & 0 & i\frac{2v_{\ast}}{\ell}&-i\frac{\sqrt{2}v_*'}{\ell}\Sigma_0 & \gamma \Sigma_1 \\
0&-i\frac{2v_*}{\ell} & \frac{J}{2} & 0& 0\\
\gamma \Sigma_0& i\frac{\sqrt{2}v_*'}{\ell}\Sigma_0& 0 & \frac{U_1}{2} & 0\\
-i\frac{\sqrt{2}v_*'}{\ell}\Sigma_1 & \gamma \Sigma_1 & 0 & 0& \frac{U_1}{2} 
\end{array}
\right).
\end{equation}

Similarly, the ansatz $\bar{\theta}^{m}_6$ yields the following $6\times 6$ Hermitian matrix for each $m$, whose eigenvectors are $c^{(6,m)}_{\alpha}$: 
\begin{eqnarray}\label{decoupledblock66,K'}
h^{-1,m}_6&=&\left(\begin{array}{cccccc}
0 & 0 & -i\sqrt{2m-2}\frac{v_*}{\ell} & 0 &\gamma \Sigma_{m-1} & i\sqrt{2m}\frac{v_*'}{\ell} \Sigma_{m}  \\
0 & 0 & 0 & i\sqrt{2m+2}\frac{v_*}{\ell} & -i\sqrt{2m}\frac{v_*'}{\ell}\Sigma_{m-1} &\gamma \Sigma_{m}\\
+i\sqrt{2m-2}\frac{v_*}{\ell} & 0 & \frac{J}{2} & 0& 0& 0\\
0 & -i\sqrt{2m+2}\frac{v_*}{\ell} & 0 & \frac{J}{2} & 0 & 0\\
\gamma \Sigma_{m-1} & i\sqrt{2m}\frac{v_*'}{\ell}\Sigma_{m-1}  & 0  & 0 & \frac{ U_1}{2} & 0\\
-i\sqrt{2m}\frac{v_*'}{\ell}\Sigma_{m} & \gamma \Sigma_{m} & 0 & 0 & 0& \frac{U_1}{2} 
\end{array}
\right).
\end{eqnarray}
The magnetic subbands within the narrow bands from the coupled modes emanate out of the $\bB\rightarrow 0$ energy eigenvalue of the above decoupled matrices, $\frac{J}{2}$, which is $2$ fold degenerate for matrix in Eq.(\ref{decoupledblock66,K'})$\forall m$ and singly degenerate for matrices in Eq.(\ref{3x3,K'}) and Eq.(\ref{5x5,K'}). Including the decoupled $c$ mode, we have $2m_{max}+3$ magnetic modes emanating out of this $\bB\rightarrow 0$ energy eigenvalue. Now recall that we have $2q-(2m_{max}+3)$ decoupled $f$ modes with energy $\frac{U_1}{2}$. Thus in total we have $2q$ magnetic modes within the narrow bands, which corresponds to 2 states per moir\'e unit cell per spin. The spectrum for flat band limit has been shown in Supplementary Fig.(\ref{fig:K' CNP}b). Note that the $\bB\rightarrow 0$ energies recovered by the decoupled matrices are the corresponding zero field energies of THFM at $\Gamma$ in mBZ.
\subsection{\texorpdfstring{$\nu=\pm 1$}{nu = pm 1} Spin \texorpdfstring{$\uparrow \downarrow$}{up down}}\label{apdx:K'nu=-1}
The interactions at $\nu=\pm 1$ for valley $\bK'$ spin $\uparrow\downarrow$\cite{song2022magic} is given as
\begin{eqnarray}
V^{\tau=-1,s=\uparrow\downarrow}_{\nu=\pm 1} &=& \nu \sum_{k}\bigg( \sum_{a=1}^{4}\sum_{m=0}^{m_{a,-1}} \sum_{r=0}^{p-1}W_{a}c^{\dagger}_{a-1krms}c_{a-1krms} - \sum_{a=3,4}\sum_{m=0}^{m_{a,-1}}\sum_{r=0}^{p-1}\frac{J}{2} c^{\dagger}_{a-1krms} c_{a-1krms} \nonumber \\
&+& \sum_{b=1,2}\sum_{r'=0}^{q-1}\left(\frac{1}{2}U_1 + 6U_2\right)f^{\dagger}_{b-1kr's}f_{b-1kr's} \bigg),
\end{eqnarray}
$W_{a\in \{1\ldots 4\}}$ and $U_2$ are mean field coefficients with $W_1=W_2$ and $W_3=W_4$ \cite{song2022magic}. The interaction for $f$ modes in the $\bar{f}$ basis can then be given as
\begin{eqnarray}
V^{f,\tau=-1,s}_{\nu=\pm1} &=& \sum_{k\in[0,1)\otimes [0,\frac{1}{q})}V^{f,\tau=-1,s,\nu=\pm 1}_{coupled} + V^{f,\tau=-1,s,\nu=\pm 1}_{decoupled}
\end{eqnarray}
where
\begin{eqnarray}
&& V^{f,\tau=-1,s,\nu=\pm 1}_{coupled} = \nu\left(\frac{1}{2}U_1 + 6U_2 \right) \left(\sum_{m=0}^{m_{max}}\bar{f}^{\dagger}_{1-1kms}\bar{f}_{1-1kms} + \sum_{m=0}^{m_{max}+1}\bar{f}^{\dagger}_{2-1kms}\bar{f}_{2-1kms}\right),\\
&& V^{f,\tau=-1,s,\nu= \pm 1}_{decoupled}= \nu\sum_{b=1}^{2}\sum_{m'=m_{max}+b}^{q-1} \left(\frac{1}{2}U_1 + 6U_2\right) \bar{f}^{\dagger}_{b-1km's}\bar{f}_{b-1km's}.
\end{eqnarray}
\begin{figure}
\includegraphics[width=7cm]{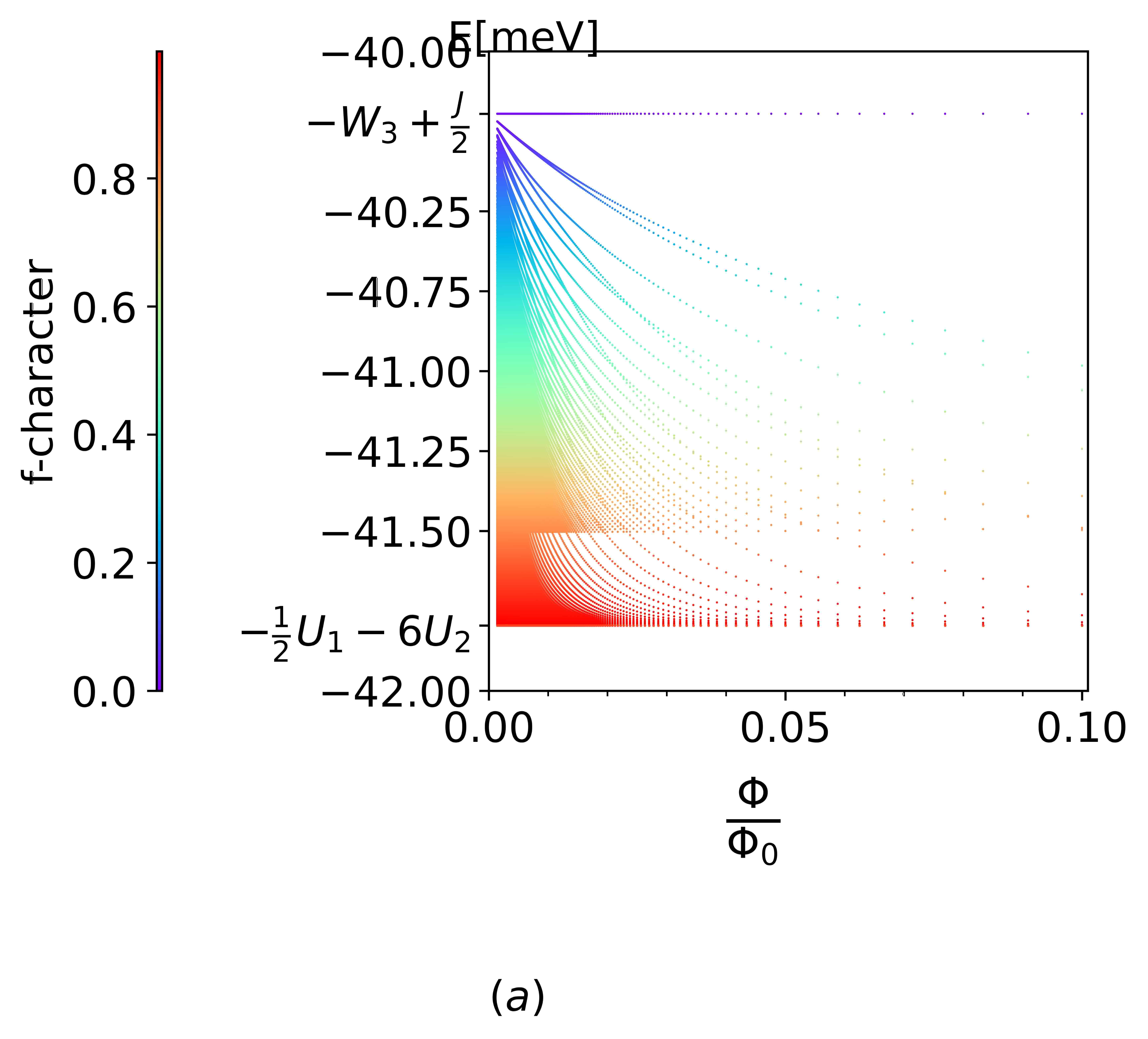}
\includegraphics[width=7cm]{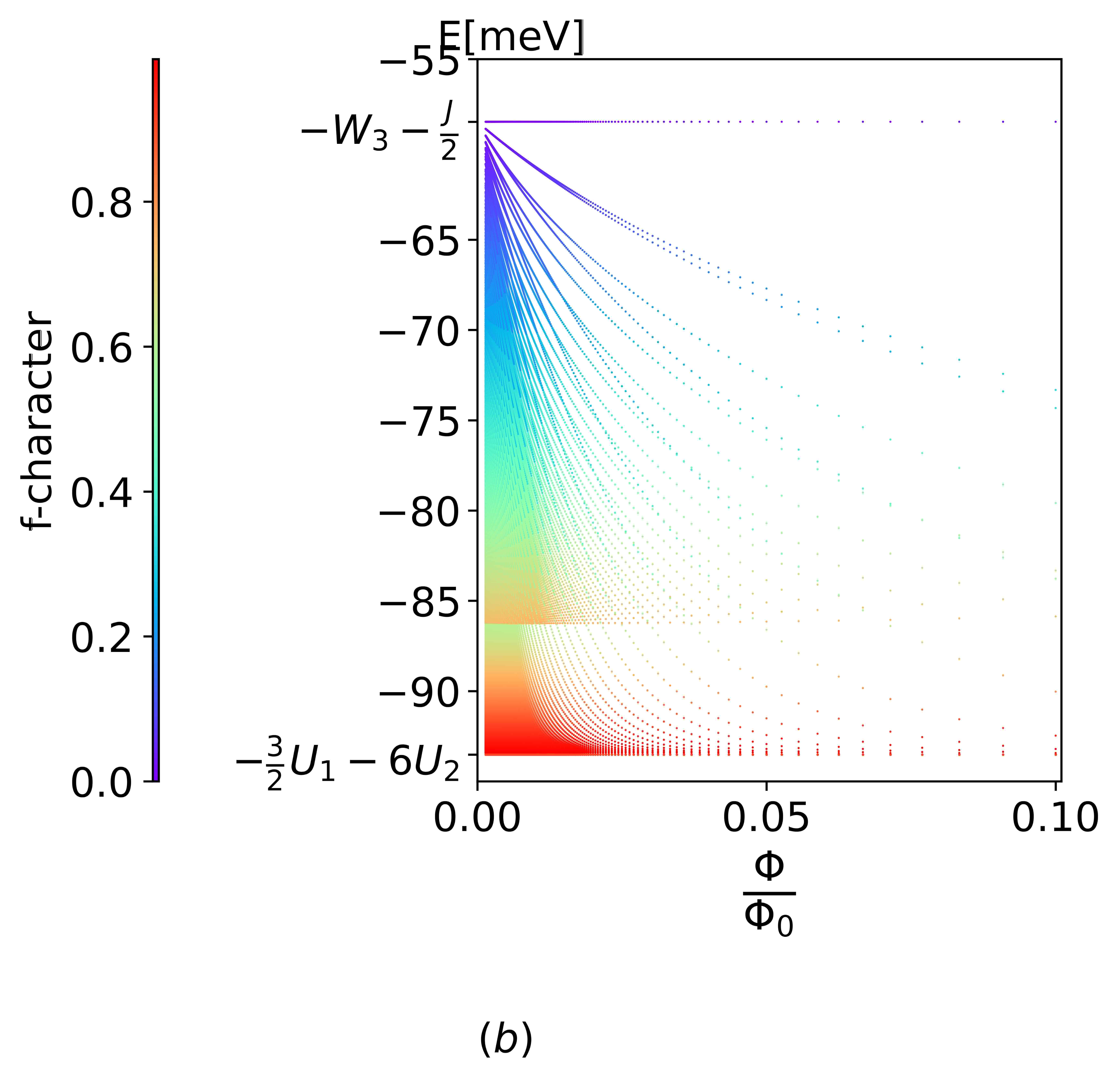}
\caption{Interacting heavy fermion Hofstadter spectra for (a) valley $\bK'$ spin $\uparrow\downarrow$ (degenerate) and (b) valley $\bK$ spin $\uparrow$ at filling $\nu=-1$ at $w_0/w_1=0.7$ in the flat band limit $M=0$. The value of parameters used are $W_1=44.05meV$, $W_3=49.33meV$ and $U_2=2.656meV$. $m_{max}=\lceil \frac{q-3}{2} \rceil$.}\label{fig:nu=-1 for Supp}
\end{figure}
Yet again there are $2q-(2m_{max}+3)$ decoupled $f$ modes for each $k$. Physically this corresponds to $2-(2m_{max}+3)/q$ states per moir\'e unit cell for each spin projection. 
The coupled modes can then be described by
\begin{equation}\label{xi-hamiltonian-1 K'}
H^{\tau=-1,s,\nu=\pm 1}_{coupled} = \sum_k\sum_{\alpha,\alpha'=1}^6
\sum_{m=0}^{m_\alpha}
\sum_{m'=0}^{m_{\alpha'}}
\bar{\Xi}^{s,\nu=\pm 2}_{m\alpha,m'\alpha'}\bar{d}^{\dagger}_{m\alpha s}(k)\bar{d}_{m'\alpha' s}(k),
\end{equation}
where $m_{\alpha=1,\ldots, 4}=m_{\alpha,-1}$, $m_{5}=m_{max}$ and $m_6=m_{max}+1$, and
\begin{eqnarray}\label{6-spinor-1,K'}
\bar{d}^{\dagger}_{m\alpha s}(k) = \left(c^{\dagger}_{1-1k0m s},c^{\dagger}_{2-1k0 m s},c^{\dagger}_{3  -1k0m s},c^{\dagger}_{4-1k0m s},\bar{f}^{\dagger}_{1-1km s},\bar{f}^{\dagger}_{2-1km s} \right)_\alpha.
\end{eqnarray}
with
\begin{eqnarray}
\bar{\Xi}^{s,\nu=\pm 1}_{m\alpha,m'\alpha'} = \langle m|\hat{h}^{-1,s,\nu=\pm 1}_{\alpha,\alpha'} |m'\rangle.
\end{eqnarray}
where the operators $\hat{h}^{-1,s,\nu=\pm 1}_{\alpha,\alpha'}$ are given as
\begin{eqnarray}\label{eq:hat h K' -1}
\hat{h}^{-1,s,\nu=\pm 1}_{\alpha,\alpha'} = \left(\begin{array}{cccccc}
\nu W_1&0&-i\sqrt{2}\frac{v_{\ast}}{\ell}\hat{a}^{\dagger}&0&\gamma \Sigma(\hat{a}^{\dagger}\hat{a})&i\sqrt{2}\frac{v'_{\ast}}{\ell}\hat{a} \Sigma(\hat{a}^{\dagger}\hat{a})  \\
0&\nu W_1&0&i\sqrt{2}\frac{v_{\ast}}{\ell}\hat{a}&-i\sqrt{2}\frac{v'_{\ast}}{\ell}\hat{a}^{\dagger} \Sigma(\hat{a}^{\dagger}\hat{a})  &\gamma \Sigma(\hat{a}^{\dagger}\hat{a}) \\
i\sqrt{2}\frac{v_{\ast}}{\ell}\hat{a}&0&\nu(W_3-\frac{J}{2})&M&0&0 \\0&-i\sqrt{2}\frac{v_{\ast}}{\ell}\hat{a}^{\dagger}&M&\nu(W_3-\frac{J}{2})&0&0\\
\gamma \Sigma(\hat{a}^{\dagger}\hat{a})    & i\sqrt{2}\frac{v'_{\ast}}{\ell}\Sigma(\hat{a}^{\dagger}\hat{a})\hat{a}&0&0&\nu(\frac{U_1}{2}+6U_2)&0\\
-i\sqrt{2}\frac{v'_{\ast}}{\ell}\Sigma(\hat{a}^{\dagger}\hat{a})\hat{a}^{\dagger} &  \gamma \Sigma(\hat{a}^{\dagger}\hat{a})&0&0&0&\nu(\frac{U_1}{2}+6U_2)
\end{array}
\right)_{\alpha,\alpha'},
\end{eqnarray}
where $\hat{a}$ is a simple h.o. lowering operator with $|m\rangle$ being a simple h.o. eigenstate and $\Sigma(m)=\Sigma_m$. The exact eigenstates for the above operator are exactly solvable in flat band limit, $M=0$. The field independent $\nu(W_3-\frac{J}{2})$ level is formed by the decoupled anomalous $c$ level given in Eq.(\ref{anam c at K'}). The rest of eigenstates can be solved using the ans\"atze given in Eq.(\ref{ansatz 3 K'})-Eq.(\ref{ansatz 6 K'}). The corresponding coefficients $c^{(3)}_{\alpha}$, $c^{(5)}_{\alpha}$ and $c^{(6,m)}_{\alpha}$ can be solved as eigenvectors of the following $3\times 3$, $5\times 5$ and $m_{max}$ $6\times 6$ Hermitian matrices respectively:
\begin{eqnarray}\label{3x3,K',-1}
&&h^{-1,\nu=\pm 1}_3=\left(\begin{array}{ccc}
\nu W_1  & i\frac{\sqrt{2}v_*}{\ell} & \gamma \Sigma_0 \\
-i\frac{\sqrt{2}v_*}{\ell} & \nu(W_3-\frac{J}{2}) & 0\\
\gamma \Sigma_0 & 0 & \nu(\frac{U_1}{2}+6U_2)
\end{array}
\right),
\end{eqnarray}

\begin{equation}\label{5x5,K',-1}
h^{-1,\nu=\pm 1}_5 = \left(\begin{array}{ccccc}
\nu W_1 & 0&0& \gamma\Sigma_0& i\frac{\sqrt{2}v_*'}{\ell}\Sigma_1 \\
0 & \nu W_1 & i\frac{2v_{\ast}}{\ell}&-i\frac{\sqrt{2}v_*'}{\ell}\Sigma_0 & \gamma \Sigma_1 \\
0&-i\frac{2v_*}{\ell} & \nu(W_3-\frac{J}{2}) &  0& 0\\
\gamma \Sigma_0& i\frac{\sqrt{2}v_*'}{\ell}\Sigma_0& 0 & \nu(\frac{U_1}{2}+6U_2) & 0\\
-i\frac{\sqrt{2}v_*'}{\ell}\Sigma_1 & \gamma \Sigma_1 & 0 & 0& \nu(\frac{U_1}{2}+6U_2)
\end{array}
\right).
\end{equation}
\begin{eqnarray}\label{decoupledblock66,K',-1}
h^{-1,m,\nu=\pm 1}_6&=&\left(\begin{array}{cccccc}
\nu W_1 & 0 & -i\sqrt{2m-2}\frac{v_*}{\ell} & 0 &\gamma \Sigma_{m-1} & i\sqrt{2m}\frac{v_*'}{\ell} \Sigma_{m}  \\
0 & \nu W_1 & 0 & i\sqrt{2m+2}\frac{v_*}{\ell} & -i\sqrt{2m}\frac{v_*'}{\ell}\Sigma_{m-1} &\gamma \Sigma_{m}\\
+i\sqrt{2m-2}\frac{v_*}{\ell} & 0 & \nu(W_3-\frac{J}{2}) & 0& 0& 0\\
0 & -i\sqrt{2m+2}\frac{v_*}{\ell} & 0 & \nu(W_3-\frac{J}{2}) & 0 & 0\\
\gamma \Sigma_{m-1} & i\sqrt{2m}\frac{v_*'}{\ell}\Sigma_{m-1}  & 0  & 0 & \nu(\frac{ U_1}{2}+6U_2) & 0\\
-i\sqrt{2m}\frac{v_*'}{\ell}\Sigma_{m} & \gamma \Sigma_{m} & 0 & 0 & 0& \nu(\frac{U_1}{2}+6U_2) 
\end{array}
\right).
\end{eqnarray}
The magnetic subbands within the narrow bands from the coupled modes emanate out of the $\bB\rightarrow 0$ energy eigenvalue of the above decoupled matrices, $\nu(W_3-\frac{J}{2})$, which is $2$ fold degenerate for matrix in Eq.(\ref{decoupledblock66,K',-1})$\forall m$ and singly degenerate for matrices in Eq.(\ref{3x3,K',-1}) and Eq.(\ref{5x5,K',-1}). Including the decoupled $c$ mode, we have $2m_{max}+3$ magnetic modes emanating out of this $\bB\rightarrow 0$ energy eigenvalue. Now recall that we have $2q-(2m_{max}+3)$ decoupled $f$ modes with energy $\nu(\frac{U_1}{2}+6U_2)$. Thus in total we have $2q$ magnetic modes within the narrow bands, which corresponds to 2 states per moir\'e unit cell for each spin projection. The spectrum for flat band limit has been shown in Supplementary Fig.(\ref{fig:nu=-1 for Supp}a). Note that the $\bB\rightarrow 0$ energies recovered by the decoupled matrices are the corresponding zero field energies of THFM at $\Gamma$ in mBZ.
\subsection{\texorpdfstring{$\nu=\pm 2$}{nu = pm 2} Spin \texorpdfstring{$\uparrow \downarrow$}{up down}}\label{apdx:K'nu=-2}
The interactions at $\nu=\pm 2$ for valley $\bK'$ spin $\uparrow\downarrow$\cite{song2022magic} is given as
\begin{eqnarray}
V^{\tau=-1,s=\uparrow\downarrow}_{\nu=\pm 2} &=& \nu \sum_{k}\bigg( \sum_{a=1}^{4}\sum_{m=0}^{m_{a,-1}} \sum_{r=0}^{p-1}W_{a}c^{\dagger}_{a-1krms}c_{a-1krms} - \sum_{a=3,4}\sum_{m=0}^{m_{a,-1}}\sum_{r=0}^{p-1}\frac{J}{4} c^{\dagger}_{a-1krms} c_{a-1krms} \nonumber \\
&+& \sum_{b=1,2}\sum_{r'=0}^{q-1}\left(\frac{3}{4}U_1 + 6U_2\right)f^{\dagger}_{b-1kr's}f_{b-1kr's} \bigg).
\end{eqnarray}
The interaction for $f$ modes in the $\bar{f}$ basis can then be given as
\begin{eqnarray}
V^{f,\tau=-1,s}_{\nu=\pm1} &=& \sum_{k\in[0,1)\otimes [0,\frac{1}{q})}V^{f,\tau=-1,s,\nu=\pm 2}_{coupled} + V^{f,\tau=-1,s,\nu=\pm 2}_{decoupled}
\end{eqnarray}
where
\begin{eqnarray}
&& V^{f,\tau=-1,s,\nu=\pm 2}_{coupled} = \nu\left(\frac{3}{4}U_1 + 6U_2 \right) \left(\sum_{m=0}^{m_{max}}\bar{f}^{\dagger}_{1-1kms}\bar{f}_{1-1kms} + \sum_{m=0}^{m_{max}+1}\bar{f}^{\dagger}_{2-1kms}\bar{f}_{2-1kms}\right),\\
&& V^{f,\tau=-1,s,\nu= \pm 2}_{decoupled}= \nu\sum_{b=1}^{2}\sum_{m'=m_{max}+b}^{q-1} \left(\frac{3}{4}U_1 + 6U_2\right) \bar{f}^{\dagger}_{b-1km's}\bar{f}_{b-1km's}.
\end{eqnarray}
\begin{figure}
\includegraphics[width=10cm]{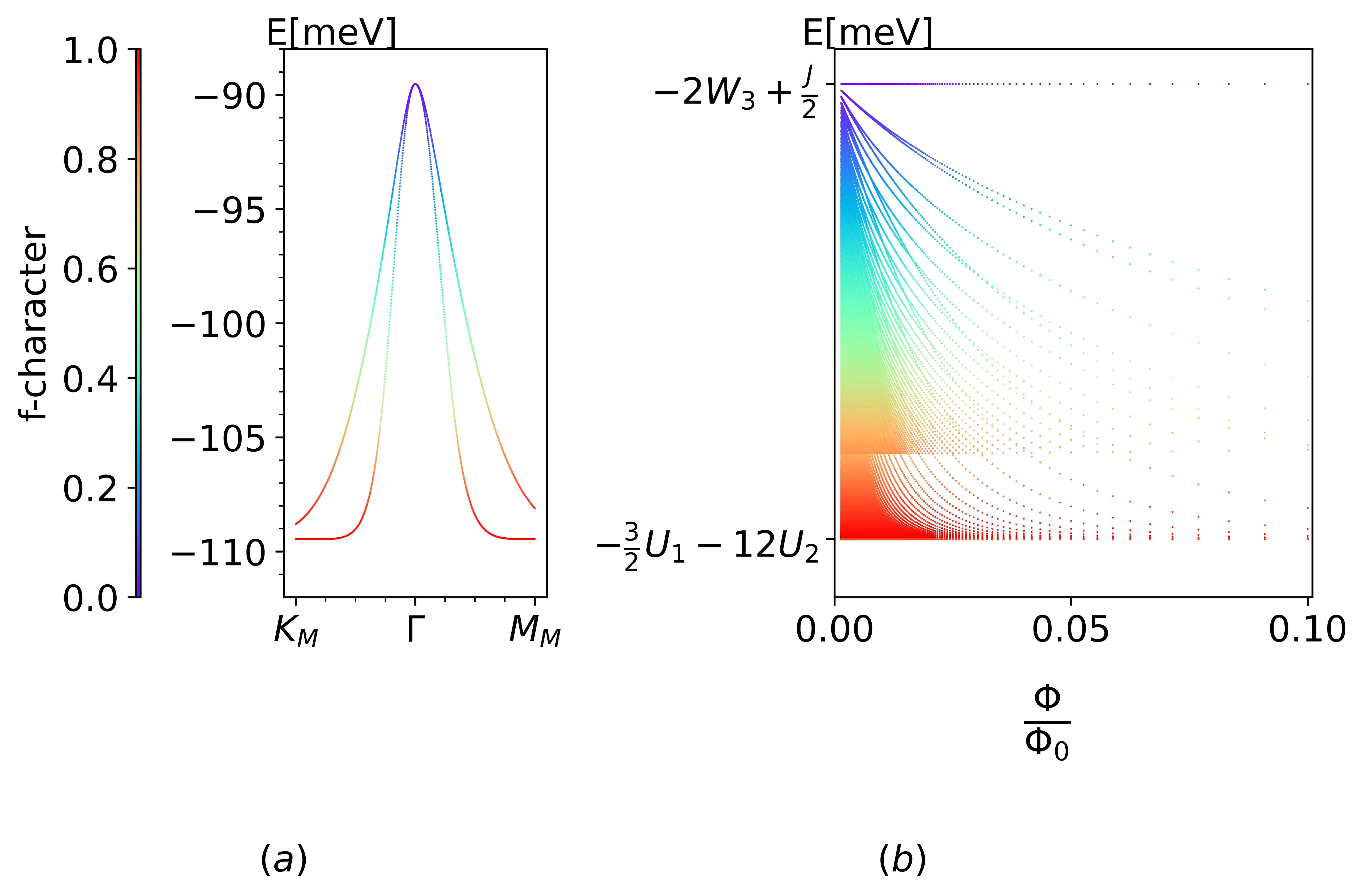}
\caption{Interacting heavy fermion Hofstadter spectra for (b) valley $\bK'$ spin $\uparrow\downarrow$ (degenerate) contrasted with (a) corresponding zero-field spectra at filling $\nu=-2$ at $w_0/w_1=0.7$ in the flat band limit $M=0$. $m_{max}=\lceil \frac{q-3}{2} \rceil$.}\label{fig:nu=-2 for Supp}
\end{figure}

Yet again there are $2q-(2m_{max}+3)$ decoupled $f$ modes for each $k$. Physically this corresponds to $2-(2m_{max}+3)/q$ states per moir\'e unit cell for each spin projection. 
The coupled modes can then be described by
\begin{equation}\label{xi-hamiltonian-2 K'}
H^{\tau=-1,s,\nu=\pm 2}_{coupled} = \sum_k\sum_{\alpha,\alpha'=1}^6
\sum_{m=0}^{m_\alpha}
\sum_{m'=0}^{m_{\alpha'}}
\bar{\Xi}^{s,\nu=\pm 1}_{m\alpha,m'\alpha'}\bar{d}^{\dagger}_{m\alpha s}(k)\bar{d}_{m'\alpha' s}(k),
\end{equation}
where $m_{\alpha=1,\ldots, 4}=m_{\alpha,-1}$, $m_{5}=m_{max}$ and $m_6=m_{max}+1$, and
\begin{eqnarray}\label{6-spinor-2,K'}
\bar{d}^{\dagger}_{m\alpha s}(k) = \left(c^{\dagger}_{1-1k0m s},c^{\dagger}_{2-1k0 m s},c^{\dagger}_{3  -1k0m s},c^{\dagger}_{4-1k0m s},\bar{f}^{\dagger}_{1-1km s},\bar{f}^{\dagger}_{2-1km s} \right)_\alpha.
\end{eqnarray}
with
\begin{eqnarray}
\bar{\Xi}^{s,\nu=\pm 2}_{m\alpha,m'\alpha'} = \langle m|\hat{h}^{-1,s,\nu=\pm 2}_{\alpha,\alpha'} |m'\rangle,
\end{eqnarray}
where the operators $\hat{h}^{-1,s,\nu=\pm 2}_{\alpha,\alpha'}$ are given as
\begin{eqnarray}\label{eq:hat h -2}
\hat{h}^{-1,s,\nu=\pm 2}_{\alpha,\alpha'} = \left(\begin{array}{cccccc}
\nu W_1&0&-i\sqrt{2}\frac{v_{\ast}}{\ell}\hat{a}^{\dagger}&0&\gamma \Sigma(\hat{a}^{\dagger}\hat{a})&i\sqrt{2}\frac{v'_{\ast}}{\ell}\hat{a} \Sigma(\hat{a}^{\dagger}\hat{a})  \\
0&\nu W_1&0&i\sqrt{2}\frac{v_{\ast}}{\ell}\hat{a}&-i\sqrt{2}\frac{v'_{\ast}}{\ell}\hat{a}^{\dagger} \Sigma(\hat{a}^{\dagger}\hat{a})  &\gamma \Sigma(\hat{a}^{\dagger}\hat{a}) \\
i\sqrt{2}\frac{v_{\ast}}{\ell}\hat{a}&0&\nu(W_3-\frac{J}{4})&M&0&0 \\0&-i\sqrt{2}\frac{v_{\ast}}{\ell}\hat{a}^{\dagger}&M&\nu(W_3-\frac{J}{4})&0&0\\
\gamma \Sigma(\hat{a}^{\dagger}\hat{a})    & i\sqrt{2}\frac{v'_{\ast}}{\ell}\Sigma(\hat{a}^{\dagger}\hat{a})\hat{a}&0&0&\nu(\frac{3U_1}{4}+6U_2)&0\\
-i\sqrt{2}\frac{v'_{\ast}}{\ell}\Sigma(\hat{a}^{\dagger}\hat{a})\hat{a}^{\dagger} &  \gamma \Sigma(\hat{a}^{\dagger}\hat{a})&0&0&0&\nu(\frac{3U_1}{4}+6U_2)
\end{array}
\right)_{\alpha,\alpha'},
\end{eqnarray}
where $\hat{a}$ is a simple h.o. lowering operator with $|m\rangle$ being a simple h.o. eigenstate and $\Sigma(m)=\Sigma_m$. The exact eigenstates for the above operator are exactly solvable in flat band limit, $M=0$. The field independent $\nu(W_3-\frac{J}{4})$ level is formed by the decoupled anomalous $c$ level given in Eq.(\ref{anam c at K'}). The rest of eigenstates can be solved using the ans\"atze given in Eq.(\ref{ansatz 3 K'})-Eq.(\ref{ansatz 6 K'}). The corresponding coefficients $c^{(3)}_{\alpha}$, $c^{(5)}_{\alpha}$ and $c^{(6,m)}_{\alpha}$ can be solved as eigenvectors of the following $3\times 3$, $5\times 5$ and $m_{max}$ $6\times 6$ Hermitian matrices respectively:
\begin{eqnarray}\label{3x3,K',-2}
&&h^{-1,\nu=\pm 2}_3=\left(\begin{array}{ccc}
\nu W_1  & i\frac{\sqrt{2}v_*}{\ell} & \gamma \Sigma_0 \\
-i\frac{\sqrt{2}v_*}{\ell} & \nu(W_3-\frac{J}{4}) & 0\\
\gamma \Sigma_0 & 0 & \nu(\frac{3U_1}{4}+6U_2)
\end{array}
\right),
\end{eqnarray}

\begin{equation}\label{5x5,K',-2}
h^{-1,\nu=\pm 2}_5 = \left(\begin{array}{ccccc}
\nu W_1 & 0&0& \gamma\Sigma_0& i\frac{\sqrt{2}v_*'}{\ell}\Sigma_1 \\
0 & \nu W_1 & i\frac{2v_{\ast}}{\ell}&-i\frac{\sqrt{2}v_*'}{\ell}\Sigma_0 & \gamma \Sigma_1 \\
0&-i\frac{2v_*}{\ell} & \nu(W_3-\frac{J}{4}) &  0& 0\\
\gamma \Sigma_0& i\frac{\sqrt{2}v_*'}{\ell}\Sigma_0& 0 & \nu(\frac{3U_1}{4}+6U_2) & 0\\
-i\frac{\sqrt{2}v_*'}{\ell}\Sigma_1 & \gamma \Sigma_1 & 0 & 0& \nu(\frac{3U_1}{4}+6U_2)
\end{array}
\right).
\end{equation}
\begin{eqnarray}\label{decoupledblock66,K',-2}
h^{-1,m,\nu=\pm 2}_6&=&\left(\begin{array}{cccccc}
\nu W_1 & 0 & -i\sqrt{2m-2}\frac{v_*}{\ell} & 0 &\gamma \Sigma_{m-1} & i\sqrt{2m}\frac{v_*'}{\ell} \Sigma_{m}  \\
0 & \nu W_1 & 0 & i\sqrt{2m+2}\frac{v_*}{\ell} & -i\sqrt{2m}\frac{v_*'}{\ell}\Sigma_{m-1} &\gamma \Sigma_{m}\\
+i\sqrt{2m-2}\frac{v_*}{\ell} & 0 & \nu(W_3-\frac{J}{4}) & 0& 0& 0\\
0 & -i\sqrt{2m+2}\frac{v_*}{\ell} & 0 & \nu(W_3-\frac{J}{4}) & 0 & 0\\
\gamma \Sigma_{m-1} & i\sqrt{2m}\frac{v_*'}{\ell}\Sigma_{m-1}  & 0  & 0 & \nu(\frac{3U_1}{4}+6U_2) & 0\\
-i\sqrt{2m}\frac{v_*'}{\ell}\Sigma_{m} & \gamma \Sigma_{m} & 0 & 0 & 0& \nu(\frac{3U_1}{4}+6U_2) 
\end{array}
\right).
\end{eqnarray}
The magnetic subbands within the narrow bands from the coupled modes emanate out of the $\bB\rightarrow 0$ energy eigenvalue of the above decoupled matrices, $\nu(W_3-\frac{J}{4})$, which is $2$ fold degenerate for matrix in Eq.(\ref{decoupledblock66,K',-2})$\forall m$ and singly degenerate for matrices in Eq.(\ref{3x3,K',-2}) and Eq.(\ref{5x5,K',-2}). Including the decoupled $c$ mode, we have $2m_{max}+3$ magnetic modes emanating out of this $\bB\rightarrow 0$ energy eigenvalue. Now recall that we have $2q-(2m_{max}+3)$ decoupled $f$ modes with energy $\nu(\frac{3U_1}{4}+6U_2)$. Thus in total we have $2q$ magnetic modes within the narrow bands, which corresponds to 2 states per moir\'e unit cell for each spin projection. The spectrum for flat band limit has been shown in Supplementary Fig.(\ref{fig:nu=-2 for Supp}b). Note that the $\bB\rightarrow 0$ energies recovered by the decoupled matrices are the corresponding zero field energy of THFM at $\Gamma$ in mBZ.
\section{Naive Minimal Substitution}\label{apdx:Naive Minimal Coupling}
\begin{figure}
    \centering
    \includegraphics[width=10cm]{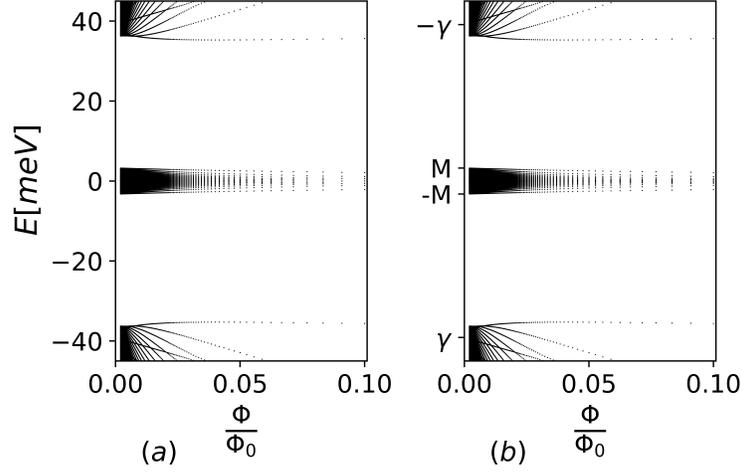}
    \caption{The spin valley degenerate non-interacting Hofstadter spectrum for $M\neq 0$ at $\omega_0/\omega_1=0.7$ (a) obtained using naive minimal coupling compared to the (b) spectrum obtained using the formalism constructed in the paper for $\bB\neq 0$ THFM. The magnetic subbands within the narrow bands are spread in the energy window of $|2M|$, which also sets the bandwidth of narrow bands at $\bB=0$ in THFM. The total number of subbands within the narrow bands i.e. in the energy widow [-M,M] at flux $1/q$ in (a) is ($2m_\ast + 1$) while in (b) is correctly $2q$.  We use $m_{max}=m_{\ast}=\lfloor\frac{q-3}{2}\rfloor$}.
    \label{fig:Non-Interacting with M}
\end{figure}
Including both the $c$-$c$ and $c$-$f$ coupling in Eq.(2) of main-text, the zero $\bB$ field THFM can be written as a $6\times 6$ matrix, say at $\tau=+1$
\begin{eqnarray}
h_0^{\tau=1} =
\left(\begin{array}{cccccc}
        0& 0& v_\ast k&0 &e^{-\frac{\bk^2\lambda^2}{2}}\gamma &e^{-\frac{\bk^2\lambda^2}{2}}v_\ast '\bar{k} \\
        0& 0& 0& v_\ast \bar{k}&e^{-\frac{\bk^2\lambda^2}{2}}v_\ast 'k &e^{-\frac{\bk^2\lambda^2}{2}}\gamma\\
        v_\ast \bar{k}&0 &0 &M &0 &0\\
        0& v_\ast k& M& 0&0 &0\\
        e^{-\frac{\bk^2\lambda^2}{2}}\gamma&e^{-\frac{\bk^2\lambda^2}{2}}v_\ast '\bar{k}& 0&0 &0 &0\\
        e^{-\frac{\bk^2\lambda^2}{2}}v_\ast 'k& e^{-\frac{\bk^2\lambda^2}{2}}\gamma&0 &0 &0 &0
    \end{array}\right)_{\alpha,\alpha'}, \label{Eq:Zero field in 6by6}\\
\end{eqnarray}
where $k=k_x+ik_y$ and $\bar{k}=k_x-ik_y$. Following a naive minimal substitution, we promote $k_x+ik_y\rightarrow -i\sqrt{2}\hat{a}/\ell$, where $\hat{a}$ is the Landau level (LL) lowering operator. Although it is completely unclear on how to perform the minimal substitution on $e^{-\frac{1}{2}\bk^2\lambda^2}$, if we were to Taylor expand it as $1-\frac{1}{2}\bk^2\lambda^2+\ldots=1-\frac{1}{2}\frac{k\bar{k}+\bar{k}k}{2}\lambda^2+\ldots$ and use the finite $\bB$ substitution for $k$ and $\bar{k}$ mentioned above we get $1-(a^\dagger a+\frac{1}{2})\frac{\lambda^2}{\ell^2}+\ldots$. If this operator acts on LL state $|m\rangle$, upto $O(\frac{\lambda^2}{\ell^2})$, we get $1-(m+\frac{1}{2})\frac{\lambda^2}{\ell^2}$. But recall that this is same as the $\bB\rightarrow 0$ limit of $\Sigma_m$ as given in Eq.(\ref{Eq:limit for singular values}). Based on this assumption, we thus promote $e^{-\frac{1}{2}\bk^2\lambda^2}$ to $\Sigma_{a^\dagger a}=\Sigma(a^\dagger a)$ at finite $\bB$. We thus have 
\begin{eqnarray}
 \hat{h}_B^{\tau=1} = \left(\begin{array}{cccccc}
0&0&-i\sqrt{2}\frac{v_{\ast}}{\ell}\hat{a}&0&\gamma \Sigma(\hat{a}^{\dagger}\hat{a})&i\sqrt{2}\frac{v'_{\ast}}{\ell}\hat{a}^{\dagger} \Sigma(\hat{a}^{\dagger}\hat{a})  \\
0&0&0&i\sqrt{2}\frac{v_{\ast}}{\ell}\hat{a}^{\dagger}&-i\sqrt{2}\frac{v'_{\ast}}{\ell}\hat{a} \Sigma(\hat{a}^{\dagger}\hat{a})  &\gamma \Sigma(\hat{a}^{\dagger}\hat{a}) \\
i\sqrt{2}\frac{v_{\ast}}{\ell}\hat{a}^{\dagger}&0&0&M&0&0 \\0&-i\sqrt{2}\frac{v_{\ast}}{\ell}\hat{a}&M&0&0&0\\
\gamma \Sigma(\hat{a}^{\dagger}\hat{a})    & i\sqrt{2}\frac{v'_{\ast}}{\ell}\Sigma(\hat{a}^{\dagger}\hat{a})\hat{a}^{\dagger}&0&0&0&0\\
-i\sqrt{2}\frac{v'_{\ast}}{\ell}\Sigma(\hat{a}^{\dagger}\hat{a})\hat{a} &  \gamma \Sigma(\hat{a}^{\dagger}\hat{a})&0&0&0&0 \label{Eq:Naive Minimal operator}
\end{array}
\right)_{\alpha,\alpha'},
\end{eqnarray}  
where we assume that the correct way of ordering $\Sigma(a^\dagger a)$ with respect to the operators $a$ and $a^\dagger$ (coming from the promotion of $k$ and $\bar{k}$ respectively) is the one we have in Eq.(\ref{eq:hat h plus 1}). The Hofstadter spectrum based on the naive minimal substitution approach can thus be obtained by solving the eigenstates of $\hat{h}^{\tau=1}_{B_{\alpha,\alpha'}}$ in the LL basis.
%four $c$ ($\alpha=\{1,\ldots 4\}$) and two $f$ fermions ($\alpha =\{5,6\}$).
The LL basis $\ket{m}$ used for computing the matrix elements for index $\alpha\in\{1,6\}$ are: $m\in\{0,m_\ast\}$ for index $\alpha=1$ ($a=1$ $c$-fermion), $m\in\{0,m_\ast-1\}$ for index $\alpha=2$ ($a=2$ $c$-fermion),  $m\in\{0,m_\ast+1\}$ for index $\alpha=3$ ($a=1$ $c$-fermion), $m\in\{0,m_\ast-2\}$ for index $\alpha=4$ ($a=1$ $c$-fermion), $m\in\{0,m_\ast\}$ for index $\alpha=5$ ($b=1$ $f$-fermion)  and $m\in\{0,m_\ast-1\}$ for index $\alpha=6$ ($b=2$ $f$-fermion). The spectrum is shown in Fig.(\ref{fig:Non-Interacting with M}a). Although the magnetic subbands within the narrow bands are well separated from the remote bands, the total number of states within the narrow bands is $(2m_\star + 1)\phi/\phi_0$ per moi\'e unit cell per spin per valley (same as the number of zero modes for $M=0$, shown in main-text), which of course is incorrect as the total number of states should rather be $2$ per moir\'e unit cell per spin per valley, independent of $\bB$.
\section{Parent Kramers inter-valley coherent state}
In this section, we discuss the Landau quantization of the one-shot Hartree Fock (HF) bands obtained for a parent Kramers inter-valley coherent (KIVC) state at fillings $\nu=0$ (CNP) and $\nu=-2$.
\subsection{CNP}\label{apdx:KIVC CNP 6x6}
\begin{figure}
    \centering
    \includegraphics[width=8cm]{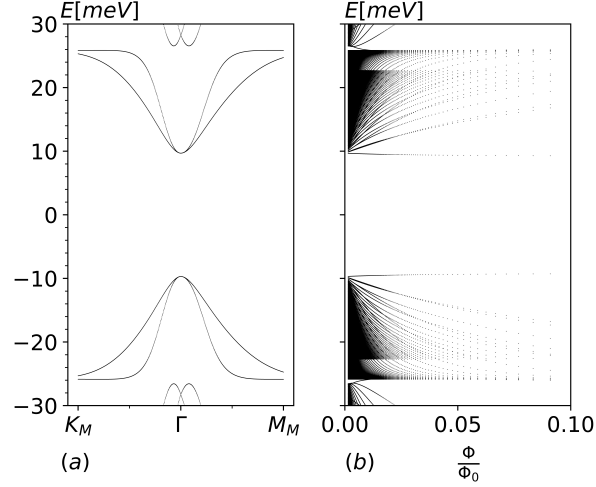}
    \caption{Spin degenerate (b) interacting heavy fermion Hofstadter spectrum for KIVC state at CNP contrasted with (a) corresponding zero-field spectrum at $w_0/w_1=0.7$. $m_{max}=\lceil \frac{q-3}{2} \rceil$.}
    \label{fig:KIVC at CNP}
\end{figure}
The spin degenerate $\bB=0$ one-shot Hartree-Fock (HF) mean-field (MF) Hamiltonian for parent KIVC state at $\nu=0$ is given as\cite{song2022magic}
\begin{eqnarray}
      H^{\nu=0,s=\uparrow,\downarrow}_{\mathbf{0},\text{KIVC}}(\bk) = \left(\begin{array}{ccc}
          0 &v_\star(k_x\sigma_0\tau_z+ik_y\sigma_z\tau_0)  &e^{-\frac{\mathbf{k}^2\lambda^2}{2}}\left(\gamma\sigma_0\tau_0 +v_\star'(k_x\sigma_x\tau_z+k_y\sigma_y\tau_0)\right)\\
           & -\frac{J}{2}\sigma_y\tau_y + M\sigma_x\tau_0  & 0\\
          h.c. & &\frac{U_1}{2}\sigma_y\tau_y
      \end{array}\right)\label{eq:KIVC B=0 CNP-0},
\end{eqnarray}
where $\bk=(k_x,k_y)\in$moir\'e BZ, the Pauli matrices $\sigma$ and $\tau$ act in the orbital and valley space respectively and h.c. represents hermitian conjugate. 

Let us begin the discussion by first analyzing the decoupled $\bar{f}$ modes at $\mathbf{B}\neq 0$. Recall that at every $k\in$magnetic BZ, $q-(m_{max}+2)$ of $\bar{f}_{11kr's}$ and $\bar{f}_{2-1kr's}$ modes, i.e. the ones with $r'\geq m_{max}+1$ decouple from the $c$s. Similarly $q-(m_{max}+1)$ of $\bar{f}_{21kr's}$ and $\bar{f}_{1-1kr's}$ modes, i.e. the ones with $r'\geq m_{max}$ decouple from the $c$'s at every $k\in$magnetic BZ. 
We see that even for the KIVC state, the decoupled $\bar{f}$ modes contribute to $2q-(2m_{max}+3)$ $\mathbf{B}$ independent energy levels at $\pm U_1/2$.
%(the eigenvalues of 
%$f$-$f$ block $\frac{U_1}{2}\sigma_y\tau_y$ of Eq.(\ref{eq:KIVC B=0 CNP-0})). 
Physically, these modes are the linear combination $\frac{1}{\sqrt{2}}\left(\bar{f}_{11kr's}\pm \bar{f}_{2-1kr's}\right)$ with $r'\geq m_{max}+1$ and $\frac{1}{\sqrt{2}}\left(\bar{f}_{21kr's}\pm \bar{f}_{1-1kr's}\right)$ with $r'\geq m_{max}$, which contribute $q-(m_{max}+2)$ and $q-(m_{max}+1)$ modes at $\pm U_1/2$, respectively. We thus have $2q-(2m_{max}+3)$ modes at $\pm U_1/2$ as motivated above.

Before discussing the Hofstadter spectrum for the coupled modes, we first perform a $U(4)$ rotation ($e^{i\frac{\pi}{2}\hat{\Sigma}_{x0}}$, where $\hat{\Sigma}_{x0}$ is defined in Eq.(4) of Ref.\cite{song2022magic})
) on the zero magnetic field Hamiltonian in Eq.(\ref{eq:KIVC B=0 CNP-0}) to re-write it in a much familiar form as 
\begin{eqnarray}\tilde{H}^{\nu=0,s=\uparrow,\downarrow}_{\mathbf{0},\text{KIVC}}(\bk) = \left(\begin{array}{ccc}
          0 &v_\star(k_x\sigma_0\tau_z+ik_y\sigma_z\tau_0)  &e^{-\frac{\mathbf{k}^2\lambda^2}{2}}\left(\gamma\sigma_0\tau_0 +v_\star'(k_x\sigma_x\tau_z+k_y\sigma_y\tau_0)\right)\\
           & -\frac{J}{2}\sigma_0\tau_z -M\sigma_z\tau_x  & 0\\
          h.c. & &-\frac{U_1}{2}\sigma_0\tau_z
      \end{array}\right)\label{eq:KIVC B=0 CNP},
\end{eqnarray}
If we were to set $M=0$ the two valleys in the above Hamiltonian decouple (the model becomes identical to that for VP state due to the $U(4)$ symmetry in flat band limit \cite{song2022magic}). Using the results in Sections.(\ref{apdx:KCNP}) and (\ref{apdx:K'CNP}), we know how to promote each of the valley block to $\mathbf{B}\neq 0$. The role of $M$ is to couple both the valleys. Hence  the Hofstadter spectrum for the coupled modes of $c$ and $\bar{f}$, for the KIVC state at $\nu=0$, can be obtained by solving the eigenvalues of the operator  \begin{eqnarray}
      \hat{h}^{\text{KIVC},\nu=0}_{\alpha,\alpha'} = \left(\begin{array}{cc}
       \hat{h}^{+1,\nu=0}    & \hat{h}_M \\
        \hat{h}_M   & \hat{h}^{-1,\nu=0}
      \end{array}\right)_{\alpha,\alpha'},\label{Eq: h for KIVC nu=0}
  \end{eqnarray}
The intra-valley operator $\hat{h}^{\tau,\nu=0}$ can be obtained by setting $M=0$ in operators given in Eqs.(\ref{eq:hat h plus 1}) and (\ref{eq:hat h -1}) for $\tau=+1$ and $-1$, respectively. The inter-valley part is given as
\begin{eqnarray}
    \hat{h}_M = \left(\begin{array}{ccc}
        0 &0  &0 \\
         & -M\sigma_z &0 \\
         h.c.& &0
    \end{array}\right)\label{eq:KIVC Mass}
\end{eqnarray}
The Hofstadter spectrum is shown in Fig.(\ref{fig:KIVC at CNP}). 
\subsection{\texorpdfstring{$\nu=-2$}{nu=-2}}
\subsubsection{Spin \texorpdfstring{$\uparrow$}{up}}
\begin{figure}
    \includegraphics[width=8cm]{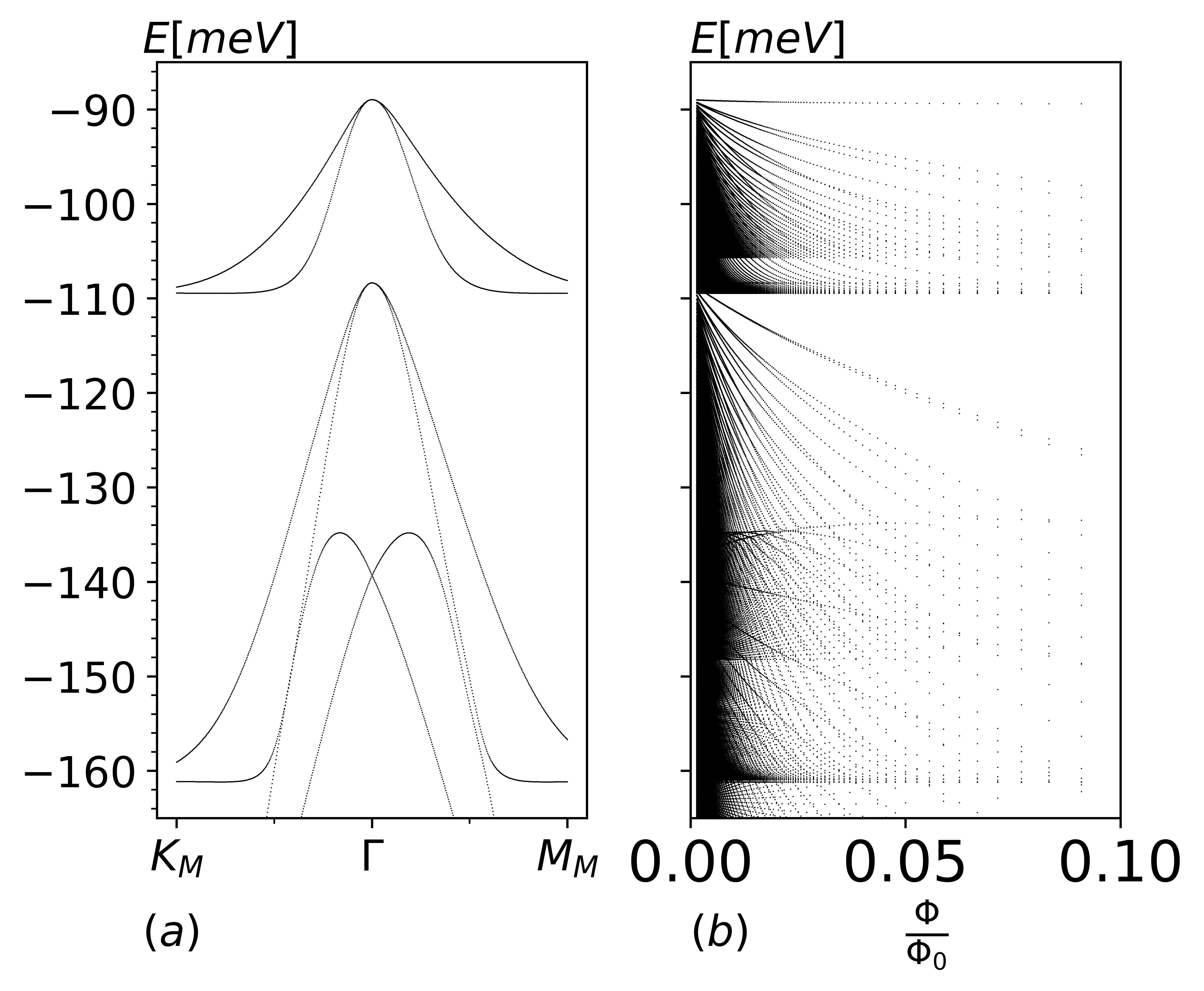}
    %\includegraphics[width=8cm]{Images for supplementary/6x6_KIVC_nu2_down_SuppFig.jpg}
    %\caption{The (b) interacting heavy fermion Hofstadter spectrum for KIVC state at $\nu=-2$ spin $\uparrow$ (left) and $\downarrow$ (Right) sector contrasted with (a) corresponding zero-field spectrum at $w_0/w_1=0.7$. $m_{max}=\lceil \frac{q-3}{2} \rceil$. The left (a,b) panel and the right (a,b) panel is for spin sector $\uparrow$ and $\downarrow$ respectively.}
    \caption{The (b) interacting heavy fermion Hofstadter spectrum for KIVC state at $\nu=-2$ spin $\uparrow$ sector contrasted with (a) corresponding zero-field spectrum at $w_0/w_1=0.7$. $m_{max}=\lceil \frac{q-3}{2} \rceil$.}
    \label{fig:KIVC at nu=-2}
\end{figure}
For filling $\nu=-2$, in this section we discuss the Landau quantization of spin flavor $s=\uparrow$ of the $\bB=0$ one-shot HF MF Hamiltonian obtained for the parent KIVC state
%the doubly degenerate lowest energy level at $\Gamma$, $-2W_3-\sqrt{M^2+J^2/4}$, 
%is contributed by the spin flavor $s=\uparrow$ of the one-shot HF MF Hamiltonian obtained for the parent KIVC state and thus has the lowest quasi-particle energy\cite{song2022magic}. 
The $\mathbf{B}=0$ Hamiltonian upto a $U(4)$ rotation (($e^{i\frac{\pi}{2}\hat{\Sigma}_{x0}}$, where $\hat{\Sigma}_{x0}$ is defined in Eq.(4) of Ref.\cite{song2022magic})
) can be given as\cite{song2022magic}
\begin{eqnarray}
      H^{\nu=-2\uparrow}_{\mathbf{0},\text{KIVC}}(\bk) = \left(\begin{array}{ccc}
          -2W_1 &v_\star(k_x\sigma_0\tau_z+ik_y\sigma_z\tau_0)  &e^{-\frac{\mathbf{k}^2\lambda^2}{2}}\left(\gamma\sigma_0\tau_0 +v_\star'(k_x\sigma_x\tau_z+k_y\sigma_y\tau_0)\right)\\
           & -2W_3\sigma_0\tau_0-\frac{J}{2}\sigma_0\tau_z -M\sigma_z\tau_x  & 0\\
          h.c. & &-(2U_1+12U_2)\sigma_0\tau_0-\frac{U_1}{2}\sigma_0\tau_z
      \end{array}\right)\label{eq:KIVC B=0 nu=-2},
\end{eqnarray}
where $\bk=(k_x,k_y)\in$moir\'e BZ, the Pauli matrices $\sigma$ and $\tau$ act in the orbital and valley space respectively and h.c. represents hermitian conjugate. The intra-valley terms in Eq.(\ref{eq:KIVC B=0 nu=-2}) can be promoted to $\mathbf{B}\neq 0$ using the results in Sections.(\ref{apdx:nu=-2 Valley K}) and (\ref{apdx:K'nu=-2}). The two valleys can then be coupled by $M$. Hence the Hofstadter spectrum for the coupled modes of $c$ and $\bar{f}$, for the KIVC state at $\nu=-2$ spin sector $\uparrow$, can be obtained by solving the eigenvalues of the operator
\begin{eqnarray}
      \hat{h}^{\text{KIVC},\uparrow,\nu=-2}_{\alpha,\alpha'} = \left(\begin{array}{cc}
       \hat{h}^{+1,\nu=-2,\uparrow}    & \hat{h}_M \\
        \hat{h}_M   & \hat{h}^{-1,\nu=-2,\uparrow}
      \end{array}\right)_{\alpha,\alpha'},\label{Eq: h for KIVC nu=-2 up}
  \end{eqnarray}
where the operator $\hat{h}^{\tau,\nu=-2,\uparrow}$ can be obtained by setting $M=0$ and $s=\uparrow$ in operators given in Eqs.(\ref{eq:pm 2 spin updown}) and (\ref{eq:hat h -2}) for $\tau=+1$ and $-1$, respectively. The inter-valley coupling $\hat{h}_M$ is given in Eq.(\ref{eq:KIVC Mass}). As argued in the previous section, the decoupled $\bar{f}$ modes give rise to the $2q-(2m_{max}+3)$ $\mathbf{B}$ independent energy levels at $-12U_2-2U_1\pm \frac{U_{1}}{2}$, which physically are linear combinations $\frac{1}{\sqrt{2}}\left(\bar{f}_{11kr's}\pm \bar{f}_{2-1kr's}\right)$ and $\frac{1}{\sqrt{2}}\left(\bar{f}_{21kr's}\pm \bar{f}_{1-1kr's}\right)$, with $r'\geq m_{max}+1$ and $r'\geq m_{max}$, respectively. The Hofstadter spectrum is shown in Fig.(\ref{fig:KIVC at nu=-2}b) 

\section{Effective Hamiltonians for Landau Quantization of light modes}
In order to better understand the Landau quantization of the zero $\mathbf{B}$ bands of interacting THFM in vicinity of $\Gamma\in$mBZ, i.e. where the lowest energy single particle excitations (light modes) reside, we derive an effective low energy Hamiltonian for THFM at $\mathbf{B}\neq 0$ in this section. Such an analysis offers us with a deeper qualitative understanding of not only the low energy but also the low $\mathbf{B}$ physics of THFM. As discussed in the main text, the decoupled $\bar{f}$ modes are only responsible for forming the $\bB$ independent higher energy level (heavy modes)
%near the boundary (opposite to CNP) of narrow band strong coupling energy window. 
and it is rather the coupled modes of $c$ and $f$ fermions which dictate the Hofstadter spectrum of THFM. Thus the results obtained for coupled modes in previous sections serves as the starting point for the following analysis.

\subsection{CNP}
In the flat band limit $M=0$, THFM is U(4) symmetric\cite{song2022magic}. The coupled modes at CNP can be described by $6\times 6$ operators given in Eq.(\ref{eq:hat h plus 1}) and Eq.(\ref{eq:hat h -1}) with $M$ set to zero, for $\tau=\pm 1$ respectively. The magnetic subbands within the narrow bands emanate out of the $\bB\rightarrow 0$ energy $\mp J/2$, while the remote subbands emanate out of $\bB\rightarrow 0$ energy $-\tau U_1/4-\sqrt{U_1^2/16+\gamma^2}$ and $-\tau U_1/4+\sqrt{U_1^2/16+\gamma^2}$, for $\tau=\pm 1$ and marked by $\pm\mathcal{E}_{\mp \tau}$ in main text Fig.(2). These energies correspond to the eigenvalues of the $\bB=0$ flat band THFM at $\Gamma$ obtained for the parent VP state. Recall that it has the form\cite{song2022magic} $\left(\begin{array}{cc}
   F^{\tau=+1}  &  \\
    & F^{\tau=-1}
\end{array}\right)$ at $\Gamma$, where $F^{\tau}=\left(\begin{array}{ccc}
  0 &0&\gamma\sigma_0\\
  & -\tau J/2\sigma_0  & 0 \\
   h.c.& & -\tau U_1/2\sigma_0
\end{array}\right)$, where Pauli matrix $\sigma_0$ acts in orbital space.
%, which is given as $H^{\nu=0}_{\mathbf{0},VP}(\bk)=\left(\begin{array}{cc}
   %H^{\nu=0}_{\tau=+1}(\bk)  &  \\
    % & H^{\nu=0}_{\tau=-1}(\bk)
%\end{array}\right)$, where $\bk=(k_x,k_y)\in$moir\'e BZ and is given in Eq.(\ref{Eq:Zero field in 6by6}).
%the above three energies correspond to the three doubly degenerate subspaces spanning the Hilbert space at $\Gamma$. %We will refer to these three subspaces by $\rho\in\{1,2,3\}$, respectively. 
The corresponding eigenstates, labelled by $|\rho,j,\tau\rangle$ with $\rho\in\{1,2,3\}$ and $j\in\{1,2\}$ are given as 
\begin{eqnarray}
    |1,1,\tau=+1\rangle &=& \left(0,0,0,1,0,0,0_{6\times 1}\right)^T \text{;} ~~~~~~~|1,1,\tau=-1\rangle = \left(0_{6\times 1},0,0,0,1,0,0\right)^T
    \label{Eq:zero B spinor 1}\\
    |1,2,\tau=+1\rangle &=& \left(0,0,1,0,0,0,0_{6\times 1}\right)^T \text{;} ~~~~~~~ |1,2,\tau=-1\rangle = \left(0_{6\times 1},0,0,1,0,0,0\right)^T
    \label{Eq:zero B spinor 2}\\
    |2,1,\tau=+1\rangle &=& \frac{1}{\sqrt{N_{X1}}}\left(0,X_1,0,0,0,1,0_{6\times 1}\right)^T \text{;} ~~~~|2,1,\tau=-1\rangle = \frac{1}{\sqrt{N_{X-1}}}\left(0_{6\times 1},0,X_{-1},0,0,0,1\right)^T
    \label{Eq:zero B spinor 3}\\
    |2,2,\tau=+1\rangle &=& \frac{1}{\sqrt{N_{X1}}}\left(X_1,0,0,0,1,0,0_{6\times 1}\right)^T \text{;} ~~~~ |2,2,\tau=-1\rangle = \frac{1}{\sqrt{N_{X-1}}}\left(0_{6\times 1},X_{-1},0,0,0,1,0\right)^T
    \label{Eq:zero B spinor 4}\\
    |3,1,\tau=+1\rangle &=& \frac{1}{\sqrt{N_{Y1}}}\left(0,Y_1,0,0,0,1,0_{6\times 1}\right)^T \text{;} ~~~~|3,1,\tau=-1\rangle = \frac{1}{\sqrt{N_{Y-1}}}\left(0_{6\times 1},0,Y_{-1},0,0,0,1\right)^T
    \label{Eq:zero B spinor 5}\\
    |3,2,\tau=+1\rangle &=& \frac{1}{\sqrt{N_{Y1}}}\left(Y_1,0,0,0,1,0,0_{6\times 1}\right)^T \text{;} ~~~~|3,2,\tau=-1\rangle = \frac{1}{\sqrt{N_{Y-1}}}\left(0_{6\times 1},Y_{-1},0,0,0,1,0\right)^T\label{Eq:zero B spinor 6}
\end{eqnarray}
where $(0_{6\times 1})$ represent 6 zero entries $(0,\ldots,0)$, 
\begin{eqnarray}
  X_\tau&=&-\frac{-\tau U_1/4+\sqrt{U_1^2/16+\gamma^2}}
  {\gamma},\label{Eq:Xtau for CNP}\\
  Y_\tau &=& -\frac{-\tau U_1/4-\sqrt{U_1^2/16+\gamma^2}}{\gamma},\label{Eq:Ytau for CNP}
\end{eqnarray}
The normalizations 
$N_{X\tau}=1+X_\tau^2$ and $N_{Y,\tau}=1+Y_\tau^2$. The energy of state $|\rho,j,\tau\rangle$ can be labelled as $E_{\rho,j,\tau}$, where 
\begin{eqnarray}
 E_{1j\tau}&=&-\tau J/2\equiv E_\tau \label{Eq:CNP VP energy label 1}, \\E_{2j\tau}&=&-\tau U_1/4-\sqrt{U_1^2/16+\gamma^2}\equiv E_{X\tau}\label{Eq:CNP VP energy label 2} \\E_{3j\tau}&=&-\tau U_1/4+\sqrt{U_1^2/16+\gamma^2}\equiv E_{Y\tau}. \label{Eq:CNP VP energy label 3}
\end{eqnarray}
As shown in the section(\ref{apdx:Naive Minimal Coupling}), the $\mathbf{B}\neq 0$ THFM for the coupled modes, upto the ambiguity of correct ordering of singular values, is the same as having naively minimally substituted in $\mathbf{B}=0$ THFM.
Hence, throughout the discussion we consider the $\mathbf{B}\neq 0$ basis to be the finite $\mathbf{B}$ $\mathbf{k}\cdot\mathbf{p}$ basis, i.e. given as $|\rho,j,\tau,m\rangle = |\rho,j,\tau\rangle |m\rangle$, where $|m\rangle$ is the $m^{th}$ LL.
%, given as $|\rho,j,\tau,m\rangle$, with $\langle \mathbf{r}|\rho,j,\tau,m\rangle=e^{i\Gamma\cdot\mathbf{r}}|\rho,j,\tau\rangle LL_{m}(\mathbf{r})=|\rho,j,\tau\rangle LL_{m}(\mathbf{r})$, where $LL_{m}(\mathbf{r})$ is the $m^{th}$ Landau level wavefunction in real space. 
%\textcolor{red}{KS: I will edit this part later}
\subsubsection{Parent Valley Polarized State}\label{apdx:CNP VP SWT}
\begin{figure}
    \centering
    \includegraphics[width=10cm]{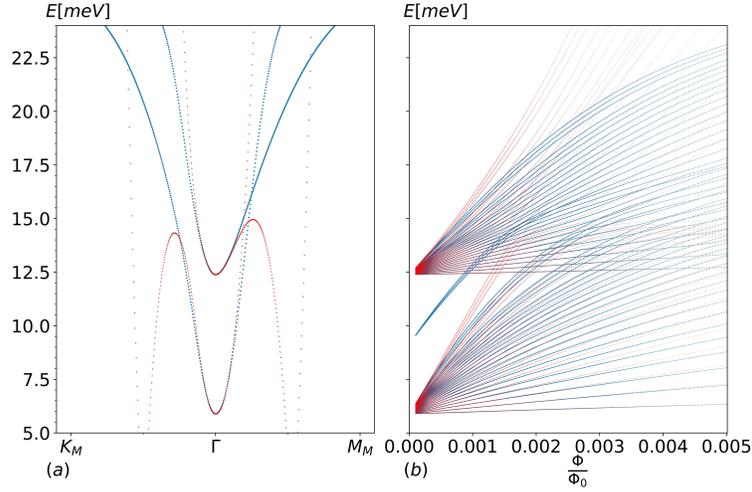}
    \caption{The comparison of (b) LL spectrum obtained using the effective Hamiltonian (\boldsymbol{\textcolor{red}{$\cdot$}}) with the exact calculation (\boldsymbol{\textcolor{blue}{$\cdot$}}) contrasted with the corresponding comparison at (a) $\mathbf{B}=0$ for valley $\mathbf{K}'$ at CNP. The spectrum at valley $\mathbf{K}$ is related by particle hole symmetry. For the comparison to be tractable we set $m_{max}=26$. Note that the above figure is spin degenerate.}
    \label{fig:VP CNP SWT}
\end{figure}
Clearly the subspace of interest, for which we want to derive an effective Hamiltonian, is the one spanned by states $|\rho=1,j,\tau,m\rangle$, i.e. the $a=\{3,4\}$ $c$ fermions for each $\tau=\pm 1$. To obtain the effective Hamiltonian we eliminate all the terms in operator Eqs.(\ref{eq:hat h plus 1}) and (\ref{eq:hat h -1}) (i.e. at $\tau=+1,-1$ respectively) which can mix the subspaces using Schrieffer Wolff Transformation (SWT). We moreover omit the contributions to effective Hamiltonian of order greater than $B^{3/2}$, i.e. $1/\ell^3$. %Since the two valleys are decoupled at CNP for VP state, let us discuss the SWT scheme for either valley $\tau$.
Below, we discuss the SWT scheme followed to obtain the effective Hamiltonian. We first rewrite the operators in Eqs.(\ref{eq:hat h plus 1}) and (\ref{eq:hat h -1}) with respect to the $\mathbf{B}=0$ eigenstates at $\Gamma$, $|\rho,j,\tau\rangle$, given in Eqs.(\ref{Eq:zero B spinor 1})-(\ref{Eq:zero B spinor 6}) as 
\begin{eqnarray}
    H = H_0 + \Delta V,
\end{eqnarray}
where $\Delta$ is an artificial parameter that helps us keeping track of the order in perturbation, to which we compute the effective Hamiltonian and is later set to 1. Along with $M$, we treat all the terms in operator in Eqs.(\ref{eq:hat h plus 1}) and (\ref{eq:hat h -1}) (i.e. at $\tau=+1,-1$ respectively) which can mix the subspaces as perturbation $V$. The unperturbed part $H_0=\left(\begin{array}{cc}
    H_0^{\tau=+1} &  \\
     & H_0^{\tau=-1}
\end{array}\right)$ where
\begin{eqnarray}
   H_0^{\tau} = \left(\begin{array}{ccc}
        E_\tau \sigma_0& 0&0 \\
        0&E_{X\tau} \sigma_0&0 \\
        0&0&E_{Y\tau} \sigma_0
   \end{array}\right). \label{Eq:H_0 for VP}
   %=\left(\begin{array}{ccc}
        %-\tau \frac{J}{2} \sigma_0& 0&0 \\
        %0&\bar{\epsilon}_\tau \sigma_0&0 \\
        %0&0&\epsilon_\tau \sigma_0
   %\end{array}\right),\label{Eq:H_0 for VP}
\end{eqnarray}
The Pauli matrix $\sigma_0$ above acts in the $j\in\{1,2\}$ space of for each $\rho$, and the energies $E$, $E_{X\tau}$, $E_{Y\tau}$ are given in Eqs.(\ref{Eq:CNP VP energy label 1})-(\ref{Eq:CNP VP energy label 3}). The perturbation is given as $V=\left(\begin{array}{cc}
    V^{\tau=+1} &0  \\
    0. & V^{\tau=-1}
\end{array}\right)$, where
\begin{eqnarray}
V^{\tau=+1}= 
\left(\begin{array}{cccccc}
        0 & M & -i\frac{\sqrt{2}}{\ell}\frac{X_1v_\star}{\sqrt{N_{X1}}}a&0&-i\frac{\sqrt{2}}{\ell}\frac{Y_1v_\star}{\sqrt{N_{Y1}}}a&0\\
         & 0&0 &i\frac{\sqrt{2}}{\ell}\frac{X_1v_\star}{\sqrt{N_{X1}}}a^\dagger &0 &i\frac{\sqrt{2}}{\ell}\frac{Y_1v_\star}{\sqrt{N_{Y1}}}a^\dagger \\
        & & 0&-i\frac{2\sqrt{2}}{\ell}\frac{X_1v_\star'}{N_{X1}}a &0 &-i\frac{\sqrt{2}}{\ell}\frac{(X_1+Y_1)v_\star'}{\sqrt{N_{X1}N_{Y1}}}a \\
         & & & 0&i\frac{\sqrt{2}}{\ell}\frac{(X_1+Y_1)v_\star'}{\sqrt{N_{X1}N_{Y1}}}a^\dagger &0 \\
         & & & & 0&-i\frac{2\sqrt{2}}{\ell}\frac{Y_1v_\star'}{N_{Y1}}a \\
         h.c.& & & & & 0
    \end{array}\right), \label{Eq:Perturbation_CNP_VP_K}
\end{eqnarray}
where $h.c.$ represents hermitian conjugate. In above matrix, the terms of $O(\frac{1}{\ell^2})$ have been opted out as they contribute terms of $O(\frac{1}{\ell^4})$ to the effective Hamiltonian. $V^{\tau=-1}$ can be obtained by replacing $X_1,Y_1\rightarrow X_{-1},Y_{-1}$, $N_{X1},N_{Y1}\rightarrow N_{X-1},N_{Y-1}$ and $a\leftrightarrow a^\dagger$ in Eq.(\ref{Eq:Perturbation_CNP_VP_K}). 
%\textcolor{red}{Note that the operators $a$ and $a^\dagger$ do not act on the $\mathbf{B}=0$ basis states. Infact, $H$ can be understood as a simple unitary transformation on the $6\times 6$ matrices in Eqs.(\ref{eq:hat h plus 1}) and (\ref{eq:hat h -1})}.
Let us now consider the SWT, generated by $S=S^\dagger$, such that it decouples the three subspaces in the transformed Hamiltonian. The transformed Hamiltonian is given as
 \begin{eqnarray}
    H' &=& e^{iS}He^{-iS}\\
    &=& H+i[S,H] + \frac{i^2}{2!}[S,[S,H]] + \frac{i^3}{3!}[S,[S,[S,H]]] + \ldots. \label{Eq:Hamiltonian expansion 1 for SWT}
\end{eqnarray}
The generator of the transformation  can be expanded in orders of $\Delta$ as
\begin{eqnarray}
    S = \Delta S_1 + \Delta^2 S_2 + \Delta^3 S_3 +\ldots \label{Eq:Generator expansion}
\end{eqnarray}
Substituting Eq.(\ref{Eq:Generator expansion}) into Eq.(\ref{Eq:Hamiltonian expansion 1 for SWT}) gives us
\begin{eqnarray}
 H' &=& H_0 + \Delta \left(V+i[S_1,H_0]\right)  + \Delta^2\left(i[S_2,H_0]+i[S_1,V] + \frac{i^2}{2!}[S_1,[S_1,H_0]]\right) +\\
 &&\Delta^3 \left(i[S_3,H_0] + i[S_2,V] + \frac{i^2}{2!}[S_2,[S_1,H_0]] + \frac{i^2}{2!}[S_1,[S_2,H_0]] +\frac{i^2}{2!}[S_1,[S_1,V]]] + \frac{i^3}{3!}[S_1,[S_1,[S_1,H_0]]\right) \\
 &&+ O(\Delta^4) + \ldots \\
 &=& H_0 + \Delta H_1' +\Delta^2 H_2' + \Delta^3 H_3' + O(\Delta^4) + \ldots \label{Eq:Hamiltonian expansion 2 for SWT}
\end{eqnarray}
Recall that the generator $S$ is defined by the condition that the SWT decouples the three subspaces. Since the two valleys are decoupled, for either valley $\tau$, we have
\begin{eqnarray}
    \langle\rho,j,\tau,m| H_n'|\rho',j',\tau,m'\rangle = \delta_{\rho,\rho'}\langle\rho,j,\tau,m| H_n'|\rho,j',\tau,m'\rangle. 
\end{eqnarray}
We thus have, for $\rho\neq \rho'$ and $n=1$ in above 
\begin{eqnarray}
    \langle\rho,j,\tau,m| \left(V+i[S_1,H_0]\right)|\rho',j',\tau,m'\rangle &=& 0\\
   \implies \langle\rho,j,\tau,m| S_1|\rho',j',\tau,m'\rangle = \begin{cases}
       i \frac{\langle\rho,j,\tau,m| V|\rho',j',\tau,m'\rangle}{E_{\rho'j'\tau} - E_{\rho j \tau}} & \text{for}~~ \rho \neq \rho',\\
      0 & \text{for}~~ \rho=\rho',
    \end{cases}\label{Eq:S_1 for VP}  
\end{eqnarray}
Similarly, for $\rho\neq \rho'$ and $n=2$, we have
\begin{eqnarray}
&&\langle\rho,j,\tau,m| \left(i[S_2,H_0]+i[S_1,V] + \frac{i^2}{2!}[S_1,[S_1,H_0]]\right)|\rho',j',\tau,m'\rangle = 0\\
&&\implies \langle\rho,j,\tau,m| S_2|\rho',j',\tau,m'\rangle =    
\frac{1}{E_{\rho j \tau}-E_{\rho'j'\tau}}\left(\sum_{\tilde{\rho}\neq\rho;\tilde{j};\tilde{m}} \langle\rho,j,\tau,m|S_1|\tilde{\rho},\tilde{j},\tau,\tilde{m}\rangle\langle \tilde{\rho},\tilde{j},\tau,\tilde{m}|V|\rho',j',\tau,m'\rangle - \right.\nonumber\\
&&\left. \sum_{\tilde{\rho}\neq\rho';\tilde{j};\tilde{m}} \langle\rho,j,\tau,m|V|\tilde{\rho},\tilde{j},\tau,\tilde{m}\rangle\langle \tilde{\rho},\tilde{j},\tau,\tilde{m}|S_1|\rho',j',\tau,m'\rangle \right. \nonumber \\
&& -\left. \sum_{\tilde{\rho}\neq\rho \rho';\tilde{j},\tilde{m}} \left(E_{\tilde{\rho}\tilde{j}\tau}-\frac{E_{\rho j \tau}+E_{\rho' j' \tau}}{2}\right)\langle\rho,j,\tau,m|S_1|\tilde{\rho},\tilde{j},\tau,\tilde{m}\rangle\langle \tilde{\rho},\tilde{j},\tau,\tilde{m}|S_1|\rho',j',\tau,m'\rangle \right)  \text{for} ~~ \rho\neq\rho', \nonumber\\
&&~~~~~~~~~~~~~~~~~~~~~~~~~~~~~~~~~~~~~~~= 0  ~~ \text{for} ~~ \rho=\rho', \label{Eq:S_2 for VP}
\end{eqnarray}
were $\tilde{j}$ is summed over $\{1,2\}$, i.e. the two states constituting each subspace. We further assume that 
\begin{eqnarray}
\sum_{\tilde{m}}|\tilde{\rho},\tilde{j},\tau,\tilde{m}\rangle\langle \tilde{\rho},\tilde{j},\tau,\tilde{m}|=|\tilde{\rho},\tilde{j},\tau\rangle\langle \tilde{\rho},\tilde{j},\tau|, 
\end{eqnarray}
which is justified in the ($\bk\cdot\bp$) continuum limit, as the upper cutoff on LLs in such a limit is unbounded. The effective Hamiltonian for either subspace $\rho$, can now be deduced in orders of $\Delta$. The O($\Delta^0$) contribution to the effective Hamiltonian in subspace $\rho$ is trivially $E_{\rho j\tau}\delta_{jj'}$ and O($\Delta^1$) terms are
\begin{eqnarray}
    \langle \rho, j, \tau,m|H_1'|\rho, j', \tau,m'\rangle &=&  \langle \rho, j, \tau,m|V|\rho, j', \tau ,m'\rangle .
\end{eqnarray}
 The O($\Delta^2$) terms are found out to be 
 \begin{eqnarray}
    \langle \rho, j, \tau,m|H_2'|\rho, j', \tau,m'\rangle &=& \frac{1}{2} \sum_{\tilde{\rho}\neq \rho;\tilde j}\langle \rho, j, \tau,m|V|\tilde{\rho},\tilde{j},\tau\rangle\langle \tilde{\rho},\tilde{j},\tau|V|\rho, j', \tau,m'\rangle\left(\frac{1}{E_{\rho j\tau}-E_{\tilde{\rho}\tilde{j}\tau}}+\frac{1}{E_{\tilde{\rho}\tilde{j}\tau}-E_{\rho j'\tau}}\right) \label{Eq:O2 contribution for VP}.
\end{eqnarray}
 The O($\Delta^3$) terms are found out to be 
 \begin{eqnarray}
   &&\langle \rho, j, \tau,m|H_3'|\rho, j', \tau,m' \rangle =   \frac{i}{2}\sum_{\tilde{\rho}\neq \rho;\tilde j}\left[\langle \rho, j, \tau,m|S_2|\tilde{\rho},\tilde{j},\tau\rangle\langle \tilde{\rho},\tilde{j},\tau|V|\rho, j', \tau,m'\rangle - \langle \rho, j, \tau,m|V|\tilde{\rho},\tilde{j},\tau\rangle\langle \tilde{\rho},\tilde{j},\tau|S_2|\rho, j', \tau,m'\rangle\right] \nonumber\\
   &&-\frac{1}{12}\sum_{\tilde{\rho}\neq \rho;\tilde {j}}\sum_{\bar{\rho}\neq \rho \tilde{\rho};\bar{j}}\left[\langle \rho, j, \tau,m|S_1|\bar{\rho},\bar{j},\tau\rangle\langle \bar{\rho},\bar{j},\tau|S_1|\tilde{\rho}, \tilde{j}, \tau\rangle \langle \tilde{\rho}, \tilde{j}, \tau|V|\rho, j', \tau,m'\rangle +\right. \nonumber \\ 
   && \left. \langle \rho, j, \tau,m|V|\bar{\rho},\bar{j},\tau\rangle\langle \bar{\rho},\bar{j},\tau|S_1|\tilde{\rho}, \tilde{j}, \tau\rangle \langle \tilde{\rho}, \tilde{j}, \tau|S_1|\rho, j', \tau,m'\rangle  -2\langle \rho, j, \tau,m|S_1|\bar{\rho},\bar{j},\tau\rangle\langle \bar{\rho},\bar{j},\tau|V|\tilde{\rho}, \tilde{j}, \tau\rangle \langle \tilde{\rho}, \tilde{j}, \tau|S_1|\rho, j', \tau,m'\rangle\right] \nonumber \\\label{Eq:O3 contribution for VP}
 \end{eqnarray}
Substituting Eqs.(\ref{Eq:S_1 for VP}) and (\ref{Eq:S_2 for VP}) in Eqs.(\ref{Eq:O2 contribution for VP}) and (\ref{Eq:O3 contribution for VP}) and further substituting them and Eq.(\ref{Eq:H_0 for VP})
 in Eq.(\ref{Eq:Hamiltonian expansion 2 for SWT}), upto O($\Delta^3$), the effective Hamiltonian for $\rho=1$ is given as
\begin{eqnarray}
  H^{eff} = \left(\begin{array}{cc}
      H^{\tau=+1}_{VP} & 0  \\
      0 & H^{\tau=-1}_{VP}
  \end{array}\right)\label{Eq:VP for CNP, final abstract},
\end{eqnarray}
where 
\begin{eqnarray}
 H^{\tau=+1} &=&  \left(\begin{array}{cc}
-\frac{J}{2}+\hbar \omega_c^{(1)} aa^\dagger & i\frac{A^{(1)}}{\ell^3}a^3\\
-i\frac{A^{(1)}}{\ell^3}{a^\dagger}^3 &-\frac{J}{2}+\hbar \omega_c^{(1)} a^\dagger a
     \end{array} \right) + M\left(1+\frac{M_c^{(1)}}{\ell^2}(aa^\dagger + a^\dagger a)\right)\sigma_x
     \label{Eq:Effective 2x2 for VP, +1},
\end{eqnarray}
where the Pauli matrix acts in the orbital space of $a\in\{4,3\}$ $c$-fermions. The cyclotron frequency $\omega_c^{\tau}$ and the other coefficients above are 
\begin{eqnarray}
    \hbar \omega_c^{(\tau)} &=& \frac{2 v_{\star}^2}{\ell^2}\left(\frac{X_\tau^2}{N_{X\tau}(E_\tau-E_{X\tau})} + \frac{Y_\tau^2}{N_{Y\tau}(E_\tau-E_{Y\tau})}\right),\\
    A^{(\tau)} &=& 4\sqrt{2}v_{\star}^2v_\star'\left(\frac{X_\tau Y_\tau(X_\tau + Y_\tau)}{N_{X\tau}N_{Y\tau}(E_\tau-E_{X\tau})(E_\tau-E_{Y\tau})}+\frac{X_\tau^3}{N_{X\tau}^2(E_\tau-E_{X\tau})^2} +\frac{Y_\tau^3}{N_{Y\tau}^2(E_\tau-E_{Y\tau})^2} \right),\\
    M_c^{(\tau)} &=& -v_\star^2\left(\frac{X_\tau^2}{N_{X\tau}(E_\tau-E_{X\tau})^2} + \frac{Y^2_\tau}{N_{Y\tau}(E_\tau-E_{Y\tau})^2}\right).
\end{eqnarray}
$H^{\tau=-1}_{VP}$ can be obtained by replacing $a\leftrightarrow a^\dagger$, $\omega_c^{(1)},A^{(1)},M_c^{(1)}\rightarrow \omega_c^{(-1)},A^{(-1)},M_c^{(-1)}$ and $-\frac{J}{2}\rightarrow\frac{J}{2}$ in Eq.(\ref{Eq:Effective 2x2 for VP, +1}). Substituting Eqs.(\ref{Eq:Xtau for CNP})-(\ref{Eq:CNP VP energy label 3}) in the coefficients, we have $\ell^2 \hbar \omega_c^{(\tau)}=-\tau (399586.49352) meV \mathring{A}^2$, $A^{(\tau)}=-\tau(42668446.86852)$ $meV \mathring{A}^3$ and $M_c^{(\tau)}= -12847.14767 \mathring{A}^2$. 
  
The above effective Hamiltonians $H^{\tau}$ describe the Landau quantization of the light single particle excitations towards CNP obtained for parent VP state. Clearly, the LLs emanate out of $-\tau J/2\pm M$. For the U(4) symmetric THFM (i.e. $M=0$), in the $\mathbf{B}\rightarrow 0$ limit, we can the drop off-diagonal O($\ell^{-3}$) terms in $H^{\tau}$. Then apart from the mode $\left(0,|0\rangle\right)^T$ and $\left(|0\rangle,0\right)^T$ at $\tau=+1$ and $-1$, respectively, all other LLs come in degenerate pair of two.
  %split by an energy $|\frac{Av_{\star}^2}{\gamma^2\ell^2}|$  
This degeneracy at $\mathbf{B}\rightarrow 0$ limit gets split as we tune back $M$. The comparison of the LL spectrum obtained using the above effective Hamiltonian with the one obtained via the exact calculation for valley sector $\mathbf{K}'$ is shown in Fig.(\ref{fig:VP CNP SWT}). The LL basis used to generate the plot is $(0,|m_1\rangle,|m_1\rangle,0)^T$, $(|m_2\rangle, |m_2+3\rangle,|m_2+3\rangle, |m_2\rangle)^T$ and $(|m_3\rangle,0,0,|m_3\rangle)^T$, where $m_1\in\{0,1,2\}$, $m_2\in\{0,\ldots,m_{max}-4\}$ and $m_3\in\{m_{max}-3,m_{max}-2,m_{max}-1\}$ respectively. We use $(|m_3\rangle,0,0,|m_3\rangle)^T$ rather than $(|m_3\rangle,|m_3+3\rangle,|m_3+3\rangle,|m_3\rangle)^T$ in order to avoid three modes emanating out of spurious $\mathbf{B}\rightarrow 0$ energy $-\tau J/2$, for each $\tau$. These three spurious modes are present in the exact calculation, as can be seen in Fig.(\ref{fig:VP CNP SWT}). However, recall that in practice we use $m_{max}=\lceil\frac{q-3}{2}\rceil$, so these three LLs are lost as $\mathbf{B}(q)$ increases(decreases).

\subsubsection{Parent Kramers Intervalley Coherent State}\label{apdx:KIVC CNP}
\begin{figure}
    \centering
    \includegraphics[width=10cm]{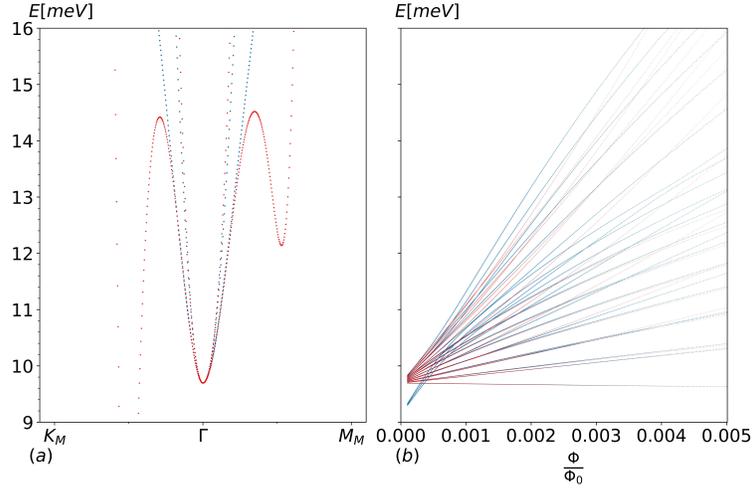}
    \caption{The comparison of (b) LL spectrum obtained using the effective Hamiltonian (\boldsymbol{\textcolor{red}{$\cdot$}}) with the exact calculation (\boldsymbol{\textcolor{blue}{$\cdot$}}) contrasted with the corresponding comparison at (a) $\mathbf{B}=0$ for KIVC state above CNP. The spectrum below CNP is related by particle hole symmetry. For the comparison to be tractable we set $m_{max}=11$. Note that the above figure is spin degenerate.}
    \label{fig:KIVC CNP SWT}
\end{figure}
In the case of the parent Kramers intervalley coherent (KIVC) state, the coupled $c$ and $f$ modes are described by the operator in Eq.(\ref{Eq: h for KIVC nu=0}). Unlike in previous section, we now break the $\bB=0$ Hilbert space at $\Gamma$ into 5 subspaces. The subspace of interest is the one spanned by states $|1,j,\tau\rangle$ with $j\in\{1,2\}$ and $\tau=\pm1$, i.e. the one spanned by $a=\{3,4\}$ $c$ fermions at each valley. The remaining four subspaces are spanned by $|2,j,+1\rangle$, $|2,j,-1\rangle$, $|3,j,+1\rangle$ and $|3,j,-1\rangle$, where $j\in\{1,2\}$. These states are given in Eqs.(\ref{Eq:zero B spinor 1})-(\ref{Eq:zero B spinor 6}). In the following discussion, we refer to these states and their energies by $|\epsilon,l\rangle$ and $E_{\epsilon l}$, respectively. Here $\epsilon\in\{1,\ldots,5\}$ labels the 5 subspaces. The index $l\in\{1,\ldots,4\}$ for $\epsilon=1$ and $l\in\{1,2\}$ for $\epsilon=\{3,4,5\}$, labels the states spanning the subspace $\epsilon$. We promote these states to $\bB\neq 0$ as $|\epsilon,l,m\rangle = |\epsilon,l\rangle|m\rangle$, where $|m\rangle$ is the $m^{th}$ LL, i.e. the finite $\bB$ $\bk\cdot\bp$ basis. Using the SWT procedure introduced in the previous section, we eliminate the terms in the operator given in Eq.(\ref{Eq: h for KIVC nu=0}) which can mix between these subspaces to obtain an effective Hamiltonian for the subspace spanned by by states $|\epsilon=1,l,m\rangle$.

To begin with, we re-write the operator in Eq.(\ref{Eq: h for KIVC nu=0}) with respect to the above $\bB=0$ eigenstates at $\Gamma$  as
\begin{eqnarray}
    H = H_0+\Delta V,
\end{eqnarray}
where $H_0=\left(\begin{array}{cc}
    H_0^{\tau=+1} &  \\
     & H_0^{\tau=-1}
\end{array}\right)$ and $V=\left(\begin{array}{cc}
    \bar{V}^{\tau=+1} &V_M  \\
    h.c. & \bar{V}^{\tau=-1}
\end{array}\right)$.
$H_0^{\tau}$ is given in Eq.(\ref{Eq:H_0 for VP}). The intra-valley perturbation $ \bar{V}^{\tau=+1}$ can be obtained by setting $M=0$ in Eq.(\ref{Eq:Perturbation_CNP_VP_K}). We can obtain $ \bar{V}^{\tau=-1}$ by replacing $X_1,Y_1\rightarrow X_{-1},Y_{-1}$, $N_{X1},N_{Y1}\rightarrow N_{X-1},N_{Y-1}$ and $a\leftrightarrow a^\dagger$ in $ V^{\tau=+1}$. The inter-valley perturbation $V_M$ is given as
\begin{eqnarray}
    V_M = \left(\begin{array}{ccc}
        M\sigma_z &0  &0\\
        0 &0  &0\\
        0 & 0 &0   
    \end{array}\right)\label{Eq:VM for KIVC SWT}
\end{eqnarray}
The effective Hamiltonian can then be expanded in orders of $\Delta$, as given in Eq.(\ref{Eq:Hamiltonian expansion 2 for SWT}). Following the condition that the transformation decouples each subspace, we have 
\begin{eqnarray}
    \langle\epsilon,l,m| H_n'|\epsilon',l',m'\rangle &=& \delta_{\epsilon,\epsilon'}\langle\epsilon,l,m| H_n'|\epsilon,l',m'\rangle\label{Eq:Step 1 KIVC,CNP}
\end{eqnarray}
For $\epsilon\neq \epsilon'$ and substituting $n=1$ in $H_n'$ gives
\begin{eqnarray}
    \langle\epsilon,l,m| \left(V+i[S_1,H_0]\right)|\epsilon',l',m'\rangle &=& 0\\
   \implies \langle\epsilon,l,m| S_1|\epsilon',l',m'\rangle = \begin{cases}
       i \frac{\langle\epsilon,l,m| V|\epsilon',l',m'\rangle}{E_{\epsilon' l'} - E_{\epsilon l}} & \text{for}~~ \epsilon \neq \epsilon',\\
      0 & \text{for}~~ \epsilon=\epsilon',
    \end{cases}\label{Eq:S_1 for KIVC}  
\end{eqnarray}
For $\epsilon\neq \epsilon'$ and substituting $n=2$ in $H_n'$ gives
\begin{eqnarray}
&&\langle\epsilon,l,m| \left(i[S_2,H_0]+i[S_1,V] + \frac{i^2}{2!}[S_1,[S_1,H_0]]\right)|\epsilon',l',m'\rangle = 0\\
&&\implies \langle\epsilon,l,m| S_2|\epsilon',l',m'\rangle =    
\frac{1}{E_{\epsilon l}-E_{\epsilon'l'}}\left(\sum_{\tilde{\epsilon}\neq\epsilon;\tilde{l};\tilde{m}} \langle\epsilon,l,m|S_1|\tilde{\epsilon},\tilde{l},\tilde{m}\rangle\langle \tilde{\epsilon},\tilde{l},\tilde{m}|V|\epsilon',l',m'\rangle - \right.\nonumber\\
&&\left. \sum_{\tilde{\epsilon}\neq\epsilon';\tilde{l};\tilde{m}} \langle\epsilon,l,m|V|\tilde{\epsilon},\tilde{l},\tilde{m}\rangle\langle \tilde{\epsilon},\tilde{l},\tilde{m}|S_1|\epsilon',l',m'\rangle \right. \nonumber \\
&& -\left. \sum_{\tilde{\epsilon}\neq\epsilon \epsilon';\tilde{l},\tilde{m}} \left(E_{\tilde{\epsilon}\tilde{l}}-\frac{E_{\epsilon l }+E_{\epsilon' l' }}{2}\right)\langle\epsilon,l,m|S_1|\tilde{\epsilon},\tilde{l},,\tilde{m}\rangle\langle \tilde{\epsilon},\tilde{l},\tilde{m}|S_1|\epsilon',l',m'\rangle \right)  \text{for} ~~ \epsilon\neq\epsilon', \nonumber\\
&&~~~~~~~~~~~~~~~~~~~~~~~~~~~~~~~~~~~~~~~= 0  ~~ \text{for} ~~ \epsilon=\epsilon', \label{Eq:S_2 for KIVC}
\end{eqnarray}
As in the previous section, we further assume that 
\begin{eqnarray}
\sum_{\tilde{m}}|\tilde{\epsilon},\tilde{l},\tilde{m}\rangle\langle \tilde{\epsilon},\tilde{l},\tilde{m}|=|\tilde{\epsilon},\tilde{l}\rangle\langle \tilde{\epsilon},\tilde{l}|, 
\end{eqnarray}
which is justified in the ($\bk\cdot\bp$) continuum limit, as the upper cutoff on LLs in such a limit is unbounded. Substituting Eqs.(\ref{Eq:S_1 for KIVC}) and (\ref{Eq:S_2 for KIVC}) into the expansion in Eq.(\ref{Eq:Hamiltonian expansion 2 for SWT}), we can now obtain the effective Hamiltonian in orders of $\Delta$. The O($\Delta^0$) contribution to the effective Hamiltonian in subspace $\epsilon$ is trivially $E_{\epsilon l}\delta_{ll'}$ and O($\Delta^1$) terms are
\begin{eqnarray}
    \langle \epsilon, l ,m|H_1'|\epsilon, l',m'\rangle &=&  \langle \epsilon, l, m|V|\epsilon, l' ,m'\rangle .
\end{eqnarray}
 The O($\Delta^2$) terms are found out to be 
 \begin{eqnarray}
    \langle \epsilon, l, m|H_2'|\epsilon, l',m'\rangle &=& \frac{1}{2} \sum_{\tilde{\epsilon}\neq \epsilon;\tilde l}\langle \epsilon, l ,m|V|\tilde{\epsilon},\tilde{l}\rangle\langle \tilde{\epsilon},\tilde{l}|V|\epsilon, l' ,m'\rangle\left(\frac{1}{E_{\epsilon l}-E_{\tilde{\epsilon}\tilde{l}}}+\frac{1}{E_{\tilde{\epsilon}\tilde{l}}-E_{\epsilon l'}}\right) \label{Eq:O2 contribution for KIVC}.
\end{eqnarray}
 The O($\Delta^3$) terms are found out to be 
 \begin{eqnarray}
   &&\langle \epsilon, l, m|H_3'|\epsilon, l', m' \rangle =   \frac{i}{2}\sum_{\tilde{\epsilon}\neq \epsilon;\tilde l}\left[\langle \epsilon, l ,m|S_2|\tilde{\epsilon},\tilde{l}\rangle\langle \tilde{\epsilon},\tilde{l}|V|\epsilon, l' ,m'\rangle - \langle \epsilon, l ,m|V|\tilde{\epsilon},\tilde{l}\rangle\langle \tilde{\epsilon},\tilde{l}|S_2|\epsilon, l' ,m'\rangle\right] \nonumber\\
   &&-\frac{1}{12}\sum_{\tilde{\epsilon}\neq \epsilon;\tilde {l}}\sum_{\bar{\epsilon}\neq \epsilon \tilde{\epsilon};\bar{l}}\left[\langle \epsilon, l, m|S_1|\bar{\epsilon},\bar{l}\rangle\langle \bar{\epsilon},\bar{l}|S_1|\tilde{\epsilon}, \tilde{l} \rangle \langle \tilde{\epsilon}, \tilde{l} |V|\epsilon, l', m'\rangle +\right. \nonumber \\ 
   && \left. \langle \epsilon, l ,m|V|\bar{\epsilon},\bar{l}\rangle\langle \bar{\epsilon},\bar{l}|S_1|\tilde{\epsilon}, \tilde{l} \rangle \langle \tilde{\epsilon}, \tilde{l} |S_1|\epsilon, l', m'\rangle  -2\langle \epsilon, l ,m|S_1|\bar{\epsilon},\bar{l}\rangle\langle \bar{\epsilon},\bar{l}|V|\tilde{\epsilon}, \tilde{l}\rangle \langle \tilde{\epsilon}, \tilde{l} |S_1|\epsilon, l',m'\rangle\right] \nonumber \\\label{Eq:O3 contribution for KIVC}
 \end{eqnarray}
Up to the O($\frac{1}{\ell^3}$), the effective Hamiltonian for subspace $\epsilon=1$ is found to be 
\begin{eqnarray}
  H^{eff} = \left(\begin{array}{cc}
      H^{\tau=+1}_{\text{KIVC}} & H_M  \\
      h.c. & H^{\tau=-1}_{\text{KIVC}}
  \end{array}\right),
\end{eqnarray}
where $H^{\tau}_{\text{KIVC}}$ can be obtained by setting $M=0$ in the expressions for $H^{\tau}_{VP}$ in Eq.(\ref{Eq:VP for CNP, final abstract}). The matrix $H_M$ couples both the valleys and is given as
\begin{eqnarray}
    H_M = 
        M\left(1+\frac{M_{c1}}{\ell^2}a^\dagger a + \frac{M_{c2}}{\ell^2}aa^\dagger\right)\sigma_z
\end{eqnarray}
where the Pauli matrix acts in the orbital space of $a\in\{4,3\}$ $c$-fermions and
\begin{eqnarray}
  M_{c1} &=& v_\star^2\left(\frac{X_{-1}^2}{N_{X{-1}}(E_{1}-E_{X{-1}})(E_{X{-1}}-E_{-1})} + \frac{Y_{-1}^2}{N_{Y{-1}}(E_1-E_{Y-1})(E_{Y-1}-E_{-1})}\right),\\ 
  M_{c2} &=& v_\star^2\left(\frac{X_{1}^2}{N_{X{1}}(E_{1}-E_{X{1}})(E_{X{1}}-E_{-1})} + \frac{Y_{1}^2}{N_{Y{1}}(E_1-E_{Y1})(E_{Y1}-E_{-1})}\right).
\end{eqnarray}
Substituting Eqs.(\ref{Eq:Xtau for CNP})-(\ref{Eq:CNP VP energy label 3}) above we get $M_{c1}= M_{c2}=-20880.79644\mathring{A}^2$. Clearly the LLs emanate out of $\mathbf{B}\rightarrow 0$ energy $\pm\sqrt{M^2+J^2/4}$. 
In the $\mathbf{B}\rightarrow 0$ limit, we can drop the O($\frac{1}{\ell^3}$) terms in the effective Hamiltonian. The LL energies in this limit take the form $\pm E^\pm_n$, where $E^\pm_n=\sqrt{(\frac{J}{2}+\hbar \omega^{(-1)}_c(n+\frac{1}{2}))^2+M^2(1+\frac{M_{c1}}{\ell^2}(2n+1))^2}\pm\frac{\hbar\omega^{(-1)}_c}{2}$, and $n\in\{0,1,2,\dots\}$. For small LL index $n$, we further have $E^-_{n+1}\approx E^+_n$.
%we have $1+(2n+1)\frac{M_{c1}}{\ell^2}\approx 1$. Moreover using the fact that $\frac{J}{2}\gg M$ and $\omega^{(-1)}_c>0$, we see that for small $n$, $E^\pm_n\approx \left(\frac{J}{2}+\hbar \omega^{(-1)}_c(n+\frac{1}{2})\right)(1+O(\frac{M^2}{J^2}))\pm \frac{\hbar\omega^{(-1)}_c}{2}$. Thus $E^-_{n+1}\approx E^+_n$. 
Thus except the anomalous mode $\pm E^-_0$, the leading LLs appear in pairs approaching double degeneracy, as can be seen in Fig.(\ref{fig:KIVC CNP SWT}) where we compare the LL spectrum obtained using the above effective Hamiltonian with the one obtained via the exact calculation. Note that such an approximate degeneracy is absent in the spectrum for VP state (see Fig.(\ref{fig:VP CNP SWT})).

The LL basis used to generate the plot is $(0,|m_1\rangle,|m_1\rangle,0)^T$, $(|m_2\rangle, |m_2+3\rangle,|m_2+3\rangle, |m_2\rangle)^T$ and $(|m_3\rangle,0,0,|m_3\rangle)^T$, where $m_1\in\{0,1,2\}$, $m_2\in\{0,\ldots,m_{max}-4\}$ and $m_3\in\{m_{max}-3,m_{max}-2,m_{max}-1\}$ respectively. We use $(|m_3\rangle,0,0,|m_3\rangle)^T$ rather than $(|m_3\rangle,|m_3+3\rangle,|m_3+3\rangle,|m_3\rangle)^T$ in order to avoid three modes emanating out of spurious $\mathbf{B}\rightarrow 0$ energy $\pm J/2$. These six spurious modes are present in the exact calculation, as can be seen in Fig.(\ref{fig:KIVC nu=-2 SWT}). However, recall that in practice we use $m_{max}=\lceil\frac{q-3}{2}\rceil$, so these three LLs are lost as $\mathbf{B}(q)$ increases(decreases).
%For $M=0$, the two valleys are decoupled and the $\mathbf{B}=0$ eigenstates at $\Gamma$ for each valley are given in Eqs.(\ref{Eq:zero B spinor 1})-(\ref{Eq:zero B spinor 1}) 

\subsection{\texorpdfstring{$\nu=-1$}{nu=-1}}
\subsubsection{Parent Valley Polarized State}
\begin{figure}
    \centering
    \includegraphics[width=12cm]{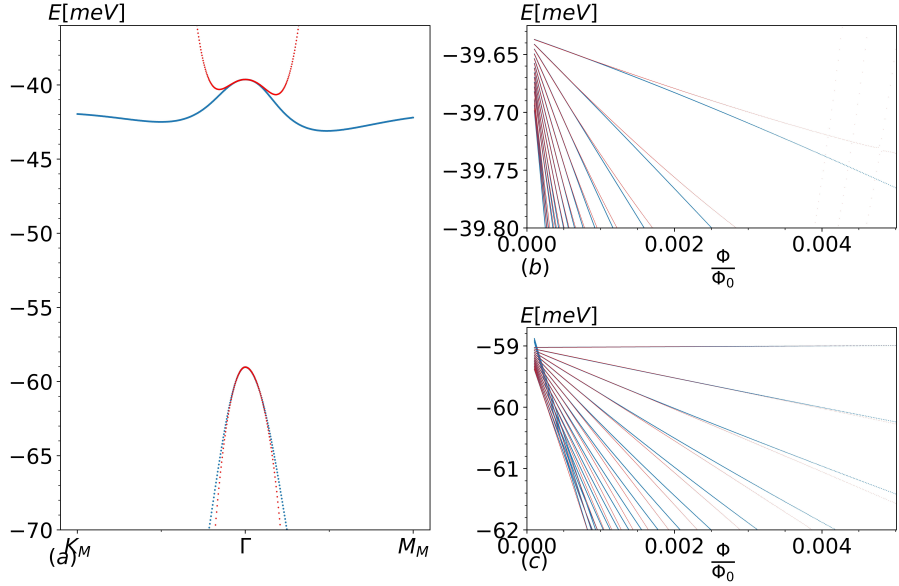}
    \caption{The comparison of (b,c) LL spectrum obtained using the effective Hamiltonian (\boldsymbol{\textcolor{red}{$\cdot$}}) with the exact calculation (\boldsymbol{\textcolor{blue}{$\cdot$}}) contrasted with the corresponding comparison at (a) $\mathbf{B}=0$ for valley spin sector $\mathbf{K} \downarrow$ at $\nu=-1$. For the comparison to be tractable we set an upper-cutoff of $25$ on the LL index. Note that the above figure is spin degenerate.}
    \label{fig:VP nu=-1 SWT}
\end{figure}
\textbf{K}\boldsymbol{$\downarrow$} : In this section we discuss the derivation of the $\bB\neq 0$ effective Hamiltonian for the spin and valley sector $\downarrow$, $\bK$ for the case of parent VP state at $\nu=-1$. The coupled modes of $c$ and $f$ fermions are described by the operator in Eq.(\ref{eq:-1 spin down}). As discussed in main text and Section(\ref{apdx:K nu=-1 spin-down}), for the above mean-field sector in flat band limit $M=0$, the magnetic subbands within the narrow bands  emanate out of the $\bB\rightarrow 0$ energy $-W_3 \mp \frac{J}{2}$. The remote subbands on the other hand emanate out of the $\bB\rightarrow 0$ energy 
$\frac{1}{2}\left(-W_1-(6U_2+\frac{U_1}{2})
\pm \sqrt{4\gamma^2 + (W_1-(6U_2+\frac{U_1}{2}))^2}\right)$ and $\frac{1}{2}\left(-W_1-(6U_2+\frac{3U_1}{2})
\pm \sqrt{4\gamma^2 + (W_1-(6U_2+\frac{3U_1}{2}))^2}\right)$. These energies correspond to the eigenvalues of the flat band $\bB=0$ THFM at $\Gamma$ at $\nu=-1$, for the same mean-field sector. Recall that at $\Gamma$, it has the form\cite{song2022magic} $\left(\begin{array}{ccc}
  -W_1\sigma_0 &0&\gamma\sigma_0\\
  & -W_3\sigma_0-\frac{J}{2}\sigma_z  &0  \\
   h.c.& & -(6U_2+U_1)\sigma_0-\frac{U_1}{2}\sigma_z
\end{array}\right)$, where Pauli matrix $\sigma$ acts in orbital space. The corresponding eigenvectors can be listed as 
\begin{eqnarray}
    |1\rangle &=& \left(0,0,1,0,0,0\right)^T \label{Eq:State1, VP, nu=-1}\\
    |2\rangle &=& \left(0,0,0,1,0,0\right)^T\label{Eq:State2, VP, nu=-1}\\
    |3\rangle &=& \frac{1}{\sqrt{N_{X}}}\left(X,0,0,0,1,0\right)\label{Eq:State3, VP, nu=-1}\\
    |4\rangle &=& \frac{1}{\sqrt{N_{Y}}}\left(Y,0,0,0,1,0\right)\label{Eq:State4, VP, nu=-1}\\
    |5\rangle &=& \frac{1}{\sqrt{N_{V}}}\left(0,V,0,0,0,1\right)\label{Eq:State5, VP, nu=-1} \\
    |6\rangle &=& \frac{1}{\sqrt{N_{Z}}}\left(0,Z,0,0,0,1\right)\label{Eq:State6, VP, nu=-1}
\end{eqnarray}
where the normalizations are $N_{X}=1+X^2$, $N_{Y}=1+Y^2$, $N_{V}=1+V^2$ and $N_{Z}=1+Z^2$ and 
\begin{eqnarray}
    X &=& -\frac{1}{2\gamma}\left(W_1-(6U_2+\frac{3U_1}{2})+\sqrt{4\gamma^2 + (W_1-(6U_2+\frac{3U_1}{2}))^2}\right)\label{Eq:Xtau for nu=-1}\\
    Y &=& -\frac{1}{2\gamma}\left(W_1-(6U_2+\frac{3U_1}{2})-\sqrt{4\gamma^2 + (W_1-(6U_2+\frac{3U_1}{2}))^2}\right)\\
    V &=& -\frac{1}{2\gamma}\left(W_1-(6U_2+\frac{U_1}{2})+\sqrt{4\gamma^2 + (W_1-(6U_2+\frac{U_1}{2}))^2}\right)\\
    Z &=& -\frac{1}{2\gamma}\left(W_1-(6U_2+\frac{U_1}{2})-\sqrt{4\gamma^2 + (W_1-(6U_2+\frac{U_1}{2}))^2}\right).
\end{eqnarray}
The energy of the state $|j\rangle$ where $j\in\{1,\ldots,6\}$, labelled by $E_j$ is given as
\begin{eqnarray}
    E_1 &=& -W_3-\frac{J}{2}\\
    E_2 &=& -W_3+\frac{J}{2}\\
    E_3 &=& \frac{1}{2}\left(-W_1-(6U_2+\frac{3U_1}{2})-\sqrt{4\gamma^2 + (W_1-(6U_2+\frac{3U_1}{2}))^2})\right)\equiv E_X\\
    E_4 &=& \frac{1}{2}\left(-W_1-(6U_2+\frac{3U_1}{2})+\sqrt{4\gamma^2 + (W_1-(6U_2+\frac{3U_1}{2}))^2}\right) \equiv E_Y\\
    E_5 &=& \frac{1}{2}\left(-W_1-(6U_2+\frac{U_1}{2})-\sqrt{4\gamma^2 + (W_1-(6U_2+\frac{U_1}{2}))^2}\right)\equiv E_V\\
    E_6 &=& \frac{1}{2}\left(-W_1-(6U_2+\frac{U_1}{2})+\sqrt{4\gamma^2 + (W_1-(6U_2+\frac{U_1}{2}))^2}\right)\equiv E_Z\label{Eq:nu=-1 VP energy label 6}
\end{eqnarray}
%{-((A - X + Sqrt[A^2 + 4 G^2 - 2 A X + X^2])/(2 G)), 0, 0, 0, 1, 0}, {-(( A - X - Sqrt[A^2 + 4 G^2 - 2 A X + X^2])/(2 G)), 0, 0, 0, 1, 0}
We break the $\bB=0$ Hilbert space at $\Gamma$ into 5 subspaces. These five subspaces are spanned by the states $\{|1\rangle,|2\rangle\}$, $|3\rangle$, $|4\rangle$, $|5\rangle$ and $|6\rangle$ respectively. We can relabel these states by $|\epsilon,l\rangle$, where $\epsilon$ labels the subspace and $l$ labels the state spanning the subspace, with $l\in\{1,2\}$ and $l=1$ for $\epsilon=1$ and $\epsilon\in\{2,\ldots,5\}$, respectively. We promote these states to $\bB\neq 0$ as $|\epsilon,l,m\rangle = |\epsilon,l\rangle|m\rangle$, where $|m\rangle$ is the $m^{th}$ LL, i.e. the finite $\bB$ $\bk\cdot\bp$ basis. Now using the SWT procedure introduced in previous section, we eliminate terms in the operator in Eq.(\ref{eq:-1 spin down}) which can mix the subspaces, to obtain an effective Hamiltonian for subspace $\epsilon=1$.

 To begin with, we rewrite the operators in Eqs.(\ref{eq:-1 spin down}) with respect to the states given in Eqs.(\ref{Eq:zero B spinor 1})-(\ref{Eq:zero B spinor 6}) as 
\begin{eqnarray}
    H = H_0 + \Delta V,
\end{eqnarray}
where $\Delta$ is an artificial parameter that helps us keeping track of the orders in perturbation, to which we compute the effective Hamiltonian and is later set to 1. Along with $M$, we treat all the terms in operator in Eq.(\ref{eq:-1 spin down}) which can mix the subspaces as perturbation $V$. The unperturbed part is given as
\begin{eqnarray}
   H_0 = \left(\begin{array}{cccccc}
        E_1& 0&0& 0&0&0 \\
        0&E_2&0&0&0& 0 \\
        0&0&E_{X}&0&0&0 \\
        0&0&0&E_{Y}&0&0 \\
        0&0&0&0&E_{V}&0 \\
        0&0&0&0&0&E_{Z}        
   \end{array}\right). \label{Eq:H_0 for VP, nu=-1}
   %=\left(\begin{array}{ccc}
        %-\tau \frac{J}{2} \sigma_0& 0&0 \\
        %0&\bar{\epsilon}_\tau \sigma_0&0 \\
        %0&0&\epsilon_\tau \sigma_0
   %\end{array}\right),\label{Eq:H_0 for VP}
\end{eqnarray}
The perturbation $V$ is given as
\begin{eqnarray}
V= 
\left(\begin{array}{cccccc}
        0 & M & i\frac{\sqrt{2}}{\ell}\frac{X v_\star}{\sqrt{N_{X}}}a^\dagger&i\frac{\sqrt{2}}{\ell}\frac{Y v_\star}{\sqrt{N_{Y}}}a^\dagger&0&0\\
        &0&0&0& -i\frac{\sqrt{2}}{\ell}\frac{V v_\star}{\sqrt{N_{V}}}a&-i\frac{\sqrt{2}}{\ell}\frac{Z v_\star}{\sqrt{N_{Z}}}a \\
        &&0&0&i\frac{\sqrt{2}}{\ell}\frac{(X+V) v_\star'}{\sqrt{N_{X}N_V}}a^\dagger&i\frac{\sqrt{2}}{\ell}\frac{(X+Z) v_\star'}{\sqrt{N_{X}N_Z}}a^\dagger \\
         & & & 0&i\frac{\sqrt{2}}{\ell}\frac{(Y+V) v_\star'}{\sqrt{N_{Y}N_V}}a^\dagger&i\frac{\sqrt{2}}{\ell}\frac{(Y+Z) v_\star'}{\sqrt{N_{Y}N_Z}}a^\dagger \\
         & & & & 0&0 \\
         h.c.& & & & & 0
    \end{array}\right), \label{Eq:Perturbation_nu=-1_VP_K}
\end{eqnarray}
Following the SWT procedure in Sec.(\ref{apdx:KIVC CNP}), i.e. the steps from Eq.(\ref{Eq:Step 1 KIVC,CNP})-(\ref{Eq:O3 contribution for KIVC}), upto O($\frac{1}{\ell^3}$), the effective Hamiltonian for subspace $\epsilon=1$ is found to be
\begin{eqnarray}
    H^{\nu=-1,VP}_{eff} &=& \left(\begin{array}{cc}
 -W_3+\frac{J}{2}+\hbar \tilde{\omega}_c a a^\dagger 
& i\frac{A}{\ell^3}{a}^3\\
-i\frac{A}{\ell^3}{a^\dagger}^3 & -W_3-\frac{J}{2}+\hbar \bar{\omega}_c a^\dagger a
     \end{array} \right) \nonumber \\
     &&+M\left(1+\frac{\bar{M}_c}{\ell^2}aa^\dagger + \frac{\tilde{M}_c}{\ell^2}a^\dagger a \right)\sigma_x
     \label{Eq:Effective 2x2 for nu=-1},
\end{eqnarray}
where the Pauli matrix acts in the orbital space of $a\in\{4,3\}$ $c$-fermions. The cyclotron frequencies and the coefficients accompanying $M$ above are
\begin{eqnarray}
   \hbar \bar{\omega}_c &=& \frac{2v_\star^2}{\ell^2}\left(\frac{X^2}{N_X(E_1-E_X)}+\frac{Y^2}{N_Y(E_1-E_Y)}\right)\\
   \hbar \tilde{\omega}_c &=&\frac{2v_\star^2}{\ell^2}\left(\frac{V^2}{N_V(E_2-E_V)}+\frac{Z^2}{N_Z(E_2-E_Z)}\right)\\
   \bar{M}_c &=& v_\star^2\left(\frac{Z^2}{N_Z(E_Z-E_2)(E_1-E_Z)} + \frac{V^2}{N_V(E_V-E_2)(E_1-E_V)}\right)\\
   \tilde{M}_c &=&v_\star^2\left(\frac{X^2}{N_X(E_1-E_X)(E_X-E_2)} + \frac{Y^2}{N_Y(E_1-E_Y)(E_Y-E_2)}\right).
   \end{eqnarray}
   The coefficient for $O(\frac{1}{\ell^3})$ term is
   \begin{eqnarray}
   A&=&2\sqrt{2}v_\star^2 v_\star'\left(\frac{XV(X+V)}{12N_XN_V}\left(\frac{1}{(E_V-E_X)(E_2-E_V)}+\frac{1}{(E_V-E_X)(E_X-E_1)}-\frac{2}{(E_X-E_1)(E_2-E_V)}\right)\right) + \nonumber \\
   &&2\sqrt{2}v_\star^2 v_\star'\left(\frac{YV(Y+V)}{12N_YN_V}\left(\frac{1}{(E_V-E_Y)(E_2-E_V)}+\frac{1}{(E_V-E_Y)(E_Y-E_1)}-\frac{2}{(E_Y-E_1)(E_2-E_V)}\right)\right) + \nonumber \\
   &&2\sqrt{2}v_\star^2 v_\star'\left(\frac{XZ(X+Z)}{12N_XN_Z}\left(\frac{1}{(E_Z-E_X)(E_2-E_Z)}+\frac{1}{(E_Z-E_X)(E_X-E_1)}-\frac{2}{(E_X-E_1)(E_2-E_Z)}\right)\right) + \nonumber \\
&&2\sqrt{2}v_\star^2 v_\star'\left(\frac{YZ(Y+Z)}{12N_YN_Z}\left(\frac{1}{(E_Z-E_Y)(E_2-E_Z)}+\frac{1}{(E_Z-E_Y)(E_Y-E_1)}-\frac{2}{(E_Y-E_1)(E_2-E_Z)}\right)\right) + \nonumber \\
&&-\frac{\sqrt{2}v_\star^2 v_\star'XV(X+V)}{N_XN_V(E_1-E_V)}\left(\frac{1}{E_X-E_1}-\frac{1}{E_V-E_X} + \frac{E_X-\frac{E_1+E_V}{2}}{(E_X-E_1)(E_V-E_X)}\right) + \nonumber 
\end{eqnarray}
\begin{eqnarray}
&&-\frac{\sqrt{2}v_\star^2 v_\star'YV(Y+V)}{N_YN_V(E_1-E_V)}\left(\frac{1}{E_Y-E_1}-\frac{1}{E_V-E_Y} + \frac{E_Y-\frac{E_1+E_V}{2}}{(E_Y-E_1)(E_V-E_Y)}\right) + \nonumber \\
&&-\frac{\sqrt{2}v_\star^2 v_\star'XZ(X+Z)}{N_XN_Z(E_1-E_Z)}\left(\frac{1}{E_X-E_1}-\frac{1}{E_Z-E_X} + \frac{E_X-\frac{E_1+E_Z}{2}}{(E_X-E_1)(E_Z-E_X)}\right) + \nonumber \\
&&-\frac{\sqrt{2}v_\star^2 v_\star'YZ(Y+Z)}{N_YN_Z(E_1-E_Z)}\left(\frac{1}{E_Y-E_1}-\frac{1}{E_Z-E_Y} + \frac{E_Y-\frac{E_1+E_Z}{2}}{(E_Y-E_1)(E_Z-E_Y)}\right) + \nonumber \\
&&\frac{\sqrt{2}v_\star^2 v_\star'XV(X+V)}{N_XN_V(E_X-E_2)}\left(\frac{1}{E_V-E_X}-\frac{1}{E_2-E_V} + \frac{E_V-\frac{E_2+E_X}{2}}{(E_V-E_X)(E_2-E_V)}\right) + \nonumber \\
&&\frac{\sqrt{2}v_\star^2 v_\star'XZ(X+Z)}{N_XN_Z(E_X-E_2)}\left(\frac{1}{E_Z-E_X}-\frac{1}{E_2-E_Z} + \frac{E_Z-\frac{E_2+E_X}{2}}{(E_Z-E_X)(E_2-E_Z)}\right) + \nonumber \\
&&\frac{\sqrt{2}v_\star^2 v_\star'YV(Y+V)}{N_YN_V(E_Y-E_2)}\left(\frac{1}{E_V-E_Y}-\frac{1}{E_2-E_V} + \frac{E_V-\frac{E_2+E_Y}{2}}{(E_V-E_Y)(E_2-E_V)}\right) + \nonumber \\
&&\frac{\sqrt{2}v_\star^2 v_\star'YZ(Y+Z)}{N_YN_Z(E_Y-E_2)}\left(\frac{1}{E_Z-E_Y}-\frac{1}{E_2-E_Z} + \frac{E_Z-\frac{E_2+E_Y}{2}}{(E_Z-E_Y)(E_2-E_Z)}\right).
\end{eqnarray}
Substituting Eqs.(\ref{Eq:Xtau for nu=-1})-(\ref{Eq:nu=-1 VP energy label 6}) in the coefficients, we have $\ell^2 \hbar \bar{\omega}_c =-692364.83538 meV \mathring{A}^2$, $\ell^2 \hbar \tilde{\omega}_c =-42241.47578$,  $A=-66946640.13364$ $meV \mathring{A}^3$, $\bar{M}_c= -15377.73807 \mathring{A}^2$ and $\tilde{M}_c= -25351.10053 \mathring{A}^2$. The comparison of the LL spectrum obtained using the above effective Hamiltonian with the one obtained via the exact calculation is shown in Fig.(\ref{fig:VP nu=-1 SWT}). 
The LL basis used to generate the plot is $(0,|m_1\rangle)^T$, $(|m_2\rangle, |m_2+3\rangle)^T$ and $(|m_3\rangle,0)^T$, where $m_1\in\{0,1,2\}$, $m_2\in\{0,\ldots,m_{max}-4\}$ and $m_3\in\{m_{max}-3,m_{max}-2,m_{max}-1\}$ respectively. We use $(|m_3\rangle,0)^T$ rather than $(|m_3\rangle,|m_3+3\rangle)^T$ in order to avoid three modes emanating out of spurious $\mathbf{B}\rightarrow 0$ energy $-(W_3+J/2)$. These three spurious modes are present in the exact calculation, as can be seen in Fig.(\ref{fig:VP nu=-1 SWT}). However, recall that in practice we use $m_{max}=\lceil\frac{q-3}{2}\rceil$, so these three LLs are lost as $\mathbf{B}(q)$ increases(decreases).
%    VS2_22 = 0.5*(V1452*(1/(Ev-Ey) -1/(E2-Ev) + (Ev-0.5*(Ey+E2))/(Ev-Ey)/(E2-Ev)) + V1462*(1/(Ez-Ey) -1/(E2-Ez) + (Ez-0.5*(Ey+E2))/(Ez-Ey)/(E2-Ez)))/(Ey-E2) 
% VS2_21 = 0.5*(V1352*(1/(Ev-Ex) -1/(E2-Ev) + (Ev-0.5*(E2+Ex))/(Ev-Ex)/(E2-Ev)) + V1362*(1/(Ez-Ex) -1/(E2-Ez) + (Ez-0.5*(E2+Ex))/(Ez-Ex)/(E2-Ez)))/(Ex-E2)

% S2V_11 = -0.5*(V1352*(1/(Ex-E1) -1/(Ev-Ex) + (Ex-0.5*(E1+Ev))/(Ex-E1)/(Ev-Ex)) + V1452*(1/(Ey-E1) -1/(Ev-Ey) + (Ey-0.5*(E1+Ev))/(Ey-E1)/(Ev-Ey)))/(E1-Ev)
% S2V_12 = -0.5*(V1362*(1/(Ex-E1) -1/(Ez-Ex) + (Ex-0.5*(E1+Ez))/(Ex-E1)/(Ez-Ex)) + V1462*(1/(Ey-E1) -1/(Ez-Ey) + (Ey-0.5*(E1+Ez))/(Ey-E1)/(Ez-Ey)))/(E1-Ez)

% d11 = (v_star**2)*(X**2/Nx/(E1-Ex) + Y**2/Ny/(E1-Ey)) #kbr k
%d22 = (v_star**2)*(V**2/Nv/(E2-Ev) + Z**2/Nz/(E2-Ez)) #k k_br
%S2V_2M = 0.5*M*(v_star**2)*(Z**2)/(Ez-E2)/Nz/(E1-Ez) #*k*k_br
%S2V_1M = 0.5*M*(v_star**2)*(V**2)/(Ev-E2)/Nv/(E1-Ev) #*k*k_br
%VS2_21M = 0.5*M*(v_star**2)*(X**2)/(E1-Ex)/Nx/(Ex-E2) #*kbrk
%VS2_22M = 0.5*M*(v_star**2)*(Y**2)/(E1-Ey)/Ny/(Ey-E2) #*kbrk

%$d11=-692364.8353793388$ $d22=-42241.47577742081$
%$A=-66946640.133643776$
%$M1=-15377.738068829087 $
%$M2=-25351.100527382063$
\begin{figure}
    \centering
    \includegraphics[width=10cm]{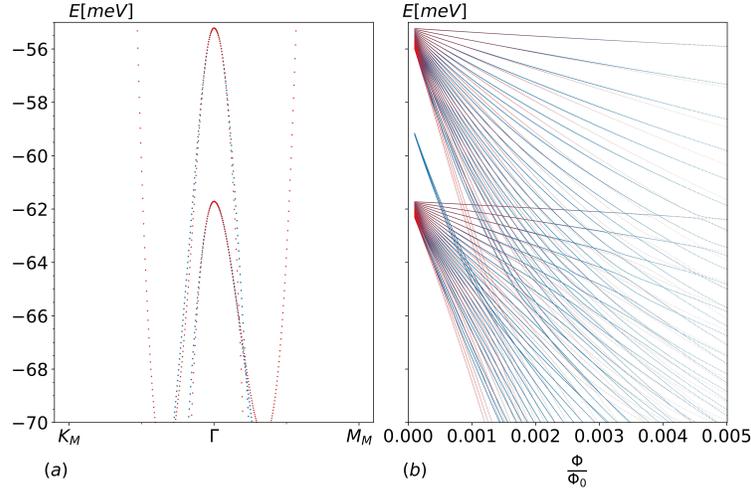}
    \caption{The comparison of (b) LL spectrum obtained using the effective Hamiltonian (\boldsymbol{\textcolor{red}{$\cdot$}}) with the exact calculation (\boldsymbol{\textcolor{blue}{$\cdot$}}) contrasted with the corresponding comparison at (a) $\mathbf{B}=0$ for VP state at $\nu=-1$ for sector $\mathbf{K}\uparrow$. For the comparison to be tractable we set $m_{max}=25$.}
    \label{fig:nu=-1 K up SWT}
\end{figure}
\textbf{K}\boldsymbol{$\uparrow$} : In this section we discuss the derivation of $\bB\neq 0$ effective Hamiltonian for the spin and valley sector $s=\uparrow$ $\bK$ for the case of parent VP state at $\nu=-1$. As discussed in Sec.(\ref{apdx:K nu=-1 spin-up}), for the above mean-field sector in flat band limit $M=0$, the magnetic subbands within the narrow bands  emanate out of the $\bB\rightarrow 0$ energy $-W_3-\frac{J}{2}$. The remote subbands on the other hand emanate out of the $\bB\rightarrow 0$ energy 
$\frac{1}{2}\left(-W_1-(6U_2+3U_1/2)
\pm \sqrt{4\gamma^2 + (W_1-(6U_2 + 3U_1/2))^2}\right)$. These energies correspond to the eigenvalues of the flat band $\bB=0$ THFM at $\Gamma$ at $\nu=-1$ for the same sector. Recall that at $\Gamma$, it has the form\cite{song2022magic} $\left(\begin{array}{ccc}
  -W_1\sigma_0 &0&\gamma\sigma_0\\
  & -(W_3+ J/2)\sigma_0  & 0 \\
   h.c.& & -(6U_2+3U_1/2)\sigma_0
\end{array}\right)$, where Pauli matrix $\sigma_0$ acts in orbital space. The corresponding eigenvectors can be listed as 
\begin{eqnarray}
    |1,1\rangle &=& \left(0,0,0,1,0,0\right)^T, 
    \label{Eq:nu-1 up VP zero B spinor 1}\\
    |1,2\rangle &=& \left(0,0,1,0,0,0\right)^T, 
    \label{Eq:nu-1 up VPzero B spinor 2}\\
    |2,1\rangle &=& \frac{1}{\sqrt{N_{X}}}\left(0,X,0,0,0,1\right)^T, 
    \label{Eq:nu-1 up VPzero B spinor 3}\\
    |2,2\rangle &=& \frac{1}{\sqrt{N_{X}}}\left(X,0,0,0,1,0\right)^T, 
    \label{Eq:nu-1 up VPzero B spinor 4}\\
    |3,1\rangle &=& \frac{1}{\sqrt{N_{Y}}}\left(0,Y,0,0,0,1\right)^T, 
    \label{Eq:nu-1 up VPzero B spinor 5}\\
    |3,2\rangle &=& \frac{1}{\sqrt{N_{Y}}}\left(Y,0,0,0,1,0\right)^T, \label{Eq:nu-1 up VPzero B spinor 6}
\end{eqnarray}
where 
\begin{eqnarray}
X &=& -\frac{1}{2\gamma}\left( W_1-(6U_2+3U_1/2) + \sqrt{4\gamma^2 + (W_1-(6U_2+3U_1/2))^2}\right), \label{Eq:Xtau for CNP,nu=-1 up}\\
Y &=& -\frac{1}{2\gamma}\left( W_1-(6U_2+3U_1/2) - \sqrt{4\gamma^2 + (W_1-(6U_2+3U_1/2))^2}\right), \label{Eq:Ytau for CNP,nu=-1 up}
\end{eqnarray}
and normalizations  $N_{X}=1+X^2$, $N_{Y}=1+Y^2$. The energy of state $\rho,j$ with $\rho\in\{1,2,3\}$ and $j\in\{1,2\}$, $E_{\rho,j}$, are given as
\begin{eqnarray}
E_{1,j} &=& -(W_3 +\frac{J}{2}) \equiv E ,\label{Eq:CNP VP energy label 1,nu=-1up}\\
E_{2,j} &=& \frac{1}{2}\left(-W_1-(6U_2+3U_1/2)
- \sqrt{4\gamma^2 + (W_1-(6U_2+3U_1/2))^2}\right) \equiv E_{X},
\label{Eq:CNP VP energy label 2,nu=-1up}\\
E_{3,j} &=& \frac{1}{2}\left(-W_1-(6U_2+3U_1/2)
+ \sqrt{4\gamma^2 + (W_1-(6U_2+3U_1/2))^2}\right) \equiv E_{Y}.\label{Eq:CNP VP energy label 3,nu=-1up}
\end{eqnarray}
We break the $\bB=0$ Hilbert space at $\Gamma$ into 3 subspaces. These three subspaces are spanned by the states $|1,j\rangle$, $|2,j\rangle$ and $|3,j\rangle$, respectively. We promote these states to $\bB\neq 0$ as $|\rho,j,m\rangle = |\rho,j\rangle |m\rangle$, where $|m\rangle$ is the $m^{th}$ LL, i.e. the finite $\bB$ $\bk\cdot\bp$ basis. Using the SWT procedure introduced in Sec.(\ref{apdx:CNP VP SWT}), we eliminate all the terms in the operator in Eq.(\ref{eq:-1 spin up}) that can mix the subspaces to obtain an effective Hamiltonian for the subspace spanned by states $|1,j,m\rangle$, i.e. the one spanned by $a=\{3,4\}$ $c$ fermions.
We first re-write the operator in Eq.(\ref{eq:-1 spin up}) with respect to the eigenstates in Eqs.(\ref{Eq:nu-1 up VP zero B spinor 1})-(\ref{Eq:nu-1 up VPzero B spinor 6}) as
\begin{eqnarray}
    H = H_0 + \Delta V,
\end{eqnarray}
where 
\begin{eqnarray}
    H_0 =\left( \begin{array}{ccc}
    E\sigma_0  & 0  & 0 \\
     0 & E_{X}\sigma_0   & 0 \\         
     0 &  0 & E_{Y}\sigma_0
    \end{array}\right),\label{Eq:H0 for SWT, nu=-1, VP}
\end{eqnarray}
where Pauli matrix $\sigma_0$ acts in the $j$ space for each $\rho$. Along with $M$, we treat all the terms in operator in Eq.(\ref{eq:-1 spin up}) which can mix the subspaces as perturbation $V$. $\Delta$ is an artificial parameter that helps us keeping track of the orders in perturbation, to which we compute the effective Hamiltonian and is later set to 1. The matrix $V$ can be obtained by replacing $X_{1}$, $Y_1$, $N_{X1}$ and $N_{Y1}$ in Eq.(\ref{Eq:Perturbation_CNP_VP_K}) by $X$, $Y$, $N_{X}$ and $N_{Y}$ respectively. Following the procedure for SWT in Sec.(\ref{apdx:CNP VP SWT}), i.e. from Eq.(\ref{Eq:Hamiltonian expansion 1 for SWT})-(\ref{Eq:O3 contribution for VP}) and replacing $|\rho,j,\tau,m\rangle$ by $|\rho,j,m\rangle$ in these steps, the effective Hamiltonian for $\rho=1$ and spin $s$ is found to be
\begin{eqnarray}
 H^{\tau=1,s=\uparrow,\nu=-1}_{VP} &=&  \left(\begin{array}{cc}
-(W_3 +\frac{J}{2})+\hbar \omega_c aa^\dagger & i\frac{A}{\ell^3}a^3\\
-i\frac{A}{\ell^3}{a^\dagger}^3 &-(W_3 +\frac{J}{2})+\hbar \omega_c a^\dagger a
     \end{array} \right) + M\left(1+\frac{M_c}{\ell^2}(aa^\dagger + a^\dagger a)\right)\sigma_x   \label{Eq:Effective 2x2 for VP, +1 up, nu=-1},
\end{eqnarray}
where the Pauli matrix acts in the orbital space of $a\in\{4,3\}$ $c$-fermions. The cyclotron frequency $\omega_c$ and the other coefficients above are 
\begin{eqnarray}
    \hbar \omega_c &=& \frac{2 v_{\star}^2}{\ell^2}\left(\frac{X^2}{N_{X}(E-E_{X})} + \frac{Y^2}{N_{Y}(E-E_{Y})}\right),\label{Eq:cyclotron freq nu=-1up CNP}\\
    A &=& 4\sqrt{2}v_{\star}^2v_\star'\left(\frac{X Y(X + Y)}{N_{X}N_{Y}(E-E_{X})(E-E_{Y})}+\frac{X^3}{N_{X}^2(E-E_{X})^2} +\frac{Y^3}{N_{Y}^2(E-E_{Y})^2} \right),\label{Eq:A nu=-1up CNP}\\
    M_c &=& -v_\star^2\left(\frac{X^2}{N_{X}(E-E_{X})^2} + \frac{Y^2}{N_{Y}(E-E_{Y})^2}\right).
\end{eqnarray}
Substituting Eqs.(\ref{Eq:Xtau for CNP,nu=-1 up})-(\ref{Eq:Ytau for CNP,nu=-1 up}) in the coefficients, we have $\ell^2 \hbar \omega_c= -692364.83538 meV \mathring{A}^2$, $A=-61125292.83865 meV \mathring{A}^3$ and $M_c=-13387.26641
 \mathring{A}^2$. For the U(4) symmetric THFM (i.e. $M=0$), in the $\mathbf{B}\rightarrow 0$ limit, we can the drop off-diagonal O($\ell^{-3}$) terms in in the above effective Hamiltonian. Then apart from the mode $\left(0,|0\rangle\right)^T$, all other LLs come in degenerate pair of two. This degeneracy at $\mathbf{B}\rightarrow 0$ limit gets split as we tune back $M$.
 The comparison of the LL spectrum obtained using the above effective Hamiltonian with the one obtained via the exact calculation is shown in Fig.(\ref{fig:nu=-1 K up SWT}). The LL basis used to generate the plot is $(0,|m_1\rangle)^T$, $(|m_2\rangle, |m_2+3\rangle)^T$ and $(|m_3\rangle,0)^T$, where $m_1\in\{0,1,2\}$, $m_2\in\{0,\ldots,m_{max}-4\}$ and $m_3\in\{m_{max}-3,m_{max}-2,m_{max}-1\}$ respectively. We use $(|m_3\rangle,0)^T$ rather than $(|m_3\rangle,|m_3+3\rangle)^T$ in order to avoid three modes emanating out of spurious $\mathbf{B}\rightarrow 0$ energy $-(W_3+J/2)$. These three spurious modes are present in the exact calculation, as can be seen in Fig.(\ref{fig:nu=-1 K up SWT}). However, recall that in practice we use $m_{max}=\lceil\frac{q-3}{2}\rceil$, so these three LLs are lost as $\mathbf{B}(q)$ increases(decreases).

\subsection{\texorpdfstring{$\nu=-2$}{nu=-2}}
\subsubsection{Parent Valley Polarized State}
\begin{figure}
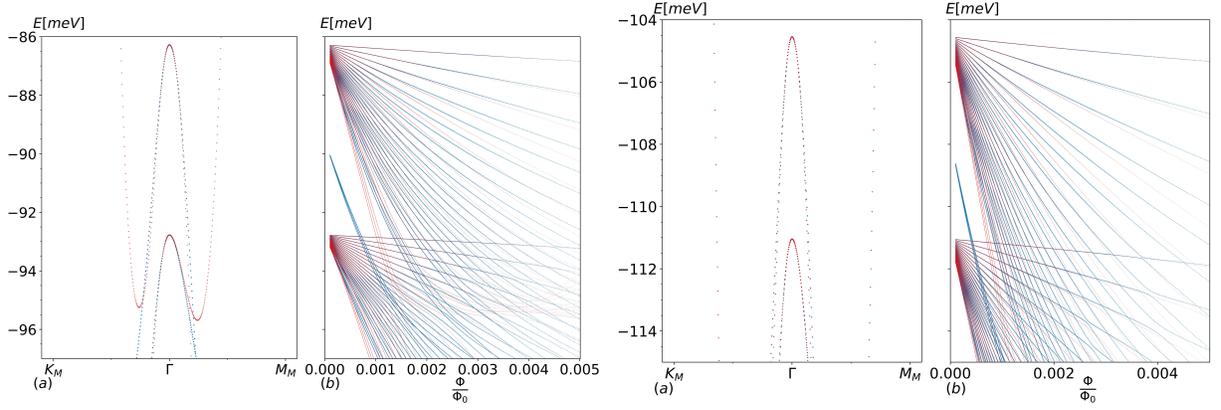

    \includegraphics[width=8cm]{Images_for_supplementary/Supp_Fig_SWT_nu2_VP1_arxiv.jpg}
    \includegraphics[width=8cm]{Images_for_supplementary/Supp_Fig_SWT_nu2_VP2_arxiv.jpg}
    \caption{(i)\textbf{Left Panel} - a,b : The comparison of (b) LL spectrum obtained using the effective Hamiltonian (\boldsymbol{\textcolor{red}{$\cdot$}}) with the exact calculation (\boldsymbol{\textcolor{blue}{$\cdot$}}) contrasted with the corresponding comparison at (a) $\mathbf{B}=0$ for valley $\mathbf{K}$ spin $\downarrow$ at $\nu=-2$. (ii) \textbf{Right Panel} - a,b : The comparison of (b) LL spectrum obtained using the effective Hamiltonian (\boldsymbol{\textcolor{red}{$\cdot$}}) with the exact calculation (\boldsymbol{\textcolor{blue}{$\cdot$}}) contrasted with the corresponding comparison at (a) $\mathbf{B}=0$ for valley $\mathbf{K}$ spin $\uparrow$ at $\nu=-2$. For the comparison to be tractable we set an upper-cutoff of $25$ on the LL index.}
    \label{fig:nu=-2 VP SWT}
\end{figure}
In this section we discuss the derivation of the case of $\bB\neq 0$ effective Hamiltonian for the spin and valley sector $s=\uparrow,\downarrow$, $\bK$ for the case of parent VP state at $\nu=-2$. As discussed in Sec.(\ref{apdx:nu=-2 Valley K}), for the above mean-field sector in flat band limit $M=0$, the magnetic subbands within the narrow bands  emanate out of the $\bB\rightarrow 0$ energy $-2W_3-\zeta_s \frac{J}{2}$, where $\zeta_s=1,-1$ for $s=\uparrow,\downarrow$ respectively. The remote subbands on the other hand emanate out of the $\bB\rightarrow 0$ energy 
$\frac{1}{2}\left(-2W_1-(12U_2+\frac{4+\zeta_s}{2}U_1)
\pm \sqrt{4\gamma^2 + (2W_1-(12U_2+\frac{4+\zeta_s}{2}U_1))^2}\right)$. These energies correspond to the eigenvalues of the flat band $\bB=0$ THFM at $\Gamma$ at $\nu=-2$ for the same mean-field sector. Recall that at $\Gamma$, it has the form\cite{song2022magic} $\left(\begin{array}{ccc}
  -2W_1\sigma_0 &0&\gamma\sigma_0\\
  & -(2W_3+\zeta_s J/2)\sigma_0  &0  \\
   h.c.& & -(12U_2+\frac{4+\zeta_s}{2}U_1)\sigma_0
\end{array}\right)$, where Pauli matrix $\sigma_0$ acts in orbital space. The corresponding eigenvectors for given spin $s$ can be listed as 
\begin{eqnarray}
    |1,1,s\rangle &=& \left(0,0,0,1,0,0\right)^T, 
    \label{Eq:nu-2 VP zero B spinor 1}\\
    |1,2,s\rangle &=& \left(0,0,1,0,0,0\right)^T, 
    \label{Eq:nu-2 VPzero B spinor 2}\\
    |2,1,s\rangle &=& \frac{1}{\sqrt{N_{Xs}}}\left(0,X_s,0,0,0,1\right)^T, 
    \label{Eq:nu-2 VPzero B spinor 3}\\
    |2,2,s\rangle &=& \frac{1}{\sqrt{N_{Xs}}}\left(X_s,0,0,0,1,0\right)^T, 
    \label{Eq:nu-2 VPzero B spinor 4}\\
    |3,1,s\rangle &=& \frac{1}{\sqrt{N_{Ys}}}\left(0,Y_s,0,0,0,1\right)^T, 
    \label{Eq:nu-2 VPzero B spinor 5}\\
    |3,2,s\rangle &=& \frac{1}{\sqrt{N_{Ys}}}\left(Y_s,0,0,0,1,0\right)^T, \label{Eq:nu-2 VPzero B spinor 6}
\end{eqnarray}
where 
\begin{eqnarray}
X_s &=& -\frac{1}{2\gamma}\left( 2W_1-(12U_2+\frac{4+\zeta_s}{2}U_1) + \sqrt{4\gamma^2 + (2W_1-(12U_2+\frac{4+\zeta_s}{2}U_1))^2}\right), \label{Eq:Xtau for CNP,nu=-2}\\
Y_s &=& -\frac{1}{2\gamma}\left( 2W_1-(12U_2+\frac{4+\zeta_s}{2}U_1) - \sqrt{4\gamma^2 + (2W_1-(12U_2+\frac{4+\zeta_s}{2}U_1))^2}\right), \label{Eq:Ytau for CNP,nu=-2}
\end{eqnarray}
and normalizations  $N_{Xs}=1+X_s^2$, $N_{Ys}=1+Y_s^2$. The energy of state $\rho,j,s$ with $\rho\in\{1,2,3\}$ and $j\in\{1,2\}$, $E_{\rho,j,s}$, are given as
\begin{eqnarray}
E_{1,j,s} &=& -(2W_3 +\zeta_s\frac{J}{2}) \equiv E_s ,\label{Eq:CNP VP energy label 1,nu=-2}\\
E_{2,j,s} &=& \frac{1}{2}\left(-2W_1-(12U_2+\frac{4+\zeta_s}{2}U_1)
- \sqrt{4\gamma^2 + (2W_1-(12U_2+\frac{4+\zeta_s}{2}U_1))^2}\right) \equiv E_{Xs},
\label{Eq:CNP VP energy label 2,nu=-2}\\
E_{3,j,s} &=& \frac{1}{2}\left(-2W_1-(12U_2+\frac{4+\zeta_s}{2}U_1)
+ \sqrt{4\gamma^2 + (2W_1-(12U_2+\frac{4+\zeta_s}{2}U_1))^2}\right) \equiv E_{Ys}.\label{Eq:CNP VP energy label 3,nu=-2}
\end{eqnarray}
We break the $\bB=0$ Hilbert space at $\Gamma$ into 3 subspaces for each $s$. These three subspaces are spanned by the states $|1,j,s\rangle$, $|2,j,s\rangle$ and $|3,j,s\rangle$, respectively. We promote these states to $\bB\neq 0$ as $|\rho,j,s,m\rangle = |\rho,j,s\rangle |m\rangle$, where $|m\rangle$ is the $m^{th}$ LL, i.e. the finite $\bB$ $\bk\cdot\bp$ basis. Using the SWT procedure introduced in Sec.(\ref{apdx:CNP VP SWT}), we eliminate all the terms in the operator in Eq.(\ref{eq:pm 2 spin updown}) that can mix the subspaces to obtain an effective Hamiltonian for the subspace spanned by states $|1,j,s,m\rangle$, i.e. the one spanned by $a=\{3,4\}$ $c$ fermions.

We first re-write the operator in Eq.(\ref{eq:pm 2 spin updown}) with respect to the eigenstates in Eqs.(\ref{Eq:nu-2 VP zero B spinor 1})-(\ref{Eq:nu-2 VPzero B spinor 6}) as
\begin{eqnarray}
    H = H_0 + \Delta V,
\end{eqnarray}
where 
\begin{eqnarray}
    H_0 =\left( \begin{array}{ccc}
    E_s\sigma_0  & 0  & 0 \\
     0 & E_{Xs}\sigma_0   & 0 \\         
     0 &  0 & E_{Ys}\sigma_0
    \end{array}\right),\label{Eq:H0 for SWT, nu=-2, VP}
\end{eqnarray}
where Pauli matrix $\sigma_0$ acts in the $j$ space for each $\rho$. Along with $M$, we treat all the terms in operator in Eq.(\ref{eq:pm 2 spin updown}) which can mix the subspaces as perturbation $V$. $\Delta$ is an artificial parameter that helps us keeping track of the orders in perturbation, to which we compute the effective Hamiltonian and is later set to 1. The matrix $V$ can be obtained by replacing $X_{1}$, $Y_1$, $N_{X1}$ and $N_{Y1}$ in Eq.(\ref{Eq:Perturbation_CNP_VP_K}) by $X_s$, $Y_s$, $N_{Xs}$ and $N_{Ys}$ respectively. Following the procedure for SWT in Sec.(\ref{apdx:CNP VP SWT}), i.e. from Eq.(\ref{Eq:Hamiltonian expansion 1 for SWT})-(\ref{Eq:O3 contribution for VP}) and replacing $|\rho,j,\tau,m\rangle$ by $|\rho,j,s,m\rangle$ in these steps, the effective Hamiltonian for $\rho=1$ and spin $s$ is found to be
\begin{eqnarray}
 H^{\tau=1,s=\uparrow,\downarrow,\nu=-2}_{VP} &=&  \left(\begin{array}{cc}
-(2W_3 +\zeta_s\frac{J}{2})+\hbar \omega_c^{(s)} aa^\dagger & i\frac{A^{(s)}}{\ell^3}a^3\\
-i\frac{A^{(s)}}{\ell^3}{a^\dagger}^3 &-(2W_3 +\zeta_s\frac{J}{2})+\hbar \omega_c^{(s)} a^\dagger a
     \end{array} \right) + M\left(1+\frac{M_c^{(s)}}{\ell^2}(aa^\dagger + a^\dagger a)\right)\sigma_x  \label{Eq:Effective 2x2 for VP, +1, nu=-2},
\end{eqnarray}
where the Pauli matrix acts in the orbital space of $a\in\{4,3\}$ $c$-fermions. The cyclotron frequency $\omega_c^{(s)}$ and the other coefficients above are 
\begin{eqnarray}
    \hbar \omega_c^{(s)} &=& \frac{2 v_{\star}^2}{\ell^2}\left(\frac{X_s^2}{N_{Xs}(E_s-E_{Xs})} + \frac{Y_s^2}{N_{Ys}(E_s-E_{Ys})}\right),\label{Eq:cyclotron freq nu=-2 CNP}\\
    A^{(s)} &=& 4\sqrt{2}v_{\star}^2v_\star'\left(\frac{X_s Y_s(X_s + Y_s)}{N_{Xs}N_{Ys}(E_s-E_{Xs})(E_s-E_{Ys})}+\frac{X_s^3}{N_{Xs}^2(E_s-E_{Xs})^2} +\frac{Y_s^3}{N_{Ys}^2(E_s-E_{Ys})^2} \right),\label{Eq:A nu=-2 CNP}\\
    M_c^{(s)} &=& -v_\star^2\left(\frac{X_s^2}{N_{Xs}(E_s-E_{Xs})^2} + \frac{Y^2_s}{N_{Ys}(E_s-E_{Ys})^2}\right).
\end{eqnarray}
Substituting Eqs.(\ref{Eq:Xtau for CNP,nu=-2})-(\ref{Eq:CNP VP energy label 3,nu=-2}) in the coefficients, we have $\ell^2 \hbar \omega_c^{(\uparrow)}= -831301.40524 meV \mathring{A}^2$, $\ell^2 \hbar \omega_c^{(\downarrow)}=-514097.23133 $ $meV \mathring{A}^2$, $A^{(\uparrow)}=-57864769.72509 meV \mathring{A}^3$, $A^{(\downarrow)}= -59278877.34125$ $meV \mathring{A}^3$, $M_c^{(\uparrow)}=-13211.61944
 \mathring{A}^2$ and $M_c^{(\downarrow)}=-15952.11343 \mathring{A}^2$. For the U(4) symmetric THFM (i.e. $M=0$), in the $\mathbf{B}\rightarrow 0$ limit, we can the drop off-diagonal O($\ell^{-3}$) terms in in the above effective Hamiltonian. Then apart from the mode $\left(0,|0\rangle\right)^T$, all other LLs come in degenerate pair of two.
  %split by an energy $|\frac{Av_{\star}^2}{\gamma^2\ell^2}|$  
This degeneracy at $\mathbf{B}\rightarrow 0$ limit gets split as we tune back $M$.
 The comparison of the LL spectrum obtained using the above effective Hamiltonian with the one obtained via the exact calculation is shown in Fig.(\ref{fig:nu=-2 VP SWT}). The LL basis used to generate the plot is $(0,|m_1\rangle)^T$, $(|m_2\rangle, |m_2+3\rangle)^T$ and $(|m_3\rangle,0)^T$, where $m_1\in\{0,1,2\}$, $m_2\in\{0,\ldots,m_{max}-4\}$ and $m_3\in\{m_{max}-3,m_{max}-2,m_{max}-1\}$ respectively. We use $(|m_3\rangle,0)^T$ rather than $(|m_3\rangle,|m_3+3\rangle)^T$ in order to avoid three modes emanating out of spurious $\mathbf{B}\rightarrow 0$ energy $-(2W_3+\zeta_sJ/2)$, for each $s$. These three spurious modes are present in the exact calculation, as can be seen in Fig.(\ref{fig:nu=-2 VP SWT}). However, recall that in practice we use $m_{max}=\lceil\frac{q-3}{2}\rceil$, so these three LLs are lost as $\mathbf{B}(q)$ increases(decreases).

\subsubsection{Parent Kramers Intervalley Coherent State}
\begin{figure}
    \centering
    \includegraphics[width=10cm]{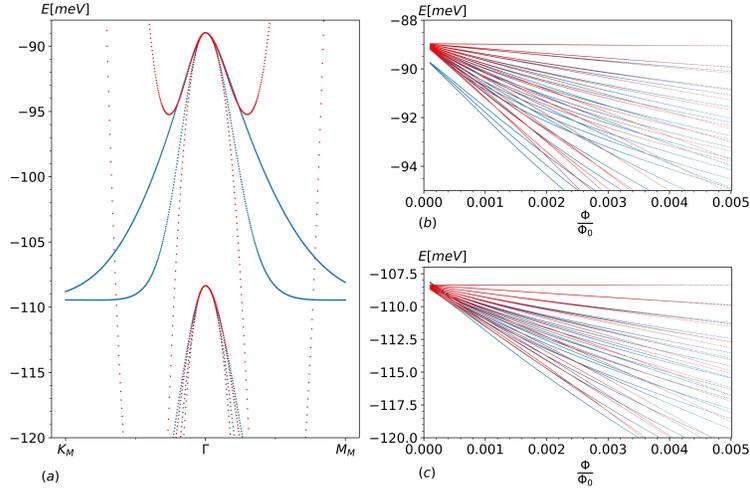}
    \caption{The comparison of (b,c) LL spectrum obtained using the effective Hamiltonian (\boldsymbol{\textcolor{red}{$\cdot$}}) with the exact calculation (\boldsymbol{\textcolor{blue}{$\cdot$}}) contrasted with the corresponding comparison at (a) $\mathbf{B}=0$ for spin sector $\uparrow$ for the KIVC state at $\nu=-2$. For the comparison to be tractable we set an upper-cutoff of $10$ on the LL index.}
    \label{fig:KIVC nu=-2 SWT}
\end{figure}
In this section, we discuss the derivation of $\bB\neq0$ effective Hamiltonian for the spin sector $s=\uparrow$ for the case of parent KIVC state at $\nu=-2$. The exact $\bB=0$ THFM for the spin sector $s=\uparrow$ at $\nu=-2$ for the parent KIVC state is presented in Eq.(\ref{eq:KIVC B=0 nu=-2}). The coupled modes of $c$ and $f$ fermions at $\bB\neq 0$ are described by the operator in Eq.(\ref{Eq: h for KIVC nu=-2 up}).
In the flat band limit at $\Gamma$, the $\bB=0$ THFM has the form  $\left(\begin{array}{cc}
    H^{\tau=1,\nu=-2} & 0 \\
   0  & H^{\tau=-1,\nu=-2}
\end{array}\right)$, where $\left(\begin{array}{ccc}
  -2W_1\sigma_0 &0&\gamma\sigma_0\\
  & -(2W_3+\tau J/2)\sigma_0  &0  \\
   h.c.& & -(12U_2+\frac{4+\tau}{2}U_1)\sigma_0
\end{array}\right)$, where the Pauli matrix $\sigma_0$ acts in the orbital space. The corresponding eigenstates, labelled by $|\rho,j,\tau\rangle$ with $\rho\in\{1,2,3\}$ and $j\in\{1,2\}$ are given as 
\begin{eqnarray}
    |1,1,\tau=+1\rangle &=& \left(0,0,0,1,0,0,0_{6\times 1}\right)^T \text{;} ~~~~~~~|1,1,\tau=-1\rangle = \left(0_{6\times 1},0,0,0,1,0,0\right)^T
    \label{Eq:zero B spinor 1,nu=-2KIVC}\\
    |1,2,\tau=+1\rangle &=& \left(0,0,1,0,0,0,0_{6\times 1}\right)^T \text{;} ~~~~~~~ |1,2,\tau=-1\rangle = \left(0_{6\times 1},0,0,1,0,0,0\right)^T
    \label{Eq:zero B spinor 2,nu=-2KIVC}\\
    |2,1,\tau=+1\rangle &=& \frac{1}{\sqrt{N_{X1}}}\left(0,X_1,0,0,0,1,0_{6\times 1}\right)^T \text{;} ~~~~|2,1,\tau=-1\rangle = \frac{1}{\sqrt{N_{X-1}}}\left(0_{6\times 1},0,X_{-1},0,0,0,1\right)^T
    \label{Eq:zero B spinor 3,nu=-2KIVC}\\
    |2,2,\tau=+1\rangle &=& \frac{1}{\sqrt{N_{X1}}}\left(X_1,0,0,0,1,0,0_{6\times 1}\right)^T \text{;} ~~~~ |2,2,\tau=-1\rangle = \frac{1}{\sqrt{N_{X-1}}}\left(0_{6\times 1},X_{-1},0,0,0,1,0\right)^T
    \label{Eq:zero B spinor 4,nu=-2KIVC}\\
    |3,1,\tau=+1\rangle &=& \frac{1}{\sqrt{N_{Y1}}}\left(0,Y_1,0,0,0,1,0_{6\times 1}\right)^T \text{;} ~~~~|3,1,\tau=-1\rangle = \frac{1}{\sqrt{N_{Y-1}}}\left(0_{6\times 1},0,Y_{-1},0,0,0,1\right)^T
    \label{Eq:zero B spinor 5,nu=-2KIVC}\\
    |3,2,\tau=+1\rangle &=& \frac{1}{\sqrt{N_{Y1}}}\left(Y_1,0,0,0,1,0,0_{6\times 1}\right)^T \text{;} ~~~~|3,2,\tau=-1\rangle = \frac{1}{\sqrt{N_{Y-1}}}\left(0_{6\times 1},Y_{-1},0,0,0,1,0\right)^T\label{Eq:zero B spinor 6,nu=-2KIVC}
\end{eqnarray}
where the normalizations are $N_{X\tau}=1+X_\tau^2$, $N_{Ys}=1+Y_\tau^2$ and  
\begin{eqnarray}
X_{\tau} &=& -\frac{1}{2\gamma}\left( 2W_1-(12U_2+\frac{4+\tau}{2}U_1) + \sqrt{4\gamma^2 + (2W_1-(12U_2+\frac{4+\tau}{2}U_1))^2}\right), \label{Eq:Xtau for KIVC,nu=-2}\\
Y_{\tau} &=& -\frac{1}{2\gamma}\left( 2W_1-(12U_2+\frac{4+\tau}{2}U_1) - \sqrt{4\gamma^2 + (2W_1-(12U_2+\frac{4+\tau}{2}U_1))^2}\right), \label{Eq:Ytau for KIVC,nu=-2}
\end{eqnarray}
The energy of state $\rho,j,\tau$ with $\rho\in\{1,2,3\}$ and $j\in\{1,2\}$, $E_{\rho,j,\tau}$, are given as
\begin{eqnarray}
E_{1,j,{\tau}} &=& -(2W_3 +\tau\frac{J}{2}) \equiv E_{\tau} \label{Eq:CNP KIVC energy label 1,nu=-2}\\
E_{2,j,{\tau}} &=& \frac{1}{2}\left(-2W_1-(12U_2+\frac{4+\tau}{2}U_1)
- \sqrt{4\gamma^2 + (2W_1-(12U_2+\frac{4+\tau}{2}U_1))^2}\right) \equiv E_{X{\tau}}
\label{Eq:CNP KIVC energy label 2,nu=-2}\\
E_{3,j,{\tau}} &=& \frac{1}{2}\left(-2W_1-(12U_2+\frac{4+\tau}{2}U_1)
+ \sqrt{4\gamma^2 + (2W_1-(12U_2+\frac{4+\tau}{2}U_1))^2}\right) \equiv E_{Y{\tau}}\label{Eq:CNP KIVC energy label 3,nu=-2}
\end{eqnarray}

We break the $\bB=0$ Hilbert space at $\Gamma$ into 5 subspaces. The subspace of interest is the one spanned by states $|1,j,\tau\rangle$ with $j\in\{1,2\}$ and $\tau=\pm1$, i.e. the one spanned by $a=\{3,4\}$ $c$ fermions at each valley. The remaining four subspaces are spanned by $|2,j,+1\rangle$, $|2,j,-1\rangle$, $|3,j,+1\rangle$ and $|3,j,-1\rangle$, respectively with $j\in\{1,2\}$. These states and their energies can now be relabelled by $|\epsilon,l\rangle$ and $E_{\epsilon l}$, respectively. Here $\epsilon\in\{1,\ldots,5\}$ labels the 5 subspaces, $l\in\{1,\ldots,4\}$ for $\epsilon=1$ and $l\in\{1,2\}$ for $\epsilon=\{3,4,5\}$ labels the states spanning the subspace $\epsilon$. We promote these states to $\bB\neq 0$ as $|\epsilon,l,m\rangle = |\epsilon,l\rangle|m\rangle$, where $|m\rangle$ is the $m^{th}$ LL, i.e. the finite $\bB$ $\bk\cdot\bp$ basis.

To begin with, we re-write the operator in Eq.(\ref{Eq: h for KIVC nu=-2 up}) with respect to the above $\bB=0$ eigenstates at $\Gamma$  as
\begin{eqnarray}
    H = H_0+\Delta V,
\end{eqnarray}
where $H_0=\left(\begin{array}{cc}
    {H_0}^{\tau=+1}_{KIVC} &  \\
     & {H_0}^{\tau=-1}_{KIVC}
\end{array}\right)$ and $V=\left(\begin{array}{cc}
    \bar{V}^{\tau=+1}_{KIVC} &V_M  \\
    h.c. & \bar{V}^{\tau=-1}_{KIVC}
\end{array}\right)$.
${H_0}^{\tau}_{KIVC}$ can be obtained by replacing $\zeta_s$ by $\tau$ in Eq.(\ref{Eq:H0 for SWT, nu=-2, VP}). The intra-valley perturbation $ \bar{V}^{\tau=+1}_{KIVC}$ can be obtained by setting $M=0$ in Eq.(\ref{Eq:Perturbation_CNP_VP_K}) and $ \bar{V}^{\tau=-1}_{KIVC}$ can be obtained by replacing $X_1,Y_1\rightarrow X_{-1},Y_{-1}$, $N_{X1},N_{Y1}\rightarrow N_{X-1},N_{Y-1}$ and $a\leftrightarrow a^\dagger$ in $ V^{\tau=+1}$. However note that the values of $Y_\tau$ and $Y_\tau$ are changed for $\nu=-2$ and given in Eq.(\ref{Eq:Xtau for KIVC,nu=-2}) and Eq.(\ref{Eq:Ytau for KIVC,nu=-2}) respectively. The inter-valley perturbation $V_M$ is given in Eq.(\ref{Eq:VM for KIVC SWT}). Following the SWT procedure in Sec.(\ref{apdx:KIVC CNP}), i.e. the steps from Eq.(\ref{Eq:Step 1 KIVC,CNP})-(\ref{Eq:O3 contribution for KIVC}), upto O($\frac{1}{\ell^3}$), the effective Hamiltonian for subspace $\epsilon=1$ is found to be
\begin{eqnarray}
  H^{eff} = \left(\begin{array}{cc}
      H^{\tau=+1,\nu=-2}_{\text{KIVC}} & H_M^{\nu=-2}  \\
      h.c. & H^{\tau=-1,\nu=-2}_{\text{KIVC}}
  \end{array}\right),
\end{eqnarray}
where 
\begin{eqnarray}
   H^{\tau=+1,\nu=-2}_{\text{KIVC}} =  \left(\begin{array}{cc}
-(2W_3 +\frac{J}{2})+\hbar \omega_c^{(1)} aa^\dagger & i\frac{A^{(1)}}{\ell^3}a^3\\
-i\frac{A^{(1)}}{\ell^3}{a^\dagger}^3 &-(2W_3 +\frac{J}{2})+\hbar \omega_c^{(1)} a^\dagger a
     \end{array} \right)\label{Eq:KIVC nu=-2 SWT11}.
\end{eqnarray}
$H^{\tau=-1,\nu=-2}_{\text{KIVC}}$ can be obtained by replacing $a\leftrightarrow a^\dagger$ and  $\omega_c^{(1)},A^{(1)},\frac{J}{2}\rightarrow \omega_c^{(-1)},A^{(-1)},-\frac{J}{2}$ in Eq.(\ref{Eq:KIVC nu=-2 SWT11}). The cyclotron frequency $\omega_c^{(\tau)}$ and coefficient $A^{\tau}$ are
\begin{eqnarray}
    \hbar \omega_c^{(\tau)} &=& \frac{2 v_{\star}^2}{\ell^2}\left(\frac{X_\tau^2}{N_{X\tau}(E_\tau-E_{X\tau})} + \frac{Y_\tau^2}{N_{Y\tau}(E_\tau-E_{Y\tau})}\right),\label{Eq:cyclotron freq nu=-2 KIVC}\\
    A^{(\tau)} &=& 4\sqrt{2}v_{\star}^2v_\star'\left(\frac{X_\tau Y_\tau(X_\tau + Y_\tau)}{N_{X\tau}N_{Y\tau}(E_\tau-E_{X\tau})(E_\tau-E_{Y\tau})}+\frac{X_\tau^3}{N_{X\tau}^2(E_\tau-E_{X\tau})^2} +\frac{Y_\tau^3}{N_{Y\tau}^2(E_\tau-E_{Y\tau})^2} \right),\label{Eq:A nu=-2 KIVC}.
\end{eqnarray}
The matrix $H_M$ couples both the valleys and is given as
\begin{eqnarray}
    H_M = 
        M\left(1+\frac{M_{c1}}{\ell^2}a^\dagger a + \frac{M_{c2}}{\ell^2}aa^\dagger\right)\sigma_z 
\end{eqnarray}
where the Pauli matrix acts in the orbital space of $a\in\{4,3\}$ $c$-fermions and
\begin{eqnarray}
  M_{c1} &=& v_\star^2\left(\frac{X_{-1}^2}{N_{X{-1}}(E_{1}-E_{X{-1}})(E_{X{-1}}-E_{-1})} + \frac{Y_{-1}^2}{N_{Y{-1}}(E_1-E_{Y-1})(E_{Y-1}-E_{-1})}\right),\\ 
  M_{c2} &=& v_\star^2\left(\frac{X_{1}^2}{N_{X{1}}(E_{1}-E_{X{1}})(E_{X{1}}-E_{-1})} + \frac{Y_{1}^2}{N_{Y{1}}(E_1-E_{Y1})(E_{Y1}-E_{-1})}\right).
\end{eqnarray}
Substituting Eqs.(\ref{Eq:Xtau for KIVC,nu=-2})-(\ref{Eq:CNP VP energy label 3,nu=-2}) in the coefficients, we have $\ell^2 \hbar \omega_c^{(1)}= -831301.40524 meV \mathring{A}^2$, $\ell^2 \hbar \omega_c^{(-1)}=-514097.23133 $ $meV \mathring{A}^2$, $A^{(1)}=-57864769.72509 meV \mathring{A}^3$, $A^{(-1)}=-59278877.34125$ $meV \mathring{A}^3$,
$M_{c1}= -12902.68825 \mathring{A}^2$ and $M_{c2}= -25550.93059
\mathring{A}^2$. The comparison of the LL spectrum obtained using the above effective Hamiltonian with the one obtained via the exact calculation is shown in Fig.(\ref{fig:KIVC nu=-2 SWT}). The LL basis used to generate the plot is $(0,|m_1\rangle,|m_1\rangle,0)^T$, $(|m_2\rangle, |m_2+3\rangle,|m_2+3\rangle, |m_2\rangle)^T$ and $(|m_3\rangle,0,0,|m_3\rangle)^T$, where $m_1\in\{0,1,2\}$, $m_2\in\{0,\ldots,m_{max}-4\}$ and $m_3\in\{m_{max}-3,m_{max}-2,m_{max}-1\}$ respectively. We use $(|m_3\rangle,0,0,|m_3\rangle)^T$ rather than $(|m_3\rangle,|m_3+3\rangle,|m_3+3\rangle,|m_3\rangle)^T$ in order to avoid three modes emanating out of spurious $\mathbf{B}\rightarrow 0$ energy $-2W_3\pm J/2$. These six spurious modes are present in the exact calculation, as can be seen in Fig.(\ref{fig:KIVC nu=-2 SWT}b) and (\ref{fig:KIVC nu=-2 SWT}c). However, recall that in practice we use $m_{max}=\lceil\frac{q-3}{2}\rceil$, so these three LLs are lost as $\mathbf{B}(q)$ increases(decreases).

\section{Review of Magnetic Bloch's Theorem at \texorpdfstring{$2\pi$}{2 pi} Flux}
\label{app:2pi_review}
In the following two appendices we detail the gauge-invariant technique of calculating the strong-coupling spectrum of TBG in flux. 
We first review our technique of calculating the band structures and topology of twisted bilayer graphene with $2\pi$ flux per unit cell, as derived in \cite{PhysRevB.106.085140, PhysRevLett.129.076401}, with the eventual goal of extending the $2\pi$ formalism to rational flux $\phi = 2\pi p/q$.  Compared to other techniques, our method explicitly defines gauge-invariant \emph{momentum eigenstates} $\ket{\kk,m}$ that are simultaneous eigenkets of magnetic translation operators with the same formal properties as Bloch states. By keeping the full momentum quantum numbers and manifestly preserving the lattice symmetries, we are able to leverage the machinery of band topology, such as Wilson loops and the quantum metric tensor as will be explored in future work, in Hofstadter systems at rational flux.

To add nonzero flux to the BM model, we perform canonical substitution:  $-i {\partial} \rightarrow {\cmom}$, with $\cmom$ the canonical momentum yielding the operators
\bea
\pi_\mu &= -i \partial_\mu - e A_\mu,~~Q_\mu = \pi_\mu - eB \epsilon_{\mu\nu} x_\nu.
\eea
where ${\bf A}$ is the vector potential $\pmb{\nabla} \times {\bf A} = B {\hat z}$, and $Q_\mu$ is the guiding center operator obeying $[\pi_\mu,Q_\nu] = 0$. We define the Landau level operators $a, b$ as
\begin{align}
a &= \dfrac{\pi_x + i\pi_y}{\sqrt{2B}},~~a^\dagger = \dfrac{\pi_x - i\pi_y}{\sqrt{2B}},~~[a, a^\dagger] = 1, \\
b &= \dfrac{(\bL_1 - i\bL_2) \cdot {\bf Q}}{\sqrt{2\phi}},~~b^\dagger = \dfrac{(\bL_1 + i\bL_2) \cdot {\bf Q}}{\sqrt{2\phi}},~~[b, b^\dagger] = 1 \ . 
\label{eq:ab1}
\end{align} In magnetic flux, the translation operators $T_\RR$ which commute with the Hamiltonian are
\begin{align}
{\hat t}_{\RR} = e^{i \RR \cdot {\bf Q}}
\label{eq:magnetic_translation}
\end{align}
where $\mbf{R}$ is a lattice vector. Because at $2\pi$ flux the magnetic translation operators $T_{\bL_1}, T_{\bL_2}$ commute with each other and the Hamiltonian $H_\text{BM}$, one can label the eigenstates of $H_\text{BM}$ by the momentum carried by the operators $T_{\bL_1}, T_{\bL_2}$. A gauge-invariant definition of the momentum eigenstates was given in \cite{PhysRevB.106.085140} as\begin{align}
\ket{\kk, m} = \frac{1}{\sqrt{{\cal N}(k_1,k_2)}} \sum_{\RR} e^{-2\pi i \kk \cdot \RR} T_{\bL_1}^{R_1} T_{\bL_2}^{R_2} \ket{m}, \qquad \braketter{\mbf{k},m|\mbf{k}',n} = \delta(\mbf{k}-\mbf{k}') \delta_{mn} \ .
\label{eq:two_pi_basis}
\end{align} 
Here momentum is $\kk = k_1 \bg_1 + k_2 \bg_2$, $k_i = \kk \cdot \bL_i \in [0,1)$, and $\RR = R_1 \bL_1 + R_2 \bL_2$ is summed over all integers $R_1, R_2$.  The ket $\ket{m}$ is the $m$th Landau level
\begin{align}
\ket{m} = \frac{1}{\sqrt{m!}} a^{\dag m} \ket{0}
\label{eq:landau}
\end{align}
and ${\cal N}(k_1,k_2)$ is the normalization factor, worked out to be \cite{PhysRevB.106.085140}
\begin{align}
{\cal N}(k_1,k_2) = {\vartheta}\left(k_1, k_2 \bigg| \Phi \right) = \sqrt{2} e^{-\pi (k_1^2 + k_2^2 - 2i k_1 k_2)} \theta \left(ik_1 + k_2 \bigg| i \right) \theta \left(k_1 + i k_2 \bigg| i \right).
\label{eq:two_pi_norm}
\end{align}  Here $\vartheta$ denotes the Siegel theta function and $\Phi$ the Riemann matrix given by
\begin{align}
\vartheta(\mbf{k}|M) &= \sum_\mbf{R} e^{2 \pi^2 \mbf{k} \cdot M \cdot \mbf{k} + 2\pi i \mbf{k} \cdot \mbf{R}}, \qquad\Phi = \frac{i}{2}\begin{bmatrix} 1 & i \\
i & 1
\end{bmatrix}.
\label{}
\end{align}

In Eq.(\ref{eq:two_pi_norm}), we have simplified the Siegel theta into a product of Jacobi theta functions $\theta$. There is a zero in the normalization at $\kk = (\pi, \pi)$ which is protected by the Chern number of the Landau level states $\ket{\mbf{k},m}$.

Within this basis, one may calculate the matrix elements at momentum $\kk$.  The matrix elements read \cite{PhysRevB.106.085140}
\begin{align}
\braketter{\kk,m|a^\dagger a|\kk',n} &= \delta(\kk-\kk') n \delta_{mn} \\
\braketter{\kk,m|e^{-i\qqq \cdot \rr}|\kk',n} &= \delta(\kk-\kk'-\qqq) e^{i\xi_\qqq(\kk)} {\cal H}_{mn}^\qqq,
\label{eq:2pi_form_fact_calc}
\end{align} where the form factor matrix and phases are defined as
\begin{align}
e^{i\xi_\qqq(\kk)} &= \dfrac{e^{-\frac{q{\bar q}}{4\phi}} \vartheta \left( \dfrac{(k_1-q/2,k_2+iq/2)}{2\pi} \bigg| \Phi \right)}{\sqrt{\vartheta \left( \dfrac{(k_1,k_2)}{2\pi} \bigg| \Phi \right) \vartheta \left( \dfrac{(k_1-q_1,k_2-q_2)}{2\pi} \bigg| \Phi \right)} },~~q = q_1 + iq_2,~~q_j = \qqq \cdot \bL_j \\
{\cal H}_{mn}^\qqq &= \braketter{m|e^{i\frac{\gamma_q a + {\bar \gamma}_q a^\dagger}{\sqrt{2\phi}}}|n},~~\gamma_q = \epsilon_{ij} q_i {\bar z}_j,~~{\bar z}_j = \dfrac{({\hat x} - i{\hat y})\cdot \bL_j}{\sqrt{\Omega}},
\label{eq:phases_and_form_factors}
\end{align}  with $\Omega = |\mbf{a}_1 \times \mbf{a}_2|$ the area of the unit cell.  These matrix elements have been worked out in detail in \cite{PhysRevB.106.085140}, and the momentum Landau levels form a complete orthonormal basis.  To evaluate the band structure of the BM model in flux, one simply evaluates the matrix elements of the Hamiltonian with respect to this basis at any momentum. Being able to block diagonalize the Hamiltonian by momentum allows for efficient calculations as well as access to the Wilson loop and quantum geometric tensor.

\section{Rational Flux}
\label{apdx:rational}
To solve the Bistritzer MacDonald model (or other continuum models) at rational flux $\phi = 2\pi p/q$, we will create new momentum eigenstates that reflect the projective representation of the translation group in flux. At flux $\phi = 2\pi p/q$, we generalize Eq.(\ref{eq:ab1}) to the following raising and lowering operators
\begin{align}
a &= \dfrac{\pi_x + i\pi_y}{\sqrt{2B}},~~a^\dagger = \dfrac{\pi_x - i\pi_y}{\sqrt{2B}} \\
b &= \dfrac{(\bL_1/p - iq\bL_2)\cdot \mbf{Q}}{\sqrt{4\pi}},~~ b^\dagger = \dfrac{(\bL_1/p + iq\bL_2)\cdot \mbf{Q}}{\sqrt{4\pi}}
\label{eq:andbblattice}
\end{align} The $a,a^\dag$ operators are unchanged and obey $[a, a^\dagger] = 1$, and we verify that  $[b, b^\dagger] = 1$ holds:
\begin{align}
[b, b^\dagger] &= 2i \frac{1}{4\pi}\frac{q}{p} [\bL_1 \cdot \QQ, \bL_2 \cdot \QQ] = eB \Omega \frac{1}{2\pi} \frac{q}{p}= 1
\end{align}
since $eB\Omega = \phi$. The translation operators $T_\RR = e^{i \RR \cdot \QQ}$ now obey
\begin{align}
{\hat t}_{\bL_1} {\hat t}_{\bL_2} = e^{-[\bL_1\cdot \QQ, \bL_2 \cdot\QQ]} {\hat t}_{\bL_2} {\hat t}_{\bL_1} = e^{i\phi} {\hat t}_{\bL_2} {\hat t}_{\bL_1}.
\label{}
\end{align}  These magnetic translation operators commute with the Hamiltonian, but do not commute with each other. However, the operators ${\hat t}_{\bL_1}, {\hat t}_{q\bL_2}$ form a commuting subgroup, $[{\hat t}_{\bL_1}, {\hat t}_{q\bL_2}] = 0$ and so label the magnetic Brillouin zone (MBZ):
\begin{align}
\kk \in \text{MBZ},~~\text{MBZ} = \{\kk | \kk = 2\pi k_1 \bg_1 + 2\pi k_2 \bg_2,~~k_1 \in [0, 1),~~k_2 \in [0, 1/q)\}.
\label{}
\end{align}
We will now provide a complete basis of states which carry the momentum quantum number $\mbf{k} \in$ MBZ. This is accomplished by adapting the basis states in $2\pi$ flux in Eq.(\ref{eq:two_pi_basis}). The completeness and orthogonality of the basis in Eq.(\ref{eq:two_pi_basis}) is an algebraic property that relies on the fact that $\hat{t}_{\mbf{L}_1}$ and $\hat{t}_{\mbf{L}_2}$ enclose $2\pi$ flux and therefore commute. So in rational flux $\phi = \frac{2\pi p}{q}$, we consider the operators $\hat{t}_{\mbf{a}_1/p},\hat{t}_{q\mbf{L}_2}$ which also enclose $2\pi$ flux and commute. Notably, $\hat{t}_{\mbf{L}_1/p}$ is not a symmetry of the Hamiltonian because it is a translation by a \emph{partial} lattice vector (although it is a well-defined operator). Thus we are led to consider the orthonormal states
\bea
\ket{\mbf{\tilde{k}},m} &= \frac{1}{\sqrt{{\cal N}(\tilde{k}_1,\tilde{k}_2)}}\sum_{R_1,R_2 \in \mathbb{Z}} e^{-2\pi i ( \tilde{k}_1 R_1 +\tilde{k}_2 R_2 )} {\hat t}_{\bL_1/p}^{R_1} {\hat t}_{q\bL_2}^{R_2} \ket{m}, \quad
\ket{m} = \dfrac{1}{\sqrt{m!}} a^{\dagger m} \ket{0} 
\label{eq:rational_basis}
\eea
which are parameterized by $\mbf{\tilde{k}} \in [0,1) \times [0,1)$. To understand the physical meaning of $\mbf{\tilde{k}}$ in terms of the momentum quantum number $\mbf{k} \in \text{MBZ}$, we compute
\bea
{\hat t}_{\mbf{a}_1} \ket{\mbf{\tilde{k}},m} &= {\hat t}_{\mbf{a}_1/p}^p \ket{\mbf{\tilde{k}},m} = e^{2\pi i p \tilde{k}_1}  \ket{\mbf{\tilde{k}},m}, \qquad \implies p \tilde{k}_1 = k_1 \mod 1 \\
{\hat t}^q_{\mbf{a}_2} \ket{\mbf{\tilde{k}},m} &= e^{2\pi i \tilde{k}_2}  \ket{\mbf{\tilde{k}},m}, \qquad \implies \tilde{k}_2 = q k_2 \mod 1  \\
\eea
and thus we identify $\tilde{k}_1 = \frac{k_1 + \kappa}{p}, \ \kappa = 0,\dots, p-1$ and $\tilde{k}_2 = q k_2$. Plugging into Eq.(\ref{eq:rational_basis}), we arrive at the following basis of magnetic translation group eigenstates
\bea
\ket{\mbf{k},\kappa,m} &= \frac{1}{\sqrt{{\cal N}((k_1 + \kappa)/p,q k_2)}}\sum_{R_1,R_2 \in \mathbb{Z}} e^{-2\pi i \big( \frac{k_1 + \kappa}{p} R_1 +q k_2 R_2 \big)} {\hat t}_{\bL_1/p}^{R_1} {\hat t}_{q\bL_2}^{R_2} \ket{m}, \nonumber \\
&\quad \mbf{k} \in \text{MBZ}, \ \kappa = 0,\dots, p-1 \mod p \ .
\label{eq:ketpq}
\eea
We see that the basis of states are labeled by $\mbf{k}$ in the MBZ and an additional flavor index $\kappa = 0,\dots, p-1$. This index has a simple physical interpretation: consider the Landau level limit of zero crystalline potential so that each momentum in the MBZ is $p$-fold degenerate due to $\kappa$. This means there are $p$ states per magnetic unit cell $\mbf{a}_1 \times q \mbf{a}_2$, leading to an electron density of $p/q = \phi/2\pi$ per unit cell. This is expected because Landau levels have a density of $\nu = p/q = \frac{\phi}{2\pi}$. For later reference, we remark that the eigenstates Eq.(\ref{eq:ketpq}) obey the embedding property
\begin{align}
\ket{\kk + \frac{1}{2\pi} \bg_1, \kappa, m} = \ket{\kk,\kappa +1, m},~~\ket{\kk + \frac{1}{2\pi q} \bg_2, \kappa, m} = \ket{\kk,\kappa , m}
\label{}
\end{align}
over the MBZ. These eigenstates are similar to the light fermions used to describe the conduction electrons in flux, Eq.(\ref{eq:conduction_basis_def}).  The orthonormality $\braketter{\mbf{\tilde{k}},m|\mbf{\tilde{k}}',n} = \delta(\mbf{\tilde{k}}-\mbf{\tilde{k}}') \delta_{mn}$ of the states Eq.(\ref{eq:rational_basis}) is used to calculate the orthonormality of the new states $\ket{\kk,\kappa , m}$ by employing the previously derived identification
\begin{align}
    \tilde{k}_1 &= \frac{k_1 + \kappa}{p}, \ \kappa = 0,\dots, p-1,~~\tilde{k}_2 = q k_2, \\
    \delta(\tilde{k}_1 - \tilde{k}_1') \delta(\tilde{k}_2 - \tilde{k}_2') &= \delta( \frac{k_1 + \kappa}{p} -  \frac{k_1' + \kappa'}{p}) \delta(qk_2 - q k_2') = \frac{p}{q} \delta(k_1-k'_1)\delta(k_2-k_2') \delta_{\kappa \kappa'},
\end{align} giving overlap
\bea
\braketter{\mbf{k},\kappa,m|\mbf{k}',\kappa',n} &= \delta((k_1-k'_1)/p) \delta_{\kappa \kappa'} \delta(q(k_2-k_2')) \delta_{mn} = \frac{p}{q} \delta(k_1-k'_1)\delta(k_2-k_2') \delta_{\kappa \kappa'} \delta_{mn}
\eea
where the factor of $p/q = \phi/2\pi$ shows that the density of the states changes with flux.  This dependence of density of states on flux is equivalent to the Streda formula $\nu = C \frac{\phi}{2\pi}$ with $C=1$ for the Landau level state $\ket{\mbf{k},\kappa,m}$.

Given a periodic Hamiltonian in magnetic flux,
\begin{align}
H = h(\cmom) + V(\mbf{r}), \qquad V(\mbf{r} + \mbf{a}_i)  = V(\mbf{r})
\label{}
\end{align} where $h$ is the kinetic energy, a function of canonical momentum, and $V$ is the lattice-periodic potential energy, one can compute the matrix elements of this Hamiltonian $H^\phi_{m\kappa,n\kappa'}(\mbf{k})$ in the basis Eq.(\ref{eq:ketpq}):

\begin{align}
    \braketter{\kk,\kappa,m|H(\rr)|\kk',\kappa',n} &= \dfrac{p}{q} \delta(k_1-k_1')\delta(k_2-k_2') H^\phi_{m\kappa,n\kappa'}(\mbf{k}).
\end{align}

The kinetic term $h(\pmb{\pi})$ has a simple action on the $\ket{\mbf{k},\kappa,n}$ states because it is $\mbf{k}$-independent:
\begin{align}
    h(\cmom) &= \hbar v_F \sqrt{2B} \begin{bmatrix}
    0 & a^\dagger \\
    a & 0
    \end{bmatrix}\\
    \braketter{\kk',\kappa',m|a^\dagger|\kk,\kappa,n} &= \frac{p}{q} \delta(k_1-k'_1)\delta(k_2-k_2') \ \sqrt{n+1} 
 \delta_{\kappa \kappa'} \delta_{mn},
\end{align} calculated by observing that $a^\dagger$ commutes with the magnetic translation operators:

\begin{align}
    a^\dagger \ket{\mbf{k},\kappa,m} &= \frac{1}{\sqrt{{\cal N}((k_1 + 2\pi \kappa)/p,q k_2)}}\sum_{R_1,R_2 \in \mathbb{Z}} e^{-i \big( \frac{k_1 + 2\pi \kappa}{p} R_1 +q k_2 R_2 \big)} T_{\bL_1/p}^{R_1} T_{q\bL_2}^{R_2} a^\dagger \ket{m}, \\
    &= \frac{1}{\sqrt{{\cal N}((k_1 + 2\pi \kappa)/p,q k_2)}}\sum_{R_1,R_2 \in \mathbb{Z}} e^{-i \big( \frac{k_1 + 2\pi \kappa}{p} R_1 +q k_2 R_2 \big)} T_{\bL_1/p}^{R_1} T_{q\bL_2}^{R_2} \sqrt{m+1} \ket{m+1} \\
    &= \sqrt{m+1} \ket{\mbf{k},\kappa,m+1}.
\end{align}

We now move on to the an explicit expression for the potential term $V(\mbf{r})$ containing $e^{ 2\pi i\mbf{b} \cdot \mbf{r}}$ operators. Let us first focus on the case where $p =1$, where many of the formulae simplify greatly.  We recall the formula at $\phi = 2\pi$ \cite{PhysRevB.106.085140}:
\bea
\label{eq:scattering2pi}
\braketter{\mbf{k},m| e^{- 2\pi i \mbf{b} \cdot \mbf{r}}|\mbf{k}',n}  &= (2\pi)^2 \delta(\mbf{k}-\mbf{k}') \exp {\lp - i \pi b_1 b_2 - i \epsilon_{ij} b_i k_j \rp} [\exp ( i \epsilon_{ij} 2\pi b_i \tilde{Z_j})]_{mn},
\eea
where the matrix $\tilde{Z}$ on the Landau level indices (derived for all $\phi$ in \cite{PhysRevB.106.085140}) is
\bea
\null  Z_j = \frac{\bar{z}_j a + z_j a^\dag}{\sqrt{2\phi}},~~[\tilde{Z}_j]_{mn} = \braketter{m|Z_j|n},  z_j = \frac{ (\hat{x}+ i \hat{y}) \cdot\mbf{a}_j}{|\mbf{a}_1 \times \mbf{a}_2|} \ .
\eea
Note that $\tilde{Z}_j$ is a sparse matrix since $a,a^\dag$ are sparse (each row has only one nonzero element). At $\phi = \frac{2\pi}{q}$, we take $\mbf{a}_2 \to q \mbf{a}_2$ in Eq.(\ref{eq:scattering2pi}), again employing our identification $k_1 \rightarrow k_1, k_2 \rightarrow qk_2, G_1 \rightarrow G_1, G_2 \rightarrow qG_2$ to obtain
\bea
\label{eq:potentialterm}
\braketter{\mbf{k},m| e^{-i \mbf{G} \cdot \mbf{r}}|\mbf{k}',n} &= \frac{(2\pi)^2}{q}  \delta(k_1-k_1') \delta(k_2-k_2') \exp {\lp - i \frac{q}{2\pi} ( \frac{1}{2} G_1 G_2 +\epsilon_{ij} G_i k_j) \rp} [ e^{i \epsilon_{ij} G_i \tilde{Z}_j}]_{mn},\nonumber \\
&k \in \text{MBZ}, \ \mbf{G} \cdot \mbf{a}_i = 0 \mod 2\pi \\   
&= (2\pi)^2 \frac{\phi}{2\pi} \delta(k_1-k_1') \delta(k_2-k_2') \exp \lp - \frac{i}{\phi} ( \frac{1}{2} G_1 G_2 +\epsilon_{ij} G_i k_j) \rp [ e^{i \epsilon_{ij} G_i \tilde{Z_j}}]_{mn}
\eea
Note that from Eq.(\ref{eq:potentialterm}), $H^\phi(k_1,k_2) = H^\phi(k_1+1,k_2) = H^\phi(k_1,k_2+1/q)$ is explicitly periodic across the magnetic BZ, and additionally $H^\phi(k_1,k_2) = H^\phi(k_1+1/q,k_2)$ which proves immediately that $\epsilon_n(\mbf{k})  = \epsilon_n(\mbf{k}+\frac{1}{q} \mbf{b}_i)$ as follows from the properties of the magnetic translation group \cite{PhysRev.134.A1602,PhysRev.134.A1607}.

The simple form for the potential for $p = 1$ becomes significantly more complex when $p \neq 1$.  One replaces the formula Eq.(\ref{eq:2pi_form_fact_calc}) as $\bL_1 \rightarrow \bL_1/p, \bL_2 \rightarrow q\bL_2$, to yield the overlap 

\begin{align}
\braketter{\kk',\kappa',m|e^{-i\qqq \cdot \rr}|\kk,\kappa,n} &= \frac{p}{q} \delta(k_1 - k_1' - q_1/2\pi + \kappa - \kappa') \delta(k_2-k_2'-q_2/2\pi) \nonumber \\
&\times e^{i\xi_{q_1/p + iqq_2}((k_1+\kappa)/p, qk_2)} {\exp (i \epsilon_{ij}q_i {\tilde Z}_j)}_{mn}.
\label{eq:form_factors_complex}
\end{align}
where the phase factor is
\begin{align}
e^{i\xi_{q_1/p + iqq_2}((k_1+ \kappa)/p, qk_2)} = \dfrac{e^{-\frac{\tilde{q}\bar{{\bf q}}}{8\pi}} \vartheta \left((k_1+\kappa)/p-\tilde{q}/4\pi,q k_2+i\tilde{q}/4\pi \bigg| \Phi \right)}{\sqrt{\vartheta \left( (k_1+\kappa)/p,qk_2 \bigg| \Phi \right)\vartheta \left((k_1+\kappa-q_1/2\pi)/p,q(k_2-q_2/2\pi) \bigg| \Phi \right)}},~{\tilde{q}} = q_1/p + iq q_2.
\end{align}

In practice, we will not need to full expression for the phase in Eq.(\ref{eq:form_factors_complex}). To evaluate the Hamiltonian, we only need an expression for the overlap at $\mbf{q} = \mbf{G}$, a reciprocal lattice vector. In this case, the expression simplifies to
\bea
\braketter{\kk',\kappa',m|e^{-i \mbf{G} \cdot \rr}|\kk,\kappa,n} &= \frac{p}{q} \delta(k_1 - k_1') \delta_{\kappa, \kappa'+G_1} \delta(k_2-k_2') \nonumber \\
&\times e^{i\xi_{G_1/p + iqG_2}((k_1+\kappa)/p, qk_2)} {\exp (i \epsilon_{ij}G_i {\tilde Z}_j)}_{mn}
\eea

\begin{figure}
\includegraphics[width=0.45\textwidth]{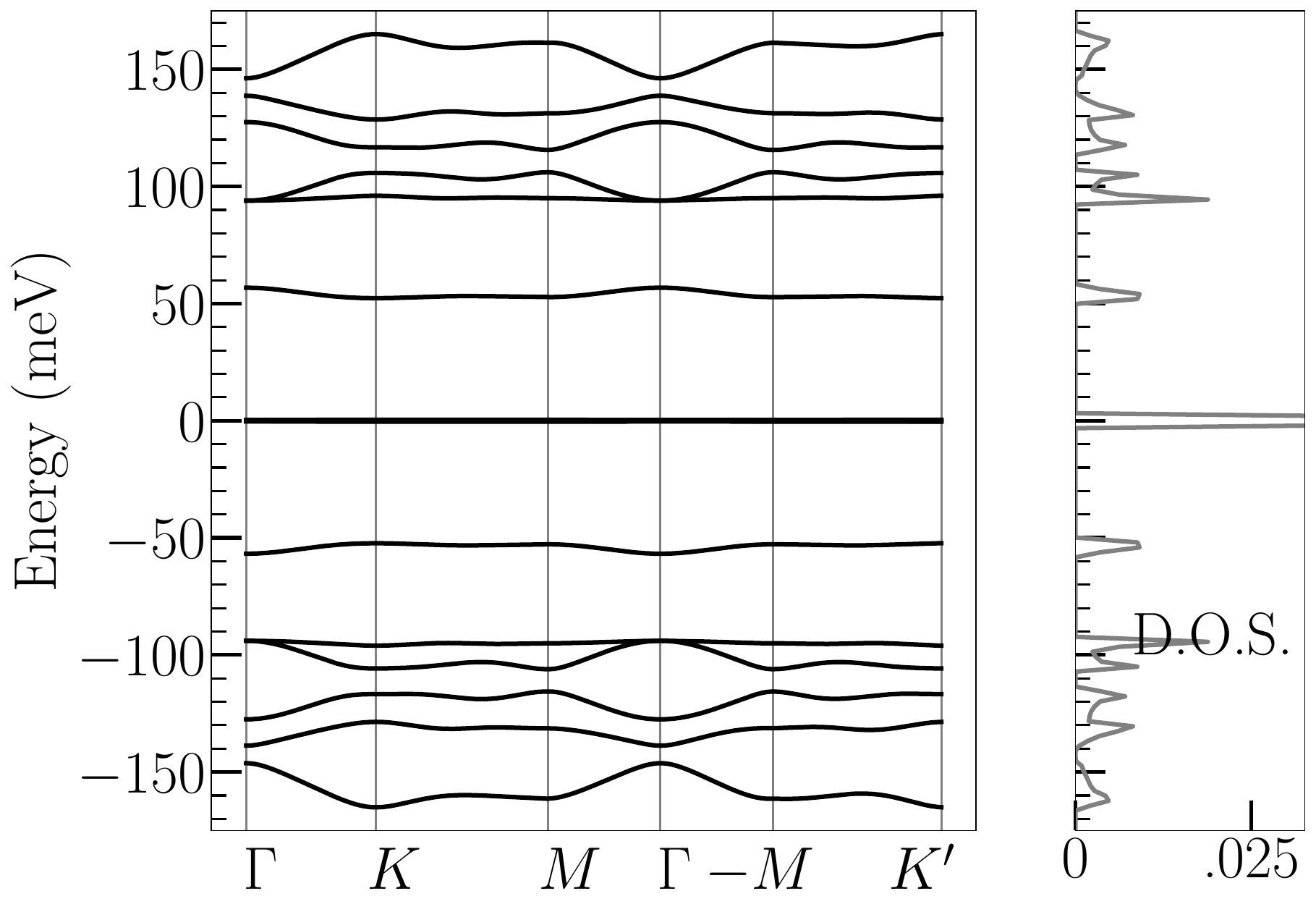}
\includegraphics[width=0.35\textwidth]{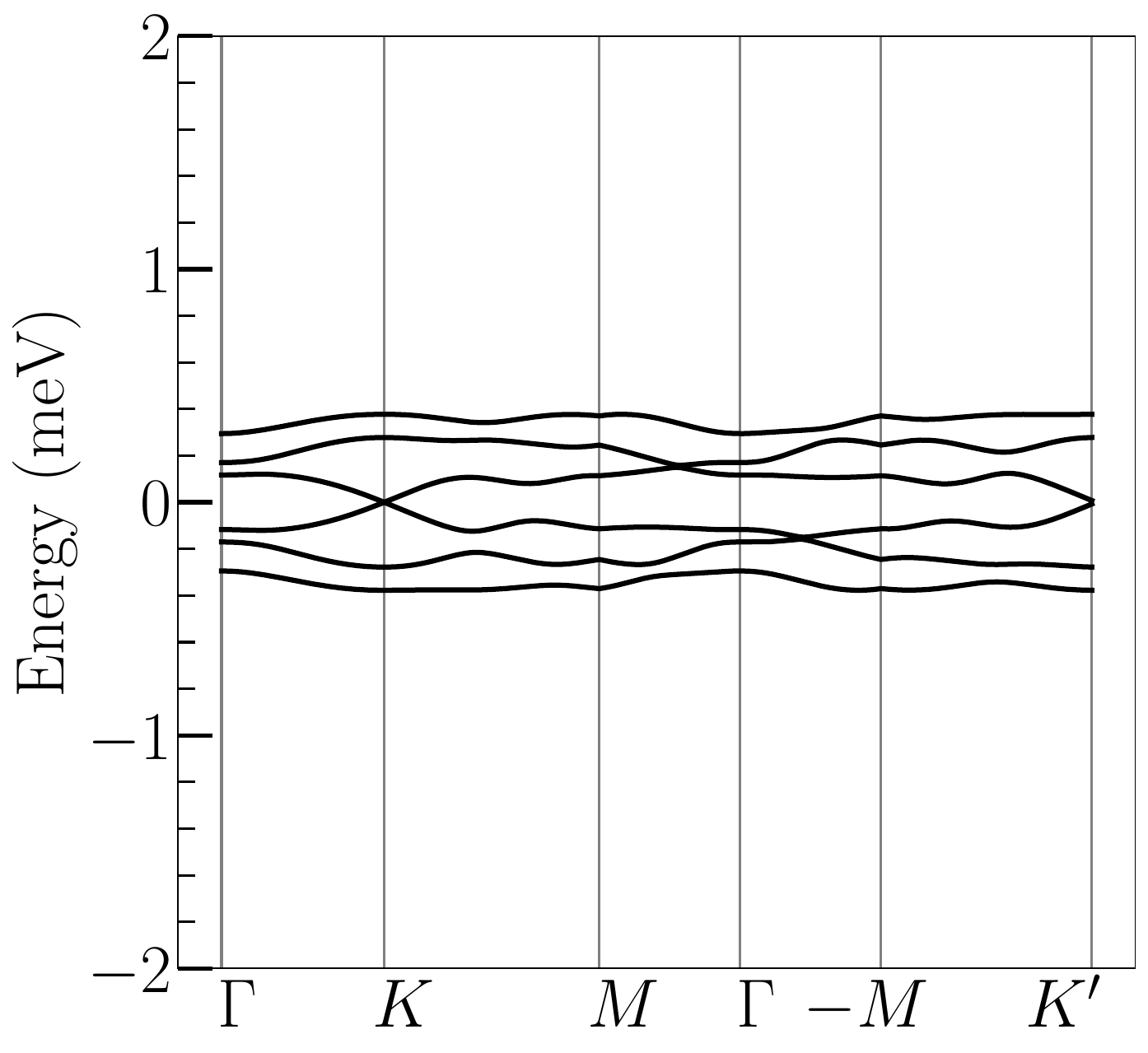}
\caption{Band structures of the BM model at flux $\phi = 2\pi/3$, calculated with our momentum eigenstates. \emph{Left}. The Hofstadter bands show significant flattening over the MBZ, leading to pronounced peaks in the density of states (\emph{Center}). \emph{Right}. Zoom-in of the  Hofstadter splitting of the flat bands.}
\label{fig:PiBS}
\end{figure}

To calculate the matrix elements of the BM model, we first perform a gauge transformation to recast the Hamiltonian into an explicitly translation-symmetric form.  Because $\qqq_1,\qqq_2,\qqq_3$ are not reciprocal lattice vectors, the BM model is only periodic up to a gauge transformation.  Conjugating $H_\text{BM}$ by $e^{-i\pi \qqq_3 \cdot \rr}$ restores translation symmetry, giving Hamiltonian

\begin{align}
{\tilde H}_\text{BM} &= e^{\pi i\qqq_3 \cdot \rr} H_\text{BM} e^{-\pi i\qqq_3 \cdot \rr} = \begin{bmatrix}
-i \hbar v_F \pmb{\sigma} \cdot \pmb{\nabla} - \pi \hbar v_F \qqq_3 \cdot \pmb{\sigma} & {\tilde T}^\dagger(\rr) \\
{\tilde T}(\rr) & -i \hbar v_F \pmb{\sigma} \cdot \pmb{\nabla} + \pi \hbar v_F \qqq_3 \cdot \pmb{\sigma}
\end{bmatrix},
\label{eq:BM}~~\\
{\tilde T}(\rr) &= \sum_{j=1}^3 e^{2\pi i \bg_1 \cdot \rr} T_1 + e^{2\pi i \bg_2 \cdot \rr} T_2 + T_3,
~~T_{j+1} = w_0 \sigma_0 + w_1 \left(\sigma_1 \cos \dfrac{2\pi j}{3} + \sigma_2 \sin \dfrac{2\pi j}{3} \right).
\end{align}. Upon canonical substitution, the Hamiltonian in flux reads
\begin{align}
    {\tilde H}_\text{BM}(\phi) &= \begin{bmatrix}
\hbar v_F \pmb{\sigma} \cdot \cmom- \pi \hbar v_F \qqq_3 \cdot \pmb{\sigma} & {\tilde T}^\dagger(\rr) \\
{\tilde T}(\rr) & \hbar v_F \pmb{\sigma} \cdot \cmom + \pi \hbar v_F \qqq_3 \cdot \pmb{\sigma}
\end{bmatrix},
\end{align}

An example band structure at flux $\phi = 2\pi/3$ calculated with our technique is depicted in Supplementary Fig.(\ref{fig:PiBS}).  

\subsection{Groundstate Charge Density}

A key feature of the HF model of TBG is the concentration of charge at the moir\'e unit cell center in the non-interacting model. This is one indication that the Wannierization must contain orbitals at the $1a$ position. To check the validity of the HF approximation in small flux, we will compute the charge density of the groundstate in flux $\phi = \frac{2\pi}{q}$, for this section restricting to $p=1$ for convenience.  We will compute the quantity
\begin{eqnarray}\label{eq:nrGS}
\braketter{GS|n(\mbf{r})|GS} &=&\sum_{l,\alpha,\eta,s}\braketter{GS|c^\dag_{l,\alpha,\eta,s}(\mbf{r})c_{l,\alpha,\eta,s}(\mbf{r})|GS} \\
&=& \frac{1}{q\Omega} \sum_{\mbf{G}_q} \int_{q_1 = 0}^{2\pi} \int_{q_2 = 0}^{2\pi/q} \frac{d^2q}{(2\pi)^2/q}e^{i (\mbf{q}+\mbf{G}_q) \cdot \mbf{r}} \braketter{GS|\rho_{\mbf{q}+\mbf{G}_q}|GS}    
\end{eqnarray}
where $\ket{GS} = \prod_j \prod_{\mbf{k},n,\eta_j,s_j} \gamma^\dag_{\mbf{k},n,\eta_j,s_j} \ket{0}$ is the Slater determinant state filling all the $2q$ Hofstadter flat bands and $n(\mbf{r}) = \sum_{l,\alpha,\eta,s} c^\dag_{l,\alpha,\eta,s}(\mbf{r})c_{l,\alpha,\eta,s}(\mbf{r})$ is the local charge operator summed over layer $l$, sublattice $\alpha$, valley $\eta$, and spin $s$.  The vectors $\GG_q = 2\pi \mathbb{Z} \bg_1 + 2\pi \mathbb{Z} \bg_2/q$ are the reciprocal lattice vectors of the MBZ (note that $\mbf{G}_q \cdot \mbf{a}_2$ can be fractional).  The momentum space density operator $\rho$ is defined as  
\begin{eqnarray}
 \rho_{\bq+\GG_q} = \int d^2 r e^{-i(\qqq + \GG_q)\cdot \rr} \sum_{l,\alpha,\eta,s} c^\dagger_{l,\alpha,\eta,s}(\rr) c_{l,\alpha,\eta,s}(\rr).   
\end{eqnarray}
 We first find the matrix elements of this operator in terms of Landau levels.  Define the creation operator $\psi^\dag_{\kk,m,\alpha,l,\eta,s}$ as the operator that creates the state $\ket{\kk,m,\alpha,l,\eta,s}$.  Following the calculations from Ref.~\cite{PhysRevB.106.085140} gives the following relations:
\begin{eqnarray}
\braketter{\kk',\alpha,l,m,\eta,s|  c^\dagger_{l,\alpha,\eta,s}(\rr) c_{l,\alpha,\eta,s}(\rr) |\kk,\alpha,l,n,\eta,s} &=& \braketter{0| \psi_{\kk',\alpha,l,m,\eta,s}  c^\dagger_{l,\alpha,\eta,s}(\rr) c_{l,\alpha,\eta,s}(\rr) \psi^\dagger_{\kk,\alpha,l,n,\eta,s}|0} \\
&=& \braketter{0| \psi_{\kk',\alpha,l,m,\eta,s}  c^\dagger_{l,\alpha,\eta,s}(\rr)|0} \braketter{0|c_{l,\alpha,\eta,s}(\rr) \psi^\dagger_{\kk,\alpha,l,n,\eta,s}|0} \\
&=& \psi_{\kk',\alpha,l,m,\eta,s}(\rr) \psi^*_{\kk,\alpha,l,n,\eta,s}(\rr),
\end{eqnarray} 
where $\psi_{\bk',\alpha,l,m,\eta,s}(\br)$ is the real-space wavefunction corresponding to the momentum ket $\ket{\bk',\alpha,l,m,\eta,s}$. These wavefunctions are orthonormal, and integrating over $\br$ gives 
\begin{align}
    \int d^2r e^{-i\qqq \cdot \rr} \psi_{\kk',\alpha,l,m,\eta,s}(\rr) \psi^*_{\kk,\alpha,l,n,\eta,s}(\rr) = \braketter{\kk',\alpha,l,m,\eta,s|  e^{-i \qqq \cdot \rr} |\kk,\alpha,l,n,\eta,s}
\end{align}
The density operator becomes
\begin{align}
    \rho_{\qqq + \GG_q} &= \int d^2 r \sum_{\alpha,l,\eta,s} \sum_{mn} \int \dfrac{d^2 k d^2 k'}{1/q^2} e^{-i(\qqq + \GG_q) \cdot \rr}  \nonumber \\
    &\times \braketter{\kk',\alpha,l,m,\eta,s|  c^\dagger_{l,\alpha,\eta,s}(\rr) c_{l,\alpha,\eta,s}(\rr) |\kk,\alpha,l,n,\eta,s} \psi^\dagger_{\kk',\alpha,l,m,\eta,s} \psi_{\kk,\alpha,l,n,\eta,s} \nonumber \\
    &=\sum_{\alpha,l,\eta,s} \sum_{mn} \int \dfrac{d^2 k d^2 k'}{1/q^2} \braketter{\kk',m,\alpha,l,\eta,s|e^{-i(\qqq+\GG_q)\cdot \rr}|\kk,n,\alpha,l,\eta,s} \psi^\dagger_{\kk',\alpha,l,m,\eta,s} \psi_{\kk,\alpha,l,n,\eta,s},
\end{align} 
where the form factor $\braketter{\kk',m|e^{-i(\qqq+\GG_q)\cdot \rr}|\kk,n}$ has been defined in Eq.(\ref{eq:form_factors_complex}).  The factor of $1/q^2$ in the integrand arises from the fact that working at $\phi = 2\pi/q$, the Brillouin zone size has reduced by a factor of $q$.

We project into the flat bands, defining the eigenoperators $\gamma^\dag_{\kk,M,\eta,s}$ as the operators that create the $M$th energy eigenstate at momentum $\kk$, with $M$ ranging from $1$ to $2q$.  To do this, expand the eigenoperators as 
\begin{eqnarray}
    \gamma_{\kk,N,\eta,s} = \sum_{\alpha,l,m} [U^{(\eta,s)}]^N_{\alpha,l,m}(\kk) \psi_{\kk,\alpha,l,m,\eta,s},
\end{eqnarray} 
so upon projection 
\begin{eqnarray}
\psi^\dagger_{\kk',\alpha,l,m,\eta,s} \psi_{\kk,\alpha,l,n,\eta,s} \rightarrow \sum_{MN} [U^{(\eta,s) *}]^M_{\alpha,l,m}(\kk') [U^{(\eta,s)}]^N_{\alpha,l,n}(\kk) \gamma^\dag_{\mbf{k}',M,\eta,s} \gamma_{\mbf{k},N,\eta,s}
\end{eqnarray}

The projected form factors read
\begin{align}
\rho_{\mbf{q}} \rightarrow \rho_{\mbf{q}+\mbf{G}_q} = \int_{[0,1) \times [0, 1/q)} \frac{d^2k}{1/q} \sum_{MN} M_{mn}^{(\eta,s)}(\mbf{k},\mbf{q}+\mbf{G}_q) \gamma^\dag_{\mbf{k}-\mbf{q}/2\pi,M,\eta_j,s_j} \gamma_{\mbf{k},N,\eta_j,s_j},
\end{align} where $M^{(\eta,s)}(\mbf{k},\mbf{q})$ are the form factor matrices given by 
\begin{eqnarray}
M^{(\eta,s)}_{MN}(\kk,\qqq+\GG_q) &=& \sum_{\alpha,l,\eta,s,mn} e^{i\xi_{\qqq+\GG_\qqq}(\kk)} [{U^{(\eta,s)}}^\dagger_{\alpha,l,n} (\mbf{k}-\mbf{q}) {\cal H}^{\qqq+\GG_q}_{mn} U^{(\eta,s)}_{\alpha,l,n}(\kk)]_{MN} ,\\
\dfrac{1}{q} \delta(k_1 - k_1') \delta(k_2 - k_2') e^{i\xi_{\qqq+\GG_\qqq}(\kk)} {\cal H}^{\qqq+\GG_q}_{mn} &=& \braketter{\kk',m|e^{-i(\qqq+\GG_q)\cdot \rr}|\kk,n}
\label{eq:form_factors}
\end{eqnarray} 
with $[U^{(\eta,s)}]^N_{\alpha,l,m}(\bk)$ the wavefunction of the $N$th energy band in the $m$th Landau level, with repeated indices summed.  We have suppressed the indices $\alpha,l,\eta,s$ in the ket $\ket{\kk,m}$ as they are unchanged by the operator $e^{-i\qqq\cdot\rr}$.

When computing many-body overlaps, it is more convenient to discretize the magnetic BZ and work with Kronecker delta normalized fermion operators instead of Dirac delta normalized operators. To do so, we discretize according to the normalization
\bea
\{\frac{1}{\sqrt{N}} \gamma_{\mbf{k},m,\eta_j,s_j}, \frac{1}{\sqrt{N}}\gamma^\dag_{\mbf{k}',n,\eta_{j'},s_{j'}} \} &= \frac{1}{N} \frac{1}{q} \delta(\mbf{k}-\mbf{k}') \delta_{mn} \delta_{jj'} \to \delta_{\mbf{k}\mbf{k}'}  \delta_{mn} \delta_{jj'}
\eea
where $N \to \delta^2(0)$ is the number of unit cells. Equivalently, $N$ is the number of magnetic unit cells which is equal to the number of $k$-points in the magnetic BZ. Working in momentum space, we discretize the momentum
\bea
\rho_{\mbf{q}+\mbf{G}_q} &= \int_{[0,1) \times [0, 1/q)} \frac{d^2k}{1/q} \sum_{MN} M_{MN}^{(\eta,s)}(\mbf{k},\mbf{q}/2\pi+\mbf{G}_q) \gamma^\dag_{\mbf{k}-\mbf{q}/2\pi,M,\eta_j,s_j} \gamma_{\mbf{k},N,\eta_j,s_j} \nonumber \\
&\to \sum_{\mbf{k}MN} M_{MN}^{(\eta,s)}(\mbf{k},\mbf{q}/2\pi+\mbf{G}_q) \gamma^\dag_{\mbf{k}-\mbf{q}/2\pi,M,\eta_j,s_j} \gamma_{\mbf{k},N,\eta_j,s_j}
\eea
where $M,N$ sum over all the Hofstadter bands and the arrow indicates a discretization of the the magnetic BZ.

\begin{figure}[h]
\includegraphics[width=0.5\textwidth]{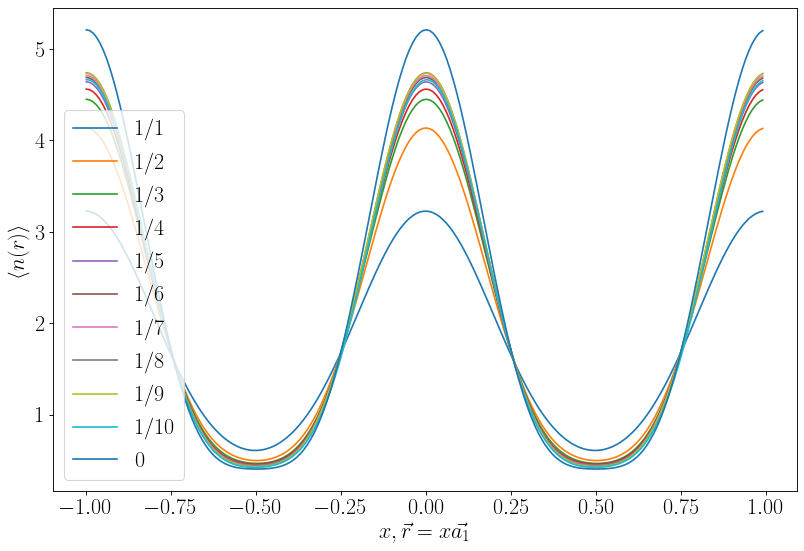}
\caption{Real space cuts along the $\bL_1$ lattice vector.  All fluxes have many-body charge densities that are peaked at the $1a$ location, but the $0$ flux is the most peaked while the $2\pi$ is the most spread.  As flux is increased, the wavefunction weight is increasingly spread, but remains strongly localized at $1a$. This validates the use of the HF Hamiltonian in flux.}
\label{fig:localization}
\end{figure}

Acting on the groundstate, we have
\bea
\label{eq:lambdaint}
\bra{GS} \rho_{\mbf{q}+\mbf{G}_q} \ket{GS} &= \delta_{\mbf{q},0}  \bra{GS} \rho_{\mbf{G}_q} \ket{GS}= \delta_{\mbf{q},0} \sum_{\mbf{k}} \Tr M(\mbf{k},\mbf{G}_q) \\
\bra{GS} \rho_{\mbf{q}+\mbf{G}_q} \ket{GS}  &= \frac{(2\pi)^2}{q} \delta(q_1) \delta(q_2) \int_{[0,1) \times [0, 1/q)} \frac{d^2k}{1/q} \Tr M(\mbf{k},\mbf{G}_q)
\eea
switching back to the continuum normalization in the second line.  As $\qqq$ ranges over the MBZ, the continuum delta function $\delta(q_1)\delta(q_2)$ appears with a prefactor of $(2\pi)^2/q$, so that the following integral over the MBZ is normalized to $1$:
\begin{align}
    \int_{[0,2\pi) \times [0, 2\pi/q)} \dfrac{d^2 q}{(2\pi)^2/q} \frac{(2\pi)^2}{q} \delta(q_1) \delta(q_2) = 1.
\end{align}
We now prove that only $\bra{GS} \rho_{\mbf{G}_q} \ket{GS}$ is nonzero, for $\mbf{G}_q$ a reciprocal lattice vector of the zero-flux (non-magnetic) BZ: that is, $\bra{GS} \rho_{\mbf{G}_q} \ket{GS}$ is nonzero only if $\GG_q = 2\pi {\mathbb{Z}} \bg_1 + 2\pi \mathbb{Z} \bg_2$.  This follows from the periodicity of the form factors:
\bea
M_{mn}(\mbf{k}+\mbf{b}_1/q,\mbf{G}_q) &= e^{i \mbf{G}_q \cdot \mbf{a}_2} M_{mn}(\mbf{k},\mbf{G}_q)
\eea
which is derived from the form factor definition:
\bea
\label{}
&\frac{1}{q}  \delta(k_1-k_1') \delta(k_2-k_2') e^{i\xi_{\GG_q}(\kk+\bg_1/2\pi q)} {\cal H}^{\GG_q}_{mn} = \braketter{\mbf{k}+\bg_1/2\pi q,m| e^{- 2\pi i \mbf{G}_q \cdot \mbf{r}}|\mbf{k}' +\bg_1/2 \pi q,n} \\
&=  \frac{1}{q}  \delta(k_1-k_1') \delta(k_2-k_2') e^{- i q (\pi (G_q)_1 (G_q)_2 +\epsilon_{ij} (G_q)_i k_j - 2\pi (G_q)_2/q)}[ e^{i \epsilon_{ij} 2\pi (G_q)_i {\tilde Z}_j}]_{mn} \\
&= e^{2\pi i \GG_q \cdot \bL_2} \braketter{\mbf{k},m| e^{- 2\pi i \mbf{G}_q \cdot \mbf{r}}|\mbf{k}',n},
\eea or
\begin{align}
    \braketter{\mbf{k}+ \bg_1/2\pi q,m| e^{- 2\pi i \mbf{G}_q \cdot \mbf{r}}|\mbf{k}' + \bg_1/2\pi q,n} = e^{2\pi i \GG_q \cdot \bL_2} \braketter{\mbf{k},m| e^{- 2\pi i \mbf{G}_q \cdot \mbf{r}}|\mbf{k}',n}
\end{align}
so that
\bea
\label{eq:efficient}
\int \frac{d^2k}{1/q} \Tr M(\mbf{k},\mbf{G}_q) &= \int_0^{1} \frac{dk_1}{2\pi} \int_0^{1/q} \frac{dk_2}{1/q} \Tr M(\mbf{k},\mbf{G}_q) \\
&= \lp \sum_{j=0}^{q-1} e^{i j \mbf{G} \cdot \mbf{a}_2} \rp \int_0^{1/q} dk_1 \int_0^{1/q} \frac{dk_2}{1/q} \Tr M(\mbf{k},\mbf{G}) \\
&= \delta_{{\mbf G}_\qqq,\mbf{G}} \int_0^{1/q} \frac{dk_1}{1/q} \int_0^{1/q} \frac{dk_2}{1/q} \Tr M(\mbf{k},\mbf{G}_q)
\eea
because $ \sum_{j=0}^{q-1} e^{i j \mbf{G}_q \cdot \mbf{a}_2} = q$ if $\mbf{G}_q \cdot \bL_2 \in \mathbb{Z}$, and otherwise vanishes: that is, only if $\GG_q$ is a reciprocal lattice vector.  This observation allows significant speedup in numerical calculations, as $\mbf{G}_q$ vectors that are not reciprocal lattice vectors do not need to be calculated in the expression for the many-body charge density.   The integrals in Eq.(\ref{eq:lambdaint}) will appear throughout the many-body calculations, so we find it convenient to define
\bea
n_{\mbf G} &= \int_{[0,1) \times [0,1/q)} \frac{d^2k}{1/q} \frac{1}{q} \Tr M(\mbf{k},\mbf G), \qquad \mbf G \in 2\pi \mbf{b}_1 \mathbb{Z} +2\pi \mbf{b}_2 \mathbb{Z}
\eea
which is the average of the form factor over the magnetic BZ and (the $2q$) occupied Hofstadter bands. Physically, $n_{\tilde{\mbf G}}$ are the Fourier modes of the charge distribution, which we see from plugging into Eq.(\ref{eq:nrGS}):
\bea
\braketter{GS|n(\mbf{r})|GS}  &= \sum_{\tilde{\mbf G}} e^{i {{\mbf G}} \cdot \mbf{r}} \ \int \frac{d^2k}{(2\pi)^2\Omega/q} \frac{1}{q} \Tr M(\mbf{k},{{\mbf G}}) = \sum_{{\mbf G}} e^{i {{\mbf G}} \cdot \mbf{r}} \ \int \frac{d^2k}{\Omega /q} \frac{1}{q} \Tr M(\mbf{k},{{\mbf G}}) = \frac{1}{\Omega} \sum_{{\mbf G}} e^{i {{\mbf G}} \cdot \mbf{r}} n_{{\mbf G}} \ .
\eea
By proving that $n_\mbf{G}$ vanishes if $\mbf{G}$ is not a reciprocal lattice vector, we have shown that the charge distribution at rational flux is periodic over the physical unit cell, not just over the (extended) magnetic unit cell.  A figure illustrating the  expectation value over a slice in real space along the $\bL_1$ direction is depicted in Supplementary Fig.(\ref{fig:localization}).  We see that even in the presence of flux, the charge density remains localized at the $1a$ position, though the spread does increase slightly with flux.

\subsection{Many-body Charge Excitations}

It was realized in \cite{wang2022narrow} that the double commutator calculation could be extended to TBG in flux.  In this appendix, we detail a variation of this calculation using our gauge invariant formalism \cite{PhysRevB.106.085140, PhysRevLett.129.076401}, which makes the derivation of the charge excitations very simple. First we recall some basic results from the strong coupling theory of TBG: the double-commutator technique of \cite{PhysRevLett.125.257602, PhysRevB.103.205415} will be used extensively in this section, along with results from \cite{PhysRevB.103.205413, PhysRevB.106.085140}.  Starting with a filled-band groundstate ansatz  \cite{PhysRevLett.122.246401, PhysRevX.10.031034, PhysRevB.103.205414} at filling $\nu$ ($\nu = 0$ is the charge neutrality point) given by
\bea
\ket{\Psi_\nu} = \prod_{\mbf{k},n} \prod_{j=1}^{\nu + 4} \gamma^\dag_{\mbf{k},n, \eta_j,s_j} \ket{0}
\eea
in spin-valley flavors $\{s_j\},\{\eta_j\}$, we compute the excitations of the Coulomb Hamiltonian projected into $2q$ flat bands (per spin-valley) defined by
\bea
H_{int} &= \frac{1}{2} \int \frac{d^2q}{(2\pi)^2} V(\mbf{q}) \bar{\rho}_{\mbf{q}} \bar{\rho}_{-\mbf{q}}  = \frac{1}{2} \sum_{\mbf{G},j=0}^{q-1} \int_{(0,2\pi) \times (0,2\pi/q)} \frac{d^2q}{(2\pi)^2} \frac{V(\mbf{q}+\frac{2\pi j}{q}\mbf{b}_2+\mbf{G})}{\Omega} \bar{\rho}_{\mbf{q}+\frac{2\pi j}{q}\mbf{b}_2+\mbf{G}} \bar{\rho}_{-(\mbf{q}+\frac{2\pi j}{q}\mbf{b}_2+\mbf{G})} \\
&\qquad \bar{\rho}_{\mbf{q}} = \sum_{\mbf{k},MN,\eta s} M^{(\eta,s)}_{MN}(\mbf{k},\mbf{q})  (\gamma^\dag_{\mbf{k}-\mbf{q},M,\eta,s} \gamma_{\mbf{k},N,\eta,s}  - \frac{1}{2} \delta_{\mbf{q},0} \delta_{MN}).
\eea
Working at $p = 1$, the equation breaks up the momentum $\qqq$ by restricting $\qqq$ into a vector in the MBZ, then adding a multiple of $\bg_2/q$, and a reciprocal lattice vector $\GG$, so that the total momentum $\qqq + \frac{2\pi j}{q} \bg_2 + \GG$.

At charge neutrality, $\nu = 0$, $\ket{\Psi_\nu}$ is a groundstate of $H_{int}$ because $H_{int}$ is positive semi-definite and
\bea
\bar{\rho}_{\mbf{q}+\frac{2\pi}{q} j\mbf{b}_2+\mbf{G}} \ket{\Psi_\nu} = \frac{\nu}{2}\frac{(2\pi)^2}{q} \delta(q_1) \delta(q_2) \delta_{j,0} q n_\mbf{G} \ket{\Psi_\nu}
\eea
and thus $H_{int} \ket{\Psi_{\nu=0}} = 0$. This convention justifies the choice of the projected interaction Hamiltonian. More generally, we observe that
\bea
\label{eq:endens}
H_{int} \ket{\Psi_\nu} &= \frac{1}{2} \sum_{\mbf{G},j=0}^{q-1} \int_{(0,2\pi) \times (0,2\pi/q)} \frac{d^2q}{(2\pi)^2} \frac{V(\mbf{q}+\frac{2\pi j}{q}\mbf{b}_2+\mbf{G})}{\Omega} (\frac{\nu}{2})^2 ((2\pi)^2 \delta(q_1) \delta(q_2))^2 \delta_{j,0} |n_\mbf{G}|^2 \ket{\Psi_\nu}\\
&=(2\pi)^2 \delta^2(0) \lp \frac{\nu^2}{8} \sum_{\mbf{G}} \frac{V(\mbf{G})}{\Omega} |n_\mbf{G}|^2 \rp \ket{\Psi_\nu} \\
\eea
where $(2\pi)^2 \delta^2(0)$ is the total number of unit cells. Thus the many-body energy density of $\ket{\Psi_\nu}$ is given by the quantity in parentheses in Eq.(\ref{eq:endens}).
\begin{figure}
\includegraphics[width=0.5\columnwidth]{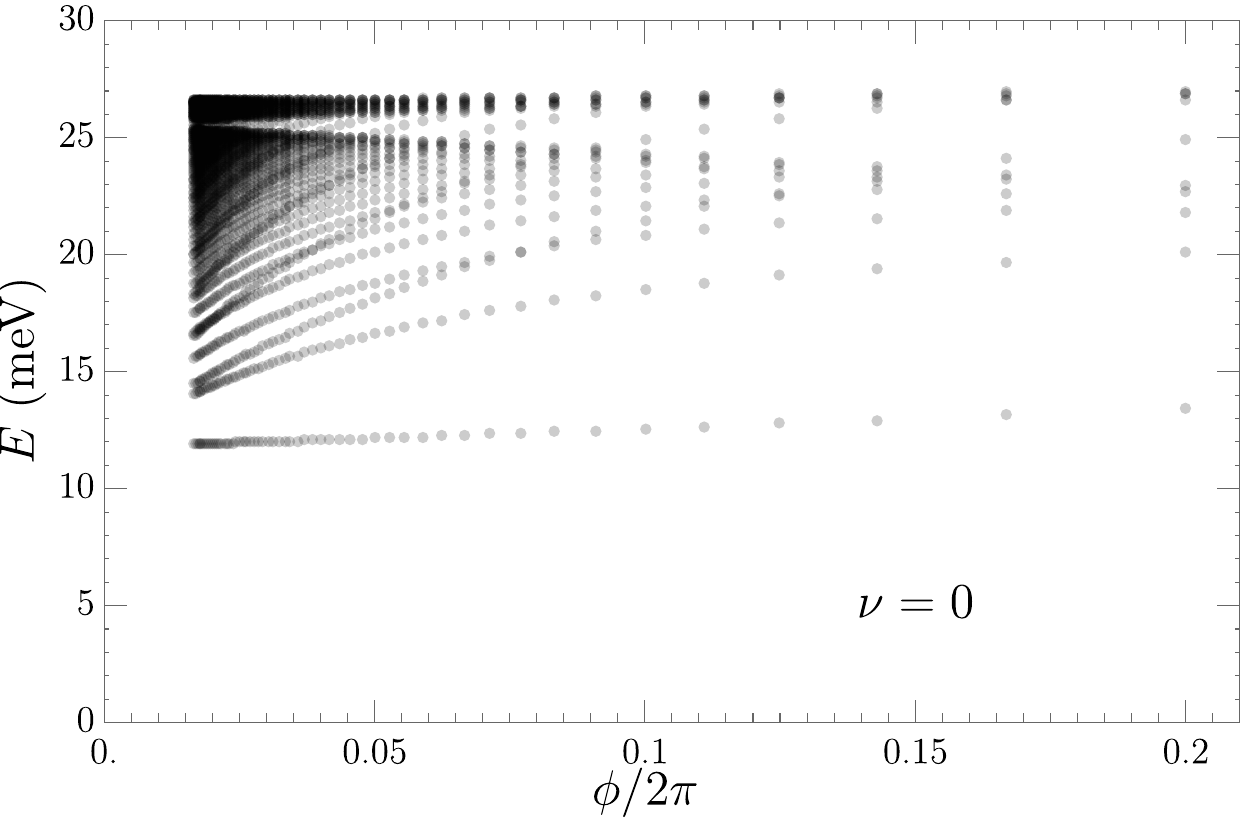}
\caption{Charge $+1$ excitations above the $\nu = 0$ VP state in flux, calculated via the strong coupling technique.}
\label{fig:strong-coupling}
\end{figure}

A double-commutator calculation \cite{PhysRevLett.125.257602,PhysRevB.103.205415} gives the charge $\pm1$ excitations in terms of the following $2q\times 2q$ effective Hamiltonians (the $\phi = 2\pi/q$ dependence is suppressed in $M(\mbf{k},\mbf{q})$ and $n_\mbf{G}$):
\bea
R^+_{MN}(\mbf{k}) &= \frac{1}{N} \sum_{\mbf{q},\mbf{G}} \frac{1}{2}\frac{V(\mbf{q}+\mbf{G})}{\Omega} [M^\dag(\mbf{k},\mbf{q}+\mbf{G})M(\mbf{k},\mbf{q}+\mbf{G})]_{MN} + \frac{\nu}{2} \sum_\mbf{G} \frac{V(\mbf{G})}{\Omega} [M(\mbf{k},\mbf{G})]_{MN} n^*_\mbf{G}  - \mu \delta_{MN}\\
R^-_{MN}(\mbf{k})^* &= \frac{1}{N} \sum_{\mbf{q},\mbf{G}} \frac{1}{2}\frac{V(\mbf{q}+\mbf{G})}{\Omega} [M^\dag(\mbf{k},\mbf{q}+\mbf{G})M(\mbf{k},\mbf{q}+\mbf{G})]_{MN} - \frac{\nu}{2} \sum_\mbf{G} \frac{V(\mbf{G})}{\Omega} [M(\mbf{k},\mbf{G})]_{MN} n^*_\mbf{G} + \mu \delta_{MN} \\
\eea
where $\mbf{q}$ sums over the $N$ states in the full Brillouin zone, and the chemical potential $\mu = (E_{min,+1}-E_{min,-1})/2$ is chosen so that the minima of the charge $\pm1$ spectra at $\mu=0$, denoted $E_{min,+\pm}$, are the same. Physically, this choice of $\mu$ corresponds to setting zero energy in the middle of the excitation gap. We refer to the first term in $R^{\pm}(\mbf{k})$ as the Fock term. Since $V(\mbf{q}) > 0$ and $M^\dag M$ is a positive-semi-definite matrix, the Fock term is positive semi-definite. It also does not depend on the filling $\nu$. Note that the phase $e^{i \xi_\mbf{q}(\mbf{k})}$ in the form factors $M(\mbf{k},\mbf{q})$ cancels in the $M^\dag M$ product, and so does not need to be evaluated. To further speed up the numerical calculations, $n_\mbf{G}$ can be precomputed since it does not depend on $\mbf{k}$. Most importantly, the computation of $M(\mbf{k},\mbf{q})$ can be carried out with sparse matrix multiplication of $e^{i \epsilon_{ij} q_i \tilde{Z}_j}$ on $U(\mbf{k})$. We find that even for very large $q$, the computations of $R^\pm(\mbf{k})$ are feasible taking the Landau level cutoff to be $30 q$ and truncating the $\mbf{G}$ sum after three shells.  A plot of the many-body charge excitations calculated with strong coupling is shown in Supplementary Fig.(\ref{fig:strong-coupling}); this is compared to the spectrum obtained via heavy fermions to excellent agreement in the main text.
\section{Notational Changes}
\begin{itemize}
    \item The notation $I^0_{[m,r']}$ is used in the supplementary to represent the
     $c$-$f$ hybridization matrix (see Eq.(\ref{I0})). In the main text, we use the notation $\Upsilon_{[m,r']}$ for the same. It is defined in Eq.(13) of main-text.
\item The notation $m_{max}$ is used in the supplementary to represent the upper cut-off on c-LL index. In the main text, we use the notation $m_{\star}$ for the same. 
\end{itemize}
%\bibliography{supp}% Produces the bibliography via BibTeX.
\end{document}